\documentclass[a4paper,12pt,twoside]{report}

\usepackage{amsfonts}
\usepackage{amssymb}
\usepackage{amsmath}
\usepackage{latexsym}
\usepackage{cite}
\usepackage[sf,sl,clearempty]{titlesec}
\usepackage{fancyhdr}
\usepackage{graphicx}
\usepackage{epsfig}

\newcommand{\beq}{\begin{equation}}
\newcommand{\eeq}{\end{equation}}
\newcommand{\be}{\begin{eqnarray}}
\newcommand{\ee}{\end{eqnarray}}
\newcommand{\bed}{\begin{displaymath}}
\newcommand{\eed}{\end{displaymath}}
\newcommand{\bea}{\begin{array}}
\newcommand{\eea}{\end{array}}
\newcommand{\bep}{\begin{split}}
\newcommand{\eep}{\end{split}}
\newcommand{\Tr}{\textrm{Tr}}
\newcommand{\tr}{\textrm{tr}}
\newcommand{\mat}{\mathcal}

\newcommand{\algebra}{\mathfrak{g}}

\newcommand{\Ad}{\textrm{Ad}}
\newcommand{\ad}{\textrm{ad}}

\newcommand{\longpage}{\enlargethispage{\baselineskip}}

\titleformat{name=\chapter}
   [display]
   {\vspace*{-15mm}\sffamily\bfseries\huge}
   {\filleft\LARGE\sffamily\thechapter}
   {0pt}
   {\titlerule[1pt]\filright}
   [\vspace*{-5mm}]
\titleformat{name=\section}
   {\filright\sffamily\LARGE}
   {\thesection}
   {1em}
   {}
\titleformat{name=\subsection}
   {\filright\sffamily\Large}
   {\thesubsection}
   {1em}
   {}
\titleformat{name=\subsubsection}
   {\sffamily \large}
   {\thesubsubsection}
   {0.4em}
   {}

\def\RR{I\!\!R}

\begin{document}


\title{\sffamily\bfseries\Huge Gauge fixing and BRST formalism in non-Abelian gauge theories}
\author{{\sf\LARGE Marco Ghiotti}\vspace*{4mm}\\
{\large Supervisors: Dr.~L.~von~Smekal and Prof.~A.~G.~Williams}
   \vspace*{5mm} \\
  {\it Special Research Centre for the}\\
  {\it Subatomic Structure of Matter}\\
  and \\
  {\it Department of Physics,}\vspace*{2mm}\\
  {\it University of Adelaide,}\\
  {\it Australia}} 
\date{January 2007} 
\maketitle

\pagestyle{empty}
\cleardoublepage{\ }
\pagenumbering{roman}
\pagestyle{plain}

\vspace*{5cm}
\begin{center}
        {\it To the morning star of my life,}\\
{\it my beloved sister Samantha}\\
\end{center}
\clearpage
\vspace*{0cm}
\begin{center}
        {\sf\Large Abstract}\\
\end{center}
In this Thesis we present a comprehensive study of perturbative and non-perturbative non-Abelian gauge theories in the light
of gauge-fixing procedures, focusing our attention on the BRST formalism in Yang-Mills theory.
We propose first a model to re-write the Faddeev-Popov quantisation method in terms of group-theoretical techniques and then
we give a possible way to solve the no-go theorem of Neuberger for lattice Yang-Mills theory with double BRST symmetry.
In the final part we present a study of the Batalin-Vilkovisky quantisation method for non-linear gauges in non-Abelian gauge theories.

\cleardoublepage{\ } 

\vspace*{5cm}
\begin{center}
        {\sf\Large Statement of Originality}\\
\end{center}
This work contains no material which has been accepted for the award
of any other degree or diploma in any university or other tertiary
institution and, to the best of my knowledge and belief, contains no
material previously published or written by another person, except
where due reference has been made in the text.

I give consent to this copy of my thesis, when deposited in the
University Library, being available for loan and photocopying.
\\
\\
\\
\\
\\
\\
\\
\\
\\
\begin{flushright}
Marco Ghiotti
\end{flushright}
\cleardoublepage{\ } 

\vspace*{2cm}
\begin{center}
        {\sf\Large Acknowledgements}\\
\end{center}
I wish to thank all the people who worked at the CSSM during these three years. 
First of all my supervisors, Anthony Williams and Lorenz von Smekal, who have been
not only incredibly competent leading figures in my Ph.D, but also because they have 
been good friends. Alongside them, I must also thank Alex Kalloniatis,
who has been my co-supervisor in the first year: I will never forget our discussions on several issues. 
I also thank Sara Boffa, Sharon Johnson and
Ramona Adorjan, three incredible and wonderful persons, who allowed me to go through the 
necessary bureocracy and the incomprehensible computer world.
Last but not the least, all the other Ph.D students who came and went at the CSSM in these 
three years. Thanks to all of you, for the amazing experience I enjoyed
and lived.
Yet, my biggest thank must go to my family spread out, geographically speaking, in the four 
cornerstones of the world. If the usual saying affirms that long-distance
relations never work, my family is exactly the perfect countexample. We became more united and 
we learned how to share our true deep feelings through our minds. 
Of course, I will ever thank Mr. Meucci for his extraordinary invention.

\clearpage
\newpage
\pagestyle{empty}
\mbox{}
\clearpage
\newpage
\pagestyle{plain}
\tableofcontents
\pagestyle{empty}
\cleardoublepage

\pagenumbering{arabic}
\pagestyle{fancy}
\fancypagestyle{plain}{\fancyhf{}
   \renewcommand{\headrulewidth}{0pt}
   \renewcommand{\footrulewidth}{0pt}
   }
\renewcommand{\chaptermark}[1]{\markboth{\thechapter.\ #1}{}}
\renewcommand{\sectionmark}[1]{\markright{\thesection\ #1}}
\fancyhf{}
\fancyhead[LE,RO]{\sf \thepage}
\fancyhead[RE]{{\sf \slshape\usefont{OT1}{phv}{m}{sl} \leftmark}}
\fancyhead[LO]{{\sf \slshape\usefont{OT1}{phv}{m}{sl} \rightmark}}
\renewcommand{\headrulewidth}{0.7pt}


\chapter{Introduction}

Since the seminal work by Yang and Mills in 1954 \cite{Yang:1954ek}, non-Abelian gauge theories have been indisputably of enormous 
importance in particle, and most generally, in theoretical physics.
It is well known that all the four forces discovered so far (electromagnetic, weak, strong and gravitation force)
are mediated by bosons or simply gauge fields (respectively photons, $W^{\pm}$ and $Z$, gluons, gravitons): 
apart from the photon,
all the other vector particles belong to a specific representation of non-Abelian gauge theory.
The fundamental property of these theories, regardless of their commutativity, lies in the fact that the Lagrangian of the model
is invariant under a local redefinition of the gauge field, known as a gauge transformation for gauge (or
general coordinate transformation for gravity), as Dirac noticed in the light
of quantum electrodynamics (QED) many years ago.

This local gauge invariance is responsible for one of the most challenging problems to solve in current
theoretical physics: the understanding of the non-perturbative regime of non-Abelian theories.
As pointed out by Gribov in the late 70-s \cite{Gribov:1977wm}, quantum chromodynamics (QCD), the most accurate model for the strong interaction,
suffers from a topological obstruction whenever one deals with low energies. This is due to the fact that the QCD Lagrangian, described by
a Yang-Mills interaction, being left unchanged by a local gauge transformation, contains an infinite overcounting of physically equivalent gauge
configurations, grouped together in different gauge orbits (equivalence classes). 
To take into account only gauge inequivalent configurations, ideally one would need to find an unambiguous procedure
to consider only one representative per gauge orbit: such a method is called gauge-fixing. Unfortunately, as Gribov discovered and
explained first in the physics language, and then Singer \cite{Singer:1978dk} in a more mathematical manner, there is no
analytic gauge-fixing procedure 
which guarantees the non-perturbative regime of QCD to be free of such an ambiguity, called the Gribov ambiguity.
The topological obstruction can be easily understood if one considers the gauge-fixing term as a hypersurface which intersects
the functional space of gauge orbits. If the fixing procedure were correct, then the surface would intersct only once per orbit.
This is the case only for high-energy or perturbative QCD. Beyond this regime, when we are entitled to move 
far from configurations around the trivial gauge field
$A_\mu=0$ (and thus infinitesimal gauge transformations), even finite gauge transformations play a decisive
role: consequently, many intersections are found along the orbit.

Surprisingly, a very interesting consequence of gauge-fixing had been already observed by four physicists three years before the paper of Gribov:
Becchi, Rouet and Stora \cite{Becchi:1975nq}, and independently Tyutin \cite{Tyutin:1975ty},
using the Faddeev-Popov quantisation method \cite{Faddeev:1967fc} to integrate out the infinite gauge
redundancy in the path integral representation of QCD, discovered a new global structure, now called BRST formalism.
To exponentiate the Faddeev-Popov operator appearing in the integrand of the path integral due to a functional change of variables, 
they introduced a pair of anticommuting fields, the ghost fields. These fields were the ones Kac and Feynman were looking for
to guarantee the unitarity of the $\mathcal S$-matrix in non-Abelian scattering processes.
Alongside the ghost fields, a Nakanishi-Lautrup field was used to rewrite the gauge-fixing term,
such that the QCD path integral was constituted by a quartet of fields.
However, this formalism was only fully understood some years later. In fact,
mainly due to the extensive works of
Atiyah \cite{Atiyah:1978mv}, Witten \cite{Witten:1982im} and Schwarz \cite{Schwarz:1978cn},
it was noticed that the theory of invariant polynomials, namely
Donaldson theory and knot theory, could be reformulated in more physical terms: this was the birth of
topological quantum field theory (TQFT). Within this theory, the BRST formalism was regarded as the most trivial example of
supersymmetry, and the BRST charge, generator of the new global symmetry, regarded as a nilpotent supersymmetric operator.

All of a sudden, QCD, and generally speaking non-Abelian gauge theories, started being studied and analyzed from many different angles:
Kugo and Ojima \cite{Kugo:1979gm} first interpreted
the BRST transformations as the quantum version of the classical gauge transformation. Then,
they discovered the generalisation to non-Abelian gauge theory of the Gupta-Bleuler formalism in QED.
They were thus able to give a prescription for confinement of quarks (known as the Kugo-Ojima criterion):
this non-perturbative mechanism is responsible for avoiding quarks as free particles at low energies
and therefore confining them  into the hadrons such as
nucleons and pions observed in nature, 
via the action of QCD self-interacting gauge fields, the gluons.

Moreover, 
mathematicians realised that the well-known theory of principal bundles could be re-expressed in terms of supersymmetric structures.
The classical geometry of gauge theory started being enriched with ghost fields, extended coordinates and Grassmann
manifolds.
In \cite{Bonora:1980pt}, \cite{Quiros:1980ec} and  \cite{Baulieu:1981sb} it was pointed out how the classical theory
of gauge geometry had as a logical implementation that of superfields and superconnections.

Schwarz and Witten constructed two different theories involving BRST formalism to study non-Abelian theories as supersymmetric ones
by means of topological path integrals: 
mathematical/geometrical theorems such the Hodge decomposition of connections, the Poincar\'e-Hopf and Gauss-Bonnet theorems
became soon familiar to many physicists. 
The connection between the Gribov ambiguity and topological field theory then naturally emerged through this massive quantity
of work: at the non-perturbative level, all of the equivalent configurations (Gribov copies) sum up to give a vanishing
topological path integral \cite{Fujikawa:1979ay, Birmingham:1991ty}. This result is regarded as the vanishing
Euler character of the gauge-group manifold on which QCD is evaluated \cite{Schaden:1999ew}.

Although much important work has been carried out, in the
understanding of non-perturbative gauge theories, some fundamental questions still remain
to be answered: in particular, the mechanism responsible for quark confinement seems to be the hardest and most challenging of all.
Nowadays the study of non-perturbative QCD is performed through different approaches: Dyson-Schwinger equations \cite{Alkofer:2000wg} and
lattice gauge theory \cite{Wilson:1974sk} perhaps are the more succesful ones for practical purposes.
In the former, functional and operator identities are pursued by the observation that the path integral of a total functional
derivative with respect to one of the fields involved vanishes. In this scenario, the BRST formalism is largely adopted
to facilitate highly complicated computations. However, the Gribov ambiguity is not fully avoided or solved:
Dyson-Schwinger calculations are performed in the region where the Faddeev-Popov operator is positive definite, and thus the complication
of going beyond this region is thus by-passed.
On the other side, in lattice QCD, the discretisation of the space-time manifold
determines a natural regularisation of the path integral, and therefore no gauge-fixing procedure is required.
It is the stochastic nature of the numerical computation of the path integral (e.g. by means of Monte-Carlo algorithms)
which self-consistently guarantees to have an insignificant probability to generate two gauge configurations on the same orbit.
It is for instance not fully understood yet how the BRST formalism can be implemented on the lattice: this is due to a no-go theorem
discovered in mid 80-s by Neuberger \cite{Neuberger:1986xz}. In this paper, he noticed that because of the
underlying BRST invariance, the gauge/fixing Y-M partition function can be shown to be independent of any gauge parameter.
The consequence of this property is that the lattice path integral of the gauge-fixing action ratios 
turns out to be exactly zero.
It seems then that the BRST formalism cannot be straightforwardly adopted at low energies. Yet, if we wish to construct
a complete and organic theory of the strong force, it is then required to comprehend why the BRST formalism apparently only works
in perturbative QCD. The motivation behind this Thesis is then to comprehend much more clearly the topological nature of the BRST algebra,
both perturbatively and non-perturbatively in non-Abelian gauge theories.

This Thesis is structured as follows:
Chapter I is dedicated to the introduction of constrained systems, first starting from the classical point of view of Lagrangian and
Hamiltonian
systems. We then adopt the covariant formalism to provide the quantum version of Maxwell theory and Y-M theory in the light
of functional contraints.
Chapter II is entirely devoted to the path integral representation of non-Abelian gauge theories.
We also analyze the structure of the functional configuration space of the Y-M path integral both from the analytic and
topological point of view. Special emphasis is given to the Gribov problem.
In Chapter III starting from the Faddeev-Popov quantisation method, we describe how the BRST formalism emerges in non-Abelian
gauge theories, focusing our attention on the supersymmetric structure of it and on the Kugo-Ojima criterion as the generalisation
of the Gupta-Bleuler formalism. We give a detailed explanation of the BRST algebra with respect to linear and non-linear gauges.
Chapter IV is dedicated to our first work conducted on the interpretation of the Faddeev-Popov quantisation method in terms of
group-theoretical techniques. We proposed a supersymmetric manner to re-write the Faddeev-Popov operator, entering the Y-M path integral
through a functional determinant, using the Nicolai map.
In Chapter V, we move to lattice gauge theory, on which we present a model to circumvent and possibly solve the Neuberger problem, which so far
prevented us from using the BRST formalism in lattice gauge theory. 
To conclude this Thesis, in Chapter VI we present the Batalin-Vilkovisky formalism both in continuum and lattice Yang-Mills theory in
the light of non-linear gauges. We will then re-propose the same methodology of Chapter V in this framework.



\thispagestyle{empty}
\cleardoublepage
\chapter{Symmetries and Constraints in Euclidian Gauge Theories}

In this chapter, we will present the theory of constrained systems, starting with a Euclidian classical theory: in this framework, we will 
first deal with the Hamiltonian formalism and then
we will move to the one commonly adopted in this Thesis, the Lagrangian formalism.
Gauge theories will be analyzed, considering first the Abelian gauge theory of electromagnetism and then its non-Abelian 
generalisation, Yang-Mills theory. 


\section{Classical constrained systems: Hamiltonian and Lagrangian formalism}

Generally speaking, all the possible physical information we require and need to extract
from a theory is encoded into the action $S$. This can be expressed either
through the Lagrangian $L$ or, by appropriate Legendre transformations, through the Hamiltonian $H$,
leading to two different formalisms: the former is of the covariant formalism, whereas the latter the canonical one.
Regardless of the particular type of formalism one wishes to adopt, one of the fundamental questions to answer in complex physical systems is how
we proceed in the case of one or more constraints affecting the action and consequently, either the Lagrangian or the Hamiltonian.
This problem can be immediately addressed at the classical level: we shall show how such constraints affect the dimensionality
of the configuration space, how they will enter the equations of motion and furthermore, how they influence Noether's theorem.
Consider for this purpose a local Euclidian Lagrangian \cite{Itzykson:1980rh, Nakanishi:1990qm}
over a finite-dimensional space of generalised commuting (bosonic) variables $q$, whose base space is $t\in\mathbb R$: 
we call the {\it free} Lagrangian, $L_0$, the part of $L$ which 
is being described only by the generalised coordinates $q(t)$, and their first derivatives $\dot q(t)$ at most quadratically  
(this for the canonical formalism to be applicable). The remaining part of $L$ contains higher-order terms in $q(t)$ and it is called the 
{\it interaction} Lagrangian, $L_I$. 
The classical action of the system is then defined as the integral over the base space of $L(t)$
\begin{align}
S&\equiv\int{\rm d}t\,L(t)=\int{\rm d}t\,L(q_n(t),\dot q_n(t)).
\end{align}
Consider now a local variation of both $q(t)$ and $\dot q(t)$ in the action
\begin{align}
\label{actionvariation}
\delta S&=\int{\rm d}t\left[\frac{\partial L}{\partial q_n}\delta q_n+\frac{\partial L}{\partial \dot q_n}\delta \dot q_n\right]\nonumber\\
&=\int{\rm d}t\left[\frac{\partial L}{\partial q_n}-\frac{{\rm d}}{{\rm d}t}\frac{\partial L}{\partial \dot q_n}\right]\delta q_n.
\end{align}
where the sum over $n$ is understood and
we performed an integration by parts to obtain the last line, once appropriate boundary conditions on $q_n(t)$ are imposed.
The action principle, or Hamilton principle of least action, states that the path satisfying the classical equations of motion
is the one extremising $S$, such that $\delta S=0$.
Because of the variations $\delta q_n$ are independent, we obtain the Euler-Lagrange equations
\beq
\label{EL}
\frac{\partial L}{\partial q_n}-\frac{{\rm d}}{{\rm d}t}\frac{\partial L}{\partial \dot q_n}=0,
\eeq
which are a set of $n$ de-coupled second-order differential equations. 
In the case when we deal with fields, they will be also called {\it field equations}.
To reduce (\ref{EL}) to first-order differential equations, we introduce the canonical momentum  conjugate
\beq
\label{momenta}
p_n\equiv \frac{\partial L}{\partial \dot q_n}\,,
\eeq
and we suppose we can invert this relation, i.e. expressing velocities in terms of positions and momenta.
The Hamiltonian is obtained through a Legendre transformation as
\beq
\label{hamilt}
H(p_n, q_n)\equiv p_n \, \dot q_n( p, q)-L(q_n, \dot q_n( p, q)).
\eeq
The Euler-Lagrange equations are being translated into the symplectic space of the Hamiltonian formalism: they now become the Hamilton equations.
To see this, insert (\ref{hamilt}) into the action and vary both momenta and positions to get
\beq
\delta S=\int{\rm d}t\left( \dot q_n\delta p_n-\dot p_n \delta q_n-\frac{\partial H}{\partial p_n}\delta p_n
-\frac{\partial H}{\partial \dot q_n}\delta q_n\right),
\eeq
with
\beq
\frac{\partial H}{\partial p_n}=\dot q_n,\qquad\frac{\partial H}{\partial q_n}=-\dot p_n.
\eeq
These are the Hamilton equations, which determine a symplectic structure\footnote{A symplectic vector 
space is a vector space V equipped with a nondegenerate, skew-symmetric, bilinear form,  called the symplectic form.
Its dimension must necessarily be even since every skew-symmetric matrix of odd size has determinant zero.}
over the configuration space: the dimensionality of this space becomes
then twice the original one because of the presence of momenta $p$. 
In this phase-space,
it is convenient to introduce a way to associate elements at different points: this is achieved by the Poisson Brackets (PB) \cite{VanHolten:2001nj}, 
defined as
\beq
\{F,G\}=\frac{\partial F}{\partial q_n}\frac{\partial G}{\partial p_n}-\frac{\partial F}{\partial p_n}\frac{\partial G}{\partial q_n},
\eeq
where repeated indices are understood to be summed.
Any transformation of the canonical variables $(p,q)$ which leaves these brackets unchanged is called canonical.
The importance of such an analytic operator lies in the fact that it determines immediately if there is a symmetry. 
In fact, any function $F(p,q)$, whose total derivative vanishes, is a constant of motion if it has
vanishing Brackets with $H$, 
\begin{align}
\label{constofmotion}
\frac{{\rm d}F}{{\rm d}t}&=\frac{\partial F}{\partial t}+\frac{\partial F}{\partial q}\dot q+\frac{\partial F}{\partial p}\dot p\nonumber\\
&=\frac{\partial F}{\partial t}+\{F,H\}\quad \Rightarrow\quad 
\{H,F\}=0=\frac{\partial F}{\partial t}. 
\end{align}
At the quantum level, (\ref{constofmotion}) describes the Heisenberg equation of motion, whereas we shall see that the Poisson Brackets will
become, according to the formalism adopted, either the Lie Brackets in the language of gauge theories or the BV Brackets in the case of
supersymmetry.
With respect to the two canonical variables, $q_n$ and $p_n$, the PB with $H$ read
\beq
\{q_n,H\}=\dot q_n\qquad \{p_n, H\}=\dot p_n.
\eeq
To introduce the fundamental concept of a constraint, we start with a pedagogical summary in the
Hamiltonian formalism, and then we will present it in the light of the
covariant Lagrangian formalism.\\
Therefore, expand the time derivative in the Euler-Lagrange equations as
\beq
\frac{\partial L}{\partial q_n}-\frac{\partial^2 L}{\partial \dot q_n\partial q^m}
\dot q^m-\frac{\partial^2 L}{\partial \dot q_n\partial \dot q^m}\ddot q^m=0.
\eeq
This is a set of $n$ coupled second-order differential equations in $q^m$ with functional coefficients, whose Lipschitz condition of solvability 
depends on the invertibility of the matrix coefficient of $\ddot q^m$ 
\beq
\label{invertability}
\det \left(\frac{\partial^2 L}{\partial \dot q_n\partial \dot q^m}\right)=\det\left(\frac{\partial p_n}{\partial \dot q^m}\right)\neq 0.
\eeq
Whenever such relation does not hold, it means that there are one or more momenta, say $m$,
which can be expressed in terms of the remaining momenta as
\beq
p_a=p_a(p_i,q_n)\qquad 1\leq a\leq m\leq i\leq n,
\eeq
and that for each momentum $p_a$, there is a velocity $\dot q^a$ which cannot be solved in terms of $p_i$ or $q_n$. We will indicate the set
of all these constraints as $\phi_a(p_n,q_n)=0$.
The corresponding constrained Hamiltonian $\overline H$ is obtained in the same formal way as for the unconstrained one: 
such Hamiltonian has the peculiarity that it does not depend on the constrained velocities
\beq
\frac{\partial {\overline H}}{\partial \dot q^a}=0,
\eeq
implying that we can always add to $\overline H$ any function of such velocities without affecting the Hamilton equations.
The unconstrained Hamiltonian will be then expressed in terms of $\overline H$ as
\beq
H=\overline H -\lambda^a(t)\phi_a,
\eeq
with $\lambda^a$ being a Lagrange multiplier, generally depending on $t$, whose role is to enforce the constraint $\phi_a=0$.
For the dynamics of such a system, in order to be consistent with the symplectic structure thus determined, not only have the canonical variables
to respect the constraints, but also the constraints themselves in a self-consistent way such that
\beq
\label{condition}
\frac{{\rm d}\phi_a}{{\rm d}t}=\{\phi_a,H\}\simeq 0,
\eeq
where by $\simeq 0$ we mean ``weakly zero'', provided the constraints are enforced.
In the case in which such a condition does not hold, then we keep going by enforcing a new constraint, say $\rho_a$, defined as
$\rho_a\equiv \{\phi_a, H\}$, until we find PB consistent with the condition (\ref{condition}). 
All these new constraints should be added {\it a posteriori}
into the Hamiltonian, associated with the appropriate Lagrange multiplier as 
$H=\overline H-\sum_a^m \lambda_a \phi_a$, for $N$ constraints.
The difference between the Hamiltonian formalism and the Lagrangian is that the Lagrange multipliers plugged into $H$ immediately
manifest the condition of a constraint, to which we must associate arbitrary trajectories to the coordinates whose dynamics
is undetermined by the equations of motion. This assigning goes under the name of gauge-fixing. Though in the Lagrangian formalism
such a procedure is unavoidable in the presence of constraints, in the case of the Hamiltonian formalism, what is actually needed is $H$
and to specify the constraints.
Even in this classical theory, we can introduce the concept of a gauge transformation:
by this we mean a map of phase space coorindates as
\begin{align}
\label{trans1}
p_n&\to p_n+\delta p_n(p_m, q^m, t)=p_n-\epsilon^a(t)\frac{\partial \phi_a}{\partial q_n}\equiv \bar p_n\nonumber\\
q_n&\to q_n+\delta q_n(p_m, q^m, t)=q_n+\epsilon^a(t)\frac{\partial \phi_a}{\partial p_n}\equiv \bar q_n,
\end{align}
such that the action $S$ (and the equations of motion) should be left invariant under these transformations.
In general, the action is not invariant under these gauges, and so we have to enforce the symmetry through an appropriate transformation
rule for all the Lagrange multipliers and the Hamiltonian, respectively
\beq
\label{trans2}
\delta \lambda^a=-\frac{{\rm d}}{{\rm d}t}\epsilon^a\qquad \delta H=\epsilon^a\{H,\phi_a\}.
\eeq
Together with (\ref{trans1}), (\ref{trans2}) leave the entire system invariant.


\subsection{Noether's theorem and charge algebra}

The Noether theorem is a crucial theorem to understand how both external and internal symmetries play a decisive role in field theories. 
It states that to any continuous one-parameter set of invariances of the Lagrangian is associated a local conserved current. Integrating
the fourth component of this current over three-space generates a conserved charge. From this charge, one can then construct a Lie charge algebra.
In the case we have also internal symmetries, as it will be the case in gauge theories, this internal charge algebra will play
a fundamental role in the quantisation procedure: it will give rise to the Lie charge algebra of BRST, a special case of a supersymmetric algebra. 
In this section we present the appearance of a charge algebra in the canonical formalism: 
consider a functional $G$ over the finite-dimensional phase-space $(p,q)$: suppose $\{G,H\}=0$ is satisfied, then
we know from previous arguments that we are dealing with a symmetry over the time, whose infinitesimal generator is the Hamiltonian.
The Hamiltonian itself is left invariant ($\delta H=0$) under the following transformations
\beq
\label{trans}
\delta q^i=\{q^i, G\}=\frac{\partial G}{\partial p_i} \qquad 
\delta p_i=\{p_i, G\}=-\frac{\partial G}{\partial q^i}.
\eeq
The difference from the Lagrangian formalism is that here constants of motion generate symmetries, and not viceversa, and moreover we have
explicit expression for the variations $(\delta q, \delta p)$. This is the inverse Noether theorem.
The algebra spanned by these symmetries is obtained by using the Jacobi identity
\beq
\{\{G_\alpha,G_\beta\},H\}+\{\{H,G_\alpha\},G_\beta\}+\{\{G_\beta,H\},G_\alpha\}=0,
\eeq
provided $\{G_\alpha\}$ is a complete set of generators. In fact,
we have the identity
\beq
\{G_\alpha, G_\beta\}=P_{\alpha\beta}=-P_{\beta\alpha},
\eeq
with $P_{\alpha\beta}$ an antisymmetric polynomial in the constants of motion $G_\alpha$
\be
P_{\alpha\beta}=c_{\alpha\beta}+f_{\alpha\beta}^{\phantom{\alpha\beta}\gamma}\, G_\gamma
+\frac12g_{\alpha\beta}^{\phantom{\alpha\beta}\gamma\delta}\, G_\gamma\, G_\delta+\ldots
\ee
All the coefficients of the Taylor expansion are constants, and therefore we have
vanishing PB's with $H$ at all orders of the expansion.
$c_{\alpha\beta}$ is called the {\it central charge} of the expansion.
Calling the generic infinitesimal variations (\ref{trans}) $\delta_\alpha$, for any functional $F$ over the phase-space
\beq
\delta_\alpha F=\{F, G_\alpha\},
\eeq
the commutation relations yield
\beq
[\delta_\alpha,\delta_\beta]F=\{\{F, G_\beta\},G_\alpha\}-\{\{F, G_\alpha\}, G_\beta\}.
\eeq
Due to the Jacobi identity, we can write
\begin{align}
\label{chargealgebra}
[\delta_\alpha,\delta_\beta] F&=\{F,\{G_\alpha, G_\beta\}\}=C_{\alpha\beta}^{\phantom{\alpha\beta}\gamma} \delta_\gamma F\nonumber\\
&=\frac{\partial P_{\alpha\beta}}{\partial G_\gamma}=
f_{\alpha\beta}^{\phantom{\alpha\beta}\gamma}
+g_{\alpha\beta}^{\phantom{\alpha\beta}\gamma\delta}\, G_\delta+\ldots.
\end{align}
where we made use of the antisymmetry of the coefficients of the Taylor expansion.
It is clear that in order to have (\ref{chargealgebra}) fulfilled in the case of local symmetries, the central charge has to vanish
identically, $c_{\alpha\beta}=0, \forall \alpha, \beta$, generating a {\it first class} constraint \cite{Henneaux:1992ig}.
We shall only deal with closed algebras, so only the linear term in the Taylor expansion will be considered, whereas for more general
algebras, such open ones, one can consider an arbitrary number of powers in $G$.\\
Algebraically speaking, whenever we have a vector space, with an antisymmetric product and Jacobi identity fulfilled, we are in the presence
of a Lie algebra (For a more formal definition see Appendix A). 
So, whenever we have a local symmetry $\dot G_\alpha =\{G_\alpha, H\}=0$ with time-dependent parameters,
the generators $\{G_\alpha\}$ turn into a constraint
\beq
G_\alpha(q, p)=0.
\eeq
Defining on the hypersurface, spun by the generators of the symmetry, a set of equivalence classes, the constraints commute with the Hamiltonian
through the corresponding PB over these classes. 
This is a necessary condition only {\it on-shell} (on the physical hypersurface), whereas {\it off-shell} this is no longer true.
When we deal with BRST algebra, we will introduce the concept of cohomology and the means to select the correct physical subspace,
thanks to the Kugo-Ojima criterion \cite{Nakanishi:1990qm}.
To anticipate such a criterion, in the light of the canonical formalism, we can assert that the condition for selecting only the physical subspace
is achieved by the two simultaneous conditions
\be
\left\{\bea{ccc}
\label{firstclass}
\{G_\alpha, G_\beta\}=P_{\alpha\beta}(G)=0\\ 
\{G_\alpha, H\}=Z_\alpha(G)=0              
\eea\right.
\quad \Leftrightarrow\quad G=0.
\ee


\section{Covariant formalism in Abelian gauge-theories: Maxwell theory}

As a further example of constrained theories, we now focus on gauge theories. First we will start with an Abelian case and then
we will treat non-Abelian theories.\\
The electromagnetic theory invented by Maxwell in 1864 can be regarded as the prototype of Abelian gauge theories. It contains two symmetries:
Lorentz invariance and gauge symmetry. The first was recognised only after the discovery of special relativity by Einstein in 1905, whereas the
second one needed to wait for quantum mechanics and general relativity to be fully appreciated. The work of Yang and Mills in mid 50's
shed light to more insights of this symmetry.
The quantisation of electromagnetism led Dirac first, and then Feynman, through the path integral representation, 
to set up the bases of quantum electrodynamics (QED). It is therefore not surprising to use this theory 
of commuting $c$-numbers as a starting framework to study and appreciate all the subtleties of gauge theories.\\
The covariant formulation of electromagnetism starts by considering 
the electric field $E$ and the magnetic field $B$ not as isolated fields, but
put together into a four-dimensional antisymmetric tensor $F_{\mu\nu}=-F_{\nu\mu}$, 
the {\it electromagnetic tensor} or {\it field-strength tensor},
such that $E_i=F_{i0}$ and $B_i=\epsilon_{ijk}F_{jk}$. The Lagrangian density of the theory is 
$\mathcal L=-\frac14 F^{\mu\nu}F_{\mu\nu}=\frac12(\vec E^2-\vec B^2)$, 
and the Maxwell equations of motion take the compact form
\begin{align}
\partial_\mu(\tilde F^{\mu\nu})&=0\nonumber\\
\partial_\mu F^{\mu\nu}&=-j^\nu,
\end{align}
with $\tilde F^{\mu\nu}=\frac12\epsilon^{\mu\nu\rho\sigma}F_{\rho\sigma}$ the dual of $F^{\mu\nu}$ and
$\epsilon^{\mu\nu\rho\sigma}$ is the four dimensional total antisymmetric tensor.
Current conservation appears as a natural compatibility condition $\partial_\mu j^\mu=0$.
To make Lorentz invariance appear more naturally, let us transform the first-order equations of motion into equivalent second-order
ones: this can be achieved by introducing a four-potential $A_\mu$, the photon field,
such that $\vec E=-\vec \nabla A^0-\frac{\partial\vec A}{\partial t}$ and $\vec B={\rm curl}\, \vec A$.
Being Maxwell theory an Abelian theory, the photon field $A_\mu$, called also the {\it gauge} or {\it vector boson potential},
does not self-couple. Therefore, $F_{\mu\nu}$ can be covariantly expressed as
\beq
F_{\mu\nu}=\partial_\mu A_\nu-\partial_\nu A_\mu,
\eeq
and the equations of motion become written in terms of $A_\mu$ only as
\beq
\Box A_\mu -\partial_\mu\partial^\nu A_\nu=-j_\mu,
\eeq
with $\Box$ being the D'Alambertian operator $\Box=\partial^\mu g_{\mu\nu}\partial^\nu$ and $g_{\mu\nu}$
being the metric. In Euclidian space, the metric is trivial, such that its signature simply reads $g_{\mu\nu}=(1,1,1,1)$.
In Maxwell theory there are two first-class constraints: first, in the Lagrangian there is no time-derivative of the gauge potential $A_0$.
Therefore, its conjugate momentum is vanishing
\beq
\pi^0=\frac{\partial L}{\partial \dot A_0}=0.
\eeq
The momentum field $\pi^i$ conjugate to $A_i$ is the electric field and it has the equal time PB with $A_i$ as follows
\beq
\{A_i(x), E_j(y)\}=\delta_{ij}\delta(x-y), \qquad x^0=y^0.
\eeq
The fields $E_j$ are not all independent, but are also subject to the Gauss law constraint
\beq
\partial_i E_i=0.
\eeq
These two constraints have also mutual vanishing PB.\\
The most important feature of this Abelian theory is the invariance of the
Lagrangian under a local redefinition of $A_\mu$, called {\it gauge transformation}, defined as
\begin{align}
A_\mu(x)\quad\to\quad A'_\mu(x)&=A_\mu(x)+\partial_\mu \Lambda(x)\nonumber\\
&\equiv{}^\Lambda\!A_\mu,
\end{align}
with $\Lambda(x)$ any smooth function defined on $\mathbb R^4$\footnote{The global gauge-invariance 
of $\mathcal L$ is straightforward, and in a more accurate language is called gauge invariance of the 
{\it first kind}, whereas the local invariance is called of the {\it second kind}.}. 
This local invariance was fully discovered in the case of QED, when the electromagnetic
field is coupled to electrons, through the Dirac Lagrangian. Though, in the course of this Thesis we will only consider pure gauge theories,
where the only degrees of freedom are the gauge potentials, and therefore we will not bother about the presence of physical fermions\footnote{We will nonetheless encounter 
other types of fermions, which will turn out to be unphysical, called {\it ghost} fields, 
generated by the quantisation procedure {\it a la} Faddeev-Popov in the case of Y-M theory. These unphysical degrees of freedom will play
a decisive role in Topological Field Theory (TFT), Supersymmetry (SUSY) and BRST.}.

Having defined our classical covariant electromagnetic theory, we would wish now to quantise it, keeping the covariance manifest.
This quantisation procedure presents three fundamental problems to overcome \cite{Nakanishi:1990qm}
\begin{enumerate}
\item the incompatibility between classical equations and quantum principles;
\item appearance of an indefinite metric;
\item subsidiary conditions to select the physical subspace.
\end{enumerate}
1.: this incompatibility derives from the fact that if one wishes to quantise Maxwell theory, 
the equations of motion have to be modified if $A_\mu$ is a nontrivial field.\\
2.: since $A_\mu$ is supposed to be a massless field, it is impossible to avoid negative norm without violating manifest covariance. This has to
do with the compactness of the little group of the Poincar\'e group. One can show \cite{Nakanishi:1990qm} the impossibility
of covariantly quantizing $A_\mu$ in the positive-metric Hilbert space $\mathcal H$. To solve this problem,
Gupta \cite{Gupta:1949rh} and Bleuler \cite{Bleuler:1950cy} proposed a formalism in the case of indefinite norm. This method will be then
translated in the language of cohomological BRST, when we will deal with BRST quantisation of non-Abelian gauge theory 
in the course of the next chapter.\\
3.: if an indefinite-metric is necessary to quantise Maxwell theory, and a positive-definite metric is indispensable for the probabilistic
interpretation of quantum states, some subsidiary conditions have to be imposed on the subspace, called ``physical'', of the entire configuration space
with indefinite metric. It is possible to select such a physical subspace, independently of the time and with positive norms.\\
\\
The fundamental postulate for these three conditions to be valid is the separability of the Hilbert physical subspace, isomorphic 
to the space of square-integrable functions, called $L^2$-space, 
i.e. functions rapidly decrease at infinity\footnote{We will moreover see that, in the case 
of non-Abelian gauge transformations, we will have to further 
restrict the functional space to be Sobolev
\cite{Semenov:1982sf, Adams:1975, Mazja:1980vg}, where even the first derivative has to be taken square-integrable.}.
As explained in \cite{Nakanishi:1990qm}, without this postulate, quantum field theory could not be properly formulated.

\subsection{Dirac quantisation method}

We wish here to briefly present a very useful quantisation procedure in the presence of first and second-class constraints, developed by Dirac
\cite{Nakanishi:1990qm}. 
As we saw in (\ref{momenta}) and (\ref{invertability}), these equations are solvable but only partially, leaving some variables undefined,
our constraints. They can be either of first-class, if the PB between the constraint $\phi_a$ and any quantity $A$ can be expressed
entirely in terms of linear combinations of $\phi_a$'s, otherwise they are second-class. In quantum field theory, it is better if we
can avoid dealing with first-class constraints: being always possible to find a quantity $\chi$, such that $(\phi'_a, \chi)\neq 0$, 
this can be interpreted as an annihilation condition for certain state vectors (i.e. $\phi'|f\rangle=0$) and consequently
inconsistent with the existence of the quantum vacuum. As we previously saw in the language of the Hamiltonian formalism,
first-class constraints are the generators of gauge transformations, and therefore they become second-class by adding gauge-fixing constraints.
In the Lagrangian formalism, if we gauge-fix, losing the local gauge-invariance, then no first-class constraint remains left, and we will
only deal with second-class constraints. To manage these second-class constraints into the quantisation procedure,
Dirac proposed a way to deal with them: he introduced a generalisation of the
Poisson Brackets, named after him as Dirac Brackets, defined as follows
\beq
(A,B)_D\equiv (A,B)_P-(A,\phi_a)_P)(\mathcal A^{-1})_{ab}(\phi_b, B)_P,
\eeq
where $\mathcal A$ is the matrix of Poisson Brackets among the constraints $\phi_i$, $A$ and $B$ are two general function.
These new brackets have the same analytic and algebraic properties of PB, such as antisymmetry, Leibniz rule and Jacobi identity, but with the additional
property that $(\phi_a, C)_D=0$, for any $C$. Quantisation is carried out by replacing the Dirac Brackets by $-i$ times a commutator
\beq
(A,B)_D\quad \to \quad -i[A,B].
\eeq
The Dirac method therefore does not bother whether or not there is constraint in the theory, and for this reason it is a very useful method
in quantizing theories with constraints.


\subsection{Covariant Quantum Theory of Maxwell Theory}

Following the works of Gupta \cite{Gupta:1949rh} and then Bleuler \cite{Bleuler:1950cy}, 
we can now define the covariant operator formalism of the electromagnetic field: 
by introducing an additional field $b$, called the Nakanishi-Lautrup field, we can quantise the theory by means of a covariant gauge-fixing.
This term, though spoiling the local gauge invariance, allows to invert the differential operator $g_{\mu\nu}\Box-\partial_\mu\partial_\nu$,
which governs the two-point function of $A_\mu$, otherwise non invertible because it is a projector operator. The gauge-fixed Lagrangian density
becomes
\beq
\mathcal L_{\rm gf}=\mathcal L_0+b\partial_\mu A^\mu+\frac12 \alpha b^2,
\eeq
where $\alpha$ is a positive real number, called {\it gauge parameter}. According to its particular value we can define different covariant gauges,
for instance $\alpha=1$ is the Feynman (or Fermi) gauge, $\alpha=0$ is the Landau gauge, and for generic positive $\alpha$ we have the Lorentz gauge.
The field equations then read
\begin{align}
\label{quantumMaxwell}
\partial_\mu F^{\mu\nu}-\partial^\nu b&=-j^\nu\nonumber\\
\partial^\mu A_\mu+\alpha b&=0,
\end{align}
also called the quantum Maxwell equations. What is remarkable in this framework is that, by taking a total divergence of the first equation in
(\ref{quantumMaxwell}),
due to the anti-symmetry of the field-strength tensor and to the conservation of the Noether current, we obtain an additional condition
\beq
\label{Gupta}
\Box b=0,
\eeq
implying $b$ is massless, despite the fact that
$A_\mu$ is an interacting field. Eq. (\ref{Gupta}) is a central feature of Abelian gauge-theory in covariant gauges,
and it is exactly the generalisation of this condition which will establish the appearance of the Ojima criterion in non-Abelian gauge theory
for selecting the appropriate Hilbert physical subspace.

The quantisation procedure in canonical formalism elevates $A_\mu$ to a canonical variable: its canonical momentum conjugate is
\begin{align}
\pi^\mu=\frac{\partial \mathcal L_{\rm gf}}{\partial \dot A_\mu}&=F^{0\mu}+g^{0\mu}b\nonumber\\
&=g^{0\mu}b,
\end{align}
where the last line is due to the antisymmetry of $F^{\mu\nu}$. The canonical commutation relation at zero-time will be 
$[\pi^\mu(x), A_\nu(y)]_0=i\delta^\mu_\nu\delta(\vec x-\vec y)$ and otherwise vanishing.
As a consequence, $\dot A_k$ ($k=1,2,3$) and $b$ are directly expressible in terms of $\pi^\mu$, whereas
$A_0$ and $\dot b$ are not. As mentioned before, the appearance of the $b$-field into covariant Maxwell theory sets up the basis for the
Gupta-Bleuler condition: it can be derived by the observation that $b$, satisfying the free-field equation (\ref{Gupta}), can be represented through
a conserved local current, from which the following integral representation of $b$ follows
\beq
\label{integralb}
b(y)=\int{\rm d}\vec z\,[\partial_0^z D(z-y)\cdot b(z)-D(y-z)\partial_0 b(z)].
\eeq
By taking the various commutation relations at equal time (ETCR) with respect to the other fields of the theory and 
making use of (\ref{integralb}), one observes that
\beq
{}[\Phi(x), b(y)]=-i\mathcal L(\Phi)^x D(x-y),
\eeq
where $\Phi$ is any local quantity and $\mathcal L$ is a differential operator. Thus, $b$ can be regarded as a generator of local gauge
transformations. Splitting the contribution of  negative/positive frequency part in $b$, the Gupta-Bleuler condition yields
\beq
\label{GB}
b^{(+)}(x)|f\rangle=0\qquad b^{(-)}=\left(b^{(+)}\right)^\dagger,
\eeq
implying that the physical subspace $\mathcal V_{\rm phys}$ (time-independent and Poincar\'e invariant) 
of the total Hilbert space is constituted by the totality of states $|f\rangle$ satisfying (\ref{GB}).


\section{Non-Abelian gauge theories: a survey into Yang-Mills theory}

Once the canonical formalism for the electromagnetic force is being translated in the quantum language, 
elevating all the canonical variables to quantum operators by means of
a suitable quantisation procedure, the theory of constrained systems with gauge symmetry that interests us the most is Yang-Mills (Y-M) theory.
Originally proposed by Yang and Mills in 1954 \cite{Yang:1954ek} in the context of isospin structures for $SU(2)$ theories,
after some initial reluctance in the physics community, 
it soon became the fundamental model to establish a connection between electromagnetism and the weak force as a unified theory.
It then circumvented parton models, and today it is believed the best candidate for the description of the strong force, especially after
the brilliant and successful work in the 70's and 80's devoted to the proof of its renormalisation at all orders
\cite{'tHooft:1972fi}, asymptotic freedom \cite{Gross:1973ju, Politzer:1974fr}
and to the understanding of non-perturbative phenomena such quark confinement \cite{Wilson:1974sk}.
The reason why gauge theories are so important in particle physics is due to the fact that the 4 fundamental forces existing in nature are
believed to be mediated by exchanging particles, called vector bosons. These integer-spin particles are subjected to dynamics described by
gauge theories, being either Abelian, as in the case of photons, or non-Abelian, as in the case of $W^{\pm}$ or $Z$ bosons, 
as well gluons and gravitons.

In this Thesis we will focus only on
Y-M theory as the preferred model for describing quantum chromo-dynamics (QCD), 
which  can be regarded as the generalisation of QED: while in the latter the underlying Lie (gauge) group 
is the Abelian Lie group $U(1)$, in the former 
the Lie group is a non-Abelian gauge group, the special unitary $SU(N)$.
This is a crucial difference, as we will see, because now the gluons, matrix-valued vector bosons of the theory, are self-interacting, differently
from the case of photons. This will be clear once we will show the Y-M field-strength tensor.
To begin with, let us introduce some concepts of Lie group and algebra theory (for more details we remind the reader
to see Appendix A): consider the semi-simple
\footnote{Semi-simplicity is equivalent to that for the Killing form \cite{Nakanishi:1990qm}, entering the Lagrangian density as 
$\mathcal L_{YM}=-\frac14\,K_{ab}$$\,F_{\mu\nu}^a F^{\mu\nu b}$, and defined as 
$K_{ab}\equiv-{\rm Tr}\,({\rm ad}(T_a)\,{\rm ad}(T_b))=-f^c_{ad}f^d_{bc}$ 
to be non-degenerate. The Killing form can be diagonalised as $K_{ab}=\delta_{ab}$, and w.r.t. this diagonalizing basis, upper and lower indexes in the 
structure constants do not make any difference any more, as long as we are concerned with compact Lie groups.}, 
compact Lie group $SU(N)$:
given a complete set of anti-Hermitian generators $X_a$ for the algebra of $SU(N)$, $su(N)$,
a group element of $SU(N)$ can be written through the local exponential (analytic) map as
\beq
\label{groupelement}
g=\exp\left(\sum_{a=1}^N\theta^a(x)\,X_a\right),
\eeq 
where the various $X_a$ satisfy the following commutation and anti-commutation relations
\beq
\label{commutation}
[X_a,X_b]=f^c_{\phantom{c}{ab}}X_c\qquad \{X_a,X_b\}=-\frac1N_c\delta_{ab}-id^c_{\phantom{c}{ab}}X_c,
\eeq
with $f^c_{\phantom{c}{ab}}=-f^c_{\phantom{c}{ba}}$ and $d^c_{\phantom{c}{ab}}=d^c_{\phantom{c}{ba}}$.
The local functions $\theta^a(x)$ are taken to be smooth over the manifold under consideration.
The normalisation condition depends on the specific representation
\footnote{We remind that a representation $\rho$ of a Lie group $G$ on a vector space $V$ is a homeomorphism of Lie groups $\rho:G\to {\rm Aut} V$.}
we use for the Lie group: for a generic representation $\rho$ we obtain
\beq
\label{1.4}
\Tr(X_a\,X_b)=-\rho\,\delta_{ab}.
\eeq
In the fundamental representation $\rho=1/2$ and in the adjoint $\rho=N_c$.
The dimension of this algebra is relevant only if gauge fields are coupled with fermions.  
Each gluon field is then a matrix-valued Lorentz vector
\footnote{According to \cite{Cotta:2003cr, Semenov:1982sf, Adams:1975, Mazja:1980vg}, the gauge fields $A_\mu$ have to be
matrix-valued functions belonging to the space $L^2$, the space of square-integrable functions, but the gauge transformations
rather to the more restrictive 
$W_1^2$, the space of Sobolev norm. Being $f\in W_1^2$,
then $|f|^2\equiv \int\frac{{\rm d}^dx}{x^2}|f|^2+\int{\rm d}^dx|{\partial}f|^2<\infty$.}, 
defined in terms of the algebra generators through the adjoint map as
\beq
A_\mu(x)=A^a_\mu(x)X_a.
\eeq
In Appendix B we shall briefly explain the geometric interpretation of the gauge field as the component of a Lie algebra-valued 
differential 1-form $\omega$, called the connection form over a principal bundle, 
which determines the profund relation between Y-M theory and the theory of principal bundles
\cite{Naber:2000bp, Cotta:2003cr}.
This algebraic and geometric structure appears in the four-dimensional free Euclidian action of YM theory, also called in a more geometric language
the Y-M functional, as
\begin{align}
\label{action}
S_{\rm YM}=\int_M{\rm d}^4x \mathcal L(x)&=\frac12\Tr\int_M{\rm d}^4x\, F_{\mu\nu}F^{\mu\nu}\nonumber\\
&=-\frac14\int_M{\rm d}^4x\, F^a_{\mu\nu}F_a^{\mu\nu},
\end{align}
where the trace over the gauge group ensures $S$ to be a scalar quantity. The manifold $M$, is generally supposed to be oriented, compact, without
boundary and endowed with a metric (in our case we have the flat, trivial Euclidian metric $\delta_{\mu\nu}$).
The field-strength tensor $F_{\mu \nu}$, component of a differential 2-form $\Omega$, is formally generated from $\omega$,
(and therefore from the potential $A_\mu$) by the Maurer-Cartan equation 
\begin{align}
\Omega&={\rm d}\omega+\frac12[\omega,\omega]\nonumber\\
&=\frac12 F_{\mu\nu}{\rm d}x^\mu\wedge{\rm d}x^\nu.
\end{align}
It can be geometrically interpreted as the Riemannian curvature in the principal bundle (also denoted as $F_A$), 
where the parallel transport along a curve is
not commutative. In Maxwell theory, being the Lie group Abelian, we have $[A,A]=0$ and this is the reason why the curvature is simply
$F_A={\rm d}A$.
According to the particular representation of $SU(N)$ we choose, $F_{\mu\nu}$ can assume different expressions: among the various 
irreducible representations, usually the fundamental and the adjoint are preferred. In the course of this Thesis we will adopt generally
the adjoint representation, unless otherwise specified 
\footnote{The difference between these two commonly used representations lies in the fact that in the fundamental representation we use
standard $N\times N$ matrices, where $N$ is the dimensionality of the Lie group and these matrices form a complete set of generators of the algebra. 
In the case of the adjoint reresentation, the matrices representing the basis elements are formed from the structure constants $f^{abc}$, 
defined through the commutation relations among the generators (\ref{commutation}). 
Therefore, each matrix has now a dimensionality $N^2-1\times N^2-1$, and this representation provides an overcomplete set of generators.},
In this representation, the covariant derivative is written as $\mathcal D_\mu=\partial_\mu + {\rm g}[A_\mu,\cdot]$, with ${\rm g}$ the coupling 
constant of the theory; consequently, the field-strength tensor becomes 
\begin{align}
\label{curvature}
F_{\mu \nu}&=\partial _{[\mu}A_{\nu ]}+{\rm g}[A_\mu, A_\nu]\nonumber\\
&=\frac1{\rm g}[\mathcal D_\mu, \mathcal D_\nu]\nonumber\\
&=\ad (F_{\mu\nu})=F_{\mu\nu}^a X_a.
\end{align}
In electromagnetism, the field-strength satisfies the relation $\mathcal P(\partial_\rho F^{\mu\nu})=0$, where $\mathcal P$ stands for cyclic 
permutation of Lorentz indices. In Y-M, however, this identity generalises to
\begin{align}
\label{jacobi}
0&=[\mathcal D_\mu, F_{\nu\rho}]+[\mathcal D_\rho, F_{\mu\nu}]+[\mathcal D_\nu, F_{\rho\mu}]\nonumber\\
&=\frac1{\rm g}\mathcal P[\mathcal D_\mu, [\mathcal D_\nu, \mathcal D_\rho]],
\end{align}
as a consequence of the Jacobi's identity and (\ref{curvature}). This is the analogue of the homogeneous Maxwell equations.
It must be stressed that if $F_{\mu\nu}$ and $A_\rho$ satisfy (\ref{jacobi}), $F_{\mu\nu}$
is not necessarily the strength tensor associated with $A_\rho$. This implies that, differently from the Abelian case, here
$F_{\mu\nu}$ does not determine uniquely all gauge-invariant quantities. 

\subsection{Local gauge invariance}

Requiring the action (\ref{action}) to be invariant under a local gauge transformation (being the pull-back of the connection form $\omega$
through a local section $\sigma$ over the principal bundle), we have a prescription for the transformation law of the gauge field as 
\beq
\label{gauge}
{}^g\!A_\mu=g^\dagger A_\mu g +\frac{1}{\rm g}g^\dagger \partial _\mu g,
\eeq
where $g=g(x)$ is a nonsingular local group element
\footnote{In a non rigorous language, we indicate $g(x)$ as the gauge transformation, but actually it is only a group element. The proper 
gauge transformation we refer to is (\ref{gauge}).}
of $SU(N)$ (cf. (\ref{groupelement})).
Given an infinitesimal group element 
$g(x) \,\substack {\simeq \\ \theta \to 0}\,{\bf 1}+\theta^a(x)X_a +\mathit o(\theta^2)$, 
the infinitesimal version of (\ref{gauge}), is reminiscent of 
the canonical formalism of symmetries $\delta_\alpha q^i=\{q^i,G_\alpha\}_{\rm Poisson}$, 
where $G_\alpha$ are the generators of the symmetry (cf. \ref{firstclass}), is
\beq
\label{infinitesimal}
\delta_{g} A^a_\mu=\partial_\mu \delta^{ab}\theta^b -{\rm g}f^{abc}\,\theta^bA^c_\mu\equiv \mathcal D_\mu^{ab}\theta^b.
\eeq
This invariance requirement is probably the most important property of Y-M theory and of any other local gauge theory. Comparing with
electromagnetism, we remember that the two fundamental physical fields of the theory, the magnetic field $B^a$ and the electric
field $E^a$ can be arbitrarily defined through the introdcution of an unphysical vector-gauge potential $A_\mu$.
If we demand the system not to change under a rotation in gauge space, then $B$ and $E$ are left unchanged by a suitable redefinition 
of the vector potential. The Lagrangian, which encodes all the physical information of the system, must not change as well: to be precise, under
the aforementioned redefinition of $A_\mu$ by (\ref{gauge}), $\mathcal L$ changes only by total derivative modulo boundary terms which vanish
if the fields vanish sufficiently fast at spatial infinity. 
The field-strength tensor transforms covariantly under the action of $g$
\beq
{}^g\!F_{\mu \nu}=g^\dagger F_{\mu \nu}g.
\eeq
An other interesting property of the field-strength tensor is that
if it is vanishing in a neighbourhood of a point (flat connections), then the gauge field is a pure gauge
\beq
F_{\mu\nu}=0\,\Leftrightarrow\,\exists\, g(x):\, A_\mu(x)=g^\dagger(x)\,\partial_\mu\, g(x).
\eeq
It follows that if $A_\mu$ is a pure gauge, then we have vanishing curvature.
In topological field theory, flat connections are studied in great details in BF theories \cite{Birmingham:1991ty}.
The field equations become
\beq
\label{fieldeq}
L^\mu_a=\partial _\nu F^{\mu \nu}_a-{\rm g}f^c_{a b}A^b_\nu F^{\nu \mu}_c=0,
\eeq
and are covariant, in the sense that if $A_\mu$ is a solution, so is ${}^g\!A_\mu$.
Because of the invariance of the action under (\ref{gauge}), the E-L equations are not independent\footnote{It must 
be stressed here that the canonical energy momentum tensor $\Theta^{\mu\nu}=-2\Tr(F^{\mu\rho}\partial^\nu A_\rho-\frac14
g^{\mu\nu}F^{\rho\sigma}F_{\rho\sigma})$ is not gauge invariant. By subtracting a term $\Delta \Theta=-2\partial_\rho\Tr(F^{\mu\rho}A^\nu)$,
we can restore the gauge invariance of $\tilde\Theta\equiv \Theta-\Delta\Theta$. This gaue-invariant energy momentum tensor
is equal to $\tilde\Theta=F_a^{\mu\rho}F^{a\nu}_{\rho}-\frac14g^{\mu\nu}F^{a\mu\sigma}F^{a\sigma\rho}$ called the Beltrami tensor \cite{Itzykson:1980rh}.}
but rather fulfill non-Abelian Bianchi identities
\be
\label{bianchi}
\left \{ \begin{array}{ll}
{\rm d}_AF=0 \\
*{\rm d}_A*F=0,
\end{array} \right.
\ee
with $*$ the Hodge operator 
\footnote{The action of $\ast$ on $F_{\mu\nu}$ determines its dual $\ast F_{\mu\nu}=\tilde F_{\mu\nu}$, fundamental to study instantons and solitons.
For instance, a more geometric expression for the Y-M action is $
S=||F_{\rm YM}||^2=-\frac14\int_M F\wedge \ast F$ \cite{Naber:2000bp}. The existence of a norm $||,||$ is guaranteed by the fact that we have a metric
on the space of gauge configurations $A$, denoted by $\mathcal A$. In the next chapter we will study the analytic and algebraic properties
of this functional space.}
and ${\rm d}_A$ the covariant derivative \cite{Naber:2000bp} (see Appendix B).
As usual, we  introduce the momenta conjugate
\beq
\Pi^\mu_a=\frac{\partial \mathcal L}{\partial \dot A_\mu^a }=F^{0\mu}_a,
\eeq
from which we immediately see, as in the case of the Abelian Maxwell theroy, that $\Pi_a^0$ is a primary constraint, and $A_0^a$ is its
canonical conjugated variable. Though, more insight can be obtained from the equations of motion directly: consider for instance
the action (\ref{action}) re-expressed in terms of the Y-M electric and magnetic fields $E$ and $B$ as \cite{Itzykson:1980rh}
\begin{align}
S&=\frac12\Tr\,\int{\rm d}^4x\,\mathcal L(x)\nonumber\\
&=\frac12\Tr\,\int{\rm d}^4x\,\left[\partial^0 \bold A\cdot \bold E+\frac12\left(\bold E^2+\bold B^2\right) 
-A^0(\bold \nabla\cdot \bold E +[\bold A, \bold E])\right],
\end{align}
where $\frac12\left(\bold E^2+\bold B^2\right)$ is the free energy. The canonical variables $p$ and $q$ become $\bold A$ and 
$\bold E$, whereas $A^0$ plays
the role of a Lagrange multiplier for the constraint identified with $\bold\nabla\cdot \bold E +[\bold A, \bold E]=\mathcal D[\bold A] \bold E$. 
This constraint comes from the equations of motion (\ref{bianchi})
by setting the Lorentz index $\nu=0$. In the Hamiltonian language we would say that, given a constraining manifold, whose constraints
having vanishing PB with $H$ or among themselves, they determine equivalent pairs of canonical variables if
\be
\label{ymconstraints}
\bea{ccc}
\frac{{\rm d}\bold A}{{\rm d}u}=\{\Gamma, \bold A\}\qquad \frac{{\rm d}\bold E}{{\rm d}u}=\{\Gamma, \bold E\}\\
\\
\Gamma=\bold \nabla\cdot \bold E +[\bold A, \bold E]=0.
\eea
\ee
Even though this has been shown in a non-covariant way, it is interesting to see how the Y-M electric field and the gauge field 
play a decisive part in the canonical formalism of constraints. The next step is then to gauge-fix the action:
in the Hamiltonian formalism, this is achieved by selecting a gauge and then taking the PB with respect to (\ref{ymconstraints}) one constructs
the appropriate path integral.
In the Lagrangian formalism, we consider the method proposed by Faddeev and Popov \cite{Faddeev:1967fc}: we will dedicate the entire next
chapter in analyzing this procedure, the appearence of the ghost fields into the Y-M path integral and of course we will give
extensive details on the Gribov problem \cite{Gribov:1977wm}.\\

\subsubsection{Euclidian solutions to the classical equations of motion: \\Instantons}
In a four dimensional manifold $M$ we can define the quantity
\beq
\label{instanton}
n=-\frac{1}{64\pi^2}\int{\rm d}^4x\epsilon^{\mu\nu\rho\sigma}F_{\mu\nu}^aF_{a\rho\sigma}=
-\frac{1}{32\pi^2}F_{\mu\nu}^a\tilde F^{\mu\nu}_a
\eeq
the {\it instanton winding number}. This quantity is called ``topological'' because, differently from the Y-M functional (\ref{action}),
it does not depend on the metric. Writing (\ref{instanton}) as $n=-\Tr\int{\rm d}^4x F\wedge F$ and comparing 
with (\ref{action}), we notice that there is no Hodge operator $\ast$: it is this operator which endowes the Y-M functional a metric and not $n$.
An other important property of $n$ is that it does not depend on $A$, but only on the algebraic structure of the manifold and of the principal bundle.
The importance of working on a four dimensional manifold lies in the fact that applying the Hodge operator to $F$, which is a differential 2-form,
gives an other differential 2-form. In particular, we have these two particular case
\beq
F_A=\pm\ast F_A,
\eeq
respectively called {\it self-dual} and {\it anti self-dual} curvatures. Moreover, $F_A$ can be decomposed into its dual and anti self-dual part
as $F_A=F_A^{\rm sd}+F_A^{\rm asd}$. It is possible to show that
\beq
\label{instanton1}
\Tr\int{\rm d}^4x F\wedge \ast F\geq \left|\Tr\int{\rm d}^4x F\wedge F\right|
\eeq
and that (\ref{instanton1}) is minimised when 
\begin{align}
\Tr\int{\rm d}^4x F_A\wedge \ast F_A&= \Tr\int{\rm d}^4x F_A\wedge F_A  &F_A^{\rm asd}&=0\nonumber\\
\Tr\int{\rm d}^4x F_A\wedge \ast F_A&= -\Tr\int{\rm d}^4x F_A\wedge F_A &F_A^{\rm sd}&=0.
\end{align}
The first case implies that $F_A=F_A^{\rm sd}$, called {\it instanton}, whereas the second case $F_A=F_A^{\rm asd}$, called {\it anti-instanton}.
These two minimizing solutions of the action are also solutions of the Y-M equations of motion (\ref{bianchi}).
The solution proposed by Belavin, Polyakov, Schwartz and Tyupkin 
\cite{Belavin:1975fg} for $n=\pm1$ reads
\beq
A_\mu=\frac{x^2}{x^2+\lambda^2}[\partial_\mu g(x)]g^\dagger(x),
\eeq
with $g(x)=\frac{x_0\pm i\bold{\sigma \cdot x}}{(x^2)^{1/2}}$.



\thispagestyle{empty}
\cleardoublepage
\chapter{Path integrals in Y-M theory}

We present here a very powerful tool in theoretical physics, known as the path integral
formalism (in Euclidean space), which we will largely adopt in the next chapters.
For a more detailed description, consult for example\cite{Itzykson:1980rh, Rivers:1987hi}.
The conceptualisation of path integrals in physics is mainly due to three scientists: Dirac, Feynman and Kac. In the 30's Dirac proposed the idea 
and Kac and Feynman established the
mathematical basis to provide us with a unified view of quantum mechanics, field theory and statistical models.
The basic idea behind the path integral approach to QFT is rather simple: at the quantum mechanical level, instead of pretending
to solve the Schr\"oedinger equation at general times $t$, one may first attempt to solve the easier problem at infinitesimal time
$\delta t$. The time-evolution operator, decomposed into its potential and kinetic part,
is divided in $N$ discrete time intervals $\delta t=t/N$. The exponential of the time-operator
can then be factorised into $N$ parts, such that the eigenstates of each component are known independently.
One then considers the amplitude of the time-evolution operator, split into $N$ time intervals, 
between initial and final state:  inserting the completeness relation of
the eigenstates of the position and momentum operator, the potential and kinetic operators thus act (to the left and to the right respectively)
on the corresponding eigenstates. In this way, the matrix element of the time-evolution operator has been expressed as $2N-1$
dimensional integrals over eigenvalues. At each time step $t_n=n\delta t$, $n=1,\ldots N$ we are integrating over coordinates
parametrising the classical phase space $(q_n, p_n)$. 
Therefore, the integral kernel (the propagator) of the time evolution operator could be expressed as a sum over all
possible paths connecting two points, $q'$ and $q''$ with a weight factor provided by the exponential of the action. 
Mathematicians, such as Wiener, Kac himself, Cameron and Martin, 
dealing with stochastic processes already knew this approach, as far as the analytic continuation is concerned.
Though, mathematicians were more reluctant to accept straightforwardly such path representation, because of its intrinsic and pathological
difficulties, mainly due to the highly non-trivial definition of an infinite functional measure (and also of the infinite sum of phases).
Historically, a semi-classical approach in solving this infinity was provided by the WKB method, in which the solution of the Feynman
kernel is based on the fact that the harmonic oscillator (the quadratic Lagrangian) is exactly solvable and its solution is only
determined by the classical path and not by the summation over all the paths.
Nonetheless, a correct mathematical interpretation and definition of the functional measure is still lacking:
there have been many attempts \cite{Unz:1985wq, Fujikawa:1979ay, Orland:1996hm, Orland:2004ve, Baez:1993id}, and references therein,
to give a definite and rigorous definition of such a quantity, trying to find a relation
with Lebesgue theory, measure theory and complex analysis in Hilbert spaces \cite{Adams:1975, Mazja:1980vg}.
In \cite{Unz:1985wq} it has been pointed out that a possible correct definition for a functional flat measure of bosonic fields could have the form of
$\label{measure1}
[\mat D\Phi]=\prod_x\left[\det\left(\frac{\delta^2 \mathcal L}{\delta(\partial_0 \Phi)\delta(\partial_0 \Phi)}\right)\right]^{\frac12}[{\rm d}\Phi],
$
whereas for fermions
$
[\mat D\Psi]=\prod_x\left[\det\left(\frac{\delta^2 \mathcal L}{\delta(\partial_0 \Psi)\delta(\partial_0 \Psi)}\right)\right]^{-\frac12}[{\rm d}\Psi]
$
\footnote{The symbol of ``$\det$'' stands for a functional determinant: a standard matrix determinant over the entire base-space $M$.}.
In the case of curved spaces, we may replace the former functional measures by
$[{\rm d}\Phi]=\prod_x(g^{00})^{1/2}(g^{\mu\nu})^{1/4}{\rm d}\Phi$,
with $g^{\mu\nu}$ the Riemannian metric tensor. 
Even these two objects can
produce some problems, especially concerning their regularity, because of the infinity arising from the number of space-times in $M$.
The most common technique to deal with such a regularisation problem is by virtue of the zeta-function \cite{Elizalde:1997nd},
also adopted in the regularisation of functional determinants.
The crucial problem is that, in any functional space, finding a converging Cauchy series
through which the underlying metric is defined, and consequently the concept of distance between two elements of the space, requires
a huge effort of mathematics techniques, not always successful. Therefore, what physicists do, and sometimes even mathematicians, is to postulate the
existence of a converging distance, a well-defined measure, and the only property openly required is the translational invariance of it, up to
boundary terms \cite{Rivers:1987hi}.
Though plagued by all these intrinsic and structural problems, path integrals are very elegant and suggestive and, they are ideally
suited to
\begin{itemize}
\item implement the symmetries of the theory directly
\item calculate correlations functions,
\item incorporate constraints simply,
\item analyze and explore field topology,
\item isolate relevant dynamical variables,
\item describe the non-zero temperature regime.
\end{itemize}
To clarify the notation we will adopt, we define as {\it functional} $F[\phi]$ of a real classical field $\phi(x)$ a rule that
associates a number (generally complex) to each real configuration $\phi(x)$. Functionals, naively speaking, include
as particular examples
integrals of functions
as $F[\phi]=\int{\rm d}x\,f(\phi(x))$. By {\it functional differentiation} we denote $\delta F[\phi]/\delta \phi(y)$ as the formal
limit (assuming the ratio exists)
$\delta F[\phi]/\delta \phi(y)=\lim_{\epsilon \to 0}\frac1\epsilon (F[\phi']-F[\phi])$, with $\phi'=\phi(x)+\epsilon \delta(x-y)$. 
The fundamental quantity in path integrals is the generating functional of the theory, formally defined as the integral of the action over
all of the possible constituent fields, taking values over all the possible space-time points
\beq
\label{path}
\mathcal Z[J]\equiv \int_{\mathcal M} [{\rm d}\Phi]\, e^{-S[\Phi]+\int_M J(x)\cdot \Phi(x)},
\eeq
where $\mathcal M$ is the functional configuration space of $\Phi$ and $M$ the space-time base space on which $\Phi$ is evaluated.
As seen in the previous chapter, in the case of the Hamilton principle of least action, 
the path configurations extremising the path integrals are the solutions of the classical equations of motion. In path integrals, though,
we consider all the possible paths represented by the functional measure: these quantum fluctuations have to be imagined as wrapped around
the classical flux tube connecting the two boundary points in the configuration space, whose contribution is weighted by the exponential of the action.
From $\mathcal Z[J]$ one can extract all the possible Green's functions of the theory by taking the appropriate functional derivative
(according to the spin-statistics of each field) 
with respect to the external sources $J(x)$.
Unfortunately, these Green's functions are cumbersome quantities to use, and in general the diagrams that constitute
a Green's function are disconnected (diagrams of two or more subdiagrams that are not linked by propagators). Moreover, being interested
in {\it one-particle irreducible} (1PI) diagrams (diagrams that cannot be separated by cutting single propagators), the isolation of connected
1PI diagrams is easily obtained by functional derivatives with respect to the external source $J(x)$
by means of a Legendre transformation of the effective action. 
To practically evaluate $\mathcal Z[J]$ from path integrals we may retain in the exponent the quadratic
part of the action, expand the rest in a Schwinger series and apply Wick's theorem: this is valid as long as the coupling constant of the theory is small.
If that is not the case, since we are only able to exactly calculate path integrals whose action is Gaussian, we may use different techniques, such as
the steepest descent or stationary space methods.

We are now ready to translate 
the quantisation procedure by means of the covariant operators we investigated in the last chapter into the language
of path integrals. This formalism is fundamental in analyzing and studying the topological nature
of Y-M theory, which constitutes the main subject of this Thesis. 
We will first introduce the Faddeev and Popov method to quantise non-Abelian
gauge theories in Euclidian space; then we will present in details the Gribov ambiguity which plagues these theories whenever one attempts to 
attack their non-perturbative nature. At last, we will concentrate on the functional configuration space $\mathcal A$,
paying particular attention to its stratification through the various Gribov regions $C_i$.


\section{Faddeev-Popov quantisation of non-Abelian gauge theories}

In late 60's, in the attempt to quantise non-Abelian theories, Faddeev and Popov \cite{Faddeev:1967fc} proposed an original method
based on covariant path integrals.  As Feynman noted in \cite{Feynman:1963ax}, in the gravitational field
and Yang-Mills theory, diagrams with closed loops depend non-trivially on the longitudinal parts of Green's functions
and scattering amplitudes are neither unitary nor transverse. Alongside Feynman, 
also De Witt \cite{DeWitt:1964yg} proposed a remedy to circumvent this problem.
Though, they were not able to give a prescription for arbitrary diagrams. The Faddeev and Popov method was developed
exactly to generalise Feynman and DeWitt's arguments.
We will follow their work in the light of Euclidian path integrals
\footnote{We choose the Euclidian metric because we will shortly analyze non-perturbative problems of Y-M, and for this purpose we will
adopt lattice gauge theories {\it a la} Wilson, in which the Euclidian metric, by means of a Wick rotation from Minkoswi space, 
is required to perform proper numerical simulations. This rotation alludes to the fact that a multiplication with the imaginary unit can
be interpreted as a $\pi/2$-rotation in the complex plane. Therefore imaginary time representations of Lagrangian actions are denoted
as Euclidean actions, whereas standard (real time) as Minkowski actions.},
emphasising the role of the functional measure and of the configuration space: these two subjects, alongside the underlying local gauge-invariance
of the Y-M action (and of the functional measure) 
will turn out to be of extreme importance
to understand the appearance of the famous Gribov ambiguity \cite{Gribov:1977wm}.

The FP quantisation procedure essentially deals with non-Abelian theories subject to constraints.
In fact,
as we noted in the previous chapter, the time component of the momentum conjugate $\Pi_\mu^a(x)$ is subjected to a vanishing constraint
due to the antisymmetry of the field-strength tensor. This condition is not consistent with the assumed commutation relations.
Thus the simple-minded application of the canonical quantisation of non-Abelian gauge theory fails.
This difficulty arises as long as we rely on a gauge invariant Lagrangian, such that $\mathcal L_{\rm YM}$ remains invariant under
an infinitesimal gauge transformation, changing $A_\mu^a$ into $A_\mu^a+\mathcal D^{ab}_\mu \theta^b$.
We previously observed how in the canonical operator formalism, the introduction of a gauge-fixing term
corresponds to a constraint that could eliminate this unnecessary gauge freedom. 
We would then wish to incorporate in the path integral only those gauge connections $A$ that are unrelated by gauge transformations.
This is a more difficult task in a non-Abelian one. 
To start with the functional quantisation of Y-M theory, we introduce in a flat (Euclidean) metric, the gauge-unfixed Y-M
generating functional
\beq
\label{YMpath}
\mathcal Z_{\rm YM}[J]\equiv \int_{\mathcal A} [{\rm d}A]\, e^{\frac12\Tr\int_M \mathcal L_{\rm YM}+\Tr\,\int_M J^\mu(x)\cdot A_\mu(x)}.
\eeq
Here, the path integral is considered over the configuration space $\mathcal A$ spanned by the gauge fields $A_\mu$,
defined on a Riemannian manifold $M$ for now left as general as possible.
In order to remove the gauge freedom of the Lagrangian density
\footnote{It is worthwhile noting that the source term $J^\mu(x)\cdot A_\mu(x)$ is not gauge invariant.}
and to preserve the manifest covariance, we may then choose the Lorentz condition
\beq
\label{Lorentz}
\partial^\mu A^a_\mu=0,
\eeq
such that the gauge freedom is eliminated because (\ref{Lorentz}) is no longer gauge invariant
\footnote{Of course in (\ref{YMpath}) the gauge invariance is already broken by the presence of the source term. Nonetheless,
as we will focus on the sourceless generating functional, it is important to stress this lost gauge invariance by means
of the gauge-fixing term.}.
There are some variety of gauges other than
the Lorentz gauge. Among them, the following noncovariant gauges are frequently used:
Coulomb (radiation) gauge $\partial^iA_i^a=0$ ($i=1,2,3$), axial gauge $A^a_3=0$ and temporal gauge $A^a_0=0$. Being interested in manifestly
covariant gauges, 
we will adopt the Lorentz gauge henceforth. To incorporate the gauge constraint (\ref{Lorentz})
into the generating
functional (\ref{YMpath}), we use the standard Lagrange multiplier method well known in analytic dynamics as follows
\beq
\label{gflagrangian}
\mathcal L_{\rm gf}=-\frac14F^a_{\mu\nu}F^{a\mu\nu}-\frac{1}{2\xi}(\partial^\mu A_\mu^a)^2.
\eeq
Though the gauge-fixing Lagrangian density so constructed manifestly breaks gauge invariance, all the physical quantities extracted from
it should of course be gauge-independent and therefore one has the freedom to fix the value of the gauge parameter $\xi$ arbitrarily
(for instance $\xi=1$ corresponds to Feynman gauge, whereas $\xi \to 0$ to Landau gauge).
According to (\ref{gflagrangian}), the equations of motion will change, and consequently the expression of the momentum conjugate as
\beq
\Pi^a_\mu=-F_{0\mu}^a-\frac1\xi\delta_{0\mu}(\partial^\nu A_\nu^a),
\eeq
which circumvents the vanishing condition for the time component of $\Pi^a_\mu$. In such a way, the commutation relations are satisfied
and the path integral quantisation is apparently well-defined. 
Yet, in the case of non-Abelian gauge theories, due to the self-interaction of the gauge field $A$,
we also have, differently from the case of QED, a three-body and a four-body interaction in the Lagrangian (\ref{gflagrangian}).
At one-loop level, the gauge-field contribution to the self-energy part $\Pi_{\mu\nu}^{ab}(q)$ for gauge fields $A_\mu^a$
does not satisfy the requirement for gauge invariance $q^\mu \Pi_{\mu\nu}^{ab}(q)=0$.
The reason why such requirement fails is due to the non correct method of extrapolating the physical polarisation for the gauge
field even with the gauge-fixing Lagrangian (\ref{gflagrangian}).
Feynman first \cite{Feynman:1963ax} 
and then De Witt \cite{DeWitt:1964yg} pointed out this difficulty in the early stages of the development of quantisation of non-Abelian theories.
To solve this problem, Faddeev and Popov tried to incorporate in (\ref{YMpath}) an appropriate gauge-fixing condition with the 
double purpose to eliminate the infinite redundancy of gauge transformed fields affecting the path integral 
and to guarantee the elimination of the unphysical polarisation states of the gauge field.

As we know well, the Y-M Lagrangian density is gauge invariant by construction, whereas the gauge-fixing and source terms are not.
The measure requires special attention: performing an infinitesimal gauge transformation we find
\begin{align}
[{\rm d}{}^g\!A]&=[{\rm d}A]\det\left(\frac{\delta {}^g\!A^a_\mu }{\delta A_\nu^b}\right)\nonumber\\
&=[{\rm d}A]\det(\delta^{ab}-f^{abc}\theta^c)\nonumber\\
&=[{\rm d}A](1+\Tr L+\ldots +\det L)\nonumber\\
&=[{\rm d}A](1+{\it O}(\theta^2)),
\end{align}
with $L=-f^{abc}\theta^c$. To analyze the problem connected with the infinite gauge measure, it is possible to disregard from
(\ref{YMpath}) the source term and dealing just with $\mathcal Z_{\rm YM}[0]$. Furthermore, in the light of covariant gauges, the
condition (\ref{Lorentz}) can be generalised to the case of a differential operator $G^\mu$
\beq
\label{Lorentz1}
G^\mu A_\mu^a(x)=\chi^a(x),
\eeq
with $\chi^a(x)$ a matrix-valued local function not depending on gauge transformations. 
The essential requirement for the gauge condition (\ref{Lorentz1}) is to be single-valued: this would guarantee the bijectivity
of the map between the space of gauge configurations $A$ and the space of gauge connections modulo gauge-transformations, $A/[{}^g\!A]$,
satisfying (\ref{Lorentz1}). Basically, it is demanded that the gauge-fixing hypersurface generated by (\ref{Lorentz1})
should intersect each gauge orbit once and only once. A single-valued gauge-fixing condition is called in the literature {\it ideal}.
Yet, it is not difficult to show that
for any field $A$ satisfying (\ref{Lorentz1}), there are many others, obtained by a gauge transformation of $A$,
satisfying the same condition
\beq
\label{Lorentz2}
G^\mu \,{}^g\!A_\mu^a(x)=\chi^a(x)\quad \Leftrightarrow\quad G^\mu A_\mu^a(x)=\chi^a(x).
\eeq
It seems then the requirement of an ideal gauge-fixing condition immediately fails: in perturbation theory, where the FP method
is discussed, this inconvenience is circumvented by considering only infinitesimal fluctuations around the trivial gauge configuration $A_\mu=0$.
In fact, any configuration ${}^g\!\,0=g^\dagger\,\partial_\mu\, g$ satisfying (\ref{Lorentz2}) lies sufficiently far from the intersection
between the hypersurface and the trivial orbit, and therefore the gauge-fixing condition can still be regarded single-valued.
However, beyond perturbation theory, when finite gauge transformations are important, single-valued gauge conditions are
considered impossible to be found (thus the adjective ``ideal'') as long as Y-M theory is evaluated on $S^4$, the standard manifold
for physical processes \cite{Singer:1978dk}. This pathological problem affecting the 
non-perturbative regime of non-Abelian theories is called the {\it Gribov ambiguity}
and will be detailed in the next section. For the moment, we will only deal with perturbation theory and consequently we
are allowed to neglect such obstruction.

Faddeev and Popov proposed a way to take into account the condition (\ref{Lorentz1}) in the sourceless generating functional $\mathcal Z_{\rm YM}[0]$,
generalising the well known formula in 
standard calculus for an appropriate change of variables, 
\begin{equation}
\label{change}
\left| \det\left({ {\partial f_i}\over {\partial x_j} }\right) 
\right|^{-1}_{{\vec f}=\vec 0}
= \int dx_1 \dots dx_n \delta^{(n)}({\vec f}({\vec x}))
\end{equation}
where the map is supposed to be bijective, i.e. single-valued and  $\det\left({ {\partial f_i}\over {\partial x_j} }\right)$
is the Jacobian of the transformation.
Being the Jacobian independent of local coordinates, we then write
\begin{equation}
\label{change1}
1= \int dx_1 \dots dx_n \,\left| \det\left({ {\partial f_i}\over {\partial x_j} }\right) 
\right|_{{\vec f}=\vec 0}\,\delta^{(n)}({\vec f}({\vec x})).
\end{equation}
The generalisation of (\ref{change1}) in the language of non-Abelian gauge theory can be cast in the form
\begin{equation}
\label{functchange1}
1
= \int [{\rm d}g] \,\left| \det\left({ {\delta F[{}^g\!A]}\over {\delta g} }
\right) \right|_{F=0}\,\delta(F[{}^g\!A]) 
\end{equation}
with $F[{}^g\!A]=0$ the local gauge condition (\ref{Lorentz1}). Some remarks are necessary here.
The functional integration $\int[{\rm d}g]$ we perform in (\ref{functchange1}) is over the group space and 
it is called the Haar measure \cite{Naber:2000bp, Nakanishi:1990qm, tilma-2004-52, Nakahara:1990th, Smit:2002ug}. 
This measure is invariant under a gauge transformation: in fact, for any functional of the gauge group $g$,
we can distinguish between left or right invariant measure, according to the kind of group action w.r.t. $g_0$ we perform on
the group element $g$. The left invariant measure satisfies for instance the following condition
\beq
\int  {\rm d}g f(g)=\int {\rm d}gf(g_0^{-1}g)=\int ({\rm d}g_0g)f(g)= \int  {\rm d}g f(g),
\eeq
and similarly for the right action.
In general, left and right invariant measures are not necessarily equal, but it is possible to
prove \cite{Muta:1998vi} (and references therein) that for compact groups, simple and semi-simple groups, and also finite groups this is the case.
According to the parametrisation for $SU(N)$ one adopts, we can give a more practical expression for the Haar measure.
For instance, instead of using group elements $g$, we can perform the integration over the local gauge functions $\theta^a(x)$
appearing through the exponential map connecting the group $SU(N)$ to its algebra $su(N)$. 
In this case, the Haar measure is proportional to $\sqrt{\det \rm g_{ab}}\prod_{a,x} {\rm d}\theta^a(x)$ \cite{Nakahara:1990th, Smit:2002ug}, 
with $\rm g_{ab}$ being the metric in $SU(N)$
\footnote{Chosen an arbitrary parametrisation for the group element, the metric in the group space can be written as 
${\rm g}_{kl}=\frac1\rho\,\Tr\,\left(\frac{\partial g}{\partial \theta^k}\frac{\partial g^\dagger}{\partial \theta^l}\right)$,
with $\rho$ the normalisation in the given representation. Under coordinate transformations $\theta^k=f^k(\theta')$,
the metric is covariant, ${\rm g}_{kl}={\rm g}_{mn}\frac{\partial \theta'^m}{\partial \theta^k}\frac{\partial \theta '^n}{\partial \theta^l}$.}.
The same result can be also achieved by parametrising the group through Euler angles, showing that
the $SU(N)$ volume can be written as $2^{N-1/2}\pi^{\frac{(N-1)(N+2)}{2}}\sqrt N \prod_{k=1}^{N-1}\frac1{k!}$ \cite{tilma-2004-52}.

The second comment is on the absolute value of the Jacobian. Faddeev and Popov did not consider it because they were
interested in quantising Y-M theory in the perturbative regime. 
The reason why in perturbation theory we can remove the absolute value lies in the positive definiteness of the Jacobian in
(\ref{functchange1}), also known as the determinant of the Faddeev-Popov (FP) operator $\mathcal M[A]$. Under an infinitesimal gauge transformation,
in local coordinates, this operator is
\begin{align}
\mathcal M_{ab}(x,y)&=\frac{\delta}{\delta \theta^a(y)}F_b[{}^g\!A(x)]\nonumber\\
&=\frac{\delta}{\delta \theta^a(y)}\left[-\frac{\delta F_b}{\delta A_c^\mu(x)}\,\mathcal D^\mu_{cd}\,\theta_d(x)\right]\nonumber\\
&=-\frac{\delta F_b}{\delta A_c^\mu(x)}\,\mathcal D^\mu_{ac}\,\delta^4(x-y).
\end{align}
If we adopt the covariant condition (\ref{Lorentz1}), we obtain
\begin{align}
\mathcal M_{ab}(x,y)&=-\partial_\mu\,\mathcal D^\mu_{ab}\,\delta^4(x-y)\nonumber\\
&=-[\Box \delta_{ab}-{\rm g}f_{abc}\partial^\mu\,A_\mu^c(x)]\delta^4(x-y).
\end{align}
It is now clear why, in the case of the trivial orbit and with appropriate boundary conditions, 
the FP operator has definite sign, because it is just the Laplacian operator multiplied tensorially with the Lie group.
It is therefore redundant to keep the absolute value in perturbation theory. We will see, however, that beyond this regime not only is the absolute
value necessary, but it is its very presence which determines the Gribov ambiguity \cite{Ghiotti:2005ih}.
According to the gauge condition (\ref{Lorentz}), the FP operator is not independent of $A$.
The determinant of $\mathcal M[A]$ has also an interesting geometric interpretation: 
in \cite{Babelon:1980uj, Dell'Antonio:1991xt, Daniel:1979ez},
it was shown how it is related to the volume of the gauge orbit.
It is worthwhile noting that
the generalisation of the linear covariant gauge condition (\ref{Lorentz}) to (\ref{Lorentz1}) does not affect the form
of the operator $\mathcal M$. 
In the case of non-covariant gauges, temporal gauge has a constant FP operator but Gauss' law is lost, whereas Coulomb gauge has
not. It is an easy task to show that in QED, the FP operator is simply the Laplacian operator and therefore $\det \mathcal M$
results in an overall constant factor for the path integral. This is the reason why in QED, in the presence of linear gauges,
there is no Gribov ambiguity. 

Inserting then (\ref{functchange1}) into the gauge-fixing generating functional, we obtain
\beq
\label{gfYM}
\mathcal Z_{\rm gf}[0]=\int [{\rm d}g] [{\rm d}A]  \,\det\left({ {\delta F[{}^g\!A]}\over {\delta g} }
\right)_{F=0}\,\delta(F[{}^g\!A])\,
e^{-\int_M \frac14F^a_{\mu\nu}F^{a\mu\nu}}.
\eeq
Making use of the local gauge invariance of the Y-M action and of the functional Haar measure, which
allows us to show that also the functional Jacobian does not depend on the gauge transformation $g$
\footnote{According to the left-invariant measure, one can show that any integral over the gauge group is
gauge invariant. In fact, as mentioned above,
it is easy to show that $\int {\rm d}g\, \Phi(g)=\int ({\rm d}g_0g)\, \Phi(g)$,
regardless whether $\Phi(g)$ is a gauge-invariant function or not.}, 
we can re-write (\ref{gfYM}) as
\beq
\label{gfYM1}
\mathcal Z_{\rm gf}[0]=\int [{\rm d}g] \int[{\rm d}{}^g\!A]  \, \det\left({ {\delta F[{}^g\!A]}\over {\delta g} }
\right)_{F=0}\,\delta(F[{}^g\!A])\,
e^{-\int_M \frac14F^a_{\mu\nu}F^{a\mu\nu}},
\eeq
where we have left the dependence of $g$ also in the measure. 
In this way we have factored out the infinite measure over the gauge group
\footnote{The Haar measure of the continuous group $\mathcal G$ of gauge transformations is infinite because, though the Lie group $SU(N)$ is compact,
a gauge transformation belongs to the functional space $\Omega^0(M, \ad P)$. Therefore $[{\rm d}g]$ also includes
an infinite product over all the $x\in M$, $[{\rm d}g]=\lim_{x\to \infty}\prod_x {\rm d}g(x)$.
In lattice Y-M theory, the discretisation of the space-time allows to make sense of such an infinity, and consequently
the Haar measure becomes regulated naturally.}.
The importance of this passage is clearly manifest when one deals with expectation values of gauge-invariant operators (thus
physical observables), 
where we need to calculate
\beq
\label{Observable}
\langle \mathcal O\rangle=
\frac{\int_{\mathcal A}[{\rm d}A]\, \mathcal O[A]\,e^{-S_{\rm YM}[A]}}{\mathcal Z_{\rm YM}}\,.
\eeq
Making use of the Faddeev-Popov method, we re-write the ratio as
\beq
\label{Observable1}
\langle \mathcal O\rangle=
\frac{\int[{\rm d}A]  \, \det\left({ {\delta F[{}^g\!A]}\over {\delta g} }
\right)\,\delta(F[{}^g\!A])\,\mathcal O[A]\,
e^{-\int_M \frac14F^a_{\mu\nu}F^{a\mu\nu}}}
{\int[{\rm d}A]  \,\det\left({ {\delta F[{}^g\!A]}\over {\delta g} }
\right)\,\delta(F[{}^g\!A])\,
e^{-\int_M \frac14F^a_{\mu\nu}F^{a\mu\nu}}},
\eeq
where $\int [{\rm d}g]$ has been cancelled both from numerator and denominator, due to the gauge-invariance of $\mathcal O[A]$.
Therefore, since gauge-invariant quantities should not be sensitive to changes of auxiliary conditions, it is possible to average over 
the local functions $\chi^a(x)$ of (\ref{Lorentz1}) with a Gaussian weight,  
substituting the delta function in the integrand of (\ref{gfYM1}) as
\beq
\int [{\rm d}\chi]\,\delta(G^\mu A_\mu^a -\chi^a)\,e^{-\frac1{2\xi}\int_M (\chi^a)^2}
=e^{-\int_M\frac{1}{2\xi}(G^\mu A_\mu^a)^2}.
\eeq
The complete gauge-fixing generating functional without sources is consequently
\beq
\label{gfYM2}
\mathcal Z_{\rm gf}[0]=\int[{\rm d}{}^g\!A]  \, \det\left({ {\delta F[{}^g\!A]}\over {\delta g} }\right)\,
e^{-\int_M \left\{\frac14F^a_{\mu\nu}F^{a\mu\nu}-\frac{1}{2\xi}(G^\mu A_\mu^a)^2\right\}}.
\eeq
The perturbative expansion of $\det \mathcal M$ 
\footnote{We remind the reader that, given a functional determinant $\det A$, this can be exponentiated as $\det A=\exp(\Tr \log A)$,
which can be diagrammatically represented as infinite non-local loops.}
leads to non local interactions between gauge fields. To express these interactions
as local ones, we perform a manipulation which takes into account Grassmann fields, unphysical and fictitious, playing only 
an algebraic role.
A more detailed explanation of Grassmann fields and algebra will be given in the next chapter, when we will introduce the BRST 
formalism. For the sake of comprehension, we just wish to remind the reader that it is possible to write a functional determinant
of any $N\times N$ matrix operator $Q$
\footnote{In the case of Grassmann fields, the operator $Q$ is not required to have special properties (apart from singularities).
On the contrary, in the case of Gaussian bosonic integration, when the field is real, $Q$ has to be positive definite.}
over a complete set of dimension $2N$ of anti-commuting generators, called Grassmann, such that
\beq
\det Q=\int\prod_{i=1}^k{\rm d}\bar\eta_k\,{\rm d}\eta_k\,e^{-\bar \eta Q \, \eta}
\eeq
In the language of Feynman diagrams, these fields have been called by Feynamn  FP ghosts: though anti-commuting, they are Lorentz scalar, and therefore
do not satisfy the spin-statistics theorem.
As far Feynman diagrams are concerned,
they are allowed to run around loops but 
not in external lines. They do not add to the spectrum of observable particles in the theory.
The indisputable importance of these unphysical fields lies in their role played to guarantee the unitarity of the $\mathcal S$-matrix.
As seen in the previous chapter, the decomposition into positive and negative frequency states
of the $B$-field led us to a subsidiary condition $B^{+}(x)|{\rm phys}\rangle=0$, whose
role was to specify and select the physical states. This condition could be associated to the Gupta-Bleuler condition, which guarantees
the unitarity of QED. In the attempt to generalise QED to the case of non-Abelian gauge theories, such as QCD for instance,
the non-linear self-interaction of the gauge fields causes the subsidiary condition on the $B$-field to fail.
Feynman pointed out that this could affect the breakdown of the unitarity of the $\mathcal S$-matrix. In fact, due to
this self-interaction, it is not guaranteed that the contribution of unphysical degrees of freedom of $A_\mu$, the longitudinal and
temporal modes, to intermediate states could cancel out. Feynman himself and De Witt found in the context of
perturbation theory that this problem concerning unitarity could be explained by the absence of massless scalar fermions
to closed loops in Feynman diagrams. It is then thanks to Faddeev and Popov that these missing unphysical particles
showed up through their quantisation method. Furthermore, in the context of Y-M theory renormalisation, it is due to the work
of Veltman and 't Hooft \cite{'tHooft:1972fi}
that we can prove now that ghosts allow exact cancellation at all orders of the longitudinally and temporally
polarised modes in the intermediate states, where the intermediate states include transverse vector particles. 
In this way unitarity is preserved. 
To insure global invariance, these ghost fields belong to the adjoint
representation of the Lie group under consideration. 

Under this manipulation, (\ref{gfYM2}) assumes the original form presented
by Faddeev and Popov
\beq
\label{gfYM4}
\mathcal Z_{\rm gf}[0]=\int[{\rm d}A][{\rm d}\bar \eta][{\rm d}\eta]  \,
e^{-\int_M \left\{\frac14F^a_{\mu\nu}F^{a\mu\nu}+\bar \eta^a\mathcal M_{ab}\eta^b+\frac{1}{2\xi}(G^\mu A_\mu^a)^2\right\}},
\eeq
and in the presence of sources
\beq
\label{gfYM4}
\mathcal Z_{\rm gf}[J, \zeta, \bar \zeta]=\int[{\rm d}A][{\rm d}\bar \eta][{\rm d}\eta]  \,
e^{-\int_M \left\{\frac14F^a_{\mu\nu}F^{a\mu\nu}+\bar \eta^a\mathcal M_{ab}\eta^b+\frac{1}{2\xi}(G^\mu A_\mu^a)^2
-J^aA_\mu^a-\bar \zeta^a\eta^a+\bar \eta^a\zeta^a\right\}},
\eeq
Our quantisation procedure {\it a la} Faddeev and Popov is then completed, as well as providing a generating functional,
through the exponentiation of the FP operator, able to generate appropriate Feynman diagrams in the context of perturbation theory.
Though, as pointed out by Gribov \cite{Gribov:1977wm} 
and then explained in the language of principal bundles by Singer \cite{Singer:1978dk},
(\ref{gfYM4}) makes only sense in the high-energy regime, but it fails to be applicable beyond it, when non-perturbative effects
have to be taken into account. This will be explained in the next section.


\section{The Gribov Ambiguity}

As we saw in the last chapter, the constraints we have to impose when quantising Maxwell's electrodynamics do not change
the energy spectrum. This is so because we can reduce the number of degrees of freedom to be quantised by taking
advantage of the gauge-invariance of the classical theory. This procedure is called {\it gauge fixing}.
The choice of a gauge fixing term is arbitrary, but it leads to different problematic. For example, Coulomb gauge $\vec \nabla \vec A=0$ 
is not compatible with the Poisson Brackets $\{A_i(\vec x), \Pi_j(\vec y)\}=\delta_{ij}\delta(\vec x-\vec y)$, because
the spatial divergence of the $\delta$-function does not vanish. This implies that the quantisation of the theory is achieved only
at the price of modifying the commutation relations. Moreover, differently from manifest covariant gauges, such as Lorentz or Landau gauges,
Coulomb gauge spoils the Lorentz invariance. Another gauge which breaks the Lorentz invariance is temporal or Weyl gauge:
in this gauge one imposes the condition $A_0=0$, which causes the Gauss Law to be lost.
We have also seen how the Faddeev-Popov method of quantising non-Abelian theories provides a way to avoid
the infinite gauge measure. Though, such a procedure is plagued by a topological obstruction, which prevents us from going beyond
perturbation theory. As we will see, this problem is a common problem in any non-Abelian theory which is evaluated on a configuration
space over equivalence classes of gauge-transformations.

In the late seventies, in fact, it was
first pointed out by V. Gribov in his seminal work \cite{Gribov:1977wm} that once Y-M theory is gauge-fixed by means of Coulomb
gauge, one has to face a degeneracy of such gauge, i.e. the gauge orbits can intersect the Coulomb gauge hypersurface at more than one point.
Following Gribov, we consider two gauge-equivalent fields $\vec A$ and $\vec A'$
\beq
\label{gaugetr}
\vec A'=g^\dagger\,\vec A\,g+g^\dagger\,\vec \nabla\,g.
\eeq
Because of the non-linearity of (\ref{gaugetr}), a transverse field potential  satisfying the Coulomb-gauge condition may actually happen
to be a pure gauge, which should not be separately counted as an additional physical degree of freedom. He explicitly constructed
such a transverse field for $SU(2)$, and showed the uncertainty in the gauge-fixing procedure
when the QCD coupling constant $\rm g$ becomes of order of unity, i.e. in the non-perturbative regime.
In terms of the FP method, this uncertainty arises when the FP operator acquires zero-mode solutions, which occurs in the infrared region,
where the vacuum enhancement of the dressed Coulomb gluon propagator becomes catastrophically large \cite{Zwanziger:2003cf}.
To see the appearance of this uncertainty, we start first with a covariant gauge, and then we limit ourselves to the three-dimensional
case.
The divergence of $A'_\mu$ is given by
\beq
\label{divergence}
\partial_\mu A'^\mu=(\partial_\mu g^\dagger)\,A^\mu\,g
+g^\dagger\,(\partial_\mu A^\mu)\,g
+g^\dagger\,A^\mu\,\partial_\mu g
+(\partial_\mu g^\dagger)\,\partial^\mu\,g
+g^\dagger\,\Box\,g.
\eeq
The requirement for the gauge transformation $g$ to not change the divergence of both $A_\mu$ and $A'_\mu$ leads to the condition
\beq
\partial_\mu(g^\dagger\,(\mathcal D^\mu[A]\,g)=0.
\eeq
This second order partial differential equation, for large values of $A_\mu$, i.e. for distant configurations from perturbation theory 
(which is evaluated around $A_\mu=0$), will produce several non-trivial (different from constant gauge transformations) solutions. Among these solutions,
we can distinguish three types: 1) solutions belonging to gauge orbits intersecting the gauge-fixing hypersurface only once, which
correspond to ideal gauge conditions. 2) solutions belonging to gauge orbits never intersecting the gauge-fixing hypersurface and
3) solutions belonging to gauge orbits intersecting the gauge-fixing hypersurface more than once.
The third case is what Gribov discovered in Coulomb gauge, which would cause the ambiguity in the gauge-fixing procedure. These redundant 
solutions have been called {\it Gribov copies}, i.e. for each configuration satisfying the gauge condition, there are gauge-equivalent
configurations that satisfy the same condition. Therefore, the gauge-fixing procedure fails in removing all the unnecessary gauge
degrees of freedom. 
It is pedagogical to show that, within perturbation theory, for an infinitesimal gauge transformation ($g\simeq 1+X^a\theta_a(x)+{\it O}(\theta^2)$)
the divergence (\ref{divergence}) becomes
\begin{align}
\Box \theta -(\partial^\mu \theta)A_\mu + A_\mu (\partial^\mu \theta)=0\quad
{\rm or}\quad
\partial^\mu(\partial_\mu\theta +[A_\mu,\theta])=0.
\end{align}
We see that the condition for the appearance of Gribov copies is equivalent to the requirement for the operator 
$-\partial^\mu(\partial_\mu\theta +[A_\mu,\theta])$, the Faddeev-Popov operator, to have zero eigenvalues (zero modes).
It is also interesting to notice \cite{Sobreiro:2005ec} that the eigenvalue equation for the FP operator 
\footnote{Performing a gauge transformation on $A^\mu$, the FP operator w.r.t. the gauge transformed field becomes
$
\mat M_{FP}[{}^g\!A]=\Box-\partial_\mu[{}^g\!A^\mu,\cdot ]=\Box-\partial_\mu[A^\mu-g^\dagger[\mat D^\mu,g],\cdot]
=\mat M_{FP}+\partial_\mu[g^\dagger[\mat D^\mu,g,\cdot]
$
Applying n-times a gauge transformation on $A$ gives 
$
{}^{g_n}\!A_\mu=(g_n)^\dagger A_\mu g_n-(g_n)^\dagger \partial_\mu g_n 
$
where $g_n=\prod_{i=1}^n g_i$.
This implies that the FP operator for the n-th gauge transformation is
$
\mat M_{FP}[{}^{g_n}\!A]=\Box-\partial_\mu[(g_n)^\dagger A_\mu g_n-(g_n)^\dagger \partial_\mu g_n,\cdot]=
\mat M_{FP}[A]+\partial_\mu[(g_n)^\dagger[\mat D^\mu,g_n],\cdot]
$.}
resembles a Schr\"odinger equation as
\beq
\label{Schro}
-\partial^\mu(\partial_\mu\alpha +[A_\mu,\alpha])=\epsilon[A]\alpha,
\eeq
with the gauge potential $A$ playing the role of a potential. Being the eigenvalue $\epsilon$ a functional of $A$
(as well as the eigenfunction $\alpha$),
we expect, for sufficient large values of $A$, the zero-energy solution ($\epsilon[A]=0$) to exist.

\subsection{Gribov pendulum}

In his work, Gribov considered the three-dimensional case with Lie group $SU(2)$ to explore explicit solutions to the Coulomb gauge.
Moreover, to simplify the parametrisation, he chose a time-independent and 
spherically symmetric gauge field $A_i$, $i=1,2,3$, such that it only depends
on the unit vector $n_i=x_i/r$, with $r=\sqrt{x^ix_i}$. Under such assumptions, he considered for the following parametrisation
\beq
A_i=f_1(r)\frac{\partial \hat n}{\partial x_i}+f_2(r)\hat n\frac{\partial \hat n}{\partial x_i}
+f_3(r)\hat n n_i,
\eeq
with $\hat n=in_i\sigma_i$ and the functions $f_i(r)$ supposed to be smooth on the domain of $r$.
Since
\beq
\frac{\partial \hat n}{\partial x_i}=\frac ir\,(\sigma_i-(\vec n\cdot \vec \sigma)n_i),
\eeq
it follows that the gauge field can be written as
\beq
\label{Gripar}
A_i=\frac ir f_1(r)\sigma_i
-i\frac irf_1(r)(\vec n\cdot \vec \sigma)n_i
-\frac irf_2(r)(\vec n\cdot \vec \sigma)\sigma_i
+\frac ir f_2(r)n_i
+if_3(r)(\vec n\cdot \vec \sigma)n_i.
\eeq
In the case in which $f_1=f_3=0$, (\ref{Gripar}) simplifies to
\beq
\label{Gripar1}
A_i=\frac i{r^2}\epsilon_{ijk}x_j\sigma_k\, f_2(r),
\eeq
which is purely transverse $\vec \nabla\vec A=\vec 0$, due to the antisymmetry of the tensor $\epsilon_{ijk}$.

\begin{figure}[ht]
\label{Gribovpendulum}
\centering \epsfig{file=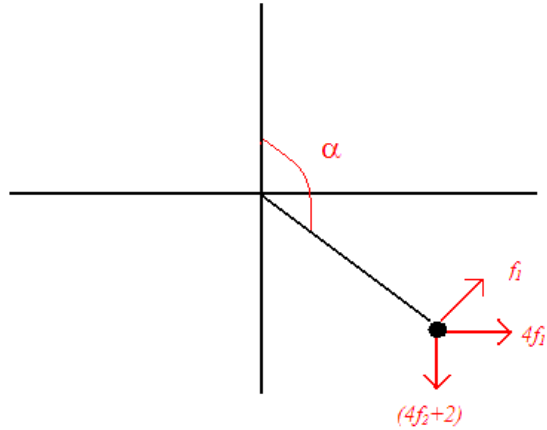,width=10cm} \caption{The Gribov
pendulum}
\end{figure}

The condition for the existence of copies
\begin{align}
\vec A'&=g^\dagger\,\vec A\,g+g^\dagger\,\vec \nabla\,g\nonumber\\
\vec \nabla \vec A'&=\vec \nabla \vec A,
\end{align}
together with a suitable parametrisation for the $SU(2)$ gauge transformation $g$ ($g=e^{\frac i2 \alpha(r)\vec n\cdot \vec \sigma}$)
determine an explicit form for the gauge transformed field $A_i'$ as follows
\begin{align}
A_i'&=\left( f_1 \cos \alpha+\left(f_2 +\frac12\right)\sin \alpha\right)\frac{\partial \hat n}{\partial x_i}
+\left(\left(f_2+\frac12\cos\alpha\right)-f_1\sin\alpha-\frac12\right)\hat n\frac{\partial \hat n}{\partial x_i}
\nonumber\\
&+\left(f_3+\frac12\frac{{\rm d}\alpha}{{\rm d} r}\right)\hat nn_i.
\end{align}
The condition on the divergence of the gauge fields determines instead a second-order differential equation in $\alpha$
\beq
\label{pendulum}
\alpha''(r)+\frac2r\alpha'(r)-\frac4{r^2}\left(\left(f_2 +\frac12\right)\sin \alpha+(f_1\cos\alpha -1)\right)=0.
\eeq
Performing a logarithmic change of variables $\tau =\log r$, (\ref{pendulum}) assumes the form of the equation for a pendulum
in the presence of a damping term, proportional to the velocity
\beq
\label{pendulum1}
\alpha''(\tau)+\alpha'(\tau)-4\left(\left(f_2 +\frac12\right)\sin \alpha+(f_1\cos\alpha -1)\right)=0.
\eeq
This equation is called the {\it Gribov pendulum}. The presence of sinusoidal functions 
makes the pendulum equation to be highly non-linear: no analytic solution
in closed form is known \cite{Canada:2001ab, Grotowski:1999ay}, but only numerical ones.
The only possibility to analyze such pendulum is to simplify the problem by imposing
particular boundary conditions and approximation. Regardless these analytic difficulties, we 
can illustrate the situation qualitatively:
at each point in the pendulum trajectory, three forces are applied: $4f_1$ and $4f_2+2$ respectively in the longitudinal
and transverse direction, whereas $f_1$ onto determines a perturbation as sketched in Figure \ref{Gribovpendulum}

To simplify further the calculation, we could adopt the pure transverse configuration (\ref{Gripar1}), such that 
the Gribov pendulum equation becomes
\beq
\label{pendulum2}
\alpha''(\tau)+\alpha'(\tau)-4\left(f+\frac12\right)\sin\alpha=0,
\eeq
in which only the force proportional to $4f+2$ is applied on the pendulum. The smooth function $f(e^\tau)$ is necessary
to preserve the regularity of solutions \cite{Sciuto:1977ti, Henyey:1978qd}:
$A_i$ is required to be regular at $r=0$, implying $\alpha(e^\tau)\to_{r\to 0} n\pi +{\it O}(r^2)$
and to go to zero at infinity faster than $1/r$, implying $\alpha(+\infty)-\alpha(-\infty)$ being either $0$ or
$\pm \pi/2$ \cite{Sciuto:1977ti}.
To be more precise, the condition at infinity can be of two types, according to the boundary conditions one
needs to impose: we distinguish between weak (WBC) or strong (SBC) boundary conditions.
Both conditions require $A_i$ to be regular at the origin, such that $f(r)\to_{r\to 0} {\it O}(r)$, but 
while WBC impose $f(r)\to_{r\to \infty}{\rm constant}$, SBC impose $f(r)\to_{r\to \infty}0$.
For more details about the various types of Gribov copies one can obtain according to WBC or SBC, we refer
the reader for \cite{Sobreiro:2005ec}.

To conclude this section, it is worthwhile addressing the solution found by Henyey in \cite{Henyey:1978qd}.
As noted in the previous chapter, a fundamental role in Euclidean Y-M theory is played by instantons: these are
classical solutions to the equations of motion of pure Euclidean Y-M theories which have finite action.
To fully understand their importance, we have to introduce some basic concepts concerning the topology of Euclidean Y-M theory.
The boundary of the four-dimensional Euclidean space-time at infinity ($r\to \infty$) is given by the three-sphere $S_\infty^3$.
The gauge field $A_\mu$, when $r\to \infty$, becomes a pure gauge, i.e. $A_\mu\to_{r\to \infty}=g^\dagger\,\partial_\mu\,g+{\it O}(1/r^2)$,
and the corresponding field-strength tensor vanishes, $F_{\mu\nu}(g^\dagger\,\partial_\mu\,g)=0$.
With such boundary condition for the gauge configuration, it is possible to show \cite{Goddard:1986sj, Naber:2000bp}
the existence of a map between $S_\infty^3$ and $SU(2)$: being the topology of the three-sphere the same of $SU(2)$ (topological equivalent) 
this map can be characterised by the {\it winding number} $\nu$ (also called the Pontryagin number) corresponding to the discrete {\it homotopy}
\footnote{Given two continuous maps from the hypersphere $S^n$ to $M$, $\phi$ and $\varphi$, 
they are said to be homotopic if there exists a map $F(x,t)$,
with $0\leq t\leq 1$, which interpolates continuously between them, namely $F(x,0)=\phi$ and $F(x,1)=\varphi$.
The homotopy between $\phi$ and $\varphi$ is denoted by the symbol $\phi\sim \varphi$. The set of homotopy classes is denoted by $\pi_n(M)$.
When $M=S^n$, the equivalence homotopy classes are labelled by the winding number $\nu$: two maps $\phi, \varphi: S^n\to S^n$
can be continuously deformed into one another iff both maps cover $S^n$ the same number of times as $x$ covers it once. }
$\pi_3(S^3)=Z$. One of the major achievements in the Yang-Mills theory 
was the discovery of the relation between
instanton solutions and their classification by the winding number $\nu$ \cite{Belavin:1975fg, 'tHooft:1974qc, Uhlenbeck:1982zm}.
The relevance of Henyey's work relies on the fact that he was able to explicitly obtain Gribov's copies with vanishing winding number
and which fall off faster than $1/r$ for $r\to \infty$. Starting with a gauge field $A_i=ia_i(r)\sigma_3$, satisfying the
Coulomb condition, and adopting a suitable parametrisation for the $SU(2)$ gauge transformation, he obtained a differential
equation of the second order similar to (\ref{pendulum}). Adopting polar coordinates for the various parameters, he showed
that the function $a$, specifying the gauge field can be put in the form
\beq
a(r,\theta)=\frac1{2r\sin\theta}-\frac1{\sin(2rb\sin\theta)}\left(b+r^2\sin^2\theta\left(b''+\frac4rb'\right)\right),
\eeq
where the function $b$ only depends on the radius and is defined as $b(r)=\frac{k}{(r^2+r_0^2)^{3/2}}$ and $k<\frac{3^{3/2}\pi}{4}r_0^2$. 
As long as such a function $b$ exists \cite{Sciuto:1978vq}, then $a(r,\theta)$ fulfills the boundary and regularity conditions
required in Euclidean Y-M theory, being regular at the origin $r=0$ and decaying at infinity faster than $1/r^2$.

\section{The functional spaces $\mathcal A$ and $\mathcal A/\mathcal G$}

The functional analysis of the Gribov ambiguity leads to examine the configuration space $\mathcal A$, the functional space of all
gauge connections $A_\mu$
\footnote{The gauge connections, as explained in the previous chapter, are the components of a 
smooth matrix-valued differential one-form $\Omega^1(M, \ad P)$.}, 
in more detail. 
This is an infinite-dimensional affine space, on which 
it is possible to fix a point and coordinate axis such that every point in the space can be represented as an $n$-tuple of its coordinates. 
Not only is $\mathcal A$ affine, but also
a Hilbert space: in fact, if we denote by $\Omega^p_k(M, \ad P)$ the $k$-Sobolev completion of $\Omega^p(M, \ad P)$ 
\cite{Uhlenbeck:1982zm,  Adams:1975, Mazja:1980vg, Naber:2000bp} 
the space of smooth sections
of degree $p$,
then $\mathcal A$ assumes the structure of a Hilbert space, as an affine space modelled 
over $\Omega^1(M, \ad P)$.

Since all physically relevant quantities are gauge invariant, the objects of interest are the families of gauge
related connections rather than the connections themselves. For this purpose we denote by $\mathcal G$ the group of all gauge transformations,
whose elements $g\in \Omega^0(M, \ad P)$ in local coordinates determine the group (adjoint) action on $A_\mu$ as
\begin{align}
\label{gauge1}
{}^g\!A_\mu(x)&=g^\dagger(x) A_\mu(x) g(x) +\frac{1}{\rm g}g^\dagger(x) \partial _\mu g(x)\nonumber\\
&=A_\mu(x)+\frac{1}{\rm g}g^\dagger(x) \mathcal D_\mu g(x).
\end{align}
The fundamental question is how we define the connections and the functions belonging to the group of gauge transformations \cite{Grotowski:1999ay}:
Uhlenbeck \cite{Uhlenbeck:1982zn}, considering Sobolev gauge connections $A_\mu\in W^{1,p}(\mathbb B^n, \mathcal G), n\geq p/2$
and gauge transformations with one more weak derivative $g\in W^{2,p}(\mathbb B^n, G)$ on the $n$-dimensional unit ball $\mathbb B^n$, was
able to prove that with such a setting, providing appropriate curvatures $F_A$, ${}^g\!A\in W^{1,p}(\mathbb B^n, \mathcal G)$
belongs to the Coulomb gauge. The restriction to Sobolev norms was also suggested in a very interesting work \cite{Semenov:1982sf}, in which
it was shown under which conditions it was possible to find the absolute minima of the Y-M functional. This issue will be
of great importance when, in the next section, we will discuss about the various Gribov regions and the so-called fundamental modular region (FMR).
On the other side, Dell'Antonio and Zwanziger \cite{Dell'Antonio:1991xt} considered less restrictive conditions on the gauge connections,
provided with a standard $L^2(\mathbb R^n)$ norm, and they proved the existence of gauge copies of $A_\mu$ in the Coulomb gauge on
the non-compact $n$-dimensional Euclidean space $\mathbb R^n$.

Regardless of these subtle distinctions, 
the set of all physically inequivalent connections is determined by the orbit space (a manifold) $\mathfrak A=\mathcal A/\mathcal G$, i.e.
the set of equivalence classes where $A$ and $A'$ are equivalent if there exists a $g \in \mathcal G$ such that $A'={}^g\!A$. 
The high non-triviality of
this Riemannian space
\footnote{A very detailed description of the Riemannian structure of the gauge configuration space in Y-M theory can be found in
\cite{Babelon:1980uj, Daniel:1979ez}.}
is the reason for which we encounter the Gribov ambiguity in non-Abelian gauge theory. Following Singer \cite{Singer:1978dk},
we try now to highlight the fundamental topological obstruction that Gribov discovered in the light of Coulomb gauge.
In the Feynman approach to quantisation of Y-M theory, one would want to make sense of
$\int\mathcal D A\,\{\cdot\}\,e^{-||F_A||^2}/\int\mathcal D A\,e^{-||F_A||^2}$, where $||F_A||^2$ is the Yang-Mills functional
(\ref{action}), and the integrand of the numerator may be constant on orbits of $\mathcal G$. These orbits are expected to have an infinite
measure though and this introduces a difficulty in evaluating the ratio. One then would like to perform the integral over $\mathfrak A$,
but this turns out to be intractable and this is the reason why we choose an arbitrary gauge-fixing condition. This procedure would
be consistent if one would be able to choose in a continuous manner one gauge connection on each orbit.
When changing variables from $\mathcal A$ to $\mathfrak A$ we introduce in the functional integral a Jacobian, which is interpreted
as the integral of a probability measure along the fibers.
Gribov observed that by choosing a Coulomb gauge with appropriate boundary conditions at $\infty$, 
there exist gauge transformed connections belonging to a trivial principal bundle $P$ over $\mathbb R^4$,  
with Lie group $SU(2)$, that intersect the Coulomb hypersurface not only in the vicinity of trivial configurations, as $A_\mu=0$,
but also at a large distance from $0$.
What Singer showed is that, this scenario is not only valid and applicable to the case of Coulomb gauge, but more generally,
if the conditions at $\infty$ amount to studying Riemannian surfaces as $S^4=\mathbb R^4\cup \infty$ (the unit sphere in $\mathbb R^5$),
then topological considerations imply that no gauge exists. Thus, the Gribov ambiguity for the Coulomb gauge will occur
in all the other gauges, and no continuous gauge fixing is possible. 
In practice, the topological obstruction occurs when one tries to invert the projective map $\sigma:\mathcal A\to \mathcal A/\mathcal G$: 
due to (\ref{gauge1}) $\sigma$ is mapping an affine space to a non-affine one, such that any $A\in \mathcal A$ is being mapped
onto the orbit $\mathcal G$ of $A$. Yet, $\sigma^{-1}$ maps back to one $A\in \mathcal A$ each representative of the same orbit without distinction, 
and therefore such a function is not bijective. 
This topological obstruction therefore prevents one from introducing affine coordinates in a global way
\footnote{This problem occurs also in General relativity when one considers diffeomorphisms
of the metric tensor \cite{Naber:2000bp, Nakahara:1990th}.}.

\section{The Gribov regions $C_i$ and the fundamental modular region $\Lambda$}

Following Gribov \cite{Gribov:1977wm}, it is possible to define on $\mathcal A/\mathcal G$ different regions $C_i$, 
according to the number of negative eigenvalues of the the Faddevv-Popov operator. 
To see this, consider the eigenvalue equation for the Faddeev-Popov
operator (\ref{Schro}): for small values of $A_\mu$ it is solvable for small and postive $\epsilon[A]$ only. More precisely, one can show that
for small $A_\mu$, $\epsilon_i[A]>0$. As the gauge field increases its magnitude, one of the eigenvalues turns out to vanish, and then
becoming negative as the field increases further. Therefore the magnitude of $A_\mu$ insures the existence of negative energy states,
i.e. bound states. Supposing to keep going with increasing $A_\mu$, some other eigenvalues will start vanishing and subsequently
changing sign. If we divide the orbit space into regions $C_i$ ($i=0,1,\ldots N$), the {\it Gribov regions}
with $i$ denoting the number of negative
eigenvalues $\epsilon[A]$ for the F-P operator $-\partial^\mu(\partial_\mu +[A_\mu,\cdot])$, we may obtain the following 
schematic picture
\begin{figure}[ht]
\centering \epsfig{file=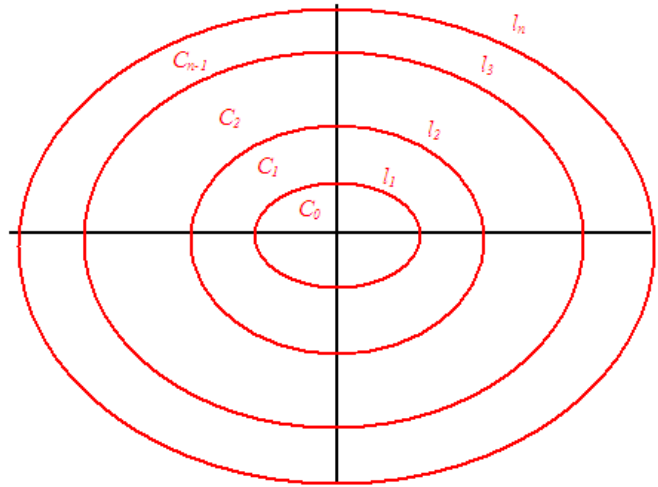,width=10cm} \caption{The Gribov regions $C_i$ and the fundamental modular region $\Lambda$}
\end{figure}
The various lines denoted by $l_i$ correspond to the so-called {\it Gribov horizons}: the label indicates the number of vanishing eigenvalues
of the corresponding F-P operator $\mathcal M$. 
Therefore, when passing from $C_i$ to $C_{i+1}$, one crosses one horizon and the overall sign 
of $\det \mathcal M[A]$ changes. 
Moreover, it is possible to show that for any configuration lying within
the region $C_{i+1}$ close to the boundary $l_{i+1}$ there is an equivalent configuration
within the region $C_i$ close to the same boundary $l_{i+1}$.
It is important to notice that in the first region $C_0$, there is no negative eigenvalue, or put in another
form, the lowest eigenvalue is positive, guaranteeing the condition for the positiveness of the Faddeev-Popov determinant such that
for a given gauge-condition $F[A]$, the zeroth Gribov region is defined as
$C_0=\{A_\mu \in \mathfrak A, F[A]=0, \epsilon_i[A]>0\,|\,-\partial^\mu(\partial_\mu +[A_\mu,\cdot])>0\}$.
As originally suggested by Gribov himself and rephrased in the language of path integrals, 
the Y-M functional integral over gauge-inequivalent configurations
should be restricted to an appropriate region, where the gauge-fixing condition
would be guaranteed unambiguously. Gribov suggested to restrict the integration over $C_0$:
yet, as proved in \cite{Semenov:1982sf}, this region wouldn't necessarily guarantee 
to find unique solutions to the gauge condition for each orbit. Therefore, it is necessary to find a better way
to evaluate the Y-M path integral in such a way that the integration region only selects one single
representative for each gauge orbit. 
It was shown in \cite{Semenov:1982sf} first and then developed in \cite{Dell'Antonio:1991xt}, 
that there is a functional method to determine such a region.
Suppose we define a covariant $L^2$-vector Morse potential 
\footnote{A smooth function $ u:M\to\mathbb{R}$ is called Morse if all its critical points are non-degenerate.
Morse functions exist on any smooth manifold, and in fact form an open dense subset of smooth functions on $M$.}
along the gauge orbit
\beq
V[{}^g\!A]\equiv||{^g A}||^2= -\int_M ~
\tr \left( \left( g^{\dagger} A_\mu g + g^{\dagger} \partial_\mu g \right)^2\right).
\label{gAnorm}
\eeq
Expanding around the minimum of eq.(\ref{gAnorm}), writing $g(x)=\exp(X(x))$,
one easily finds:
\begin{align}
||{^g A}||^2 &= ||A||^2+2\int_M \tr(X
\partial_\mu A_\mu)+\int_M \tr (X^\dagger \mathcal M[A] X) \nonumber \\
&+\frac{1}{3}\int_M\tr\left(X\left[[A_\mu,X],\partial_\mu X\right]\right)
+\frac{1}{12}\int_M\tr\left([D_\mu X,X][\partial_\mu X,X]\right)+{\it O}(X^5).
\label{Xexpan}
\end{align}
At any local minimum the vector potential is therefore transverse,
$\partial_\mu A_\mu=0$, and $\mathcal M[A]$ is a positive operator. The
set of all these vector potentials is by definition the Gribov region
$C_0$. Using the fact that $\mathcal M[A]$ is linear in $A$, $C_0$ is
seen to be a convex subspace 
\footnote{If $A_1$ and $A_2$ are two gauge fields inside $\Omega$, they by definition satisfy the Faddeev-Popov eigenvalue equation,
$-\mat M(A_{1,2})\phi_n = \lambda_n(A_{1,2}) \phi _n$,
such that $\lambda_n >0, \forall n$ ($\mat M \equiv \partial^\mu (\partial_\mu - [A_\mu, \cdot])$). 
To prove convexity it suffices to show that, given a real
parameter $s\in[0,1]$, through which we relate $A_1$ to $A_2$ as
$sA_1+(1-s)A_2 =\bar A$,
the field $\bar A$ always belongs to $\Omega$, regardless $s$ and the particular choice of the two starting fields. 
This is easy to see because, by definition, $A_1$
and $A_2$ belong to $C_0$, and so they have positive eigenvalues. This implies their combination by $s$ is always positive, 
whichever value for the parameter 
we pick up and therefore $\bar A \in \Omega$.
In addition to this convexity, there's another theorem which claims there's always an equivalent field in the second Gribov region, 
$C_1$, for a field $A_\mu$ inside 
$C_0$ and close to the Gribov horizon $l_1$. These two geometric properties of $C_0$ ensure us that the common cartoon which 
displays this region is correct.}
of the set of transverse connections $\Gamma$.
Its boundary $\partial C_0$ is called the Gribov horizon. At the Gribov
horizon, the {\it lowest} eigenvalue of the Faddeev-Popov operator
vanishes, and points on $\partial C_0$ are hence associated with coordinate
singularities. Any point on $\partial C_0$ can be seen to have a finite
distance to the origin of field space and in some cases even
uniform bounds can be derived \cite{Zwanziger:1993dh, Dell'Antonio:1991xt}.
The Gribov region is then defined as the set of {\it local} minima of the
norm functional (\ref{gAnorm})
and needs to be further restricted to the
{\it absolute} minima to form a fundamental domain, which will be denoted
by $\Lambda$. The fundamental domain is clearly contained within the
Gribov region and therefore $\Lambda$ is proven to be convex too. 
We can define $\Lambda$ in terms of the absolute minima 
over $g \in \mathcal G$ of $||{}^g\!A||^2-||A||^2=\langle g, \mathcal M[A]g\rangle$ as
\beq
\Lambda =
\{ A \in {\mat A} | \min_{g \in \mathcal G} \langle g, \mathcal M[A]g\rangle=0 \}.
\label{Lmdef}
\eeq
A different approach in restricting the integration region to $\Lambda$ may come from
stochastic quantisation as explained in \cite{Zwanziger:2003cf}.
A detailed overview of the analytic properties of the Gribov and fundamental modular region
can be found in \cite{vanBaal:1991zw, Zwanziger:1993dh, Zwanziger:2003cf} and references therein.
For the purpose of this thesis we only focus on some elementary properties
of these regions. 
As $\Lambda$ is contained in $C_0$, this means $\Lambda$ is also 
bounded in each direction and has a boundary $\partial \Lambda$. 
Convexity of $\Lambda$ allows us to consider rays extending from
the origin of $\Lambda$, set to $A_\mu=0$ out to $C_0$, crossing
the common boundary, 
such that at some point along the ray, this absolute minimum has to pass
the local minimum. At the point they are exactly degenerate, there are
two gauge equivalent vector potentials with the same norm, both at the 
absolute minimum. As in the interior the norm functional has a unique
minimum, again by continuity, these two degenerate configurations have
to both lie on the boundary of $\Lambda$. This is the generic situation.
If the degeneracy at the boundary is continuous along non-trivial directions
one necessarily has at least one non-trivial zero eigenvalue for $\mathcal M[A]$ and 
the Gribov horizon will touch the boundary of the fundamental domain at these 
so-called singular boundary points.
It is interesting to note in the case of stochastic quantisation in \cite{Zwanziger:2003cf},
it was suggested that in the thermodynamic limit, as the number of configurations tends to increase,
the Y-M functional integral would be dominated by configurations lying on the common boundary of
$\Lambda$ and $C_0$.

The final comment we would like to point out here concerns the practical realisation of such a fundamental
modular region. As the Gribov region is associated with the local minima, and since the space 
of gauge transformations resembles that of a spin model, the analogy with spin 
glasses makes it unreasonable to expect that the Gribov region is free of 
further gauge copies \cite{vanBaal:1991zw}.
Unfortunately restrictions to a subset of the transverse gauge fields is 
a rather non-local procedure. This cannot be avoided since it reflects the 
non-trivial topology of field space \cite{Singer:1978dk}.
Early after the discovery of Gribov of the degeneracy in quantising non-Abelian gauge theories,
within the context of U gauge, in \cite{Balachandran:1978uc} an unambiguous way was proposed 
to select single representatives for each orbit. Though, this gauge fails in being covariant and
hard to put in a close analytic form.
Further proposals of eliminating the Gribov ambiguity are very frequent in literature, for instance see
\cite{Manjavidze:2004qn, Grotowski:1999ay, Zwanziger:2003cf}. Nonetheless, the most rigorous
scenario in which Gribov copies can be consistently and practically avoided is Lattice Gauge Theory (LET),
according to Wilson's procedure \cite{Wilson:1974sk}. In the course of the next chapters we will often deal
with such formalism: for now, it suffices to say that
in LGT, it is well known that due to the discretisation
procedure adopted, no gauge-fixing is required. The Y-M path integral is calculated over an ensemble
of links $U_{x,\mu}$ randomly generated by appropriate Monte Carlo algorithms.
It is then possible to show that the probability to generate two link configurations 
lying on the same gauge orbit is statistically negligible. 
The fundamental difference between continuous and lattice gauge theories lies 
therefore in the fact that in the former
formalism it is possible to simulate numerically the dependence of the gluon and ghost propagators on Gribov
copies \cite{Alexandrou:2000ja, Alexandrou:2001fh, Leinweber:1998im, Bonnet:2000kw, Bowman:2002bm}.
Appearance of Gribov copies are studied in Landau gauge in \cite{Marinari:1991zv, Sharpe:1984vi} whereas,
for instance, a proposal to eliminate the ambiguity of gauge-equivalent configurations can be found
in \cite{Testa:1998az} in the light of a simple toy model using BRST arguments.
It is however a difficult task
to show how the continuous limit of LGT can be obtained, maintaining the theory free of Gribov copies.

\cleardoublepage
\chapter{BRST formalism in Yang-Mills Theory}

Soon after the work of Faddeev and Popov \cite{Faddeev:1967fc}, the attention of the physics community was focused greatly
on the appearance of these fictitious and unphysical particles, called FP ghosts. As Feynman  suggested early on, these particles
were meant to be necessary to guarantee the unitarity of the $\mathcal S$-matrix in non-Abelian gauge theories.
De Witt had also suggested that this breakdown of unitarity was due to missing contributions of a pair of massless scalar (or vector in
the case of the gravitational field) fermions to closed loops in Feynman diagrams.
It was further realised
that the Ward-Takahashi identities for Abelian theories, as well as Slavnov-Taylor identities for the non-Abelian case, 
both indispensable to prove renormalisabilty of the respective theories, should necessarily involve these
unphysical ghosts.
Though ghost particles were thus the missing particles physicists were after, the geometric structure of gauge theory seemed
to be plagued by unphysical modes which do not follow the spin-statistics for fermions.
This chapter will be then entirely dedicated to the BRST formalism, introduced independently 
in \cite{Becchi:1975nq} and \cite{Tyutin:1975ty} in the mid-1970s: this quantisation method will
be analyzed in the light of covariant Y-M theory, firstly with linear gauges, such as Landau gauge
and successively with a more general class of non-linear gauges, such as the Curci-Ferrari gauge. 
We will also present the Kugo-Ojima criterion for selecting the appropriate physical
states.


\section{Faddeev-Popov ghosts and the birth of a new symmetry}

In the last chapter we saw how FP ghosts appeared in the path integral representation of Y-M theory, through the introduction
of two Grassmann fields in order to exponentiate the determinant of the FP operator. 
As we know, the Lagrangian appearing in (\ref{gfYM4}) has lost its local gauge invariance by the
introduction of a gauge-fixing term: it would be nonetheless
desirable to maintain the infinitesimal gauge invariance of the theory. The extension of this symmetry to the case
of finite transformations can be understood heuristically by performing the same transformation many times
as $\Delta A=\lim_{n\to \infty}\delta_n A$ with $\delta A= \mathcal D[A] \theta$.
This infinite repetition of infinitesimal variations can be avoided by introducing a Grassmann parameter in the definition of
$\delta A$ in the following way: in the previous chapter we saw that by the exponential map we can define a local relation between
the group and its algebra as $g(x)=e^{X^a \theta_a(x)}$. Suppose now we introduce in the exponent a parameter $\epsilon$ as
$g(x)=e^{\epsilon\,X^a \theta_a(x)}$ and we expand the exponential in a Taylor series, 
$g(x)=e^{\epsilon\,X^a \theta_a(x)}=\mathbb I+\epsilon\,X^a \theta_a(x)+\frac12(\epsilon\,X^a \theta_a(x))^2+\ldots
\frac1{n!}(\epsilon\,X^a \theta_a(x))^n$.
If we are allowed to take $(\epsilon\,X^a)^2=0$ not as an approximate relation, but as an exact one, then we notice that
the infinitesimal form becomes exact by itself being identical with its finite one.
This constraint mimics the infinitesimal form of the original local gauge invariance, whereas it does not reproduce its finite
form which has been broken by the gauge-fixing procedure. It is well known that in differential geometry, an object which is endowed
with such nilpotency condition is a differential form \cite{Naber:2000bp, Nakahara:1990th}: 
these forms constitute a finite-dimensional Grassmann algebra equipped with exterior
product (see Appendix). This anti-commutating nature underlying classical gauge transformations led in mid 70's
Becchi, Rouet and Stora \cite{Becchi:1975nq} and independently Tyutin \cite{Tyutin:1975ty} to construct
a coherent formalism in covariant non-Abelian theories to solve in more algebraic way  Slavnov-Taylor identities
and to prove the renormalisability of the theory. Moreover, 
the canonical quantisation of Yang-Mills theory and its correct application to the Fock space of instantaneous 
field configurations were elucidated by Kugo and Ojima. Later works by many authors, notably Thomas 
Suchcker and Edward Witten, have clarified the geometric significance of the BRST operator and related fields and 
emphasised its importance to topological quantum field theory and string theory.

The BRST formalism is based on the use of the Faddeev-Popov ghosts to construct a nilpotent operator $\delta$ and its associated 
Noether charge $Q_B$, 
the generator of quantum gauge transformations. Furthermore, the Grassmann nature of $\delta$ identifies it as
a supersymmetric operator and consequently the BRST formalism is considered an example of a superymmetric
theory (SUSY). Another important property of this formalism is its understanding in terms of differential geometry and
fiber bundles. 
In \cite{Bonora:1980pt} it was pointed out how the gauge-fixing procedure by means of FP ghosts
would enlarge the Riemannian structure of the principal bundle inherited by Y-M theory into a supersymmetric space, extended
to include Grassmann degrees of freedom \cite{Delbourgo:1982iq, Delbourgo:1981cm}.  
In \cite{Quiros:1980ec, Hoyos:1981pb, Baulieu:1981sb}, these ideas were confirmed and expanded in the context of superfield formalism 
and covariant quantisation for Y-M theory. 
Topologically speaking, the central idea of the BRST construction is to identify the solutions of the gauge constraints with
the cohomology classes of a certain nilpotent operator, the BRST operator $\delta$ \cite{Kalau:1991tg}, 
generated by a pair of anitcommuting Lorentz scalar fields, the FP ghosts.
 
Following the original works of Becchi, Rouet and Stora \cite{Becchi:1975nq} and Tyutin
\cite{Tyutin:1975ty}, we want to show how the Grassmann structure shows up naturally
in gauge transformations: consider for this purpose the linearly covariant gauge-fixed Lagrangian
\footnote{We start the BRST formalism without the Nakanishi-Lautrup field $b$, also called the {\it on-shell} BRST,
i.e. when we consider the equations of motion for the $b$-field. 
Then we will show how the auxiliary field plays the role of insuring the BRST invariance {\it off-shell}.}
\beq
\label{lag1}
\mathcal L=\mathcal L_{\rm YM}+\mathcal L_{\rm gf}+\mathcal L_{\rm FP}
=\frac14F^a_{\mu\nu}F^{a\mu\nu}-\frac{1}{2\xi}(\partial_\mu A_\mu^a)^2+i\bar c^a\mathcal M_{ab}c^b,
\eeq
and a local infinitesimal gauge transformation
\beq
\label{gaugetrans}
\delta A_\mu^a(x)=\partial_\mu \theta^a(x)-{\rm g}\, f^{abc}A_\mu^c(x)\,\theta^b(x)\equiv \mathcal D_\mu^{ab}[A]\,\theta^b(x).
\eeq
Substituting $\theta^a(x)=c^a(x)$ in (\ref{gaugetrans}), with $c^a$ being a local Grassmann 
field, we obtain
\beq
\label{gaugetrans1}
\delta A_\mu^a(x)=\mathcal D_\mu^{ab}[A]\,c^b(x).
\eeq
It is worthy noting the role played by the ghost field: it replaces the classical $\theta^a$ gauge function to provide the quantum version of
(\ref{gaugetrans}).
Under such a transformation,
it is rather trivial to show the invariance 
of $\mathcal L_{\rm YM}$
under (\ref{gaugetrans1}),
because the ghost field does not affect the original gauge invariance. 
Conversely, the variation  of the Faddeev-Popov Lagrangian yields
\beq
\delta(\mathcal L_{\rm FP})=i\partial^\mu \bar c^a(\mathcal D_\mu^{ab}c^b).
\eeq
So (\ref{lag1}) is not invariant under the local infinitesimal gauge transformation (\ref{gaugetrans1}) with an arbitrary gauge function $\theta^a$.
In \cite{Becchi:1975nq, Tyutin:1975ty} it was proposed to ``gauge'' transform also the two ghosts. For this purpose
consider the following analogy with differential geometry in the case of an infinite-dimensional Lie group:
in Y-M theory we deal with an infinite-dimensional Lie group $\mat G$ of gauge transformations, together with
its Lie algebra $\mathfrak g$. On $\mathfrak g$, we can define a {\it Maurer-Cartan} differential form $\omega$, which
is a left-invariant 1-form, whose functional nature is due to the fact that
$\mathfrak g=\mat C^\infty(M, su(N))$, or a section of the fiber bundle $\Gamma(M\times_{SU(N)}su(N))$.
On this algebra, we can define a coboundary operator $\delta$ (dual of a derivative operator), which acts on
elements of $\mathfrak g$ according to the following anti-derivation rule
\beq
\delta(\varphi_1\wedge \varphi_2)=\delta(\varphi_1)\wedge \varphi_2+(-1)^{{\rm deg}\varphi_1}\varphi_1\wedge \delta(\varphi_2),
\eeq
and satisfies the nilpotency condition
\beq
\label{nilpo}
\delta^2=0.
\eeq
From the last chapter, we know that such a Maurer-Cartan form $\omega$ satisfies $\delta\omega=-\frac12\omega\times \omega=-\frac12[\omega, \omega]$.
By rewriting $\omega(x)= {\rm g} c(x)$, we chose the following on-shell transformations
\begin{align}
\label{gaugetrans2}
\delta c^a(x)&=-  \frac{\rm g}2(c(x)\times c(x))^a=- \,\frac{\rm g}2\,f^{abc}c^b(x)c^c(x)\nonumber\\
\delta \bar c^a(x)&=- \,\frac{1}{\xi}(\partial_\mu A_\mu(x)^a).
\end{align}
Under the transformations (\ref{gaugetrans1}) and (\ref{gaugetrans2}), called the 
{\it BRST transformations}, we can prove that the Lagrangian (\ref{lag1}) is left invariant.
In fact, the variations of the gauge-fixing and Faddeev-Popov read
\footnote{From now on we will not make the space-time dependence explicit in the expression of the various fields.} 
\begin{align}
\label{BRSTvar}
\delta\mathcal L_{\rm gf}&= \frac{1}{\xi}(\partial^\mu A_\mu^a)(\partial^\nu\mathcal D_\nu c^a)\nonumber\\
\delta\mathcal L_{\rm FP}&= -(\delta\bar c^a)\partial^\mu\mathcal D_\mu c^a-\bar c^a\partial^\mu\delta (\mathcal D_\mu c)^a.
\end{align}
Concentrating first on the variation of the covariant derivative we notice that
\begin{align}
\delta (\mathcal D_\mu c)^a&=\delta\left[(\partial_\mu \delta^{ab}-{\rm g}f^{abc}A_\mu^c)c^b\right]\nonumber\\
&= \left[-\frac{\rm g}2\delta^{ab}f^{bmn}\partial_\mu(c^mc^n)-{\rm g}f^{abc}(-\mathcal D_\mu^{cm}c^m)c^b
-{\rm g}f^{abc}A_\mu^c(-\frac{\rm g}2 f^{bmn}c^mc^n)\right].
\end{align}
Expanding the covariant derivative, terms linear in $\rm g$ and ${\rm g}^2$ separately cancel
\beq
- {\rm g}\delta^{ab}f^{bmn}\partial_\mu(c^m)c^n+ {\rm g}\delta^{ab}f^{bmn}\partial_\mu(c^m)c^n=0,
\eeq
and because of the Jacobi identity
\footnote{The Jacobi identity holds for any Lie algebra and its expressed through the Lie brackets as
$[X_a,[X_b, X_c]]+[X_c,[X_a, X_b]]+[X_b,[X_c, X_a]]=0$ or equivalently through the corresponding structure constants $f^{abc}$
$f^{abm}f^{cmn}+f^{cam}f^{bmn}+f^{acm}f^{bmn}=0$.}
\beq
 {\rm g}^2A_\mu^c(f^{abm}f^{cmn}+f^{cam}f^{bmn}+f^{acm}f^{bmn})c^mc^n=0.
\eeq
The terms remaining in (\ref{BRSTvar}) vanish because they can be written as a total space-time derivative
\begin{align}
\delta(\mathcal L_{\rm gf}+\mathcal L_{\rm FP})&=\int_M\left[
 \frac{1}{\xi}(\partial^\mu A_\mu^a)\partial^\nu(\mathcal D_\nu c^a)
+\partial^\mu\left(\frac{ }{\xi}\partial^\nu A_\nu^a\right)(\mathcal D_\mu^{ab}c^b)
\right]\nonumber\\
&=\int_M\partial^\mu\left(\frac1{\xi}(\partial^\nu A_\nu^a)(  \mathcal D_\mu^{ab}c^b)\right)=0.
\end{align}
This completes the proof that the Lagrangian (\ref{lag1}) is invariant under the BRST transformations 
(\ref{gaugetrans1}) and (\ref{gaugetrans2}).
As a further check, we can see if the BRST transformations are nilpotent as required.
Keeping in mind the Jacobi identity, acting twice on $A_\mu^a$ we obtain
\beq
\delta^2A_\mu^a=(\mat D_\mu(s c+\frac{\rm g}2c\times c))^a=0.
\eeq
For $c^a$, we get
\beq
\delta^2c^a=\frac{\rm g^2}6(f^{bce}f^{eda}+f^{cde}f^{eba}+f^{dbe}f^{eca})c^bc^cc^d=0.
\eeq
Yet, when applying $\delta$ twice on $\bar c^a$, we notice an inconsistency, because we obtain 
$\delta^2\bar c^a=-\frac{\rm g}{\xi}\partial^\mu(\mat D_\mu c)^a\neq 0$. It is this problem which forces us to introduce
here the $b$-field, in order to guarantee such nilpotency condition (\ref{nilpo}).
We then change (\ref{gaugetrans2}) as follows
\begin{align}
\label{gaugetrans3}
\delta c^a&=- \,\frac{\rm g}2\,f^{abc}c^bc^c\nonumber\\
\delta \bar c^a&= b^a\nonumber\\
\delta b^a &=0,
\end{align}
and it is rather trivial to prove the nilpontency on $\bar c$ and $b$. These transformations 
are called off-shell BRST transformations.
\footnote{They are called off-shell because we do not use the equations of motion of the $b$-field. Hence,
the BRST transformations (\ref{gaugetrans2}) are called on-shell.}
This all shows that the FP ghosts are to be interpreted as components of Maurer-Cartan 1-forms, as well as the gauge field $A_\mu$
\cite{Bonora:1980pt, Baulieu:1981sb}. The anticommuniting properties of the ghosts therefore are consequences
of their differential-form nature, forming a Grassmann algebra of left-invariant forms on $\mathfrak g$ and represent
all infinitesimal local gauge transformations in $\mat G$ in a generic way. With $c$ and $\bar c$ 
not identified with any particular $g\in \mat G$,
the BRST invariance of (\ref{lag1}) can be then regarded as the lost gauge invariance under infinitesimal local
gauge transformations. It must be stressed that in the case of linear covariant gauges, such as Landau gauge for instance,
the Lagrangian (\ref{lag1}) is not invariant under the interchange of ghosts into anti-ghosts and vice-versa. 
This symmetry is only
generated in the case of non-linear gauges, such as the Curci-Ferrari gauge \cite{Curci:1976ar}.

To conclude this section, we wish to make a remark
on the Hermiticity properties of the ghosts: if we demand the requirement for the
Lagrangian density to be Hermitian and for the $\mat S$-matrix to be (pseudo-)unitary
\beq
\mat L^\dagger=\mat L, \qquad \mat S^\dagger \mat S=\mat S\mat S^\dagger=\mathbb I,
\eeq
the only permissible choice for the ghosts \cite{Nakanishi:1990qm} is for them to be both Hermitian as
\beq
c^{a^\dagger}=c^a\qquad \bar c^{a^\dagger}=\bar c^a,
\eeq
and hence the factor $i$ in front of $\bar c^a\mathcal M_{ab}c^b$ is necessary. 
If we had adopted the wrong hermiticity assignment
\beq
c^{a^\dagger}=i\bar c^a\qquad \bar c^{a^\dagger}=i c^a,
\eeq
then not only would the hermiticity of the Lagrangian density (\ref{lag1}) be violated
\beq
\mat L^\dagger-\mat L=i{\rm g}\partial^\mu A_\mu^c\,f^{abc}c^a\bar c^b-i{\rm g}\partial^\mu(A_\mu^cf^{abc}c^a\bar c^b)\neq 0,
\eeq
but also it would affect the hermiticity of the BRST and Faddeev-Popov charge operators.


\section{BRST Noether's charges and algebra}

According to the Noether theorem, whenever there is a continuous symmetry in the theory, there must be a conserved current $j^\mu$, whose
associated charge is generated by the space integral of the current's temporal component. 
Making use of the Euler-Lagrange equations for (\ref{lag1}), the conserved BRST Noether charge is then
\begin{align}
j_{\mu}&=\sum_{\{\Phi\}}\frac{\partial \mat L}{\partial (\partial_\mu \Phi)}\delta\Phi\nonumber\\
&=b^a\mat (D_\mu c)^a -\partial_\mu b^a c^a+i\frac12{\rm g}f^{abc}\partial_\mu \bar c^a c^bc^c,
\end{align}
with $\{\Phi\}$ the set of all fields present in the Lagrangian 
\footnote{To be precise, we generate a conserved current up to a total divergence $\partial^\mu(F_{\mu\nu}^a c^a)$,
which should vanish provided appropriate boundary conditions, {\it unless} such a term does not generate a
massless bound-states spectrum.}.
The BRST Noether current is consequently
\beq
Q_B=\int{\rm d}\vec x \left(b^a\mat (D_0 c)^a -\dot b^a c^a+i\frac12{\rm g}f^{abc}\dot{\bar c}^a c^bc^c\right).
\eeq
Under the hermiticity properties of the ghost fields we assigned in the last section, we then check the hermiticity
of $Q_B$, $Q_B=Q_B^\dagger$, which implies that such a charge operator has real eigenvalues.
The BRST transformations (\ref{gaugetrans1}) and (\ref{gaugetrans3}) can be put in the form of BRST commutators, i.e.
as Lie derivatives of fields w.r.t. the current $Q_B$ as
\begin{align}
\delta A_\mu^a&=[iQ_B, A_\mu^a]=(\mat D_\mu c)^a\nonumber\\
\delta c^a&=[iQ_B, c^a]_{+}=-\frac{\rm g}2(c\times c)^a\nonumber\\
\delta \bar c^a&=[iQ_B, \bar c^a]_{+}=b^a\nonumber\\
\delta b^a&=[iQ_B, b^a]=0,
\end{align}
with $+$ indicating the anti-commutator. The reason for this lies at the very heart of the BRST formalism, due to its supersymmetric
nature, and hence all the operations must be understood to be Grassmann graded.
\footnote{Grassmann grading defines the {\it even} or {\it odd} character of a field under product exchange.
This rule is also applied on functional derivatives w.r.t. ghost fields: it is custom to define left (L) and right (R)
derivatives as follows $
\frac{\delta}{\delta \bar c^a}\equiv\frac{\delta^L}{\delta \bar c^a}\qquad
\frac{\delta}{\delta c^a}\equiv\frac{\delta^R}{\delta c^a}.
$}
The fundamental difference between the BRST symmetry (and its charge operator in particular) 
and the underlying infinitesimal local gauge invariance stands in the {\it global} nature of the former.
This property  allows us to interpret the BRST procedure as a topological operation.

Another conserved current emerging from this formalism is the so-called Faddeev-Popov current, which interchanges 
ghosts in anti-ghosts and vice-versa. To it, we associate a FP ghost number, resembling of the fermion number,
which is a conserved quantity too. Unlike the usual case of fermion number conservation, however, the FP ghost number
is not due to the invariance under a phase shift in the ghost fields, because this would lead to an incompatibility
with the hermiticity requirements for both $c$ and $\bar c$. Instead, the conservation of this new quantum number
is due to an invariance under a scale shift as
\beq
c^a\to e^\alpha\,c^a\qquad \qquad \bar c^a\to e^{-\alpha}\bar c^a, \qquad \alpha \in \mathbb R,
\eeq
The action of the FP charge operator $Q_C$ on the Nakanishi-Lautrup field and on the gauge connection is trivial.
The corresponding conserved Noether current reads then
\beq
J_{C\mu}=i(\bar c^a(\mat D_\mu c)^a -\partial_\mu\bar c^a c^a),
\eeq
which generates the conserved charge
\beq
Q_C=i\int{\rm d}\vec x(\bar c^a(\mat D_0 c)^a-\dot{\bar c}^ac^a)=Q_C^\dagger.
\eeq
In terms of BRST brackets, we get the following variations
\begin{align}
[iQ_C, c^a]&=c^a\nonumber\\
[iQ_C, \bar c^a]&=- \bar c^a,
\end{align}
with a minus sign to preserve the hermiticity of $i\bar c^a (\mat D_\mu c^a)$ under the action of $Q_C$.
Being hermitian, the FP charge operator, as well as the BRST one, has real eigenvalues: though, the FP ghost number $N_{FP}$
requires to be identified with the eigenvalue of $Q_C$ multiplied by a factor $i$ to be consistent with the existence 
of an indefinite-metric Hilbert space. These pure imaginary eigenvalues come in pairs with their complex conjugate,
providing the norm-cancellation necessary to isolate unphysical ghost modes with negative norms \cite{Kugo:1979gm, Nakanishi:1990qm}. 
Together with the BRST charge $Q_B$, they form the {\it BRST algebra}, which is a simple example of a superalgebra
\footnote{A Lie superalgebra is a generalisation of a classical Lie algebra to include a $\mathbb Z_2$-grading. 
Lie superalgebras are important in theoretical physics where they are used to describe the 
mathematics of supersymmetry. In most of these theories, the even elements of the superalgebra 
correspond to bosons and odd elements to fermions (but this is not always true; for example, the BRST supersymmetry is the other way around).}
\begin{align}
\label{brst1}
[Q_B, Q_B]_{+}&=(Q_B)^2=0\nonumber\\
[iQ_C, Q_B]&=Q_B\nonumber\\
[Q_C, Q_C]&=0.
\end{align}
This algebra should correspond to the superalgebra extension of the Lorentz group $SO(1,1)$, which is a non-compact Lie group,
whose only generator is a boost. BRST algebra in fact can be regarded as a Lie superalgebra whose even part is zero-dimensional
and whose odd-part is one-dimensional. The superalgebra structure of the BRST algebra will become more manifest
and complex  when we will introduce the anti-BRST operator $\bar\delta$.
Notice the presence of the factor $i$ in the second line of (\ref{brst1}) which associates the FP charge with the correct FP ghost number and
which leads to the fact that the FP charge behaves as a bosonic operator, hence the use of an even-graded commutator.
A remarkable aspect of these two charges is that they can be connected via a BRST variation
\begin{align}
Q_B&=\int {\rm d}\vec xJ_{B0}=-\int {\rm d}\vec x(\delta J_{C0}+\partial^i(F_{0i}^ac^a))\nonumber\\
&=-\delta Q_C=-[iQ_B, Q_C].
\end{align}
Therefore, the coboundary operator $\delta$ generates the BRST charge operator as a quantum gauge transformation on the FP charge.
This property of an object to be equal to the BRST variation of an other one is called {\it exactness}, states that are annihilated
by $Q_B$ are called {\it closed}.
To the reader familiar with differential geometry this terminology is reminescent of the De Rham cohomology: in classical differential geometry
the set of smooth, differential k-forms on any smooth manifold M forms an Abelian group (a real vector space) called
$\Omega^k(M)$. The exterior derivative ``${\rm d}$'' maps ${\rm d}:\Omega^k(M) \to \Omega^{k+1}(M)$.
The use of De Rham cohomology is to classify the different types of closed forms on a manifold. 
One performs this classification by saying that two closed forms $\alpha$ and $\beta$ in $\Omega^k(M)$ are cohomologous if they differ 
by an exact form, that is, if $\alpha - \beta$ is exact. This classification induces an equivalence relation on the space 
of closed forms in $\Omega^k(M)$. One then defines the $k$-th de Rham cohomology group $H^k_{\rm dR}(M)$
to be the set of equivalence classes, that is, the set of closed forms in $\Omega^k(M)$ modulo the exact forms.
In the BRST formalism one then wishes to generalise such an argument to the case of infinite-dimensional Lie algebra-valued
differential forms, by replacing $d$ with $\delta$, whose BRST De Rham cohomology 
(or simply BRST cohomology)
becomes $H_\delta(M\times_{\mat G} \mathfrak g)={\rm Ker}\, \delta/{\rm Im}\, \delta$.
In \cite{Kalau:1991tg} it was pointed out that to prove the consistency of the BRST quantisation procedure,
the BRST cohomology has to define physical states. For this purpose the authors studied the use of harmonic gauge fixing procedure
in the context of indefinite-metric Hilbert spaces. These concepts will be discussed within the Kugo-Ojima criterion.

Finally, consider the Lagrangian
\beq
\label{lag2}
\mathcal L=\mathcal L_{\rm YM}+\mathcal L_{\rm gf}+\mathcal L_{\rm FP}
=\frac14F^a_{\mu\nu}F^{a\mu\nu}+\frac{\xi}{2}(b^a)^2+ib^a(\partial_\mu A_\mu^a)+i\bar c^a\mathcal M_{ab}c^b.
\eeq
It is very important to notice that, according to the BRST transformations (\ref{gaugetrans3}), 
$\mathcal L_{\rm gf}+\mathcal L_{\rm FP}$ can be written in terms of a total BRST variation
as
\begin{align}
\mathcal L_{\rm gf}+\mathcal L_{\rm FP}&=\frac{\xi}{2}(b^a)^2+ib^a(\partial_\mu A_\mu^a)+i\bar c^a\mathcal M_{ab}c^b\nonumber\\
&=i\delta\left((\partial^\mu \bar c^a) A_\mu^a-\frac\xi2\bar c^a b^a\right).
\end{align}
Therefore, being the artificial Lagrangians  $\mathcal L_{\rm gf}+\mathcal L_{\rm FP}$ 
appearing in (\ref{lag2}) a BRST-cobaundary term (BRST exact), they do not contribute to the overall invariance of
$\mat L$ under the action of $\delta$, due to its nilpotency.
This is a consequence (and a confirmation) of the local gauge invariance.
The coboundary term can also be cast in a more general form as
\begin{align}
\mathcal L_{\rm gf}+\mathcal L_{\rm FP}=
i\delta(\bar c^a( F^a(A)-\frac\xi2b^a)),
\end{align}
with $F^a[A]$ being a general covariant gauge fixing.


\section{Kugo-Ojima criterion and Slavnov-Taylor identities}

In covariant gauge theories, negative norm states
appear naturally, and the Hilbert space of configurations $\mat V$ has consequently an indefinite metric. As in Abelian theory,
it is required to select a subspace of $\mat V$ such that, inside this Fock subspace $\mat V_{\rm phys}\subseteq \mat V$,
time invariance and norm-positivity are being guaranteed. In Abelian theory, the condition on the
positive-frequency $b$-modes $b^{(+)}|{\rm phys}\rangle=0$
satisfies such a requirement. Yet, in Y-M theory, due to the non linearity of the gauge connection $A_\mu$,
the same condition on the $b$-field cannot select straightforwardly the physical subspace $\mat V_{\rm phys}$. Thanks to Kugo and Ojima
\cite{Kugo:1979gm}, we can impose a subsidiary condition such to overpass this topological obstruction
\beq
\label{kugo}
Q_B|{\rm phys}\rangle=0 \qquad \mat V_{\rm phys}\equiv \{|\Phi\rangle;\quad Q_B|\Phi\rangle=0\}.
\eeq
The condition (\ref{kugo}), known in the literature as the Kugo-Ojima criterion, 
describes the gauge invariance of all the physical states belonging to $\mat V_{\rm phys}$
as the conserved charge associated to the $b$-field in Abelian theories represents the generator of local
gauge transformations, the BRST charge represents the generator of quantum gauge transformations. It is possible 
to show that the condition (\ref{kugo}) reduces to the Gupta-Bleuler condition for QED \cite{Nakanishi:1990qm}.  
The importance of this criterion for the selection of physical states can be also seen in the calculation of Slavnov-Taylor identities.
Due to the underlying BRST symmetry of the Lagrangian (\ref{lag2}), the corresponding Slavnov-Taylor identities
of Y-M theory are now derived from a more general argument than from local gauge invariance.
We know that these identities are indispensable to guarantee the renormalizability of the theory: though the local gauge
invariance has been broken through the gauge-fixing procedure, the BRST formalism, as we saw, provides not only the quantum version
of it, but detemines a new global symmetry. Originally, the ST identities were derived in a very complicated diagrammatic
way \cite{'tHooft:1972fi}. To translate these in the language of the BRST formalism, we assign
the condition for the vacuum to be BRST closed
\beq
Q_B|0\rangle=0,
\eeq
as long as there is {\it no} dynamical symmetry breaking. It then follows that for any physical quantity $\mat O$
\beq
\label{st}
\langle 0 |\delta \mat O|0\rangle=\langle 0 |[iQ_B, \mat O]_{\pm}|0\rangle=0.
\eeq
The subsidiary condition (\ref{kugo}) then allows us to generalise (\ref{st}) in order to include the physical states
belonging to $\mat V_{\rm phys}$ as
\beq
\langle n |\delta \mat O|m\rangle=\langle n |[iQ_B, \mat O]_{\pm}|m\rangle=0 \qquad \{|m\rangle, |n\rangle \in \mat V_{\rm phys}\}.
\eeq
The ST identities follows from defining a generating function $\Gamma$, being the effective action of the theory
of one-particle-irreducible vertices, defined through the Legendre transformation
\begin{align}
\mat S[J,K]&\equiv\int_M\left(J^{\mu a}A_\mu^a+J_c^a c^a +J_{\bar c}^a \bar c^a +J_b^ab^a\right. \nonumber\\
&\qquad\quad\left.+K^{\mu a}(\mat D_\mu c)^a-\frac{\rm g}2 K_c^a (c \times c)^a -K_{\bar c}b^a\bar c^a )\right)\nonumber\\
\exp(i W[J,K])&\equiv \langle 0| \mat T \exp(i\mat S[J,K])|0\rangle\nonumber\\
\Gamma[\Phi, K]&\equiv W[J,K]- J_i \Phi_i,
\end{align}
with $\Phi$ being $\Phi\equiv (\delta/\delta J_i)_{\pm} W[J,K]$, where $\pm$ reminds us of the correct Grassmann grading.
The main difference with standard ST identities is easily
appreciated by the new sources $K$ associated to the BRST variations (\ref{gaugetrans1}). Due to the vanishing variation for $b$,
$K_b$ does not enter $\Gamma$.
In short-notation, the ST identities then read
\beq
\label{st1}
\frac{\delta \Gamma}{\delta A_\mu^a}\frac{\delta \Gamma}{\delta K^{\mu a}}+
\frac{\delta \Gamma}{\delta c^a}\frac{\delta \Gamma}{\delta K_c^a}+
\frac{\delta \Gamma}{\delta \bar c^a}\frac{\delta \Gamma}{\delta K_{\bar c}^a}=0.
\eeq
In the case of non-linear gauges (as well as for a more symmetric form of (\ref{gaugetrans1}) as far as the $b$-field goes), we will see that 
(\ref{st1}) will also incorporate the $b$-field term.


\section{Another BRST operator}

If we take a closer look at the Lagrangian (\ref{lag2}), we notice that FP ghosts and anti-ghosts
do not play a symmetric role. Though FP ghosts are interpreted as Maurer-Cartan 1-forms,
and the operator $\delta$ is recognised as the generator of translations in the $c$-direction,
as first pointed out in \cite{Thierry-Mieg:1979kh}, we do not have at this stage an analogous interpretation
for the FP anti-ghosts. These fields are being introduced in the BRST formalism as Lagrange multipliers
for keeping the gauge-fixing condition unchanged under the BRST transformations (\ref{gaugetrans2}).

The attempt to discover an appropriate and coherent interpretation also for the anti-ghost fields
can be traced back to late 70's and early 80's.
As noted first in \cite{Thierry-Mieg:1979kh}, the anti-commuting nature of the ghost fields was associated to elements
(Maurer-Cartan 1-form of connection)
embedded in an extended principal fiber bundle.
Yet, it was realised in \cite{Quiros:1980ec} and \cite{Bonora:1980pt}
that the principal bundle needed to be enlarged to correctly correlate the classical gauge and
the new quantum global symmetry.
In particular, in \cite{Quiros:1980ec}, it was proposed, starting off the classical Maurer-Cartan 1-form for gauge connections,
how to write down the appropriate Maurer-Cartan 1-form to include ghosts and anti-ghosts in an extended principal bundle.
On the other side, in \cite{Bonora:1980pt} the idea of interpreting
the BRST transformations and charges as proper supersymmetric quantities was expressed in terms
of superprincipal bundles and superfields \cite{DeWitt:1992cy}.
However it is due to Ojima \cite{Ojima:1980da} who discovered another global symmetry in the context of the BRST formalism:
this new symmetry, called {\it anti-BRST}, behaves as the ``almost'' mirror image
\footnote{We call the anti-BRST symmetry ``almost'' mirror image of the standard one because, as we will see later, the
$b$-field breaks such symmetry. Moreover, this field is also responsible to break the superalgebra $osp(4|2)$
as discovered by Thierry-Mieg \cite{Thierry-Mieg:1979kh}.}
of the standard BRST one.
The purpose of this new operator $\bar \delta$ (and its associated charge $\overline{Q}_B$) 
is to make the geometry in the extended
ghost-space more symmetric. If we interpret $\delta$ as the generator of translations in the $c$-direction, then it would seem
appropriate, if not necessary, to construct an analogous operator for translations in the $\bar c$-direction.

For this purpose, consider the following operator identity
\beq
\partial_\mu \mat D^\mu-\mat D^\mu\partial_\mu={\rm g}[\partial_\mu A^\mu, \cdot].
\eeq
Only in the special case of Landau gauge ($\xi=0$)
\footnote{In this gauge, in fact, $\lim_{\xi\to 0}\int [{\rm d}b]\,e^{-\int_M ib^a\partial^\mu A_\mu^a 
+\frac\xi2(b^a)^2}\propto \delta(\partial^\mu A_\mu^a)$.}, 
this identity vanishes. In such a gauge, it is possible to show
that the Lagrangian (\ref{lag2}) remains invariant under the FP conjugation operator $\mat C_{\rm FP}$
\begin{align}
\mat C_{\rm FP}A_\mu^a&=A_\mu^a \nonumber\\
\mat C_{\rm FP}b^a&=b^a-i{\rm g}(\bar c\times c)^a \nonumber\\
\mat C_{\rm FP}c^a&=\bar c^a \nonumber\\
\mat C_{\rm FP}\bar c^a&=-c^a.
\end{align}
The apparent strange transformation of $b$ under $\mat C_{\rm FP}$ is necessary to cancel the term coming from 
$\mat C_{\rm FP}(\mat L_{\rm FP})$. Following \cite{Curci:1976ar, Ojima:1980da}, combining $\delta$
with $\mat C_{\rm FP}$, we can construct a new BRST operation $\bar \delta$ as
\beq
\bar \delta =\mat C_{\rm FP}\,\delta\,\mat C_{\rm FP}^{-1},
\eeq
such that the BRST and anti-BRST operators transform covariantly under FP conjugation.
Applying this identity to the BRST transformations (\ref{gaugetrans3}) we generate
the {\it anti-BRST transformations}
\begin{align}
\label{gaugetrans4}
\bar \delta A_\mu^a&=(\mat D_\mu \bar c)^a\nonumber\\
\bar \delta \bar c^a&=- \,\frac{\rm g}2(\bar c\times \bar c)^a\nonumber\\
\bar \delta c^a&= -b^a-{\rm g}(\bar c\times c)^a\nonumber\\
\bar \delta b^a &=-\frac{\rm g}2(\bar c\times b)^a.
\end{align}
It has to be pointed out that, though only in Landau gauge $\mat C_{\rm FP}$ is unbroken, the invariance the Lagrangian
under (\ref{gaugetrans4}) is preserved even for $\xi \neq 0$. Due to the nilpotency of $\bar \delta$, $\bar \delta^2=0$,
and to the following operator identity
\beq
\delta\bar \delta+\bar \delta\delta=0,
\eeq
then not only the gauge-fixing but also FP Lagrangian can be written as a total BRST variation and even more importantly as
\begin{align}
\label{variation1}
\mat L_{\rm gf}+\mat L_{\rm FP}&=i\delta(\partial^\mu \bar c^a A_\mu^a-\frac\xi2\bar c^a b^a)\nonumber\\
&=-i\bar \delta(\partial^\mu c^a A_\mu^a-\frac\xi2c^a b^a)\nonumber\\
&=\frac i2\delta \bar\delta(A^{a\mu} A^a_\mu)+i\frac\xi2\bar \delta(b^a c^a).
\end{align}
We then see how the gauge-fixing and FP Lagrangians can also be expressed as boundary terms of $\bar \delta$, though,
in the attempt to write these Lagrangians as a whole BRST--anti-BRST variation, we fail due to the presence of
the term $i\frac\xi2\bar \delta(b^a c^a)$.
Demanding the vacuum be left invariant under the action of $\overline{Q}_B$
\beq
\overline{Q}_B|0\rangle=0,
\eeq
we can generate ST identities on the same line of the previous section. Furthermore, the Kugo-Ojima criterion,
due to the anti-commutativity between $Q_B$ and $\overline{Q}_B$, becomes in terms of the anti-BRST charge
\beq
\overline{Q}_B|{\rm phys}\rangle=0.
\eeq
One difficulty in this formalism is to make (\ref{gaugetrans3}) and (\ref{gaugetrans4}) look more symmetric.
For this purpose, in \cite{Nakanishi:1990qm}, a new Nakanishi-Lautrup field is being introduced:
by demanding $\bar b^a= -b^a-{\rm g}(\bar c\times c)^a$, then we have
\beq
\bar \delta c^a=\bar b^a \qquad\bar \delta \bar b^a=0, 
\eeq
with FP conjugation
\beq
\mat C_{\rm FP} b^a=-\bar b^a \qquad \mat C_{\rm FP} \bar b^a=-b^a.
\eeq
In an interesting paper in 1982 \cite{Thierry-Mieg:1982un} it has been shown that the most general form
of Lorentz invariant renormalizable Lagrangian density can be written as
\begin{align}
\label{lag4}
\tilde{\mat L}&= \mat L+\frac\beta2\bar b^a\bar b^a\nonumber\\
&=\mat L_{\rm YM}+i\delta(\partial^\mu \bar c^a A_\mu ^a +\frac\xi2 b^a\bar c^a+\frac\beta2 \bar b^a\bar c^a)\nonumber\\
&=\mat L_{\rm YM}-i\bar\delta(\partial^\mu c^a A_\mu ^a +\frac\xi2 b^a c^a+\frac\beta2 \bar b^a c^a).
\end{align}
This new parameter $\beta$ will be of great importance when we will treat the massive Curci-Ferrari gauge and the invariance
of the Lagrangian under the related massive BRST transformations.


\section{BRST superalgebra for linear and non-linear gauges}

This abundance of new operators and charges may look awkward: it is therefore a considerable advantage to manage
these in a short-form, according to the following {\it double BRST algebra}
\begin{align}
i\{Q_B, \overline{Q}_B\}&=\bar \delta Q_B=\delta \overline{Q}_B=0\nonumber\\
{}[i Q_C, Q_B]&=\delta Q_C=Q_B\nonumber\\
{}[i Q_C, \overline{Q}_B]&=\bar \delta Q_C=-\overline{Q}_B.
\end{align}
In a series of papers \cite{Schaden:1999ew, Schaden:2001xu, Dudal:2002ye} and references therein, following the
work in \cite{Nakanishi:1980dc} it has been addressed that this double BRST algebra actually hides a more general
algebra. We define the operators $\delta_{cc}$ and $\delta_{\bar c\bar c}$ as
\begin{align}
\delta_{cc}\bar c^a&=c^a &\delta_{\bar c\bar c}c^a&=\bar c^a\nonumber\\
\delta_{cc} b^a&=\frac{\rm g}2(c\times c)^a&\delta_{\bar c\bar c}b^a&=\frac{\rm g}2(\bar c\times \bar c)^a\nonumber\\
\delta_{cc} A_\mu^a&=\delta_{cc}c^a=0& \delta_{\bar c\bar c}A_\mu^a&=\delta_{\bar c\bar c}\bar c^a=0,
\end{align}
and their conserved Hermitian charges are respectively $Q_{cc}$ and $Q_{\bar c\bar c}$.
Together with the Faddeev-Popov ghost number charge $Q _{c}$, $%
Q_{cc} $ and $Q_{\bar c\bar c}$ generate an $sl(2,R)$ algebra
\footnote{The Lie group $SL(n,\mathbb R)$ is the special linear group of real matrices with unit determinant.
Its corresponding Lie algebra $sl(n,\mathbb R)$ has an irreducible representation by square matrices with null trace.
It is important to notice that $SL(2, \mathbb R)$ is the set 
of orientation-preserving isometries of the Poincaré half-plane $SO(2,1)$, isomorphic to $SU(1,1)$ and $Sp(2,R)$ and
also it is a non-compact group since its universal cover has no finite-dimensional representations.}. 
This algebra
is a subalgebra of the algebra generated by $Q _{c}$, $Q_{cc}$, 
$Q_{\bar c\bar c}$ and the BRST and anti-BRST charges $Q_B$ and $\overline{Q}_B$. 
The algebra 
\begin{align}
\label{brst3}
Q_B^2&=0& \overline{Q}_B^2&=0\nonumber\\
\{Q_B, \,\overline{Q}_B\}&=0&[iQ_{cc}/2, \,Q_{\bar c\bar c}/2]&=-Q_c\nonumber\\
{}[iQ_c/2, \,Q_{cc}/2]&= Q_{cc}/2&[iQ_c/2, \,Q_{\bar c\bar c}/2]&=-Q_{\bar c\bar c}/2\nonumber\\
{}[iQ_c/2, \,Q_B]&= Q_{B}/2&[iQ_c/2, \,\overline{Q}_B]&=-\overline{Q}_B/2\nonumber\\
{}[iQ_{cc}/2, Q_B]&=0&[iQ_{cc}/2, \overline{Q}_B]&=-Q_B\nonumber\\
{}[iQ_{\bar c\bar c}/2, Q_B]&=\overline{Q}_B&[iQ_{\bar c\bar c}/2, \overline{Q}_B]&=0,
\end{align}
is known as the Nakanishi-Ojima (NO) algebra \cite{Nakanishi:1980dc, Dudal:2002ye}.
The remarkable aspect of this algebra, composed by these five BRST charges, is that it constitutes
the contracted superalgebra extension of the Lie algebra of 3-dimensional Lorentz group \cite{Nakanishi:1990qm}\cite{Frappat:1996pb}\cite{DeWitt:1992cy},
whose representation is $osp(1,2)$. This orthosymplectic superalgebra is denoted in the literature \cite{Frappat:1996pb} as
$B(m,n)$ or generally $osp(2m+1, 2n)$: it is defined for $m\geq 0$ and $n\geq 1$, has as its even part the Lie algebra $so(2m+1)\oplus sp(2n)$
and as its odd part $(2m+1, 2n)$ representation of the even part. It has rank $m+n$ and dimension $2(m+n)^2+m+3n$.

The BRST formalism so far has been presented in the context of a linear covariant gauge: however, as Baulieu and Thierry-Mieg showed in
\cite{Thierry-Mieg:1982un}, to achieve a more general scenario, in which ST identities are still preserved (and therefore the renormalizability
of the theory), we must incorporate into the gauge-fixing Lagrangian a quartic ghost interaction. This can be obtained in different ways.
For instance, consider the Lagrangian (\ref{lag4}): the quadratic interaction $\bar b^a\bar b^a=(b-i{\rm g}\bar c\times c)^2$ already
contains such a desired quartic ghost interaction. Furthermore, from the algebraic point of view, as demonstrated in
\cite{Thierry-Mieg:1979kh}, if we perform a shift of the $b$-field as
\beq
b^a\to b^a+\frac{\rm g}2(\bar c\times c)^a,
\eeq
then the BRST and anti-BRST transformations become
\begin{align}
\label{gaugetrans5}
\delta A_\mu^a&=(\mat D_\mu c)^a &\bar \delta A_\mu^a&=(\mat D_\mu \bar c)^a\nonumber\\
\delta c^a&=- \,\frac{\rm g}2(c\times c)^a &\bar \delta \bar c^a&=- \,\frac{\rm g}2(\bar c\times \bar c)^a\nonumber\\
\delta \bar c^a&= b^a-\frac{\rm g}2(\bar c\times c)^a&\bar \delta c^a&= -b^a-\frac{\rm g}2(\bar c\times c)^a\nonumber\\
\delta b^a &=-\frac{\rm g}2(c\times b)^a-\frac{\rm g^2}8((c\times c)\times \bar c)^a&
\bar \delta b^a &=-\frac{\rm g}2(\bar c\times b)^a+\frac{\rm g^2}8((\bar c\times \bar c)\times c)^a.
\end{align}
Though these new BRST transformations appear more complicated, the striking advantage comes from the observation that
$\mat L_{\rm gf}$ and $\mat L_{\rm FP}$ are now expressed as a proper total double BRST variation
\begin{align}
\label{variation2}
\mat L_{\rm gf}+\mat L_{\rm FP}&=\frac i2\delta \bar\delta(A^\mu A_\mu-i\xi\bar c^ac^a)\nonumber\\
&=i b^a\partial^\mu A_\mu^a+\frac\xi2(b^a)^2+i\frac12\bar c^a\,\mat M^{ab}[A]\,c^b+\frac{\rm g^2}8\xi(\bar c\times c)^2,
\end{align}
where the term $\bar \delta(b^a c^a)$ in (\ref{variation1}) has been cancelled by the shift on $b$. 
In fact, due to the triple FP ghost terms in $\delta b$ and $\bar \delta b$,
the action of $\delta \bar \delta (\bar cc)$ produces
\begin{align}
i\frac12\delta \bar \delta (-i\xi\bar c^ac^a)&=\frac12\delta(\bar c^a b^a)\nonumber\\
&=\frac{\xi}2(b^a)^2+\frac{\rm g^2}8\xi(\bar c\times c)^2\nonumber\\
&=\frac{\xi}2(b^a)^2+\frac{\rm g^2}8\xi f^{bca}f^{amn}\bar c^b c^c\bar c^mc^n
\end{align}
The reason why this gauge is non-linear is easily understood: there is no linear procedure to
reproduce the Faddeev-Popov method out of this Lagrangian as seen by the following
path integral representation
\begin{align}
\mat Z&=\int[{\rm d}A][{\rm d}c][{\rm d}\bar c][{\rm d}b]\,
e^{-\int_M\left(i b^a\partial^\mu A_\mu^a+\frac\xi2(b^a)^2+i\frac12\bar c^a\,\mat M^{ab}[A]\,c^b+\frac{\rm g^2}8\xi(\bar c\times c)^2\right)}.
\end{align}
In Appendix C we will see how to deal with such a non-linear gauge by means of the semi-classical approximation. Moreover, we will provide
the linearization of the quartic term which will allow us to construct the Hubbard-Stratonovich transformations
of this non-linear gauge. We will also determine the relative BRST algebra.
The renormalizability of such a theory has been studied for instance in \cite{Delbourgo:1987np} and checked in \cite{Gracey:2002yt} up to three loops. 
It is known from topological field theory arguments \cite{Birmingham:1991ty} that the four-fermion interactions are
governed by the Riemann tensor of the manifold and by topological considerations we can derive from them 
the Euler characteristic of the target manifold
\footnote{These topological quantities will be explained and used in great details in successive sections. For now,
it is only necessary to know that they are related to the curvature of the manifold. In particular, the Euler character
is a topological invariant, i.e. does not depend on the Riemannian metric, which ``counts'' the number of holes in the manifold $M$.}.
Such interactions are fundamental terms in supersymmetric quantum mechanics, supersymmetric Y-M theory and above all string theory.
The presence of $f^{bca}f^{amn}\bar c^b c^c\bar c^mc^n$ is particularly important to give rise to an effective potential
whose vacuum configuration favors the formation of off-diagonal ghost-condensates \cite{Schaden:1999ew, Kondo:2000ey}.
The ghost condensation has been observed in others gauges, namely
in the Curci-Ferrari gauge and in the Landau gauge \cite{Dudal:2002ye} and references therein.
In these gauges the ghost condensates do not give rise to any
mass term for the gauge fields. The existence of these condensates turns out
to be related to the dynamical breaking of a $SL(2,R)$ symmetry which is
known to be present in both Curci-Ferrari and Landau gauge since long time
\cite{Ojima:1980da, Delbourgo:1981cm, Ojima:1981fs}. We will return to this issue when we will discuss one of our works
on Extend Double Lattice BRST later on.
The idea of giving a mass to gauge fields is stricly connected to the topological nature of the BRST operators:
the nilpotency condition for both $\delta$ and $\bar \delta$ is necessary to guarantee the confining of physical states.
It would then be interesting to find to what extent such a condition can be violated and how it can be controlled.
Suppose we introduce a bare mass term of the field $A_\mu^a$ which damages the aforementioned nilpotency.
Following the seminal work \cite{Curci:1976ar}, a renormalizable covariant Lagrangian can be written as
\begin{align}
\label{lag6}
\mat L_{\rm gf}+\mat L_{\rm FP}&=\frac i2(\delta \bar\delta-im^2)(A^\mu A_\mu-i\xi\bar c^ac^a)\nonumber\\
&=i b^a\partial^\mu A_\mu^a+\frac\xi2(b^a)^2+i\frac12\bar c^a\,\mat M^{ab}[A]\,c^b+\frac{\rm g^2}8(\bar c\times c)^2\nonumber\\
&\quad-\frac{m^2}2A^a_\mu A^{a\mu}-i\xi m^2\bar c^a c^a=\mat L_{\rm mCF}.
\end{align}
We call this Lagrangian the massive Curci-Ferrari Lagrangian. Notice how the mass term enters (with the correct
factor $i$ to preserve the overall hermiticity of $\mat L_{\rm mCF}$) the double BRST variation
and how in the third line, together with the expected gluon mass term, there is also a ghost--anti-ghost one.
This Lagrangian is left invariant under the following {\it extended double BRST transformations}
\begin{align}
\label{gaugetrans6}
\delta A_\mu^a&=(\mat D_\mu c)^a &\bar \delta A_\mu^a&=(\mat D_\mu \bar c)^a\nonumber\\
\delta c^a&=- \,\frac{\rm g}2(c\times c)^a &\bar \delta \bar c^a&=- \,\frac{\rm g}2(\bar c\times \bar c)^a\nonumber\\
\delta \bar c^a&= b^a-\frac{\rm g}2(\bar c\times c)^a&\bar \delta c^a&= -b^a-\frac{\rm g}2(\bar c\times c)^a\nonumber\\
\delta b^a &=im^2 c^a-\frac{\rm g}2(c\times b)^a&
\bar \delta b^a &=im^2\bar c^a-\frac{\rm g}2(\bar c\times b)^a\nonumber\\
&\quad-\frac{\rm g^2}8((c\times c)\times \bar c)^a&&\quad+\frac{\rm g^2}8((c\times c)\times \bar c)^a.
\end{align}
Though, these transformations do not satisfy the nilpotent condition; in fact
\beq
\delta^2 \bar c^a=i\delta b^a=-im^2 c^a\qquad\bar \delta^2 \bar c^a=-im^2 \bar c^a,
\eeq
or in general
\beq
\delta^2=\bar \delta^2\sim im^2.
\eeq
The fundamental consequence of such manipulation is the unitarity breakdown of 
the physical $\mat S$-matrix. Topologically speaking we can picture this
problem as having a singularity in the domain of the exterior differential operator $d$, which fails to maintain its nilpotency.
Though the unitarity is lost, FP conjugation still reamins valid, and this is the reason why we can construct
a superalgebra irrespective  of the validity of the nilpotency of $Q_B$ and $\overline{Q}_B$ as
\begin{align}
\label{brst4}
\{Q_B, Q_B\}&= 2 Q_B^2=-i\delta Q_B\nonumber\\
&=-m^2 Q_{cc}=-m^2 Q^\dagger_{cc}\nonumber\\
\{\overline{Q}_B, \overline{Q}_B\}&= 2 \overline{Q}_B^2=-i\bar \delta \overline{Q}_B\nonumber\\
&=-m^2 Q_{\bar c\bar c}=-m^2 Q^\dagger_{\bar c\bar c}\nonumber\\
\{Q_B, \overline{Q}_B\}&= -i\bar \delta Q_B=-i\delta \overline{Q}_B\nonumber\\
&=-m^2 Q_{c},
\end{align}
with the other algebra in (\ref{brst3}) intact. This superalgebra constitutes the {\it group decontraction}
of $osp(2,1)$ for finite $m$. 
In group theory terms \cite{Nakahara:1990th}, 
the group contraction is strongly related to the existence of a little group.
In \cite{Wigner:1939cj} Wigner constructed the maximal subgroup of the Lorentz group whose 
transformations leave the four-momentum of the given particle invariant. This subgroup is called 
Wigner's little group. This little group dictates the internal space-time symmetry of relativistic particles.
In \cite{Kim:2001hn} the reader can find an extensive overview of little group theory in the Lorentz group.
For a relativistic particle, we then wish to find what the maximal subgroup of $SO(3,1)$ is leaving invariant
the first Casimir operator $C_1=-p_\mu p^\mu$. 
For the purpose of this thesis, we are interested in considering the {\it light-like} case ($C_1=0$) and 
the {\it space-like} case ($C_1<0$). In the case of the Poincar\'e group, therefore including also space-time translations
there is also the second Casimir operator to take into account $C_2=W_\mu W^\mu$, with $W^\mu=\epsilon^{\mu\lambda\sigma\nu} J_{\lambda} P_\nu/2$.
It suffices here to say that the internal space-time symmetries of massive
and massless particles (massive and massless BRST algebra) are dictated by $O(3)$-like and $E(2)$-like
little groups respectively. $O(3)$ is locally isomorphic to the three-dimensional rotational group, whereas
the Euclidean group $E(2)$ is a two-dimensional group constituted by a translation and a rotation over a flat space.

It would be also possible to include in the NO algebra other $3n$ charges (conserved in Landau gauge) following
from the equations of motion of $b^a$, $c^a$ and $\bar c^a$ respectively. It is argued in \cite{Nakanishi:1980dc, Delbourgo:1981cm}
that this new {\it extended} algebra would correspond to $osp(4,2)$, i.e. enlarging the Lorentz group to the Poincar\'e
group. 
Though, as shown in \cite{Thierry-Mieg:1985yv}
the $b$-field would create an anomaly in the algebra, and therefore $osp(4,2)$ is broken, at least on-shell.

\thispagestyle{empty}
\cleardoublepage
\chapter{Faddeev-Popov Jacobian in non-perturbative Y-M theory}
\label{chap4}

The elevation of Faddeev-Popov (FP) gauge-fixing of
Yang-Mills theory beyond the
realm of perturbation theory has been intensely pursued in
recent years for many reasons.
Nonperturbative gauge-fixed calculations on the lattice
are being compared to analogous solutions of Schwinger-Dyson
equations \cite{Alkofer:2000wg, Bowman:2002bm}. 
As well, the long-term goal of simulating the
full Standard Model using lattice Monte Carlo requires
the Ward-Takahashi identities associated with BRST symmetry
\cite{Becchi:1975nq} in order to control the lattice renormalisation. 
The main impediment to nonperturbative gauge-fixing
is the famous Gribov ambiguity \cite{Gribov:1977wm}: 
gauges such as Landau and Coulomb
gauge do not yield unique representatives on gauge-orbits 
once large scale field fluctuations are permitted.  
To some extent one could live with such non-uniqueness
if one could incorporate all Gribov copies in a computation. However
the no-go theorem of Neuberger \cite{Neuberger:1986xz}
obstructs even this: (a naive generalisation
of) BRST symmetry forces a complete cancellation of all Gribov copies 
in BRST invariant observables giving $0/0$ for expectation values.
In particular, Gribov regions contribute with alternating
sign of the FP determinant.

Here we shall propose an approach which takes seriously
that gauge-fixing when seen as a change of variables involves
a Jacobian being the absolute value of the 
Faddeev-Popov determinant. Usually the absolute
value is dropped either because of an {\it a priori} restriction
to perturbation theory or because of the identification
of the determinant in terms of an invariant of a
topological quantum field theory \cite{Birmingham:1991ty} 
such as the Euler character \cite{Hirschfeld:1978yq, Baulieu:1996rp}. 
In the latter case the Neuberger problem is encountered.

The approach we describe in the following is not restricted to
perturbation theory. Moreover, because it will be seen to
involve a gauge-fixing Lagrangian density that is not
BRST exact it falls outside the scope of the preconditions
for the Neuberger problem. 
In the next section we shall derive the Jacobian associated
with gauge-fixing in the presence of Gribov copies. We shall
give a representation of the ``insertion of the identity'' in this
case in terms of a functional integral over an enlarged set
of scalar and ghost fields. The extended BRST symmetry of this
new gauge-fixing Lagrangian density will be described
though we will see that the final form of the gauge-fixing
Lagrangian is not BRST exact. 


\section{Field theoretic representation for the Jacobian of FP gauge fixing}

In the following we shall formulate the problem in the
continuum approach to gauge theory.
  
Our aim is to generalise the standard formula from calculus
for a change of variable:
\beq
\label{change}
\left| \det\left({ {\partial f_i}\over {\partial x_j} }\right) 
\right|^{-1}_{{\vec f}=0}
= \int dx_1 \dots dx_n \delta^{(n)}({\vec f}({\vec x})).
\eeq
Here one is changing from integration variables ${\vec x}$ to
those satisfying the condition ${\vec f}({\vec x})=0$
and where, for Eq.(\ref{change}) to be valid,
in the domain of integration of $\vec x$ there 
must be only one such solution. 
In the context of gauge-fixing of Yang-Mills theory the
generalisation of Eq. (\ref{change}) is
\beq
\left| \det\left({ {\delta F[{}^gA]}\over {\delta g} }
\right) \right|^{-1}_{F=0}
= \int \mathcal{D}g \,\delta[F[{}^g\!A]] 
\label{functchange}
\eeq
where $A_{\mu}$ represents the gauge field, $g$ is an element
of the $SU(N)$ gauge group, 
$\mathcal{D}g$ is the functional integration measure in the group and 
\beq
F[{}^g\!A]=0
\label{gfcondition}
\eeq
is the gauge-fixing condition. We shall be interested 
in Landau gauge $F[A]=\partial_{\mu} A_{\mu}$. 
As in the calculus formula, here
Eq.(\ref{functchange}) is only valid as long as 
Eq.(\ref{gfcondition}) has a unique solution.
This is known not to be the case for Landau gauge. 
The FP operator nevertheless is
$M_F[A]= (\delta F[{}^gA]/\delta g)|_{F=0}$ and its determinant is
$\Delta_F[A]=\det(M_F)$.
For the Landau gauge 
$M_F[A]^{ab} = \partial_{\mu} D^{ab}_{\mu}[A]$ with $D^{ab}_{\mu}[A]$ the
covariant derivative with respect to $A_{\mu}^a$ in the 
adjoint representation.
Now the standard FP trick is
the insertion of unity in the measure of the
generating functional of Yang-Mills theory realised via
the identity (which follows from the above definitions):
\beq
1=\int \mathcal{D}g \Delta_F[{}^g\!A] \delta[F[{}^g\!A]].
\label{resolveunity}
\eeq
By analogy with standard calculus, in the presence of multiple solutions to the gauge-fixing condition
Eq.(\ref{resolveunity}) must be replaced by
\beq
\label{newidentity}
N_{F}[A] =\int \mathcal Dg\,\delta(F[{}^g\!A])\,\Big|\det
M_{F}[{}^g\!A]\Big|,
\eeq
where $N_{F}[A]$ is the number of different solutions for the
gauge-fixing condition $F[{}^g\!A]=0$ on the orbit
characterised by $A$, where $A$ is any configuration on the 
gauge orbit in question for which $\det M_F\neq 0$.
It is known that Landau gauge
has a fundamental modular region (FMR), namely a set of
unique representatives of every gauge orbit which is moreover
convex and bounded in every direction 
\cite{Dell'Antonio:1991xt,Semenov:1982sf}. 
The following discussion can be found in more
detail in \cite{vanBaal:1995gg}.
Denoted $\Lambda$, the FMR is defined as the
set of absolute minima of the functional $V_A[g]=\int d^4x ({}^gA)^2$
with respect to gauge transformations $g$. The stationary points of 
$V_A[g]$ are those $A_{\mu}$ satisfying the
Landau gauge condition. 
The boundary of the FMR, $\partial \Lambda$,
is the set of degenerate absolute minima
of $V_A[g]$. $\Lambda$ lies within
the Gribov region $C_0$ where the FP operator
is positive definite. The Gribov region is comprised of all of the local minima of $V_A[g]$.
The boundary of $C_0$,
the Gribov horizon $\partial C_0$, is where the FP operator
$M_F$ (which corresponds to the second order variation
of $V_A[g]$ with respect to infinitesimal $g$) acquires zero modes. 
When the degenerate absolute minima of $\partial\Lambda$
coalesce, flat directions
develop and $M_F$ develops zero modes. Such orbits cross the
intersection of $\partial\Lambda$ and $\partial C_0$. 
The interior of the fundamental modular region is a smooth
differentiable and everywhere convex manifold. Orbits crossing the boundary 
of the FMR on the other hand will cross that boundary again at
least once corresponding to the degenerate absolute
minima. 
Though, at present, there is no practical computational 
algorithm for constructing the FMR, it exists and we will make use of it 
for labelling orbits, i.e., $A_{\rm u}$ are defined to be configurations
in the FMR, $A_{\rm u}\in \Lambda$.
Since every orbit crosses the fundamental modular
region once we are guaranteed to have $N_{F}\geq 1$. 
In turn the ${}^gA_{\rm u}$
fulfilling the constraint of Eq. (\ref{gfcondition}) would be every other gauge copy
of $A_{\rm u}$ along its orbit. Eq.(\ref{newidentity})
is equal to the number of Gribov copies
on a given orbit, $N_{GC}=N_{F}-1$, except 
that copies lying on any of the Gribov horizons ($\Delta_{F}=0$) 
do not contribute to $N_{F}$. 
The finiteness of $N_F$ in the presence of a regularisation 
leading to a finite number of degrees of freedom 
(such as a lattice formulation) can be argued as follows.
Consider two neighboring Gribov copies corresponding to
a single orbit. If they contribute to $N_F$ they cannot
lie on the Gribov horizon. Therefore they do not
lie infinitesimally close to each other along a flat direction,
namely they have a finite separation. This is true then for all
copies on an orbit contributing to $N_F$: all copies contributing
to $N_F$ have a finite separation.  
But the $g$ which create the copies of $A_{\rm u}$
belong to $SU(N)$ which has a finite group volume. 
Thus for each space-time point there is a finite number of
such $g$.
We conclude then for a regularised formulation that $N_{F}$ is finite. 
Consider then the computation
of the expectation value of a gauge-invariant operator $O[A]$ over an ensemble
of gauge-field configurations ${ A_{\rm u} }$ which is this set of unique
representatives of gauge orbits discussed above.  
Note that for a gauge-invariant observable, it makes no difference whether $A_{\rm u}\in \Lambda$
or if the $A_{\rm u}$'s are any other unique representatives of the orbits.
The expectation value on these configurations
\beq
\langle O[A] \rangle = 
{   {\int {\mathcal D}A_{\rm u} O[A_{\rm u}] e^{-S_{YM}}   } 
  \over  {\int {\mathcal D}A_{\rm u} e^{-S_{YM}}   }  } 
\label{expecvalue}
\eeq
is well-defined.
Since in any regularised formulation $N_F$ is a finite positive integer,
we can legitimately
use Eq.(\ref{newidentity}) to resolve the identity
analogous to the FP trick
and insert into the measure of
integration for an operator expectation value. We thus have
\beq
\langle O[A]\rangle=\frac{\int \mathcal DA_{\rm u}\frac{1}{N_{F}[A_{\rm u}]}
\int \mathcal Dg\,\delta(F[{}^g\!A])\,\Big|\det 
M_{F}[{}^g\!A]\Big|\,O[A]\,
e^{-S_{YM}[A]}} {\int \mathcal DA_{\rm u}\frac{1}{N_{F}[A_{\rm u}]}
\int \mathcal Dg\,\delta(F[{}^g\!A])\,\Big|\det M_{F}[{}^g\!A]\Big|
\,e^{-S_{YM}[A]}}.
\label{expectvalGC}
\eeq
We can now pass $N_{F}[A_{\rm u}]$ under the group
integration ${\mathcal D}g$ and combine the latter with ${\mathcal D}A_{\rm u}$
to obtain the full measure of all gauge fields ${\mathcal D}({}^gA_{\rm u})$
which we can write now as ${\mathcal D}A$.
$N_{F}$ is certainly gauge-invariant: it is a property
of the orbit itself. 
So $N_{F}[A_{\rm u}]=N_{F}[{}^gA_{\rm u}]
=N_{F}[A]$.
Thus we can write
\beq
\langle O[A]\rangle=\frac{\int \mathcal DA\,
\frac{1}{N_{F}[A]}
\delta(F[A])\,\Big|
\det M_{F}[A]\Big|\,O[A]\,e^{-S_{YM}[A]}}
{\int \mathcal DA\,
\frac{1}{N_{F}[A]}
\delta(F[A])\,\Big|\det
M_{F}[A]\Big|\,e^{-S_{YM}[A]}}.
\label{expecvaluewithcopies}
\eeq
Perturbation theory can be recovered from this of
course by observing that only $A$ fields near the trivial orbit, containing 
$A=0$ and for which $S_{YM}[A]=0$, contribute significantly in the perturbative regime: the curvature of the
orbits in this region is small so that the different orbits in
the vicinity of $A=0$ intersect the gauge-fixing hypersurface
$F=0$ the same number of times.
Then the number of Gribov copies is the same for each orbit, 
$N_{F}$ is independent of $A_{\rm u}$ and
we can cancel $N_{F}$ out of the expectation value. In that case
\beq
\label{expecvaluenocopies}
\langle O[A]\rangle=\frac{\int \mathcal DA\,\delta(F[A])\,\Big|
\det M_{F}[A]\Big|\,O[A]\,e^{-S_{YM}[A]}}
{\int \mathcal DA\,\delta(F[A])\,\Big|\det 
M_{F}[A]\Big|\,e^{-S_{YM}[A]}}.
\eeq
In turn, observing that fluctuations near the trivial orbit
cannot change the sign of the determinant, the modulus can
also be dropped and one recovers the usual starting point
for a standard BRST invariant formulation of 
Landau gauge perturbation theory.
Note that perturbation theory is built on the gauge-fixing surface
in the neighbourhood of $A=0$, which for a gauge-invariant quantity
will be equivalent to averaging over the Gribov copies of $A=0$ as in Eq. (\ref{expecvaluenocopies}).
For the non-perturbative
regime, the orbit curvature increases significantly and
in general there is no reason to expect that $N_{F}$ would be the same for each
orbit. Moreover the determinant can change sign. 
Let us focus on the partition function appearing in 
Eq. (\ref{expecvaluewithcopies})
\beq
{\mathcal Z}_{\rm{gauge-fixed}} = \int {\mathcal D}A
\, N_{F}^{-1}[A]
\Big| \det(M_F[A]) \Big| \delta(F[A])\, e^{-S_{YM}}
\label{partitionfn1}
\eeq
The objective is to generalise the BRST formulation of 
Eq.(\ref{partitionfn1}) such that it is valid beyond perturbation theory
taking into account the modulus of the determinant.
We thus start with the following representation:
\beq
\Big|\det (M_F[A])\Big| = {\rm sgn}(\det (M_F[A]))\, \det (M_F[A])
\label{moddet}. 
\eeq
This representation goes under the name of the {\it Nicolai map} \cite{Nicolai:1979nr, Birmingham:1991ty}.

\subsection{The Nicolai map and Topological Field Theory}

Soon after the seminal work of Gribov \cite{Gribov:1977wm}, the attention on gauge-fixing procedures in non-Abelian gauge theories \cite{Singer:1978dk}
led physicists to examine more in depth the strong relation between these theories and Topological Field Theory (TFT).
It was immediately realised that in 4-dimensional gauge theory certain topological aspects play an important role: 
in late 70's and early 80's an enormous amount of work, mainly due to Donaldson, Schwarz and Witten, allowed the physics community
to discover how Y-M theory could be explained in terms of topologically invariant quantities, such as polynomials and knots
\cite{Schwarz:1978cn, Witten:1982im, Donaldson:1983wm, Donaldson:1990kn, Witten:1988hf}. 
The discovery of solutions to the Y-M classical equations of motion, called instantons, \cite{Belavin:1975fg} and 
the analysis of monopole structures in gauge theory \cite{'tHooft:1974qc, 'tHooft:1976fv} spread light into a world dense of
interesting topological properties \cite{Uhlenbeck:1982zm, Polyakov:1976fu, Atiyah:1978mv}, as well as
a better understanding of the geometrical/mathematical background of low dimensional manifolds of Yang-Mills theory in terms of the 
well known theory of principal bundles \cite{Babelon:1980uj, Daniel:1979ez, Atiyah:1982fa}.
The study of these relations among mathematics, topology and physics has become known as Topological Quantum Field Theory (TQFT).
In the following sections we will largely adopt \cite{Birmingham:1991ty} as a leading guide: as TQFT is a considerable subject to 
cover, we will try to focus on those parts which directly concerns supersymmetric aspects of Yang-Mills theory, the 
coherent and appropriate scenario on which the BRST formalism resides.

In topological quantum field theories we are only interested in those observables that only depend on global
features of the space on which these theories are defined. Consequently, the observables are independent of any 
Riemannian metric characterising the underlying manifolds. The study of quantities which are topological
invariant was first started by Euler, in 1736 when he published a paper on the solution of 
the Königsberg bridge problem entitled {\it Solutio problematis ad geometriam situs 
pertinentis} which translates into English as {\it The solution of a problem relating to the geometry of position}
\footnote{The paper not only shows that the problem of crossing the seven bridges in a single journey is impossible, 
but generalises the problem to show that, in today's notation,
{\it A graph has a path traversing each edge exactly once if exactly two vertices have odd degree}.}.
The quantum version of Euler's ideas deals largely with the path integral representation of topoligcal invariants such as
the Ray-Singer torsion \cite{Schwarz:1978cn} or Morse theory and its relations with supersymmetric quantum mechanics \cite{Witten:1982im}.
As we will see, in the construction of TQFT, one can adopt two main different frameworks, the Witten-type or Schwarz-type:
for the Witten type theories, also called {\it cohomological}, 
one combines certain topological shift symmetry with any other local symmetry, whereas in Schwarz
type models, called {\it quantum}, the attention is focused on the usual gauge symmetry.
We will only consider Witten-type theory throughout this work.
The necessary ingredients to construct a topological field theory are 
\begin{itemize}
\item a collection of Grassmann-graded
\footnote{Grassmann grading defines the {\it even} or {\it odd} character of a field under product exchange.}
 fields $\{\Phi\}$ defined on a Riemannian manifold $M$ with metric $\rm g$,
\item a nilpotent operator $Q$, $Q^2=0$, odd w.r.t. Grassmann grading,
\item the physical Hilbert space is defined by the condition $Q|phys\rangle=0$, and its physical states are defined to be $Q$-cohomology classes
\footnote{A state annihilated by $Q$ is said to be $Q$-closed, while a state of the form $Q|\chi\rangle$ is called $Q$-exact;
this equivalence relation partitions the physical Hilbert space into $Q$-cohomology classes, states which are $Q$-closed modulo $Q$-exact states.},
\item an energy-momentum tensor which is $Q$-exact, i.e.
corresponds to the variation of a functional $V_{\alpha,\beta}$ of fields w.r.t. $Q$
\footnote{The variation $\delta \mathcal O=\{Q, \mathcal O\}$ corresponds to a Grassmann-graded commutator of fields.} 
\beq
T_{\alpha\beta}=\{Q,V_{\alpha,\beta}(\Phi,g)\}.
\eeq
\end{itemize}

The existence of  a a Nicolai map is admitted in a Witten-type theory, such that
the path integral can be restricted to the moduli space of classical solutions (instantons).
Nicolai has proven that for theories with a global supersymmetry
there exists a non-linear and, in general, non-local mapping of the bosonic
fields which trivialises the bosonic part of the action, and whose determinant
cancels the Pfaffian
\footnote{The Pfaffian of an even $2n$-dimensional antisymmetric matrix $M_{ij}$ is defined
as $Pf(M)=\epsilon_{i_1,\ldots i_{2n}} M^{i_1}_{i_2}\cdots M^{i_{2n-1}}_{i_{2n}}$, with the property that
the determinant of the matrix $M$ is equal to the square of the Pfaffian.}
(in the case of Majorana spinors) of the fermionic fields present.

Consider the following action
\begin{align}
S&=\int{\rm d}\tau\left[i\left(\frac{{\rm d}\phi^i}{{\rm d}\tau}+s g^{ij}(\phi)\frac{\partial V}{\partial \phi} \right)B_i
+\frac12g^{ij}(\phi)B_iB_j-\frac14 R_{ijkl}\bar \psi^i \psi^k\bar \psi^j \psi^l\right.\nonumber\\
&\left.-i\bar\psi_i\left(\delta^i_j \frac{\mat D}{\mat D\tau}+s g^{ik}(\phi)\frac{\mat D^2 V}{\mat D\phi^k\mat D\phi^j} \right)\psi^j\right].
\end{align}
After integrating out the fermions, the partition function takes the form
\beq
Z=\int_\phi e^{-S(\phi)}\, Pf[\phi].
\eeq
The existence of a Nicolai map for such a theory tells us that there exists a map $\phi\to \xi(\phi)$ such that the Jacobian of
the transformation compensates the Pfaffian (up to a sign). The partition function $Z$ then assumes the topological form
\beq
Z=\int_\xi e^{-\int\frac12\ \xi^2}\times (\textrm{winding number of the mapping}),
\eeq
where the winding number is the number of times $\xi$ runs over its range as $\phi$ is varied. In \cite{Nicolai:1979nr}
Nicolai was able to show this map only up to third order in the coupling constant for $N=1$ super Yang-Mills theory in 4-dim.
This approximation is due to the highly non-local character of the map, which can be found analytically only in low dimensional cases.
In the above case, suppose we use the following change of variables, showing the instanton sector of the theory
\beq
\phi\to \xi=\frac{{\rm d}\phi}{{\rm d}\tau}+s\frac{\partial V}{\partial \phi}.
\eeq
With this change of variables we get
\beq
Z=\int_\xi e^{-\frac12\oint {\rm d}\tau\, \xi^2}\,\det\left(\frac{\delta \xi}{\delta \phi}\right)
\left[\det\left(\frac{\delta \xi}{\delta \phi}\right)\right]^{-1}=\pm 1.
\eeq
The ratio of functional determinants is then $\pm 1$, which can be regarded, when dealing with Y-M theory, 
as a topological manner to consider the Gribov problem.
In the next chapter, we will analyze this ratio problem from the topological point of view of the Poincar\'e-Hodge and Gauss-Bonnet theorem.
There, we will discuss the fundamental topological obstruction in non-perturbative Y-M theory which determines the {\it exact}
cancellation of this ratio, also known as the Neuberger problem.


\section{The Nicolai map in the Faddeev-Popov Jacobian}

As mentioned, the factor ${\rm {det}}(M_F[A])$ in Eq.(\ref{moddet}) is  
represented as a functional integral via the usual Lie algebra valued 
ghost and anti-ghost fields
in the adjoint representation of $SU(N)$. 
Let us label these as $c^a, {\bar c}^a$. It is usual also 
(see for example \cite{Nakanishi:1990qm}) 
to introduce a Nakanishi-Lautrup auxiliary field $b^a$.
Thus the effective gauge-fixing Lagrangian density
\footnote{We will use throughout this Chapter a different convention from the one we will adopt in the next Chapter. In this Chapter, in fact
we adopt Hermitian generators for the algebra, and Hermitian ghost fields.}
\beq
{\mathcal L}_{\rm{det}} =
-b^a \partial_{\mu} A^a_{\mu} + \frac{\xi}{2} b^a b^a
+ {\bar c}^a M_F^{ab}c^b
\label{usualLag}
\eeq
yields \cite{Nakanishi:1990qm} 
\beq
\lim_{\xi\rightarrow 0}
\int {\mathcal D}{\bar c}^a {\mathcal D}c^a {\mathcal D}b^a
e^{-\int d^4x {\mathcal L}_{\rm{det}} }
= \delta(F[A]) \det(M_F[A]). 
\eeq
In order to write the factor ${\rm {sgn}}(\det(M_F[A]))$   
in terms of a functional integral
weighted by a local action, we consider the following Lagrangian density
\beq
\mathcal{L}_{{\rm{sgn}}} = i B^a M_F^{ab} {\varphi}^b - i {\bar d}^a M_F^{ab} d^b
+ \frac{1}{2} B^a B^b
\label{Lagsgn1}
\eeq
with ${\bar d}^a,d^a$ being new Lie algebra valued Grassmann fields 
and $\varphi^a,B^a$ being new auxiliary commuting fields.
Consider in Euclidean space the path integral 
\beq
\mathcal{Z}_{{\rm{sgn}}}= 
\int \mathcal{D}{\bar d}^a \mathcal{D}d^a \mathcal{D}\varphi^a \mathcal{D}B^a 
e^{-\int d^4 x \mathcal{L}_{{\rm{sgn}}}}\,.
\label{sgnpartition}
\eeq
Completing the square in the Lagrangian density of Eq.(\ref{Lagsgn1}), 
the $B$ field can be integrated out in the partition function
leaving an effective Lagrangian density
\beq
\mathcal{L}'_{{\rm{sgn}}} = \frac12 \varphi^a ((M_F)^T)^{ab} M_F^{bc} \varphi^c 
- i {\bar d}^a M_F^{ab} d^b\,,  
\label{sgnLag2}
\eeq
where $(M_F)^T$ denotes the transpose of the FP operator.
Integrating all remaining fields now 
it is straightforward to see that the partition function  
Eq.(\ref{sgnpartition}) amounts to just 
\beq
\mathcal{Z}_{\rm sgn} = 
\frac{\det(M_F)}{\sqrt{\det((M_F)^T M_F)}}= {\rm sgn}(\det (M_F)). 
\eeq
Thus the representation Eq.(\ref{sgnpartition}) can be used
for the first factor of Eq.(\ref{moddet}). The Lagrangian density of 
Eq.(\ref{Lagsgn1}) therefore combines with   
the standard BRST structures of Eq.(\ref{usualLag})
coming from the determinant itself in Eq.(\ref{moddet})
so that an equivalent representation for the partition function
based on Eq.(\ref{partitionfn1}) is
\beq
{\mathcal Z}_{\rm{gauge-fixed}} =
\int {\mathcal D}A^a_{\mu} {\mathcal D}{\bar c}^a {\mathcal D}c^a
{\mathcal D}{\bar d}^a {\mathcal D}d^a {\mathcal D}b^a {\mathcal D}\varphi^a
(N_{F}[A])^{-1}\, e^{-S_{\rm YM}-S_{\rm \det}-S_{\rm sgn}}
\eeq
with $S_{\det}$ and $S_{\rm sgn}$ the actions corresponding to
the above Lagrangian densities Eqs. (\ref{usualLag},\ref{Lagsgn1}).

 
\section{A new extended BRST}

The symmetries of the new Lagrangian density, $\mathcal L_{\rm sgn}$, 
are essentially a boson-fermion
supersymmetry and can be seen from Eq.(\ref{Lagsgn1}). 
In analogy to the standard BRST transformations typically denoted by $s$,
we shall denote them by the Grassmann graded operator $t$
\begin{eqnarray}
t \varphi^a &=& d^a \nonumber \\
t d^a &=& 0 \nonumber \\
t {\bar d}^a &=& B^a \nonumber \\
t B^a &=& 0\,,  
\label{t-alg}
\end{eqnarray}
such that
\begin{equation}
t\mathcal L_{\rm sgn}=0.
\end{equation}
Eqs.(\ref{t-alg}) realise the infinitesimal form of shifts in
the fields. The operation
$t$ is nilpotent: $t^2=0$. Using Eqs.(\ref{t-alg}) we can
give the following form for the Lagrangian density $\mathcal{L}_{\rm{sgn}}$,
\begin{equation}
\mathcal{L}_{\rm{sgn}}= 
t \left( {\bar d}^a (i M_F^{ab} \varphi^b + \frac{1}{2} B^a) \right) .
\end{equation}
The question now is how to combine this with the standard BRST
transformations 
\begin{eqnarray}
s A_{\mu}^a &=& D_{\mu}^{ab} c^b \nonumber \\
s c^a &=& - \frac12 g f^{abc} c^b c^c \nonumber \\
s{\bar c}^a &=& b^a \nonumber \\
s b^a &=& 0 .
\end{eqnarray}
The transformations due to $t$ and $s$ are completely
decoupled except that the latter also act on the gauge field
on which the FP operator $M_F$ depends. We propose the
following unification of these symmetry operations.
Consider an operation $\mathcal{S}$ {\it block-diagonal}
in $s$ and $t$: $\mathcal{S}={\rm{diag}}(s,t)$.
The operator acts on the following multiplet fields:
\be
\mathcal{A}^a = 
\left( \bea{ccc} A^a_{\mu} \\\varphi^a \eea\right),
\mathcal{C}^a= 
\left( \bea{ccc} c^a \\d^a \eea\right),
\bar{ \mathcal{C}}^a= 
\left( \bea{ccc} {\bar c}^a \\{\bar d}^a \eea\right), 
\mathcal{B}^a=
\left( \bea{ccc} b^a \\B^a \eea \right).
\ee
We see that these fields transform under $\mathcal{S}$ completely
analogously to the standard BRST operations
\begin{eqnarray}
\label{newalgebra}
\mathcal{S} \mathcal{A}^a &=& \mathcal{D}^{ab} \mathcal{C}^b \nonumber \\  
\mathcal{S} \mathcal{C}^a_i &=& 
\mathcal{F}^{abc}_{ijk} \mathcal{C}^b_j \mathcal{C}^c_k 
\nonumber \\
\mathcal{S} \bar{\mathcal{C}}^a &=& \mathcal{B}^a \nonumber \\
\mathcal{S} \mathcal{B}^a &=& 0\,, 
\end{eqnarray}
where $i,j,k=1,2$ label the elements of the multiplets, and  
\begin{eqnarray}
 \mathcal{D}^{ab} &=& {\rm{diag}} (D_{\mu}^{ab}, \delta^{ab}) \nonumber \\
\mathcal{F}^{abc}_{111} &=& -\frac12 g f^{abc} ,\quad \mathcal F_{ijk}^{abc}=0 \quad {\rm for} \quad ijk\neq 111.  
\end{eqnarray}
Note that nilpotency is satisfied, $\mathcal{S}^2=0$. 
We shall refer to this type of operation as an {\it extended} BRST 
transformation which we distinguish from the BRST--anti-BRST
or double BRST algebra of the Curci-Ferrari model
\cite{Curci:1976ar,Thierry-Mieg:1979kh}.
We can thus formulate the gauge-fixing Lagrangian density  
for the Landau gauge as 
\begin{equation}
\mathcal{L}_{\rm gf} = {\rm Tr} \, \mathcal{S} 
\left( \begin{array}{ccc}
\bar c^a F^a & 0 \\
0 & \bar d^a(i M_F^{ab} \varphi^b + \frac{1}{2} B^a) \end{array} \right) .
\label{gfLag}
\end{equation}
This approach admits also an extended anti-BRST operation:
\begin{eqnarray}
{\bar {\mathcal{S}}} \mathcal{A}^a &=& 
\mathcal{D}^{ab} \bar {\mathcal{C}}^b \nonumber \\
{\bar {\mathcal{S}}} \bar{ \mathcal{C}}^a_i &=& \mathcal{F}^{abc}_{ijk} 
{\bar {\mathcal{C}}}^b_j \bar{ {\mathcal{C}}}^c_k \nonumber \\
{\bar {\mathcal{S}}} \mathcal{C}^a &=& -\mathcal{B}^a \nonumber \\
{\bar {\mathcal{S}}} \mathcal{B}^a &=& 0\,.
\end{eqnarray}
Writing $\bar{ \mathcal{S}} = {\rm{diag}} ({\bar s}, {\bar t})$
we can extract the standard anti-BRST ${\bar s}$-operations, in Landau gauge, 
\cite{Thierry-Mieg:1982un, Baulieu:1981sb} 
\begin{eqnarray}
{\bar s} A_{\mu}^a &=& D_{\mu}^{ab} {\bar c}^b \nonumber \\
{\bar s} {\bar c}^a &=& - \frac12 g f^{abc} {\bar c}^b {\bar c}^c \nonumber \\
{\bar s} c^a &=& -b^a \nonumber \\
s b^a &=& 0 
\end{eqnarray}
and those corresponding to $\bar t$:
\begin{align}
\bar t \varphi^a &= \bar d^a \nonumber \\
\bar t \bar d^a &= 0 \nonumber \\
\bar t d^a &=-B^a \nonumber \\
\bar t B^a &= 0.
\end{align}
Moreover, the ghosts and anti-ghosts in this extended structure
also fulfill the criteria for being Maurer-Cartan one-forms,
\begin{equation}
\mathcal{S} \bar{ \mathcal{C}} + \bar{ \mathcal{S}} \mathcal{C} =0.
\end{equation}
However there is no extended 
BRST--anti-BRST (or double) symmetric
form of the gauge-fixing Lagrangian density Eq. (\ref{gfLag}),
unlike the two pieces of which it consists.
Such a representation exists
in the $s-$sector of Landau gauge:
\begin{equation}
\mathcal{L}_{{\rm gf}, s} = \frac12s {\bar s} A^a_{\mu} A^a_{\mu}\,.
\end{equation}
In the $t-$sector, the corresponding structure is
\begin{equation}
\mathcal{L}_{{\rm gf},t} = \frac12t {\bar t} 
\left [ \varphi^a M_F^{ab} \varphi^b +  {\bar d}^a d^a
\right] .
\end{equation}
However the complete Landau gauge-fixing Lagrangian density
can only be expressed via a trace, namely as 
\begin{equation}
{\mathcal L}_{\rm{gf}}= \frac12 {\rm Tr}
\mathcal{S}\bar{ \mathcal{S}} \mathcal{W}
\end{equation}
with
\begin{equation}
\mathcal{W} = {\rm{diag}} 
\left( A^a_{\mu} A^a_{\mu}, 
\varphi^a M_F^{ab} \varphi^b +  {\bar d}^a d^a 
\right)\,.
\end{equation}
Nevertheless this 
compact representation formulates the modulus of the determinant
in Landau gauge fixing in terms of a local Lagrangian density
and follows as closely as possible the standard BRST
formulation without the modulus.


\thispagestyle{empty}
\cleardoublepage
\chapter{Decontracted Double Lattice BRST, the Curci-Ferrari Mass and the Neuberger Problem}


In 1974, Wilson \cite{Wilson:1974sk} formulated Euclidian gauge theories on the lattice in order to shed light 
on the confinement mechanism in QCD and to study the non-perturbative regime of non-Abelian gauge theories.
To construct the proper lattice gauge theory of QCD, we need first to discretize the space-time, then the transcription
of the gauge and fermion fields\footnote{Dealing only with pure gauge theories we will not consider fermion fields in this introduction to lattice gauge theory.}
succesively the action and the re-definition of the functional measure. 
Finally the transcription of the operators to
probe the physics.
For a detailed analysis of lattice gauge theory we refer to \cite{Gupta:1997nd} and \cite{Smit:2002ug}.
Here we just wish to give the basic properties and definitions of this theory which will be used in the following sections.
To start with, we need to stress that the lattice procedure provides a cutoff which naturally regularizes
the ultraviolet divergences of quantum field theories.
As with any regulator, it must be removed after renormalisation: the continuum version of any lattice theory is
provided by taking the adopted lattice spacing to zero. In the wide range of possible lattice regularisation, the
simplest one consists in taking the isotropic cubic grid, where there is no distinction between the space lattice spacing
$a_S$ and the time one $a_T$.
Moreover, on the lattice, we sacrifice Lorentz invariance, but all the other internal symmetries are preserved, particularly
local gauge invariance\footnote{Requiring gauge invariance at all $a$ is necessary otherwise one would have many more parameters to tune
(such as the gluon mass for instance) and there would arise many more operators at any given order in $a$.
The lattice action will be also invariant under charge conjugation $\mathcal C$, parity $\mathcal P$ and time
reversal $\mathcal T$.}. 
Having said that, any four-dimensional integral can be written in terms of the lattice spacing $a$ as
\beq
\int{\rm d}^4x\to a^4\sum_n,
\eeq
where the space-time coordinate $x_\mu$ has been replaced by a set of integers $n_\mu$, such that $x_\mu=a n_\mu$
and $\sum_n$ corresponds to a finite sum over the lattice sites $n$.
The construction of gauge fields is somewhat tricky and requires some attention: observing that a particle moving on a contour
picks up a phase factor, Wilson formulated gauge fields on a space-time lattice introducing the concept of link variables $U_\mu(x)$,
connected to this phase factor.
These links are the fundamental variables on the lattice, they live in the Lie group $G$ of the theory, connecting  $x$
to $x+{\hat \mu}$, defined as
\begin{align}
U_\mu(x)&={\rm P}\exp\left\{{\rm g}\int^{x+{\hat \mu}}_{x_\mu}\, X^a A^a_\nu(x){\rm d}z^\nu\right\}\nonumber\\
&=U_\mu(x, x+{\hat \mu}),
\end{align}
with $X^a$ the $N^2-1$ anti-hermitian generators of the Lie algebra $\mathfrak g$.
${\rm P}$ denotes the path ordering, such that
\beq
U_\mu(x, x-{\hat \mu})\equiv U_{-\mu}(x)=U^\dagger_\mu(x-{\hat \mu}, x)
\eeq
Under a gauge transformation $g(x)$, the link variable transforms as
\beq
{}^g\!U_\mu(x)\equiv g(x)\,U_\mu(x)\,g^\dagger (x+{\hat \mu}).
\eeq
With these definitions, there are two types of gauge invariant objects (which can be of arbitrary size and shape and
over any representation of the Lie group)
that one can construct on the lattice
1) a string of path-ordered product of links capped by a fermion and an antifermion;
2) closed Wilson loops, whose simplest example is the {\it plaquette}, a $1\times 1$ loop
\beq
\label{wilson}
W^{1\times 1}_{\mu\nu}={\rm Re} {\rm Tr}\Big( U_\mu(x)U_\nu(x+{\hat \mu})U^\dagger_\mu(x+{\hat \nu})U^\dagger_\nu(x)\Big).
\eeq
A gauge invariant action has to build up out of loops and strings, with the physical constraint that in the limit
$\lim_{a\to 0}$, we recover the continuum theory (in the case of QCD, the Y-M action).
Consider for this purpose the Wilson loop (\ref{wilson}), where the average field $A_\mu$ is defined at the midpoint
of the link
\begin{align}
W^{1\times 1}_{\mu\nu}&={\rm Re} {\rm Tr}\Big( U_\mu(x)U_\nu(x+{\hat \mu})U^\dagger_\mu(x+{\hat \nu})U^\dagger_\nu(x)\Big)\nonumber\\
&=e^{a{\rm g}A_\mu(x+\frac{{\hat \mu}}{2})+A_\nu(x+\frac{{\hat \nu}}{2})-A_\mu(x+\frac{{\hat \mu}}{2})-A_\nu(x+\frac{{\hat \nu}}{2}))}.
\end{align}
Expanding about $x+\frac{{\hat \mu}+{\hat \nu}}{2}$ gives
\begin{align}
W^{1\times 1}_{\mu\nu}&=\exp\Big\{a^2{\rm g}(\partial_\mu A_\nu-\partial_\nu A_\mu)+\frac{a^4{\rm g}}{12}
(\partial^3_\mu A_\nu-\partial^3_\nu A_\mu)+\ldots\Big\}\nonumber\\
&=1+a^2{\rm g} F_{\mu\nu}-\frac{a^4{\rm g^2}}{2} F_{\mu\nu}F^{\mu\nu} +{\it O}(a^6)+\ldots
\end{align}
Summing over the Lorentz indices we can write
\begin{align}
S[U]&=\frac6{\rm g^2}\sum_{x}\sum_{\mu<\nu}{\rm Re}{\rm Tr}\frac13\left(1-W^{1\times 1}_{\mu\nu}\right)\nonumber\\
&=\frac{a^4}{4}\sum_{x}\sum_{\mu<\nu}F_{\mu\nu}F^{\mu\nu}\to \frac14\int{\rm d}^4x\,F_{\mu\nu}F^{\mu\nu}.
\end{align}
Historically, lattice calculations are generally presented in terms of the coupling $\beta=6/{\rm g^2}$ (for $SU(3)$).


\section{Double BRST on the lattice}
\label{DBL}

In the covariant continuum formulation of gauge theories, in terms of
local field systems, one has to deal with the redundant degrees of
freedom due to gauge invariance. Within the language of local quantum
field theory, the machinery for that is based on the so-called
Becchi-Rouet-Stora-Tyutin (BRST) symmetry which is a global symmetry
and can be considered the quantum version of local gauge
invariance \cite{Nakanishi:1990qm, Alkofer:2003jr}. In short, one starts out
from the representations of a BRST algebra on indefinite metric spaces
with assuming the existence (and completeness) of a nilpotent BRST
charge $Q_B$. The physical Hilbert space can then be defined as the
equivalence classes of BRST closed (which are annihilated by $Q_B$)
modulo exact states (which are BRST variations of others). In QED this
machinery reduces to the usual Gupta-Bleuler construction. For the
generalisation thereof, in non-Abelian gauge theories, all is well in
perturbation theory also. Beyond perturbation theory, however, there
is a problem with such a construction that has not been fully and
comprehensively addressed as yet. It relates to the famous Gribov
ambiguity \cite{Gribov:1977wm}, the existence of so-called Gribov copies
that satisfy the Lorenz condition \cite{Jackson:2001ia} 
(or any other local gauge fixing condition) but are related by gauge
transformations, and are thus physically equivalent. As a result of
this ambiguity, the usual definitions of a BRST charge fail to be
globally valid.   

A rigorous non-perturbative framework is provided by lattice gauge
theory. Its strength and beauty derives from the fact that
gauge-fixing is not required. However, in order to arrive at a
non-perturbative definition of non-Abelian gauge theories in the
continuum, from a lattice formulation, we need to be able to perform
the continuum limit in a formally watertight way. And there is the gap
in our present understanding. The same problem as described above
comes back to haunt us in another dress when attempting to fix a gauge
via BRST formulations on the lattice. There it is known as the
Neuberger problem which asserts that the expectation value of any
gauge invariant (and thus physical) observable in a lattice BRST
formulation will always be of the indefinite form 0/0
\cite{Neuberger:1986xz}. 

The BRST algebra requires the introduction of further unphysical
degrees of freedom. These are the Faddeev-Popov ghosts and anti-ghosts
which violate the Spin-Statistics Theorem of local quantum field
theory on positive definite metric (Hilbert) spaces. Contrary to what
the name anti-ghost might suggest, however, in the usual linear
covariant gauges the treatment of ghosts and anti-ghosts is completely
asymmetric. On the other hand, it is also known for many years that it
is possible to extend the BRST algebra to be entirely symmetric
w.r.t. ghosts and anti-ghosts. This additional symmetry arises
naturally in the Landau gauge but can also be extended to more general
gauges the so-called Curci-Ferrari gauges at the expense of quartic
ghost self-intertactions. The most interesting feature of these gauges
four our purpose, however, is that they allow the introduction of a
mass term for ghosts \cite{Curci:1976ar}. While such a Curci-Ferrari mass
$m$ breaks the nilpotency of the BRST and anti BRST charges, which is
known to result in a loss of unitarity and which therefore meant that
this relatively old model received little attention for many years, it
also serves to regulate the Neuberger zeroes in a lattice formulation. 
In \cite{Kalloniatis:2005if} this was exemplified in a simple Abelian toy-model
where the zeroes in the numerator and denominator of expectation
values become proportional to $m^2$ and allow to compute a 
finite value for $m^2\to  0 $ via l'Hospital's rule. 

For the $SU(N)$ gauge theory on a finite four-dimensional lattice things
are naturally much more complicated than in the toy model. In this
Chapter we developed a full lattice formulation of the time-honored
model by Curci and Ferrari with its decontracted double BRST/anti-BRST and
ghost-mass term, as announced in \cite{Ghiotti:2006pm}. We first extend
Neuberger's no-go-theorem to include the ghost/anti-ghost symmetric
case of the non-linear covariant Curci-Ferrari gauges for $m^2=0$, a
case originally excluded by Neuberger. At non-vanishing
Curci-Ferrari mass the partition function of the model used as the
gauge-fixing device can be shown to be polynomial in $m^2$ and thus
non-vanishing. In this way regularising the Neuberger zeroes, the
leading power of that polynomial can be extracted from a suitable
number of derivatives (w.r.t.~$m^2$) before the limit $m^2\to 0$ is
taken, in the spirit of l'Hospital's rule. This gives rise to a
modified lattice BRST model without Neuberger problem. 

For the topological lattice formulation of the double BRST symmetry of
the ghost/anti-ghost symmetric covariant gauges we start out from the
standard gauge-fixing functional $V_U[g]$ of covariant gauges which
here assumes the role of a Morse potential on a gauge orbit,
\begin{equation}
  V_U[g] \, = \, - \frac{1}{2\rho} \, \sum_i \sum_{j\sim i} \mbox{tr}\,
  U^g_{ij} \, = \, - \frac{1}{\rho} \sum_{x,\, \mu} \mbox{Re tr}\,
  U_{x,\mu}^g \; . 
\end{equation}
Here, in the first form, $U_{ij} \in SU(N)$ is the directed link variable 
connecting nearest neighbour sites $i$ and $j$.  The sum $j\sim i$
denotes summation over all nearest neighbours $j$ of site $i$. We assume
periodic boundary conditions.  
The double sum thus runs twice over all links $\langle ij\rangle$, and
with $U_{ij}^\dagger = U_{ji}$ it is therefore equivalent to the
simple sum over links in the second form, where $U_{x,\mu} $ stands
for the same link field $U$ at position $x$ in direction $\mu$. The
constant $\rho$ is the normalisation of the $SU(N)$ generators $X$. We use
anti-Hermitian   $ [ X^a,X^b] = f^{abc} X^c$ with tr$X^a X^b= -\rho \,
\delta^{ab}$. We explicitly only need the fundamental representation,
where $\rho=\rho_{\mbox{\tiny fund}} = 1/2$.

As usual, under gauge transformations the link variables $U$ transfrom
\begin{equation} 
   U_{ij} \, \to \,  U_{ij}^g \, = \, g_i^\dagger U_{ij} g_j \; .
\end{equation}
BRST transformations $s$ and anti-BRST transformations $\bar s$ in the
topological setting do not act on the link variables $U$ directly,
but on the gauge transformations $g_i$ like infinitesimal 
right translations in the gauge group 
with real ghost and anti-ghost Grassmann fields
$c_i^a$, $\bar c_i^a$ as parameters, respectively,
\begin{equation}
   s g \, = \, g \, X^a c^a \, = \, g c  \; , \;\;    
   \bar s g  \, = \, g \, X^a \bar c^a \, = \, g \bar c \; , 
\end{equation}
where we introduced Lie-algebra valued, anti-Hermitian ghost
fields $c_i \equiv X^a c_i^a$ with $c_i^\dagger = - c_i$,
and analogous anti-ghost fields $\bar c_i\equiv X^a \bar c_i^a$.
For consistency, we furthermore require
\begin{equation}
   s g^\dagger \, = \, (s g)^\dagger \, = \, - c g^\dagger  \; , \;\;    
   \bar s g^\dagger  \, = \, (\bar s g)^\dagger \, = \, - \bar c
   g^\dagger 
\; . 
\end{equation}
For the gauge-transformed link variables this then implies
\begin{equation} 
s U^g_{ij} \, = \, - c_i U_{ij}^g + U_{ij}^g c_j \; , \;\; 
\bar s U^g_{ij} \, = \, - \bar c_i U_{ij}^g + U_{ij}^g \bar c_j \; .
\label{BRSUg}
\end{equation}

\vspace*{-1pt}
\noindent
The BRST transformations for (anti)ghosts and Naka\-nishi\--Lautrup fields $b$
are straightforward lattice analogues (per site) of their continuum
counterparts,
\begin{eqnarray}
  s c^a &=& - \frac{1}{2} (c \times c)^a  \; , \label{BRSc} \\
  s \bar c^a &=& b^a - \frac{1}{2} (\bar c \times c)^a \; ,\label{BRScb}\\
  s b^a &=&  - \frac{1}{2} (c \times b)^a  
 - \frac{1}{8} \big((c \times c) \times \bar c  \big)^a \;
  . \label{BRSb}  
\end{eqnarray}
The relatively obvious notation of using the ``cross-product'' 
herein refers to the structure constants for $SU(N)$, {\it e.g.}
$(\bar c \times c)^a \equiv f^{abc} \bar c^b c^c$.

In the ghost/anti-ghost symmetric gauges as considered here, the
anti-BRST variations are obtained by  substituting $c \to \bar c$ and
$\bar c \to - c$ according to Faddeev-Popov conjugation. Thus,
\begin{eqnarray}
  \bar s c^a &=& - b^a - \frac{1}{2} (\bar c \times c)^a \; ,
    \label{aBRSc} \\
  \bar s \bar c^a &=&  - \frac{1}{2} (\bar c \times \bar c)^a  \;
  ,  \label{aBRScb} \\
  \bar s b^a &=&  - \frac{1}{2} (\bar c \times b)^a  
 + \frac{1}{8} \big((\bar c \times \bar c) \times c  \big)^a \; .
  \label{aBRSb} 
\end{eqnarray}
The action of the topological lattice model for gauge fixing a la
Faddeev-Popov with double BRST invariance can then be written in
compact form as
\begin{equation} 
S_{\mbox{\tiny GF}} \, = \, i \, s \bar s \, \Big( V_U[g] + i
\frac{\xi}{2\rho}  
\sum_i \mbox{tr}\, \bar c_i c_i \Big) \; . \label{S_GF}
\end{equation}
This is the lattice counterpart of the continuum gauge-fixing 
Lagrangian
\begin{equation}
 \mathcal{L}_{\mbox{\tiny GF}} \, = \, \frac{i}{2} \, s \bar s
 \big( A_\mu^a A_\mu^a  - i \xi \bar c^a c^a \big)\;\; \mbox{with} \;\; 
S_{\mbox{\tiny GF}} \, =  \int d^Dx \,  \mathcal{L}_{\mbox{\tiny GF}}
 \label{L_GFcont}
\end{equation}
in $D$ Euclidean dimensions.

Performing the anti-BRST variation first, we obtain
\begin{eqnarray}
  \bar s \, V_U[g] &=& 
 \frac{1}{2\rho} \, \sum_i \sum_{j\sim i} \mbox{tr}\,\big(
 \bar c_i  U^g_{ij} - \bar c_j  U^g_{ij} \big) \\
 &=&   \frac{1}{2\rho} \, 
\sum_i \sum_{j\sim i}\, \bar c_i^a \, \mbox{tr}\,\big(
 X^a(  U^g_{ij} -   U^g_{ji}) \big) \nonumber \\
 &=&  -  \sum_i  \bar c_i^a  F_i^a(U^g) \; , \nonumber
\end{eqnarray}
where
\begin{equation} 
 F^a_i(U^g) \, = \,  -\frac{1}{2\rho} \,  \sum_{j\sim i}
 \mbox{tr}\,\big( X^a(  U^g_{ij} -   U^g_{ji}) \big) 
\end{equation}
is, of course, the standard gauge-fixing condition of covariant gauges
which reduces in the continuum limit to
\begin{equation} 
 F^a_i(U^g) \, \stackrel{a\to 0}{\longrightarrow} 
 \,  a^2\, \partial_\mu {A_\mu^g}^a  \, +\, \mathcal{O}(a^4) \; .
\end{equation}
As we know, the gauge-fixing condition is derived by considering the first derivative
of the Morse potential $V_U[g]$ with repsect to the group element $g$. However, since this functional
derivation takes into account matrix elements, its computation requires special attention to
the matrix-order, and so it can turn out to be ambiguous. To avoid this complication
we define the one-parameter subgroup of the Lie group as
$g_t(x)=e^{t\omega_x}$, with $t \in \mathbb R$. Through this parameter $t$ we then have to compute
a simple 1-dimensional derivation, bypassing the non-commuting nature of the matrix elements.
Adopting standard notation $(x,\mu)$, we write
\begin{align}
\frac{{\rm d}}{{\rm d}t}V_U[g]&=-\frac1{\rho}\sum_{x,\mu}\,\mbox{Re tr}[(\omega_x-\omega_{x+\mu}){}^{g_t}\! U_{x,\mu}]\nonumber\\
&=-\frac1{2\rho}\sum_{x,\mu}\,\mbox{tr}[\omega_x({}^{g_t}\! U_{x,\mu}-{}^{g_t}\! U_{x,\mu}^\dagger)
-\omega_{x+\mu}({}^{g_t}\! U_{x,\mu}-{}^{g_t}\! U_{x,\mu}^\dagger)]\nonumber\\
&=-\frac1{\rho}\sum_{x,\mu}\,\mbox{Re tr}\left\{\omega_x\left[\frac12({}^{g_t}\! U_{x,\mu}-{}^{g_t}\! U_{x,\mu}^\dagger)
-\frac12({}^{g_t}\! U_{x-\mu,\mu}-{}^{g_t}\! U_{x-\mu,\mu}^\dagger)
\right]\right\}.
\end{align}
From now on we will drop $g_t$ from the link notation. Defining $A_{x,\mu}=\frac12(U_{x,\mu}-U_{x,\mu}^\dagger)_{\rm traceless}=A_{x,\mu}^a X^a$,
we then have
\begin{align}
\frac{{\rm d}}{{\rm d}t}V_U[g]&=-\frac1{\rho}\sum_{x}\omega^a_x\,\sum_\mu{\rm tr}\{X^a(A_{x,\mu}-A_{x-\mu,\mu})\}\nonumber\\
&=\sum_x\omega^a_x\,\sum_\mu(A^a_{x,\mu}-A^a_{x-\mu,\mu})\nonumber\\
&=(\omega, \nabla \cdot A).
\end{align}

Turning back to the BRST variations, with Eqs.~(\ref{aBRSc}), (\ref{aBRScb}) we furthermore have
\begin{equation}
\bar s \, \big( \bar c^a c^a \big) \, = \, \bar c^a b^a \; ,
\label{sbcbc}
\end{equation}
and therefore, for the gauge-fixing action, we obtain the alternative form
\begin{equation}
S_{\mbox{\tiny GF}} \, = \, -i \, \sum_i \,
s  \Big(  \bar c_i^a  \, \big( F_i^a(U^g) \, + \frac{i\xi}{2} \, 
  b_i^a \big)  \Big) \; . \label{S_GF-standard}
\end{equation}
As in the continuum formulation, in this form it looks exactly like
the gauge-fixing action of standard Faddeev-Popov theory for the
linear covariant gauge. The specific features of the ghost/anti-ghost
symmetric framework show when working out the remaining BRST
variation. From the first term we have ({\it i}),
\begin{align}
\sum_i \big( s \bar c_i^a \big) \, F_i^a  &=  
-  \frac{1}{2\rho} \, \sum_i \sum_{j\sim i}\, \mbox{tr}\,\big(   
 b_i  (  U^g_{ij} -   U^g_{ji}) \big) \label{var1} \nonumber\\
&+
\frac{1}{4\rho} \, \sum_i \sum_{j\sim i}\, \mbox{tr}\,\big(   
 \{ \bar c_i,\, c_i\}\,  (  U^g_{ij} -   U^g_{ji}) \big) \;.
\end{align}
The first term implements the gauge-fixing condition as in standard
Faddeev-Popov theory. The second term, containing the anticommutator $
\{ \bar c,\, c\}$, is characteristic of ghost/anti-ghost symmetry
because it combines with the remaining quadratic ghost terms to produce
a Hermitian Faddeev-Popov operator (for any gauge parameter $\xi$). To
see this explicitly, consider ({\it ii}),
\begin{align}
\sum_i \bar c_i^a  \, \big( s F_i^a )  &= 
\frac{1}{2\rho} \, \sum_i \sum_{j\sim i}\, \mbox{tr}\,\big(   
 \bar c_i c_i  U^g_{ij} \label{stdFP}\nonumber\\
 &-   \bar c_i  U^g_{ij} c_j 
 +\,  c_j U^g_{ji} \bar c_i -  c_i \bar c_i  U^g_{ji} \big) \; , 
\end{align}
so that the difference ({\it i}) - ({\it ii}) yields
\begin{align}
\label{sMFP}
\sum_i s \big( \bar c_i^a  \, F_i^a \big) &=  
-  \frac{1}{2\rho} \, \sum_i \sum_{j\sim i}\, \mbox{tr}\,\big(   
 b_i  (  U^g_{ij} -   U^g_{ji}) \big) \nonumber\\
&+\frac{1}{2\rho} \sum_i \sum_{j\sim i}\, \mbox{tr}\,\big(   
\bar c_i  U^g_{ij} c_j   
 -   c_i U^g_{ij} \bar c_j -  [\bar c_i,\, c_i ] \,  \frac{1}{2} (  U^g_{ij} +   U^g_{ji} ) 
+ c_i \bar c_i \,  \frac{1}{2} (  U^g_{ij} +   U^g_{ji} )
 \big) \nonumber \\
&\equiv \sum_i b_i^a F_i^a \, + \, \sum_{i,\,j} \bar c_i^a \,
     {M_{\mbox{\tiny FP}}}^{ab}_{ij} \,  c_j^b  
\end{align}
which defines the lattice Faddeev-Popov operator $M_{\mbox{\tiny FP}}$
of the ghost/anti-ghost symmetric Curci-Ferrari gauges.
Following the same method we used to derive the gauge-fixing condition, we want here to show how to derive the Faddeev-Popov operator
$M_{\mbox{\tiny FP}}$, which is obtained by the second derivative with respect the real parameter $t$ of the Morse potential $V_U[g]$
as follows
\begin{align}
\frac{{\rm d}^2}{{\rm d}t^2}V_U[g]&=\sum_x\omega_x^a\,\sum_\mu\frac{{\rm d}}{{\rm d}t}(A^a_{x,\mu}-A^{a\dagger}_{x,\mu})\nonumber\\
&=\frac1{\rho}\sum_x\omega^a_x\,\sum_\mu\left\{\frac1{4\rho}\left({\rm tr}(U_{x,\mu}+U^\dagger_{x,\mu})(\omega^a_{x+\mu}-\omega^a_x)\right.
-{\rm tr}(U_{x-\mu,\mu}+U^\dagger_{x-\mu,\mu})(\omega^a_x-\omega^a_{x-\mu})\right)\nonumber\\
&-\frac{\rho}2f^{abc}\left(A^c_{x,\mu}(\omega^b_{x+\mu}+\omega^b_x)-A^c_{x-\mu,\mu}(\omega^b_x+\omega^b_{x-\mu})\right)\nonumber\\
&\left.+\frac{i}4d^{abc}\left({\rm tr}(X^c(U_{x,\mu}+U^\dagger_{x,\mu}))(\omega^b_{x+\mu}-\omega^b_x)
{\rm tr}(X^c(U_{x-\mu,\mu}+U^\dagger_{x-\mu,\mu}))(\omega^b_{x}-\omega^b_{x-\mu})
\right)\right\}.
\end{align}
Changing notation to $U_{ij}\equiv U_{x,\mu}$ and $A_{ij}\equiv A_{x,\mu}$ we can write
\begin{align}
\frac{{\rm d}^2}{{\rm d}t^2}V_U[g]&=\frac1{\rho}\sum_i
\omega_i^a\,\sum_{j\sim i}\frac1{4\rho}\left\{{\rm tr}(U_{ij}+U_{ji})(\omega^a_j-\omega^a_i)
-\frac{\rho}2f^{abc}\,A^c_{ij}(\omega^b_{j}+\omega^b_i)\right.\nonumber\\
&\left.+\frac{i}4d^{abc}\,{\rm tr}(X^c(U_{ij}+U_{ji}))(\omega^b_{j}-\omega^b_i)\right\}.
\end{align}
Therefore, the Faddeev-Popov operator can be obtained as a double derivative with respect to the gauge function $\omega^a$
of $\frac{{\rm d}^2}{{\rm d}t^2}V_U[g]$
\beq
(M_{\mbox{\tiny FP}})_{ij}^{ab}=\frac{\partial}{\partial \omega^b_j}\frac{\partial}{\partial \omega^a_i}\frac{{\rm d}^2}{{\rm d}t^2}V_U[g].
\eeq
Note that the terms in (\ref{stdFP}) can be written in the form
\begin{align}
\sum_i \bar c_i^a  \, \big( s F_i^a )  &=
\frac{1}{4\rho} \, \sum_{i,\,j\sim i}\, \bar c_i^a \, \Big\{ \,
\mbox{tr}\,\big(   [X^a,X^b] (  U^g_{ij} -  U^g_{ji} ) \big) \, (c^b_i
+ c^b_j )  \nonumber\\
& +\, \mbox{tr}\,\big(   \{X^a,X^b\} (  U^g_{ij} + U^g_{ji} ) \big) \, (c^b_i
- c^b_j )   \, \Big\}
\end{align}
This, of course, corresponds to the widely used Faddeev-Popov operator
of lattice Landau gauge, as first derived in \cite{Zwanziger:1993dh}. It
differs by the quadratic ghost terms in (\ref{var1}) from the
ghost/anti-ghost symmetric one, $M_{\mbox{\tiny FP}}$  in
(\ref{sMFP}), which can be written in the alternative form,
\begin{align}
\label{edBRSFP}
  \sum_{i,\,j} \bar c_i^a \,
     {M_{\mbox{\tiny FP}}}^{ab}_{ij} \,  c_j^b &=
     - \frac{1}{4\rho} \sum_{x,\, \mu} \, \Big\{
 \mbox{tr} \big( \{X^a,X^b\}  (
     U^g_{x,\,\mu} + U^{g\,\dagger}_{x,\,\mu} )\big) \,\times
  (\bar c_{x+\hat \mu}^a - \bar c_x^a ) (c_{x+\hat \mu}^b - c_x^b ) \nonumber\\
&+ \,  \mbox{tr} \big( [X^a,X^b]  (
     U^g_{x,\,\mu} - U^{g\,\dagger}_{x,\,\mu} )\big) \,\times
  \big(\bar c_{x}^a  (c_{x+\hat \mu}^b - c_x^b )  -    
     (\bar c_{x+\hat \mu}^a - \bar c_x^a )  c_x^b \big)
   \Big\}\,,
\end{align}
In the continuum limit this reduces to the ghost/anti-ghost
symmetric Faddeev-Popov operator
\begin{equation} 
 {M_{\mbox{\tiny FP}}}^{ab}_{ij} \, \stackrel{a\to 0}{\longrightarrow} 
 \,  - a^2\frac{1}{2} \,\big(\partial  D^{ab} + D^{ab}
 \partial \big) \, \delta(x-y)  \, +\, \mathcal{O}(a^4) \; .\nonumber
\end{equation}
To complete the derivation of the gauge-fixing action in the
ghost/anti-ghost symmetric framework, we furthermore need work out the
BRST variation of $s \bar s (\bar c^a c^a) = s (\bar c^a b^a)$ from
(\ref{BRSc})-(\ref{BRSb}). This, however, is done in exactly the same
away as in the continuum, the result is ({\it iii}),
\begin{equation}
  s \big( \bar c^a b^a \big) \, = \, b^a b^a \, +\, \frac{1}{4} 
\, (\bar c \times c )^2 \; . \label{scbb}
\end{equation}
Putting together all terms from ({\it i}) to ({\it iii}) we obtain the
full gauge-fixing action with extended double BRST invariance on the
lattice in the form,
\begin{align}
\label{SGFexpl}
S_{\mbox{\tiny GF}} = \sum_i \,  \Big\{ - i b^a_i  F_i^a(U^g)
\, - {i}\,\bar c^a_i  {M_{\mbox{\tiny FP}}}_i^a[c]
+\, 
\frac{\xi}{2} b_i^a b_i^a \, + \, \frac{\xi}{8}\, 
(\bar c_i \times c_i)^2 \, \Big\} \; ,
\end{align}
where we introduced the short-hand notation that
\begin{align}
{M_{\mbox{\tiny FP}}}_i^a[c] &\equiv  -\frac{1}{4\rho}       
 \, \sum_{j\sim i}\,  \Big\{ \,
\mbox{tr}\,\big(   [X^a,X^b] (  U^g_{ij} -  U^g_{ji} ) \big) \, c^b_j
 \nonumber\\ 
& +\, \mbox{tr}\,\big(   \{X^a,X^b\} (  U^g_{ij} + U^g_{ji} ) \big) \, (c^b_i
- c^b_j )   \, \Big\}  \; , 
\end{align}
which corresponds to the ghost/anti-ghost symmetric
Faddeev-Popov operator in (\ref{edBRSFP}), in particular, we have
\begin{equation}
 \sum_{i} \bar c_i^a \,
     {M_{\mbox{\tiny FP}}}^{a}_{i}[c]  \, = \, 
 \sum_{i,\,j} \bar c_i^a \,
     {M_{\mbox{\tiny FP}}}^{ab}_{ij} \,  c_j^b \, .
\end{equation}
The full symmetry of the ghost/anti-ghost symmetric Curci-Ferrari
gauges \cite{Curci:1976ar, Thierry-Mieg:1979kh} 
is given by a semidirect product of a global $SL(2,\RR)$, which
includes ghost number and Faddev-Popov conjugation, with
the BRST/anti-BRST symmetries as used above\footnote{Also see Appendix A of
  Ref.~\cite{Alkofer:2003jr}.}.  
This is the global symmetry of the Landau gauge, and it is sometimes 
referred to as extended BRST symmetry. 

Among the general class of all
covariant gauges \cite{Thierry-Mieg:1982un}, with a Lagrangian which is 
polynomial in the fields, Lorentz, globally gauge and BRST invariant,
and renormalisable in $D=4$, the ghost/anti-ghost symmetric case is
special and interesting in that it allows to smoothly connect to the
Landau gauge for $\xi \to 0$, without changing the global symmetry
properties.    

In particular, introducing with \cite{Thierry-Mieg:1982un} a second gauge
parameter $\beta \in [0,\, 1] $, to interpolate between the various
generalised covariant gauges,  the linear covariant gauges of standard Faddeev
Popov theory correspond to the line $\beta = 0$ in the two
gauge-parameter plane $(\xi,\,\beta)$. Along this line, the global
symmetry changes abruptly when reaching the Landau gauge limit; and for
$\beta = 1$, one obtains a mirror image of standard Faddeev-Popov 
theory with the roles of ghosts and anti-ghosts interchanged. The
ghost/anti-ghost symmetric gauges discussed here 
then correspond to the line $\beta = 1/2$. For $\xi = 0$ the
distinction is an illusion. The whole interval for $\beta \in [0,1]$
at $\xi =0$ is equivalent and corresponds to the Landau gauge. The important
difference is, however, that the $SL(2,\RR) $ symmetric line at
$\beta = 1/2$ provides a unique class of covariant gauges which share the
full extended BRST symmetry of the Landau gauge for any value of
$\xi$. The limit $\xi \to 0$ is thus a smooth one, as far as the
symmetries are concerned, only along this line. The price to pay are
the quartic ghost self-interactions in (\ref{SGFexpl}) which again
vanish only in the Landau gauge limit. 

For a further discussion of the general ghost creating gauges, and their
geometrical interpretation, see \cite{Thierry-Mieg:1979kh}. The one-loop
renormalisation was first discussed in \cite{Thierry-Mieg:1982un}, for
explicit calculations of renormalisation constants and anomalous
dimensions of the ghost/anti-ghost symmetric case up to including
the three-loop level, see \cite{deBoer:1995dh, Gracey:2002yt}.
The Dyson-Schwinger equations of these gauges were studied in
\cite{Alkofer:2003jr}.

\section{The Neuberger problem}

Following Neuberger, we introduce an auxiliary parameter $t$ in the 
Euclidean partition function to be used as the gauge-fixing device via the
Faddeev-Popov procedure of inserting unity into the unfixed partition
function of $SU(N)$ lattice gauge theory. The gauge-fixing action of 
the double BRST invariant model given by (\ref{S_GF}) consists of two
terms both of which are separately BRST (and anti-BRST)
exact. Multiplying the $1^\mathrm{st}$ term in (\ref{S_GF}) by the
real parameter $t$ amounts to a mere redefinition of the Morse
potential which should have no further effect. We can therefore write the
gauge-fixing partition function with double BRST,
\begin{eqnarray}
Z_{\mbox{\tiny GF}}(t) &=& \int d[g, b,\bar c, c]  
\label{Z_GF}
\exp\Big\{-  i s \bar s
\Big( t \, V_U[g] \, + \,  i\frac{\xi}{2\rho} 
 \sum_i \mbox{tr}\, \bar c_i c_i \Big)  
\Big\} \; ,
\end{eqnarray}
which is independent of the set of link variables $\{U\}$ and the
gauge parameter $\xi$ because of its topological nature. Moreover, the $t$
independence is really not different from the $\xi$ independence here,
and it is thus rather obvious. Explicitly, the derivative with respect
to $t$ (or $\xi$) produces the expectation value of a BRST exact
operator which vanishes, {\it i.e.}, 
\begin{align}
\label{t-indep}
   Z'_{\mbox{\tiny GF}}(t) &=\int d[g, b,\bar c, c]
\left(-is\bar s V_U[g]\right)
\exp\Big\{-  i s \bar s
\Big( t \, V_U[g] \, + \,  i\frac{\xi}{2\rho} 
 \sum_i \mbox{tr}\, \bar c_i c_i \Big)  
\Big\}
\nonumber\\
&=0,  
\end{align}
provided the BRST operators are nilpotent (property that we will see lost in the case of the massive Curci-Ferrari gauge).
At $t\!=\!0$ on the other hand, we obtain with (\ref{sbcbc}) and
(\ref{scbb}),  
\begin{eqnarray} 
\label{GaussBonnet}
Z_{\mbox{\tiny GF}}(0) &=& \mathcal{N} \int d[b,\bar c, c]  \, \times
\exp\Big\{-  \sum_i \,  \left( 
\frac{\xi}{2} b_i^a b_i^a \, + \, \frac{\xi}{8}\, 
(\bar c_i \times c_i)^2 \right) \,
\Big\} \; ,
\end{eqnarray}
where the volume of the gauge group on the lattice, from
the invariant integrations $\prod_i dg_i$ via the Haar measure over
$g_i\in SU(N)$ per site $i$, is absorbed in the constant $\mathcal{N}$.
The Gaussian integrations over the Nakanishi-Lautrup fields $b$ are also
well-defined and produce a factor $(2\pi/\xi)^{(N^2\!-\!1)/2}$ per site.

One might be tempted to conclude at this point that the quartic ghost
self-interactions in (\ref{GaussBonnet}) might remove the
uncompensated Grassmann integrations of the linear covariant gauges
where no such self-interactions occur. The ghost/anti-ghost 
integrations at $t=0$ also factorise into independent 
integrations $d\bar c^a_i d c^a_i$  over $2(N^2-1)$ Grassmann
variables per site. For $N=3$, for example, the $4^\mathrm{th}$ order
term of the exponential in (\ref{GaussBonnet}) produces a monomial in
$\bar c^a_i$, $c^a_i$ which contains each of these $16$ Grassmann
variables exactly once, so that their integration might produce 
a non-vanishing result. This is not the case, however. Working out the 
prefactor of this monomial, as we will do explicitly in the more
general case with including a non-vanishing Curci-Ferrari mass $m$
below, one finds that the prefactor of this term in
(\ref{GaussBonnet}) vanishes in the masssless case
and thus,
\begin{equation}
  Z_{\mbox{\tiny GF}}(0) \, = \, 0\;. \label{Z-zero}
\end{equation}
Because of the $t$-independence (\ref{t-indep}), this implies the
vanishing of the gauge-fixing partition function (\ref{Z_GF}) of the
ghost/anti-ghost or $SL(2,\RR)$ symmetric formulation with double BRST
invariance in the same way as that of standard Faddeev-Popov theory 
observed in \cite{Neuberger:1986xz}. As for the latter, the sign-weighted
sum over all Gribov copies, as originally proposed to generalise the
Faddeev-Popov procedure in presence of Gribov copies
\cite{Hirschfeld:1978yq, Fujikawa:1979ay}, vanishes. 
 
This cancellation of Gribov copies is well-known \cite{Sharpe:1984vi}. 
The fact that it also arises here, in the ghost/anti-ghost symmetric
formulation with its quartic self-interactions, directly relates to the
topological interpretation \cite{Baulieu:1996rp, Schaden:1999ew} 
of the Neuberger zero: $Z_{\mbox{\tiny GF}}$ can be viewed as the
partition function of a Witten-type topological model 
to compute the  Euler character $\chi$  of the gauge
group. On the lattice the gauge group is a direct product of $SU(N)$'s
per site, and because the Euler character factorises,
\[
Z_{\mbox{\tiny GF}} = \chi(SU(N)^{\#{\rm sites}}) =
\chi(SU(N))^{\#{\rm sites}} = 0^{\# {\rm sites}}\; . 
\]
For $t=0$ the action in (\ref{Z_GF}) decouples from the link-field
configuration and $Z_{\mbox{\tiny GF}}(0) $, albeit computing the same
topological invariant, has of course no effect in terms of fixing a
gauge. In the present formulation, with $Z_{\mbox{\tiny GF}}(0) $ in
(\ref{GaussBonnet}), the independent Grassmann integrations per site
of the quartic-ghost term which contains the curvature of $SU(N)$ each 
compute its Euler character via the Gauss-Bonnet theorem
\cite{Birmingham:1991ty}. This explicitly produces one factor of zero per
site on the lattice. And it provides the topological explanation for the 
vanishing of the prefactor of the corresponding monomial of degree
$2(N^2-1)$ in the Grassmann variables  $\bar c$, $c$, which could
otherwise exist in the expansion of the exponential in
(\ref{GaussBonnet}) for all odd $N$. For $N=3$, for example, the  
zero in this prefactor arises, upon normalordering, from a cancellation of     
368 non-vanishing individual terms when expanding the square of the
square of the quartic ghost self-interaction. This cancellation
would be rather unnatural to arise accidentally, without such explanation.

The vanishing of the gauge-fixing partition function
at $t=0$ part in Neuberger's argument, in the ghost/anti-ghost
symmetric gauges with $SL(2,\RR) \rtimes $ double BRST symmetry,
therefore most directly reflects the topological origin of the
Neuberger zero. Eq.~(\ref{GaussBonnet}) precisely represents a product
of one Gauss-Bonnet integral expression for $\chi(SU(N))$ per
site of the lattice. 

Note that the gauge parameter $\xi$  can be removed completely from 
the expression for $Z_{\mbox{\tiny GF}}(0)$ in
Eq.~(\ref{GaussBonnet}) by a rescaling $\sqrt{\xi}\,  b \to b$ and
$\sqrt[4]{\xi}\, \bar c \to \bar c$, $\sqrt[4]{\xi}\, c \to c$, which
leaves the integration measure unchanged. The same rescaling for the
full gauge-fixing partition function $Z_{\mbox{\tiny GF}}(t)$ in
(\ref{Z_GF}), which amounts to replacing the action in $S_{\mbox{\tiny
    GF}}$ in (\ref{SGFexpl}) by
\begin{eqnarray}
S_{\mbox{\tiny GF}}(t) = \sum_i \,  \Big\{ - i t  b^a_i  F_i^a(U^g)
\, - {i} t \,\bar c^a_i  {M_{\mbox{\tiny FP}}}_i^a[c]
 \label{SGFtexpl}
+\, 
\frac{\xi}{2} b_i^a b_i^a \, + \, \frac{\xi}{8}\, 
(\bar c_i \times c_i)^2 \, \Big\} \; ,
\end{eqnarray}
furthermore shows that $t$ and $\xi$ really represent a single
parameter $t/\sqrt{\xi} $. Setting $t=0$ in Neuberger's argument
is therefore the same as the $\xi \to \infty $ limit which is usually
what is considered as the Gauss-Bonnet limit in topological quantum
field theory \cite{Birmingham:1991ty}.  As mentioned above, there is no
gauge-fixing in this limit, but it provides a simple way to compute
the value (zero here) of the partition function which is independent
of $t/\sqrt{\xi}$. 

In the opposite limit, that of the Landau gauge $\xi \to 0$ or
$t/\sqrt{\xi} \to \infty $, of course, $Z_{\mbox{\tiny GF}}(t)$ still
reduces to the sign-weighted sum over all Gribov copies as usual
\cite{Hirschfeld:1978yq, Fujikawa:1979ay},   
\begin{equation} 
  Z_{\mbox{\tiny GF}}(t) \to \sum_{\mbox{\tiny copies $\{g^{(i)}\!\}$}} 
  \mbox{sign} \,\big( \mbox{det} \, M_{\mbox{\tiny FP}}(U^{g^{(i)}}) \big) \; ,
\end{equation}
which because of the $t$ (and $\xi$) independence (\ref{t-indep})
thus computes the same topological zero
\cite{Sharpe:1984vi, Baulieu:1996rp, Schaden:1999ew}, in this case via the 
Poincar\'e-Hopf theorem \cite{Birmingham:1991ty}. 

\bigskip

\section{The massive Curci-Ferrari model on the lattice}

In the previous section we have seen that the quartic ghost
self-interactions of the  $SL(2,\RR) \rtimes $ double BRST symmetric
Curci-Ferrari gauges have no effect on the disastrous conclusion of the
0/0 problem in lattice BRST. They rather serve to reveal most clearly 
the topological origin of this problem. 

We will demonstrate explicitly below that this zero can be
regularised, however, by introducing a Curci-Ferrari mass $m$, as
proposed in \cite{Kalloniatis:2005if, Ghiotti:2006pm}. The gauge-fixing 
action $S_{\mbox{\tiny GF}}$ is thereby once more replaced by 
\begin{eqnarray}
\label{S_mGF}
S_{\mbox{\tiny mGF}}(t) \, = \, i \,( s \bar s-im^2) \, \Big( t\, 
 V_U[g] \, + \, i\xi \sum_i \mbox{tr}\, \bar c_i c_i \Big) 
\end{eqnarray}
(where we dropped in the $2^\mathrm{nd}$ term 
the factor $1/(2\rho) = 1$, in the  fundamental representation).   
The BSRT and anti-BRST transformations of $U^g$, $\bar c $ and $c$ in
Eqs.~(\ref{BRSUg}), (\ref{BRSc}), (\ref{BRScb}) and (\ref{aBRSc}),
(\ref{aBRScb}) of Sect.~\ref{DBL} remain unchanged. Those for the
Nakanishi-Lautrup $b$-fields, Eqs.~(\ref{BRSb}) and (\ref{aBRSb}), are
replaced by \cite{Thierry-Mieg:1979kh},
\begin{eqnarray}
  s b^a &=& im^2 \, c^a \, - \frac{1}{2} (c \times b)^a  
 - \frac{1}{8} \big((c \times c) \times \bar c  \big)^a \;
  , \label{mBRSb}  \\
  \bar s b^a &=& im^2 \, \bar c^a \, - \frac{1}{2} (\bar c \times b)^a  
 + \frac{1}{8} \big((\bar c \times \bar c) \times c  \big)^a \; .
  \label{maBRSb} 
\end{eqnarray}
In the derivation of the explicit form for $S_{\mbox{\tiny mGF}}(t)$,
using these modified anti-BRST transformations, the only
modification in comparison to Sect.~\ref{DBL}, arises from $s(\bar c^a
b^a)$ in (\ref{scbb}), which now becomes,
\begin{equation}
  s \big( \bar c^a b^a \big) \, = -i m^2 \, \bar c^a c^a \,
   + \, b^a b^a \, +\, \frac{1}{4} 
\, (\bar c \times c )^2 \; . \label{mscbb}
\end{equation}
The additional first term on the right contributes an additional
term $-i\frac{\xi m^2}{2}\bar c^a_i c^a_i$ to the gauge-fixing Lagrangian,
{\it c.f.}, Eq.~(\ref{S_GF-standard}). Together with the same
contribution from the explicit mass term in (\ref{S_mGF}) we therefore
obtain {\em twice that} as the ghost mass-term of the massive
Curci-Ferrari model (this subtlety will be worth remembering for later).    
The action of the massive Curci-Ferrari model therefore becomes,
explicitly,
\begin{align}
S_{\mbox{\tiny mGF}}(t) &=  m^2 t\, V_U[g] \, + \label{mSGFtexpl}
\sum_i \,  \Big\{ - i t  b^a_i  F_i^a(U^g)
\, - {i} t \,\bar c^a_i  {M_{\mbox{\tiny FP}}}_i^a[c]
 \nonumber\\
&+\, 
\frac{\xi}{2} b_i^a b_i^a \, - i m^2 \xi\, \bar c_i^a c_i^a  \, +
\, \frac{\xi}{8}\,  (\bar c_i \times c_i)^2 \, \Big\} \; .
\end{align}
BRST and anti-BRST transformations are no-longer nilpotent at finite
$m^2$, but we have  \cite{Nakanishi:1990qm, Curci:1976ar, Thierry-Mieg:1982un}
\begin{eqnarray} 
    s^2 = i m^2 \sigma^+ \; , && \bar s^2 = -i m^2 \sigma^- \; ,
            \nonumber \\
            s\bar s + \bar s s &=& -i m^2 \sigma^0 \; , \label{superalgebra}
\end{eqnarray}
where $\sigma^\pm $ and $\sigma^0$ generate the global $SL(2,\RR)$
including ghost number and Faddeev-Popov conjugation. The
Curci-Ferrari mass decontracts the $sl(2,\RR) \rtimes$ double BRST
algebra of the massless case to the $osp(1|2)$ superalgebra extension
of the Lie algebra of the 3-dimensional Lorentz group
$SL(2,\RR)$. Conversely, the $m^2\to 0$ limit   
is interpreted as a Wigner-Inonu contraction of the simple
superalgebra  $osp(1|2)$ \cite{Nakanishi:1990qm, Thierry-Mieg:1979kh}.
The BRST and anti-BRST invariance of the massive Curci-Ferrari action
in (\ref{S_mGF}) itself follows readily from this algebra as given in
(\ref{superalgebra}), noting that only $\bar c$ and $c$ transform
non-trivially under the $SL(2,\RR)$. 
 
We emphasise that this algebra decontraction has from the very
beginning been known to lead to a breakdown of unitarity when
attempting a BRST cohomology construction of a physical Hilbert space
in analogy to the massless case \cite{Curci:1976ar}.
In fact, explicit examples exist for states of negative norm surviving
in any such construction \cite{deBoer:1995dh, Ojima:1981fs}. They do not
belong to BRST quartets and can therefore not be removed by the
quartet mechanism \cite{Nakanishi:1990qm}. Only through the algebra
contraction by $m^2\to 0$ do these states reduce to zero norm components
which have no effect on the physical $S$-matrix elements. 

Here we deliberately do not want to interpret the mass parameter by
Curci and Ferrari as a physical mass. It rather serves to meaningfully
define a limit $m^2\to 0$ on the lattice, perhaps in parallel with the
continuum limit, to recover nilpotent (anti-)BRST transformations. 

To study the parameter dependence, we first define the partition
function of the massive Curci-Ferrari model, explicitly listing all
three parameters (even though these again really only represent 2
independent ones as we will show below),  
\begin{equation} 
Z_{\mbox{\tiny mGF}}(t,\xi,m^2) \,=\, \int d[g, b,\bar c, c]  \, 
 \exp\big\{- S_ {\mbox{\tiny mGF}}(t) \big\} \; ,
\label{Z_mGF}
\end{equation}
with $S_ {\mbox{\tiny mGF}}(t) $ from (\ref{S_mGF}) or (\ref{mSGFtexpl}). 
We note in passing that the terms proportional to $m^2$ in
the massive Curci-Ferrari action (\ref{mSGFtexpl}) are given by  
\begin{equation}
 \mathcal{O} (t,\xi) \equiv  t V_U[g] - i \xi \sum_i \bar c_i^a c_i^a
 \; , \label{LatKondOp}
\end{equation}
or, in the continuum,
\begin{equation}
  \mathcal{O} (t,\xi) =  \int d^Dx  \left( \frac{t}{2} A_\mu^a(x)
  A_\mu^a(x) - i \xi   \bar c^a(x)  c^a(x) \right) \, .
\end{equation}
For $t=1$ this coincides with the on-shell BRST invariant (at $m^2=0$) 
operator proposed by Kondo as a possible candidate for a dimension 2
condensate \cite{Kondo:2000ey}.
The doubling of the explicit ghost mass-term in (\ref{S_GF}), by the BRST
variation of $\bar c b$ in (\ref{mscbb}) as mentioned above, is
crucial here. Without this difference in the relative factor of 2 between the
two terms in $\mathcal O(t,\xi)$ and the gauge fixing functional
\begin{equation}
   -i W_{\mbox{\tiny GF}} 
     =  t V_U[g] -i  \frac{\xi}{2}  \sum_i \bar c_i^a c_i^a \; ,
\end{equation}
one could not have both, the on-shell BRST invariance of $\mathcal O$
and the gauge-fixing action in (\ref{S_GF}) from the double BRST variation $
S_{\mbox{\tiny GF}}  = s \bar s \, W_{\mbox{\tiny GF}} $, at the same time.
 
The observation that the mass terms in (\ref{mSGFtexpl}) are given by
$m^2 \mathcal O(t,\xi) $ could in principle be used to obtain the 
expectation value of Kondo's operator from the derivative 
\begin{equation} 
        \langle \mathcal{O}(t,\xi) \rangle = -\frac{\partial}{\partial
        m^2} \, \ln Z_{\mbox{\tiny mGF}}(t,\xi,m^2) \Big|_{m^2=0} \; ,
\label{mderiv}
\end{equation}
upon insertion into the unfixed partition function of lattice gauge
theory, {\it i.e.}, with taking the additional expectation value in
the gauge-field ensemble. 
As any other observable at $m^2 =0$ this expectation value
as it stands, unfortunately, of course also suffers from Neuberger's
$0/0$ problem of lattice BRST.  

In order to demonstrate that the Curci-Ferrari mass
regulates the Neuberger zero, for $t=0$ we will verify by explicit
calculation that  
\begin{equation}
  Z_{\mbox{\tiny mGF}}(0,\xi,m^2) \, \not= \, 0\;. \label{Zm-nzero} 
\end{equation}
In fact, from (\ref{Z_mGF}), (\ref{mSGFtexpl}),
\begin{align} 
\label{qGaussBonnet}
Z_{\mbox{\tiny mGF}}(0,\xi,m^2) = \mathcal{N} \int d[b,\bar c, c]
\, \exp\Big\{-  \sum_i  \left( 
\frac{\xi}{2} b_i^a b_i^a \, - i m^2 \xi\, \bar c_i^a c_i^a  \,
+ \, \frac{\xi}{8}\,  
(\bar c_i \times c_i)^2 \right)
\Big\} ,
\end{align}
which again factorises into independent Grassmann (and $b$-field)
integrations per site on the lattice. Using the same 
rescaling $\sqrt{\xi}\,  b \to b$ and $\sqrt[4]{\xi}\, \bar c \to \bar
c$, $\sqrt[4]{\xi}\, c \to c$ as mentioned in the last section, we obtain,
\begin{eqnarray}
  Z_{\mbox{\tiny mGF}}(0,\xi,m^2) & =& \label{Zm}
\left( V_{N}\,
  (2\pi)^{(N^2\!-1)/2}\, I_N\big(m^2\sqrt{\xi}\,\big) \right)^{\# {\rm sites}} ,
\end{eqnarray}

\vspace*{.1cm}

\noindent
where $V_{N}$ is the group volume of $SU(N)$, and 
\begin{eqnarray} 
 I_N(\widehat m^2) & = & \int \prod_{a=1}^{N^2\!-1} d(i\bar c^a) d c^a
 \, \label{cbc-siteint}
 \exp\left\{ i \widehat m^2\,  \bar c\! \cdot \! c \, - \,
 \frac{1}{8}\,   (\bar c \times c)^2 \right\} \; ,
\end{eqnarray}
where we used the rather obvious abbreviations $\bar c \!\cdot\!c = \bar c^a
c^a $, $(\bar c \times c)^a = f^{abc} \bar c^b c^c$, and $\widehat m^2
= m^2 \sqrt{\xi}$. Note that we define the Grassmann integration
measure to include the imaginary unit $i$ with the real anti-ghosts $\bar c$ so
as to reproduce the result of integrating over complex conjugate 
Grassmann variables $c^a \pm i \bar c^a$. 
Expanding the exponential
and collecting the relevant powers in the ghost/anti-ghost variables,
for $SU(2)$ we obtain
\begin{eqnarray}
 I_N(\widehat m^2) & = & \int \prod_{a=1}^{N^2\!-1} d(i\bar c^a) d c^a
 \, 
 \left\{\frac{\widehat m^6}{3!}\,  (i\bar c\! \cdot \! c)^3 \, + \,
 \frac{1}{8}\,   (\bar c \times c)^2\,\widehat m^2\,(i\bar c\! \cdot \! c) \right\} \;.
\end{eqnarray}
Due to the anti-symmetry of the ghost fields we notice that $(i\bar c\! \cdot \! c)^3 =6\prod_a(i\bar c^a\! \cdot \! c^a) $
and $\epsilon^{abc}\bar c^b\,c^c\,\epsilon^{ade}\bar c^d\,c^e=+(\bar c\,c)^2$. Therefore
the term combining the quartic and the quadratic interaction simply becomes 
$(\bar c \times c)^2\,(i\bar c\! \cdot \! c)=(i\bar c\,c)^3$.
According to these considerations $I_N(\widehat m^2)$ becomes for $SU(2)$
\begin{equation} 
I_2(\widehat m^2) \, = \frac{3}{4} \widehat m^2 \left(
1+\frac{4}{3} \widehat m^4\right) \; . 
\end{equation}
For $SU(3)$ the computation is a bit more tedious. First of all notice that the quartic term can be written
$f^{abc}\bar c^b\,c^c\,f^{ade}\bar c^d\,c^e=\frac{2}{N_c}(\bar c^a c^c)^2\,+\,d^{abc}(\bar c^b c^c)\,d^{ade}(\bar c^d c^e)$
or equivalently, adopting the fundamental representation, with Hermitian generators
$f^{abc}\bar c^b\,c^c\,f^{ade}\bar c^d\,c^e=2\,{\rm tr}\left((T^a f^{abc}\bar c^b c^c)^2\right)$.
This last term can also be cast into a more convenient form as
\begin{align}
2\,{\rm tr}\left((T^a f^{abc}\bar c^b c^c)^2\right)&=-2\,{\rm tr}\left([T^b, T^c]\bar c^b c^c\right)^2\nonumber\\
&=-2\,{\rm tr}\left(\bar c c+c\bar c\right)^2\nonumber\\
&=-2\,{\rm tr}\left(\{\bar c,c\}\right)^2.
\end{align}
Also, using the Jacobi Identity
\begin{align}
f^{abc}\bar c^b c^c\,f^{ade}\bar c^d c^e+f^{acd}c^c\bar c^d\,f^{abe}\bar c^bc^e+f^{adb}\bar c^d \bar c^b\,f^{ace}c^cc^e=0\nonumber\\
\end{align}
we can write
\begin{align}
(\bar c\times c)^2=-\frac12(\bar c \times \bar c)(c\times c).
\end{align}
Therefore, the quartic interaction, in the fundamental representation reads
\begin{align}
(\bar c\times c)^2&=-\frac12\,f^{abc}\bar c^b\bar c^c\,f^{ade}c^dc^e\nonumber\\
&=-{\rm tr}\left(T^a f^{abc}\bar c^b\bar c^c\,T^hf^{hde}c^dc^e\right)\nonumber\\
&={\rm tr}\left(\{\bar c,\bar c\}\{c,c\}\right)\nonumber\\
&=4{\rm tr}\left(\bar c^2\,c^2\right).
\end{align}
Similarly, for the quadratic term, we have
\beq
\bar c^a c^a=2\,{\rm tr}(\bar c c).
\eeq
The integral over the ghost fields then assumes the compact form
\beq
I_N(\widehat m^2)\,=\,\int\prod_{a=1}^{N^2-1} d(i\bar c^a)dc^a\,e^{2\widehat m^2\,{\rm tr}(\bar c c)
+\frac{{\rm g^2}}{2}{\rm tr}\left(\bar c^2\,c^2\right)}.
\eeq
The result for $SU(3)$, using Mathematica to compute all the possible combinations, is
\begin{equation}
I_3(\widehat m^2) \, = \frac{45}{64} \widehat m^4 \left(
1+ 4 \widehat m^4 + \frac{64}{15} \widehat m^8 +
\frac{64}{45} \widehat m^{12} \right) \, . \label{I_3}
\end{equation}
In both cases we factored the leading power for $\widehat m^2\to 0$.
$I_N(\widehat m^2)$ is polynomial in $\widehat m^2 = m^2\sqrt{\xi} $
of degree $N^2\!-1$, for all $N$. The successively lower
powers of $\widehat m^2$ decrease by 2 in each step
in this polynomial, reflecting an increasing power of the quartic
ghost self-interactions contributing to each term. Therefore, the
polynomials $I_N(\widehat m^2)$ are odd/even in $\widehat m^2$  for $N$
even/odd.

Because the polynomial is odd for all even $N$ there can thus not be
an order-zero term in the first place. The powers of the quartic
interactions alone never match the number of independent Grassmann
variables, and the Neuberger zero at $\widehat m^2=0$ arises rather 
trivially  for even $N$ (for the same reason that the Euler character
of an odd-dimensional manifold necessarily vanishes).

For $N$ odd, $I_N(\widehat m^2) $ is an even polynomial which could in
principle have an order zero, constant term. The fact that this term
is absent, {\it e.g.}, as explicitly verified for $SU(3)$ in (\ref{I_3}),
reflects the vanishing of the Euler character of $SU(N)$ also for odd
$N$, as mentioned above (the even dimension  $N^2\!-1$ of
the algebra is deceiving in this case as, for example, the parameter
space of  $SU(3)$ can roughly be thought of consisting of odd-dimensional $S^3$
and $S^5$).   

In any case, the polynomials $I_N(\widehat m^2)$ do not have 
a constant term and therefore vanish with $\widehat m^2\to
0$, {\it i.e.},  $I_N(0)=0$, as expected. Moreover, the scaling argument used
here and in the last section shows that the partition function
(\ref{Z_mGF}) of massive Curci-Ferrari model can only depend on two of
the three parameters, 
\begin{equation} 
 Z_{\mbox{\tiny mGF}}(t,\xi,m^2) \,=\, f\big(t/\sqrt{\xi},\,\xi m^4\big) \; .
\label{Z_mGF-scale}
\end{equation}
An independent route of deriving this generic form, from the equations
of motions, will be presented below. In this section we explicitly 
obtained $f(0,y)$ with $y=\widehat m^4$ to constrain this function $f(x,y)$ of
two variables along the $x=t/\sqrt{\xi}=0$ line, and verified that 
\[
 Z_{\mbox{\tiny mGF}}(0,\xi,m^2) = f\big(0,\,\xi m^4\big) 
  \propto  \left\{ \begin{array}{ll} 
               (\xi m^4)^{\#  {\rm sites}/2} , & N=2 \\[4pt]
               (\xi m^4)^{\#  {\rm sites}} \, , & N=3 
                      \end{array} \right. 
\]
for $m^2 \to 0$. Because of the topological explanation of the zero
obtained in this limit, {\it i.e.}, $f(0,0)=0$, as discussed in the
last section, this actually constrains $f$ to vanish along the entire
$y=0$ line, $f(x,0)=0$ for all $x=t/\sqrt{\xi}$. 

For $x=0$ we could furthermore define a non-vanishing, finite limit
\begin{equation} 
    \lim_{m^2\to 0} (\xi m^4)^{-N_\mathrm{tot}}  Z_{\mbox{\tiny
    mGF}}(0,\xi,m^2) \, = \, \mbox{const.} \label{naivelim}
\end{equation}
with an appropriate power $N_\mathrm{tot} = \# $ of sites on a 
finite lattice for odd $N$, or half that for even $N$. This constant 
could in principle be inserted into the unfixed lattice gauge theory
measure without harm, {\it i.e.}, avoiding the zero in (\ref{Z-zero}). 
Because $x=0$, however, this still has no effect
in terms of gauge-fixing by the Faddeev-Popov procedure either. We
need to get away from $x=0$, at least by a small amount, to suppress
those parts of the gauge orbits with large violations of the Lorenz
condition. At finite Curci-Ferrari mass $m^2$ we no-longer have the
$t$-independence (or $x$-independence) of (\ref{t-indep}). We can
therefore not conclude at this point yet that the constant in
(\ref{naivelim}) will essentially remain unchanged when going to some
finite $x\not =0$ as we must. 

We are not quite there yet, and
we will therefore have to have a closer look at the parameter
dependence of the massive Curci-Ferrari model in the next section.

\section{Parameter Dependences}

From Eqs.~(\ref{Z_mGF}) and (\ref{S_mGF}) or (\ref{mSGFtexpl}) we
immediately obtain the following (logarithmic) derivatives,
\begin{eqnarray} 
  t \frac{\partial}{\partial
        t } \, \ln Z_{\mbox{\tiny mGF}}(t,\xi,m^2) &=& -i 
                   \big\langle\, (s \bar s - i m^2) \, t\,  V_U[g]\, 
          \big\rangle_{m^2} \, , \nonumber \\
  2\xi  \frac{\partial}{\partial
        \xi } \, \ln Z_{\mbox{\tiny mGF}}(t,\xi,m^2) &=&
             -i 
                   \big\langle \, (s \bar s - i m^2) \big(-i\xi \sum_i \bar
        c^a_i c^a_i \big) \, \big\rangle_{m^2}
        \, , \nonumber \\
  m^2  \frac{\partial}{\partial
        m^2} \, \ln Z_{\mbox{\tiny mGF}}(t,\xi,m^2) &=&  -
          \big\langle \, m^2 \, \mathcal{O}(t,\xi) \, \big\rangle_{m^2}  \; , 
\label{Z-derivs}
\end{eqnarray}
where the subscripts $m^2$ on the right denote expectation values within the
Curci-Ferrari model at finite mass. In particular, the derivative
w.r.t.~$m^2$ in the last line differs from (\ref{mderiv}) only in that
$m^2$ has not been set to zero here yet. All these expectation values
can, in general, depend on the link-field configuration $\{U\}$ which
acts as a background field to the model. Independence of $\{U\}$ is
only guaranteed to hold in the topological limit $m^2\to 0$.

From the definition of $\mathcal O$ in (\ref{LatKondOp}), we thus find that
\begin{eqnarray} 
\left(  t \frac{\partial}{\partial
        t }   +   2\xi  \frac{\partial}{\partial
        \xi }  -   m^2  \frac{\partial}{\partial
        m^2}  \right) \, \ln Z_{\mbox{\tiny mGF}}(t,\xi,m^2) &=&
 - i \big\langle \, s\bar s \, \mathcal{O}(t,\xi) \, \big\rangle_{m^2}  \; .
\nonumber
\end{eqnarray}
The standard argument that the expectation value of an (anti-)BRST
exact operator vanishes does not hold at finite $m^2$. Neither are
BRST and anti-BRST variations nilpotent, nor is $\mathcal O$ invariant
under the BRST or anti-BRST transformations. However, the equations of
motion for (anti-)ghost and Nakanishi-Lautrup fields on the lattice,
{\it i.e.}, their lattice Dyson-Schwinger equations, can be used to
show that, indeed, 
\begin{equation}
\label{DSzero}
\big\langle \, s\bar s \, \mathcal{O}(t,\xi) \, \big\rangle_{m^2}  = 0
\; ,
\end{equation}
even at finite $m^2$. 
In fact, consider the variation of the massive Curci-Ferrari action w.r.t. the $b$-field
\begin{align}
\frac{\delta S_{\mbox{\tiny mGF}}}{\delta b^a_i}&=it F^a_i(U^g)+\xi b^a_i=\xi (b')^a_i,
\end{align}
such that
\begin{align}
\label{DSbfield}
\left\langle \frac{\delta S_{\mbox{\tiny mGF}}}{\delta b^a_i} b_j^b\right\rangle_{m^2}  &=\xi\left\langle
(b')^a_i \left((b')^a_i -\frac{it}{\xi}F_j^b(U^g)\right)\right\rangle_{m^2}  \nonumber\\
&=-i(N_c^2-1) N_{\#  {\rm sites}}\delta^{ab}\delta_{ij},
\end{align}
where we used the fact that $\langle(b')^a_i(b')^a_i\rangle_{m^2} $ corresponds to a Gaussian integral, whose result is simply 
$\frac{1}{\xi}\delta^{ab}\delta_{ij}$.
For the anti-ghost field we have, where all the functional derivatives are understood left graded, we write
\begin{align}
\frac{\delta S_{\mbox{\tiny mGF}}}{\delta \bar c^a_i}&=it {M_{\mbox{\tiny FP}}}_i^a[c]-im^2 c_i^a+\xi\frac{g^2}{4}f^{abc}
c_i^b\,f^{cde}\bar c^d_ic^e_i\nonumber\\
&=it {M_{\mbox{\tiny FP}}}_i^a[c]-im^2 c_i^a-\xi\frac{g^2}{4}((\bar c\times c)\times c)^a_i.
\end{align}
Consequently we have the lattice DS equations as follows
\begin{align}
\label{DSghost}
\left\langle \frac{\delta S_{\mbox{\tiny mGF}}}{\delta \bar c^a_i} \bar c_j^b\right\rangle_{m^2}  &=it 
\left\langle{M_{\mbox{\tiny FP}}}_i^a[c]\bar c_j^b\right\rangle_{m^2} -
im^2\left\langle c^a_i\bar c_j^b\right\rangle_{m^2} 
-\left\langle\xi\frac{g^2}{4}((\bar c\times c)\times c)^a_i\bar c^b_j\right\rangle_{m^2}  \nonumber\\
&=i(N_c^2-1) N_{\#  {\rm sites}}\delta^{ab}\delta_{ij}.
\end{align}
Putting together Eq. (\ref{DSbfield}) and (\ref{DSghost}), we exactly obtain Eq. (\ref{DSzero}).
Therefore,
\begin{equation}
\left(  t \frac{\partial}{\partial
        t }   +   2\xi  \frac{\partial}{\partial
        \xi }  -   m^2  \frac{\partial}{\partial
        m^2}  \right) \, Z_{\mbox{\tiny mGF}}(t,\xi,m^2) \, =\, 0 \; .
\end{equation}
This differential equation entails that we can write the partition
function of the model in the generic form (\ref{Z_mGF-scale}).

As we already did in the previous sections, we therefore continue to
use the new parameters  $x=t/\sqrt{\xi} $ and $\widehat m^2 =
m^2\sqrt{\xi}\,$ from now on, writing
\begin{equation}
   Z_{\mbox{\tiny mGF}} \, \equiv \,    Z_{\mbox{\tiny
   mGF}}(x,\widehat m^2)  \; .
\end{equation}
Again using rescaled fields $\sqrt{\xi}\,  b \to b$, $\sqrt[4]{\xi}\,
\bar c \to \bar c$, $\sqrt[4]{\xi}\, c \to c$ and
 $\sqrt[4]{\xi}\, \bar s \to \bar s$, $\sqrt[4]{\xi}\,
s \to s$, the (anti-)BRST transformations of Eq.~(\ref{BRSc}) --
(\ref{aBRSb}) remain formally unchanged, and $m^2 $ is replaced by
$\widehat m^2$ in those of the massive model in Eqs.~(\ref{mBRSb}),
  (\ref{maBRSb}).  Correspondingly, all other relations above are
then converted by the formal replacements $\xi\to 1$, $t\to x$ and
$m^2 \to \widehat m^2$. In particular,
\begin{eqnarray}
S_{\mbox{\tiny mGF}}(x) &=& i \,( s \bar s-i\widehat m^2) \, \Big( x\, 
 V_U[g] \, - \, \frac{i}{2}  \sum_i \bar c_i^a c_i^a \Big) \\
   &=& \sum_i \,  \Big\{ - i x  \, b^a_i  F_i^a(U^g)
\, - {i} x \,\bar c^a_i  {M_{\mbox{\tiny FP}}}_i^a[c]\nonumber \\[-4pt]
&& \hskip 1cm +\, 
\frac{1}{2} b_i^a b_i^a \, + \, \frac{1}{8}\, 
(\bar c_i \times c_i)^2 \, \Big\} \, + \widehat m^2\, \mathcal{O}(x)
 \; , \nonumber 
\end{eqnarray}
with
\begin{equation}
\mathcal{O}(x)\, = \,  x\, 
 V_U[g] \, - \, {i}  \sum_i \bar c_i^a c_i^a  
\end{equation}
The two independent derivatives left, are readily  
read off 
in an analogous way to give 
\begin{eqnarray} 
   \frac{\partial}{\partial
        x } \, \ln Z_{\mbox{\tiny mGF}}(x,\widehat m^2) &=& -i 
                   \big\langle\, (s \bar s - i \widehat m^2) \,  V_U[g]\, 
          \big\rangle_{\widehat m^2} \, , \nonumber \\
    \frac{\partial}{\partial
       \widehat m^2} \, \ln Z_{\mbox{\tiny mGF}}(x,\widehat m^2) &=&  -
          \big\langle \, \mathcal{O}(x) \,
        \big\rangle_{\widehat m^2}  \; . 
\label{indep-Z-derivs}
\end{eqnarray}
In absence of a topological argument for the gauge parameter
independence at finite Curci-Ferrari mass, the best we can do to
achieve independence of $x=t/\sqrt{\xi}$ is to allow an $x$ dependent 
mass parameter $\widehat m^2 \equiv \widehat m^2(x)$. In particular, the 
$x=0$ results of the previous section are then to be interpreted as
being expressed in terms of $\widehat m^2(0)$. These results will
remain unchanged for $x\not= 0$, if we adjust the mass function
$\widehat m^2(x) $ with $x$ in the partition function $ Z_{\mbox{\tiny
   mGF}}$, accordingly. That is, if
\begin{eqnarray}
  0 &=&  \frac{d}{dx} \,    Z_{\mbox{\tiny
   mGF}}\big(x,\widehat m^2(x)\big) \\ &=&
  \left( \frac{\partial}{\partial x} + 
\frac{d \widehat m^2}{dx} \, \frac{\partial}{\partial
   \widehat m^2} 
 \right) \,    Z_{\mbox{\tiny mGF}}\big(x,\widehat m^2(x)\big) \; . \nonumber
\end{eqnarray}
From Eqs.~(\ref{indep-Z-derivs}) we see that this requires that
\begin{equation}
\frac{d \widehat m^2}{dx} \, = \, -i \frac{   \big\langle\, (s
          \bar s - i \widehat m^2) \,  V_U[g]\ \big\rangle_{\widehat
          m^2}}{ \big\langle \, \mathcal{O}(x) \, \big\rangle_{\widehat m^2}} 
           \; . 
\end{equation}
This is not a very profound insight. The crucial question at this
point is, whether the tuning of the Curci-Ferrari mass parameter with
$x$ is possible indpendent of the link configuration $\{U\}$ which is
far from obvious here. Otherwise we would have to choose a different
trajectory in the parameter space $(x,\widehat m^2)$ for different
gauge orbits which would be of little use then, as far as the 
Faddeev-Popov gauge-fixing procedure is concerned. If it is possible,
on the other hand, we can then use the value of the mass $\widehat
m^2_0 = \widehat m^2(0)$ at $x=0$ to regulate the Neuberger zero and
use the $x$ and $\{ U\}$ independent, non-vanishing and finite constant
\begin{equation} 
    \lim_{\widehat m_0^2\to 0} (\widehat m_0^4)^{-N_\mathrm{tot}}
    Z_{\mbox{\tiny 
    mGF}}(x,\widehat m^2(x)) \, = \, \mbox{const.} \label{suitlim}
\end{equation}

\vspace*{.1cm}

\noindent
as the starting definition of Faddeev-Popov gauge fixing on the lattice.
Then, of course, we would also expect that there should be a
topological meaning to this constant which is so far, however,
unfortunately unknown to us. 

All we can offer at the moment is to verify that all is well at $x=0
$, where we can do the explicit calculations. It is relatively
straight-forward to show in this way that
\begin{equation}
      \frac{d\widehat m^2}{dx}  \Big|_{x=0} \, = \, \mbox{const}\,x+{\it O(x^2)}    
\end{equation}
with the constant independent of $\{U\}$. While this is merely necessary, but not
sufficient, it demonstrates that we can get away from $x=0$, at least
infinitesimally. This is of qualitative importance as a non-zero value
of $x=t/\sqrt{\xi}$, no matter how small, corresponds to a large but
finite $\xi$ at $t=1$ and thus  eliminates the gauge freedom.
The study and result for the second derivative will be presented in the next publication.

\thispagestyle{empty}
\cleardoublepage
\chapter{Batalin-Vilkovisky Formalism In Y-M Theory}

We have shown so far how the quantisation of Y-M theory can be pursued by means of several formalisms: canonical, covariant operator,
path integral and BRST formalism. Yet, the most general algebraic way to quantise a gauge theory is achieved by
the Batalin-Vilkovisky (BV) formalism
\cite{Batalin:1981jr, Batalin:1984jr}. 
This formalism provides a Grassmann-graded canonical formalism, by means of a new canonical structure, called {\it anti-bracket},
which generalises and elevates the standard Hamiltonian formalism to a more algebraic scenario.
Furthermore, the Batalin-Vilkovisky method encompasses the Faddeev-Popov quantisation and can be entirely formulated in the light
of BRST and anti-BRST symmetry.
The new fields introduced in this formalism, the {\it anti-fields}, necessary to build up the global canonical structure,
are identified with functional derivatives of an anti-fermion gauge-fixing term.
We will first introduce the main ingredients of this formalism, and after that we will present the BV construction
of Euclidian 4-dimensional Y-M theory in the framework of non-linear gauges.
At last, we will provide the lattice version of our model.

\section{The Appearance Of Anti-Fields}

To introduce a canonical formalism, graded with respect to the Poisson brackets first we have to deal with the 
concept of Grassmann parity, which defines the {\it even} or {\it odd} character of a field under product exchange. Given a set of fields
$\Phi^A(x)$ of Grassmann parity $\epsilon(\Phi^A)=\epsilon_A$, then we associate to them the corresponding anti-fields $\Phi^*_A(x)$ 
of opposite parity, as $\epsilon(\Phi^*_A)=\epsilon_A+1$. Fields and anti-fields play the role of conjugate variables in the Hamiltonian framework,
and therefore it is natural to define a canonical conjugation as
\beq
(\Phi^A,\Phi^*_B)=\delta^A_B\qquad (\Phi^A,\Phi^B)=(\Phi^*_A,\Phi^*_B)=0,
\eeq
where the conjugation is defined through the BV brackets
\beq
(F,G)=\frac{\delta^rF}{\delta \Phi^A}\frac{\delta^lG}{\delta \Phi^*_A}-\frac{\delta^rF}{\delta \Phi^*_A}\frac{\delta^lG}{\delta \Phi^A},
\eeq
where $r$ and $l$ denote respectively right and left derivative.
These brackets determine therefore a canonical structure onto the BV formalism, providing it a Jacobi identity 
\beq
\sum_{\mat P(F,G,H)}(-1)^{(\epsilon_F+1)(\epsilon_H+1)}(F,(G,H)=0,
\eeq
and a Leibnitz rule
\begin{align}
(F.GH)&=(F,G)H+(-1)^{\epsilon_G(\epsilon_F+1}G(F,H)\nonumber\\
(FH.G)&=F(H,G)H+(-1)^{\epsilon_H(\epsilon_G+1}(F,G)H.
\end{align}
To the antifields we can also associate a ghost number as
\beq
{\mathcal G}h(\Phi^*_A)=-{\mathcal G}h(\Phi^A)-1
\eeq
and it is constrained such that the quantum action carries total ghost number zero.
The BV quantisation prescription amounts to solve the {\it quantum Master Equation}
\beq
\frac12 (W,W)=i\hbar \Delta W,
\eeq
where $\Delta$  a second-order odd differential (nilpotent) operator defined as
\beq
\Delta=(-1)^{\epsilon_A+1}\frac{\delta^r}{\delta \Phi^A}\frac{\delta^r}{\delta \Phi^*_A}.
\eeq
and $W=W[\Phi, \Phi^*]$ a generic quantum action, that is supposed to be expandable in powers of $\hbar$
\beq
W=S_{\rm cl}+\sum_{n=0}^\infty \hbar^n\,S^{(n)}_{\rm qu}
\eeq
Here, $S_{\rm cl}$ is the classical action obtained by setting all the anti-fields to zero,
and $S_{\rm qu}$ its quantum fluctuation. 
To the lowest order in $\hbar$, the ``classical'' Master Equation reduces to
\beq
\label{master}
(S,S)=0.
\eeq
The path integral representation of the BV formalism starts from the consideration that the classical action, as previously observed,
is generated while setting $\Phi^*_A=0$. Therefore, by defining the correct gauge-fixing anti-fermion, we can write
\beq
\mathcal Z=\int[{\rm d}\Phi_A][{\rm d}\Phi^*_A]\delta\left(\Phi^*_A-\frac{\delta^r \overline{ \Psi}}{\delta \Phi_A}\right)
\exp\left[-\frac1\hbar W\right].
\eeq
After integrating out the anti-field $\Phi^*_A$ using the delta-function, one can verify that the action is left invariant under usual
BRST transformations. 

\section{Non-linear gauges in BV formalism}

It is well known that the pure Yang-Mills action $S[A]=\frac12  \int_M F_{\mu\nu}F^{\mu\nu}$, in a certain irreducible
representation $\rho$ of $SU(N)$, is left invariant under a gauge transformation ${}^g\!A=g^\dagger \,A g+g^\dagger\,\partial g$.
An interesting question to ask ourselves is what happens if the gauge field $A$ is being shifted as $A\to A-\tilde A$: does the gauge symmetry still remain
and moreover, how does this shift-symmetry affect the underlying BRST structure?
It is this background gauge manipulation of the Y-M action which poses the bases of the Batalin-Vilkovisky method of quantisation.
The appearance of the anti-ghost fields in $S[A-\tilde A]$ can be observed by gauge-fixing iteratively the gauge-symmetry and the shift-symmetry,
as done for instance in \cite{Alfaro:1993ua}. Though pedagocically interesting, we prefer to give a more heuristic approach to it.
Suppose to gauge fix with a non-linear gauge the Euclidian Y-M action: we then insert in the path integral representation 
the following covariant, non-linear gauge-fixing Euclidean Lagrangian with ghost/anti-ghost symmetry
\beq
\mathcal L_{\rm gf}[A,b,c,\bar c]=ib^a\partial A^a+\frac{\xi}2b^2+\frac i2\bar c^a\{\partial, \mathcal D\}^{ab} c^b+\frac\xi8(\bar c\times c)^2.
\eeq
where all the fields are in the adjoint representation of $SU(N)$.
This Lagrangian is left invariant under the BRST and anti-BRST transformations (\ref{gaugetrans5}) and can be written either as
a BRST coboundary term
\beq
\mathcal L_{\rm gf}[A,b,c,\bar c]=s\overline{\Psi},\qquad
\overline{\Psi}= i\bar c^a\left(\partial_\mu A^{a\mu}-i\frac\xi2 b^a\right),
\eeq
or a double BRST coboundary term
\beq
\mathcal L_{\rm gf}[A,b,c,\bar c]=s\bar s{W},\qquad
W= \frac i2\left( A_\mu^aA^{a\mu}-i\xi \bar c^ac^a\right).
\eeq
Performing now a shift in all the fields
\footnote{From now on we will leave component notation implicit, unless needed in a specific computation.}
\beq
\label{shift}
A\to A-\tilde A\qquad c\to c-\tilde c\qquad \bar c\to \bar c-\tilde{ \bar c}\qquad b\to b-\tilde b,
\eeq
we notice that the quantum action $S[A, c, \bar c, b]$ becomes also invariant under the shift symmetry
\begin{align}
s \Phi(x)&=\alpha(x)&\bar s\Phi(x)&=\gamma(x)\nonumber\\
s \widetilde \Phi(x)&= \alpha(x)-\beta(x)&\bar s\widetilde \Phi(x)&=\gamma(x)-\chi(x),
\end{align}
with $\Phi(x)$ the set of all fields and $\widetilde \Phi(x)$ the set of shifted ones. Here $\beta(x)$ and $\chi(x)$
represent the original BRST and anti-BRST variations of $\Phi(x)$, whereas $\alpha(x)$ and $\gamma(x)$
correspond to some collective fields generating the field-shift. Let us focus first on the BRST construction of the Batalin-Vilkovisky Lagrangian.
The BRST transformations (\ref{gaugetrans5}), according to the shift (\ref{shift}), assume the form
\begin{align}
\label{BV1}
sA_\mu&=\psi_\mu & s\tilde A_\mu&=\psi_\mu-\mathcal D_\mu^{(A-\tilde A)}(c-\tilde c)\nonumber\\
sc&=\epsilon& s\tilde c &=\epsilon+\frac12[c-\tilde c,c-\tilde c] \nonumber\\
s\bar c&=\bar \epsilon  & s\tilde{\bar c}&=\bar \epsilon -(b-\tilde b)+\frac12[\bar c-\tilde {\bar c},c-\tilde c] \nonumber\\
sb&=\epsilon_b & s\tilde b&=\epsilon_b+\frac12[c-\tilde {c},b-\tilde b] +\frac18[[ c-\tilde { c},c-\tilde c],\bar c-\tilde {\bar c}] .
\end{align}
A few remarks here are needed: first of all, the choice of
transformations we have made is consistent with the hermiticity of the ghost fields, chosen to be real, to satisfy the hermiticity
of the Lagrangian and therefore the unitarity of the 
$\mathcal S$-matrix. Moreover, to generate a ghost/anti-ghost gauge-fixing action, we have shifted the $b$-field as $b\to b-\frac12[\bar c,c] $:
such a symmetry is different from standard covariant linear gauges, as Landau gauge, where the action is not symmetric under the exchange of
ghost into anti-ghost fields.  
This operation also affects the linearity of the gauge chosen, and as a consequence we obtain a nonlinear gauge, whose main feature is to
generate a quartic ghost interaction in the action. This term is required from topological considerations to produce the most general renormalisable
covariant action with an underlying BRST symmetry, as showed in \cite{Baulieu:1981sb}.
Finally, the choice we made to associate the covariant derivative only to the shifted field makes the original gauge symmetry
of the original gauge field to be carried entirely by the collective field. The transformation of the original gauge field is then taken always
just as a shift.

To enforce the overall invariance under $s$ of the quantum action, we require more fields: among some new Nakanishi-Lautrup fields,
it is important to notice the appearance of the anti-fields, denoted with an asterisk
\begin{align}
\label{BV2}
s\psi_\mu&=0 & s\epsilon&=0\nonumber\\
s\bar \epsilon&=0 & s\epsilon_b&=0\nonumber\\
sA_\mu^*&=B_\mu &sB_\mu&=0\nonumber\\
sc^*&=B&sB&=0\nonumber\\
s\bar c^*&=\overline{B}& s\overline{B}&=0\nonumber\\
sb^*&=B_b&sB_b&=0,
\end{align}
where the multiplet $(\psi_\mu,\epsilon,\bar \epsilon,\epsilon_b)$ is the ghost multiplet associated with the shift symmetry for $(A_\mu,c,\bar c,b)$,
$(A_\mu^*,c^*,\bar c^*,b^*)$ are the anti-ghosts
\footnote{These antifields are the usual antighosts of the collective fields enforcing the Dyson-Schwinger equations through
shift symmetries.}
and $(B_\mu,B,\overline{B}, B_b)$ the corresponding auxiliary fields. 
Having such an abundance of fields with different ghost number, it is worthwhile to provide the table
\begin{align}
{\mathcal G}h(A_\mu)&={\mathcal G}h(\tilde A_\mu)=0&{\mathcal G}h(A_\mu^*)&=-1&{\mathcal G}h(\psi_\mu)&=1&{\mathcal G}h(B_\mu)&=0\nonumber\\
{\mathcal G}h(c)&={\mathcal G}h(\tilde c)=1&{\mathcal G}h(c^*)&=0&{\mathcal G}h(\epsilon)&=2&{\mathcal G}h(B)&=1\nonumber\\
{\mathcal G}h(\bar c)&={\mathcal G}h(\tilde{\bar c})=-1
&{\mathcal G}h(\bar c^*)&=-2&{\mathcal G}h(\bar\epsilon)&=0&{\mathcal G}h({\overline B})&=-1\nonumber\\
{\mathcal G}h(b)&={\mathcal G}h(\tilde b)=0
&{\mathcal G}h(b^*)&=-1&{\mathcal G}h(\epsilon_b)&=1&{\mathcal G}h(B_b)&=0\nonumber.
\end{align}
It is interesting to notice that in the BV formalism, the Nakanishi-Lautrup fields associated to the FP ghosts are Grassmann, with Ghost number
respectively $\pm1$.
We can therefore write in close-form the various Ghost number and Grassmann parity for the general field $\Phi$ as
\beq
{\mathcal G}h(\Phi^*)={\mathcal G}h(\Phi)-1\qquad \epsilon(\Phi^*)=\epsilon(\Phi)+1.
\eeq
We may also construct the following table which summarises the relations among the BV fields
\be
\bea{ccc}
\begin{tabular}{|c|c|c|c|c|}
\hline
Field & Collective Field & $\bea{ccc}\textrm{Anti-Ghost} \\
\textrm{(anti-field)}\eea$ & 
Ghost& 
$\bea{ccc}\textrm{Nakanishi-Lautrup}\\
\textrm{Field}\eea$\\
\hline
$A_\mu$  & $\tilde A_\mu$   & $A^*_\mu$  & $\psi_\mu$      & $B_\mu$\\
$c$      & $\tilde c $      & $c^*$      & $\epsilon$      & $B$\\
$\bar c$ & $\tilde{\bar c}$ & $\bar c^*$ & $\bar \epsilon$ & $\bar B$\\
$b$      & $\tilde b$       & $b^*$       & $\epsilon_b$    & $B_b$\\
\hline
\end{tabular}
\eea\nonumber
\ee
The physical requirement for the gauge-fixing Lagrangian is obtained by demanding all the fields
associated to the shift symmetry to vanish. We thus recover the original theory, by choosing for instance the following 
shift-symmetry gauge-fixing Lagrangian
\begin{align}
\label{shiftlagrangian}
\widetilde {\mathcal L}_{\rm GF}=i&\left[B_\mu \tilde A_\mu-A^*_\mu\Big(\psi^\mu-\mathcal D_\mu^{(A-\tilde A)}(c-\tilde c)\Big)
+\overline{B}\tilde c+\bar c^*\left(\epsilon+\frac12[c-\tilde c,c-\tilde c] \right)\right.\nonumber\\
&+B\tilde{\bar c}+c^*\left(\bar \epsilon -(b-\tilde b)+\frac12[\bar c-\tilde {\bar c},c-\tilde c] \right)\nonumber\\
&\left.+B_b\tilde b-b^*\left(\epsilon_b+\frac12[c-\tilde {c},b-\tilde b] +\frac18[[ c-\tilde { c},c-\tilde c],\bar c-\tilde {\bar c}] \right)\right].
\end{align}
It is an easy task to check that the Lagrangian $\widetilde {\mathcal L}_{\rm GF}$ is left invariant under the BRST transformations
(\ref{BV1}) and (\ref{BV2}). Moreover, the requirement for $\widetilde {\mathcal L}_{\rm GF}$ to be Hermitian
is guaranteed by the factor $i$ in front of the trace. 
By performing the integration over the $B$-fields, we generate Delta functions in the shift-symmetry fields, which will set them to zero.

All the field couplings in (\ref{shiftlagrangian}) amount to an overall vanishing Ghost number as required.
Suppose the gauge-fixing anti-fermion $\overline{\Psi}$ to depend upon only the original fields as
\begin{align}
\mathcal L_{\rm GF}=is\overline{\Psi}&=i \left(sA_\mu\frac{\delta \overline{\Psi}}{\delta A_\mu}+sc\frac{\delta \overline{\Psi}}{\delta c}
+s\bar c\frac{\delta \overline{\Psi}}{\delta \bar c}+sb\frac{\delta \overline{\Psi}}{\delta b}\right)\nonumber\\
&= i\left(\frac{\delta \overline{\Psi}}{\delta  A_\mu}\psi^\mu+\frac{\delta \overline{\Psi}}{\delta c}\epsilon
+\frac{\delta \overline{\Psi}}{\delta \bar c}\bar \epsilon+\frac{\delta \overline{\Psi}}{\delta b}\epsilon_b\right).
\end{align}
Thus, integrating out the auxiliary fields we set all the fields associated
to the shift symmetry to zero. The remaining Lagrangian is the {\it BV gauge-fixing Lagrangian}
\begin{align}
\mathcal L_{\rm BV}&=\widetilde{\mathcal L}_{\rm GF}+\mathcal L_{\rm GF}\nonumber\\
&= i\left\{A^*_\mu \mathcal D^\mu c+\frac12\bar c^*[c,c] -c^*\left(-b+\frac12[\bar c, c] \right)
-b^*\left(\frac12[c,b] +\frac18[[ c,c],\bar c] \right)\right.\nonumber\\
&\quad-A^*_\mu\psi^\mu+\bar c^*\epsilon+c^*\bar \epsilon-b^*\epsilon_b\Big\}
+ i\left(\frac{\delta \overline{\Psi}}{\delta  A_\mu}\psi^\mu+\frac{\delta \overline{\Psi}}{\delta c}\epsilon
+\frac{\delta \overline{\Psi}}{\delta \bar c}\bar \epsilon+\frac{\delta \overline{\Psi}}{\delta b}\epsilon_b\right)\nonumber\\
&=  i\left\{A^*_\mu\mathcal D^\mu c+\frac12\bar c^*[c,c] -c^*\left(-b+\frac12[\bar c,c] \right)
+b^*\left(\frac12[c,b] +\frac18[[ c,c],\bar c] \right)\right.\nonumber\\
&\quad\left.-\left(A_\mu^*-\frac{\delta \overline{\Psi}}{\delta A_\mu}\right)\psi^\mu
+\left(\bar c^*+\frac{\delta \overline{\Psi}}{\delta c}\right)\epsilon
+\left(c^*+\frac{\delta \overline{\Psi}}{\delta \bar c}\right)\bar \epsilon
-\left(b^*-\frac{\delta \overline{\Psi}}{\delta b}\right)\epsilon_b\right\}.
\end{align}
To obtain the conditions on the anti-fields, it is sufficient to integrate out the ghosts associated with the shift symmetry
\begin{align}
\label{antifield1}
A^*_\mu&=\frac{\delta \overline{\Psi}}{\delta A_\mu}&sA_\mu&=i\frac{\delta \mathcal L_{\rm BV}}{\delta A_\mu^*}\nonumber\\
\bar c^*&=-\frac{\delta \overline{\Psi}}{\delta c}&sc&=-i\frac{\delta \mathcal L_{\rm BV}}{\delta c^*}\nonumber\\
c^*&=-\frac{\delta \overline{\Psi}}{\delta \bar c}&s\bar c&=-i\frac{\delta \mathcal L_{\rm BV}}{\delta \bar c^*}\nonumber\\
b^*&=\frac{\delta \overline{\Psi}}{\delta b}& sb&=i\frac{\delta \mathcal L_{\rm BV}}{\delta b^*}
\end{align}
Once we define the explicit form of the anti-fermion gauge-fixing $\overline{\Psi}$, according to (\ref{antifield1})
we automatically
determine the identification of the anti-fields with BRST fields as
\begin{align}
\label{antifield2}
A^*_\mu&=-i\partial_\mu \bar c\nonumber\\
\bar c^*&=0\nonumber\\
c^*&=i\left(\partial_\mu A^\mu-i\frac\xi2b\right)\nonumber\\
b^*&=\frac\xi2\bar c.
\end{align}
Equations (\ref{antifield2}) clarify the geometric interpretation of the anti-fields on the line of Maurer-Cartan 1-forms.
We are now able to write the total action as a BRST variation of a gauge-fixing anti-fermion function as a proper Witten-type theory
\begin{align}
\mathcal L&=\mathcal L_0+s\overline{\Psi}\nonumber\\
&=\mathcal L_0+ is\Big(A_\mu^* \tilde A^\mu+\bar c^* \tilde c+c^*\tilde{\bar c}+b^*\tilde b\Big)\nonumber\\
&=\mathcal L_0+ is\Big(\Phi^*\tilde \Phi\Big),
\end{align} 
such that ${\mathcal G}h(A_\mu^* \tilde A^\mu+\bar c^* \tilde c-\tilde{\bar c}c^*+b^*\tilde b)=-1$ as expected.
It is worth noting the difference with the ordinary BRST-exact gauge-fixing term
\begin{align}
\overline{\Psi}&=i\bar c\left(\partial_\mu A^\mu-i\frac\xi2 b\right)
=i\bar c \,c^*.
\end{align}


\section{Including Double BRST Algebra}

As noted by Nakanishi and Ojima \cite{Nakanishi:1980dc,Ojima:1980da} 
the quantum action is also invariant under an additional symmetry, known as anti-BRST, whose relation with the
BRST operator is given by Faddeev-Popov conjugation
\beq
\bar s= \mathcal C_{\rm FP}\,s\,\mathcal C^{-1}_{\rm FP}.
\eeq
Following the structure of (\ref{BV1}), we demand that $\bar s (\Phi-\tilde{\Phi})$ reproduces the anti-BRST variations of ordinary fields
($A_\mu, c,\bar c, b$): for instance we might have this algebra
\begin{align}
\label{BV3}
\bar s A_\mu&=A^*_\mu+\mathcal D_\mu^{(A-\tilde A)}(\bar c-\tilde{\bar c})& \bar s \tilde A_\mu&=A^*_\mu\nonumber\\
\bar s c&= c^*-(b-\tilde b)-\frac12[\bar c-\tilde{\bar c},c-\tilde c]&\bar s \tilde c&=c^*\nonumber\\
\bar s \bar c&=\bar c^*-\frac12[\bar c-\tilde{\bar c}, \bar c-\tilde{\bar c}]&\bar s \tilde{\bar c}&=\bar c^*\nonumber\\
\bar s b&=b^*-\frac12[\bar c-\tilde{\bar c},b-\tilde b] +\frac18[[\bar c-\tilde{\bar c},\bar c-\tilde{\bar c}],c-\tilde c]
&\bar s \tilde b&=b^*.
\end{align}
Imposing the condition of invariance on $\widetilde {\mathcal L}_{\rm GF}$, we generate the other variations:
consider for example
\begin{align}
\bar s\left[B_\mu \tilde A_\mu-A^*_\mu\Big(\psi^\mu-\mathcal D_\mu^{(A-\tilde A)}(c-\tilde c)\Big)\right]
&=B_\mu \bar s\tilde A_\mu+A^*_\mu\bar s\Big(\psi^\mu-\mathcal D_\mu^{(A-\tilde A)}(c-\tilde c)\Big)\nonumber\\
&=B_\mu A^*_\mu+A^*_\mu\bar s\psi_\mu-A_\mu^*\bar s\Big(\mathcal D_\mu^{(A-\tilde A)}(c-\tilde c)\Big).
\end{align}
Here we have imposed vanishing variation on $B_\mu$, as suggested in \cite{Alfaro:1993ua} and \cite{Braga:1994dv}.
In order to have invariance under $\bar s$ we need to impose this variation on the field $\psi_\mu$
\begin{align}
\bar s\psi_\mu&\equiv-B_\mu +\bar s\Big(\mathcal D_\mu^{(A-\tilde A)}(c-\tilde c)\Big)\nonumber\\
&=-B_\mu-\mathcal D_\mu^{(A-\tilde A)}(b-\tilde b)-\frac12\mathcal D_\mu^{(A-\tilde A)}[(\bar c-\tilde {\bar c}),(c-\tilde c)]
+[\mathcal D_\mu^{(A-\tilde A)}(\bar c-\tilde {\bar c}),(c-\tilde c)].
\end{align}
Since the ghost fields are anti-commuting, then the following identity holds $[\bar c, c]=[c,\bar c]$,
\footnote{In fact, using component notation $[\bar c,c]^a=f^{abc}\bar c^b c^c$. Changing the index $b$ into $c$, and using the antisymmetry
of the structure constants we get
$f^{abc}\bar c^b c^c=f^{acb}\bar c^c c^b=-f^{acb}c^b \bar c^c=f^{abc}c^b \bar c^c\equiv [c,\bar c]$.}
implying that
\begin{align}
\bar s\psi_\mu=-B_\mu-\mathcal D_\mu^{(A-\tilde A)}(b-\tilde b)+\frac12[\mathcal D_\mu^{(A-\tilde A)}(\bar c-\tilde {\bar c}),(c-\tilde c)]
-\frac12[\mathcal D_\mu^{(A-\tilde A)}(c-\tilde c), (\bar c-\tilde {\bar c})],
\end{align}
where we used the fact that $-\frac12[(\bar c-\tilde {\bar c}),\mathcal D_\mu^{(A-\tilde A)}(c-\tilde c)]=
-\frac12[\mathcal D_\mu^{(A-\tilde A)}(c-\tilde c), (\bar c-\tilde {\bar c})]$.
According to \cite{Braga:1994dv}, the anti-variation $\bar s\psi_\mu$ in a linear gauge is
\beq
\bar s\psi_\mu=-B_\mu-\mathcal D_\mu^{(A-\tilde A)}(b-\tilde b)-[\mathcal D_\mu^{(A-\tilde A)}(c-\tilde c), (\bar c-\tilde {\bar c})].
\eeq
Therefore, in the presence of non-linear gauges, this variation becomes more symmetric, as far as the action of the covariant
derivative onto the FP ghosts.
It is worth checking the nilpotency of this transformation:
\begin{align}
\bar s^2\psi_\mu&=\bar s\left(
-B_\mu-\mathcal D_\mu^{(A-\tilde A)}(b-\tilde b)-\frac12\mathcal D_\mu^{(A-\tilde A)}[(\bar c-\tilde {\bar c}),(c-\tilde c)]
+[\mathcal D_\mu^{(A-\tilde A)}(\bar c-\tilde {\bar c}),(c-\tilde c)]\right)\nonumber\\
&=\bar s\left[\mathcal D_\mu^{(A-\tilde A)}\left(
-(b-\tilde b)-\frac12[(\bar c-\tilde {\bar c}),(c-\tilde c)]\right)\right]
+\bar s\left([\mathcal D_\mu^{(A-\tilde A)}(\bar c-\tilde {\bar c}),(c-\tilde c)]\right).
\end{align}
Using the identity $\bar s\left(\mathcal D_\mu^{(A-\tilde A)}(\bar c-\tilde {\bar c})\right)=0$ we obtain
\begin{align}
\bar s^2\psi_\mu&=
(\bar s \mathcal D_\mu^{(A-\tilde A)})\left(-(b-\tilde b)-\frac12[(\bar c-\tilde {\bar c}),(c-\tilde c)]\right)\nonumber\\
&-\Big[\mathcal D_\mu^{(A-\tilde A)}(\bar c-\tilde {\bar c}), -(b-\tilde b)-\frac12[(\bar c-\tilde {\bar c}),(c-\tilde c)]\Big]\nonumber\\
&=\Big[\mathcal D_\mu^{(A-\tilde A)}(\bar c-\tilde {\bar c}), -(b-\tilde b)-\frac12[(\bar c-\tilde {\bar c}),(c-\tilde c)]\Big]\nonumber\\
&-\Big[\mathcal D_\mu^{(A-\tilde A)}(\bar c-\tilde {\bar c}), -(b-\tilde b)-\frac12[(\bar c-\tilde {\bar c}),(c-\tilde c)]\Big]\nonumber\\
&=0
\end{align}
For the FP ghosts, we apply the same procedure, supposing the variations with respect to the two auxiliary fields to vanish:
\begin{align}
&\bar s\left\{\overline{B}\tilde c+\bar c^*\left(\epsilon+\frac12[c-\tilde c,c-\tilde c] \right)
+B\tilde{\bar c}+c^*\left(\bar \epsilon -(b-\tilde b)+\frac12[\bar c-\tilde {\bar c},c-\tilde c]\right) \right\}\nonumber\\
&=-B \bar c^*+\bar c^*\bar s \epsilon +\bar c^*\frac12\bar s [c-\tilde c,c-\tilde c]\nonumber\\
&\quad-\overline{B}c^*+c^*\bar s\bar \epsilon+c^*\left(-\bar s(b-\tilde b)+\frac12\bar s[\bar c-\tilde {\bar c},c-\tilde c]\right).
\end{align}
We separate the two contributions for $\epsilon$ and $\bar \epsilon$: for the second line we obtain
\footnote{It is worth noting that
$[c,[\bar c,c]]=f^{abc}f^{cmn}c^b\bar c^mc^n$. Using the Jacobi Identity $f^{abc}f^{cmn}=-f^{mac}f^{cbn}-f^{bmc}f^{can}$, we obtain
$f^{abc}f^{cmn}c^b\bar c^mc^n=-f^{mac}f^{cbn}c^b\bar c^mc^n-f^{bmc}f^{can}c^b\bar c^mc^n$. Rearranging the indices in the third term
we get
$f^{abc}f^{cmn}c^b\bar c^mc^n=-f^{mac}f^{cbn}c^b\bar c^mc^n-f^{abc}f^{cmn}c^b\bar c^mc^n$ and thus
$2f^{abc}f^{cmn}c^b\bar c^mc^n=-f^{mac}f^{cbn}c^b\bar c^mc^n$ or similarly
$2[c,[\bar c,c]]=+[\bar c,[ c,c]]=-[[c,c],\bar c]$.}
\begin{align}
\bar c^*\bar s\epsilon&=B \bar c^* -\bar c^*\frac12\bar s [c-\tilde c,c-\tilde c]\nonumber\\
&=\bar c^*B -\bar c^*\frac12[\bar s (c-\tilde c)](c-\tilde c)
+\bar c^*\frac12(c-\tilde c)\bar s(c-\tilde c)\nonumber\\
&=\bar c^*B -\bar c^*[(c-\tilde c),(b-\tilde b)]
-\frac12\bar c^*[(c-\tilde c),[(\bar c-\tilde {\bar c}),(c-\tilde c)]],
\end{align}
implying that
\begin{align}
\bar s\epsilon&=B -[(c-\tilde c),(b-\tilde b)]
-\frac12[(c-\tilde c),[(\bar c-\tilde {\bar c}),(c-\tilde c)]]\nonumber\\
&=B -[(c-\tilde c),(b-\tilde b)]
-\frac14[[(c-\tilde c),(c-\tilde c)],(\bar c-\tilde {\bar c})].
\end{align}
Checking the nilpotency we find an inconsistency: in fact
\begin{align}
\bar s^2\epsilon&=\bar s\left(B -[(c-\tilde c),(b-\tilde b)]
-\frac12[(c-\tilde c),[(\bar c-\tilde {\bar c}),(c-\tilde c)]]\right)\nonumber\\
&=\frac14[[(\bar c-\tilde {\bar c}),(c-\tilde c)],[(\bar c-\tilde {\bar c}),(c-\tilde c)]]\neq 0.
\end{align}
This means that we have to replace $\bar s B=0$ with
\beq
\bar s B=-\frac14[[(\bar c-\tilde {\bar c}),(c-\tilde c)],[(\bar c-\tilde {\bar c}),(c-\tilde c)]],
\eeq
which is nilpotent
\begin{align}
\bar s^2B=0,
\end{align}
because of the following identities
\begin{align}
[[[\bar c,\bar c],c],[\bar c,c]]&=-[[\bar c,c],[[\bar c,\bar c],c]]\nonumber\\
[[\bar c,b],[\bar c,c]]&=-[[\bar c,c],[\bar c,b]].
\end{align}
Similarly for $\bar \epsilon$ we obtain
\begin{align}
c^*\bar s\bar \epsilon&=\overline{B}c^*-c^*\left(-\bar s(b-\tilde b)+\frac12\bar s[\bar c-\tilde {\bar c},c-\tilde c]\right)\nonumber\\
&=c^*\overline{B}-c^*[(\bar c-\tilde {\bar c}), (b-\tilde b)]
+\frac1{4}c^*[[\bar c-\tilde{\bar c},\bar c-\tilde{\bar c}],c-\tilde c].
\end{align}
It is clear that this transformation is nilpotent because
\beq
\bar s^2b=\bar s\left(-\frac12[\bar c,b]+\frac18[[\bar c,\bar c],c]\right)=0.
\eeq
For the $b$ and $\tilde b$ fields we have, according to the identities $[c,[\bar c,b]]=[[\bar c,b],c]$
and $[c,[\bar c,b]]=[[\bar c,c],b]+[[c,b],\bar c]$, derived from the anti-symmetric property of the structure constants and the Jacobi Identity,
\begin{align}
b^*\bar s\epsilon_b&=-B_bb^*-b^*\bar s\left(\frac12[(c-\tilde c),(b-\tilde b)]\right)
-b^*\bar s\left(\frac18[[c-\tilde c, c-\tilde c],\bar c-\tilde {\bar c}]\right)\nonumber\\
b^*\bar s\epsilon_b&=-b^*B_b+\frac14b^*[[\bar c-\tilde {\bar c}, c-\tilde c],b-\tilde b]
+\frac1{16}b^*[[\bar c-\tilde {\bar c},\bar c-\tilde {\bar c}], [c -\tilde c, c-\tilde c]].
\end{align}
Therefore, we generate the following anti-transformations on the ghosts
\begin{align}
\label{BV4}
\bar s \psi_\mu&=-B_\mu-\mathcal D_\mu^{(A-\tilde A)}(b-\tilde b)+\frac12[\mathcal D_\mu^{(A-\tilde A)}(\bar c-\tilde {\bar c}),(c-\tilde c)]
-\frac12[\mathcal D_\mu^{(A-\tilde A)}(c-\tilde c), (\bar c-\tilde {\bar c})]
\nonumber\\
\bar s \epsilon&=B -[(c-\tilde c),(b-\tilde b)]
+\frac14[[(c-\tilde c),(c-\tilde c)],(\bar c-\tilde {\bar c})]\nonumber\\
\bar s \bar \epsilon&=
\overline{B}-[(\bar c-\tilde {\bar c}), (b-\tilde b)]
+\frac14[[(\bar c-\tilde{\bar c}),(\bar c-\tilde{\bar c})],(c-\tilde c)]\nonumber\\
\bar s\epsilon_b&=-B_b+\frac14[[\bar c-\tilde {\bar c}, c-\tilde c],b-\tilde b]
+\frac1{16}[[\bar c-\tilde {\bar c},\bar c-\tilde {\bar c}], [c -\tilde c, c-\tilde c]]
\end{align}
and on the anti-fields
\begin{align}
\label{BV5}
\bar s A^*_\mu&=0  & \bar s B_\mu&=0  \nonumber\\
\bar s c^*&=0      & \bar s B&=0 \nonumber\\
\bar s \bar c^*&=0 & \bar s \overline{B}&=0 \nonumber\\
\bar s b^*&=0      &  \bar s B_b&=0.      
\end{align}
We may notice an usual structure for the anti-variations of the fields associated to the shift-symmetry. These equations look similar
to the usual form we saw for the anti-BRST transformations (\ref{gaugetrans4}), $\bar s c= -b -\frac12[\bar c,c]$.
The vector ghost $\psi_\mu$ behaves as a gauge field due to the presence of the covariant derivative: yet,
due to its Ghost number (${\mathcal G}h(\psi_\mu)=1$), the ghost field appearing in (\ref{gaugetrans1}) ($s A_\mu=\mat D_\mu^{(A)}c$) has been replaced
by the Nakanishi-Lautrup field in the second term, plus an additional coupling of the covariant derivative with the FP ghost in
the adjoint representation. The anti-transformations for $\epsilon$ and $\bar \epsilon$ are obtained one from the other by ghost exchange operation,
whereas for $\epsilon_b$ we notice an additional coupling with ${}\times (c-\tilde c)$ with respect to terms appearing in $\bar s b$.

In order to achieve a more symmetric for (\ref{BV3}), (\ref{BV4}) and (\ref{BV5}), we can adopt 
the same procedure used to symmetrise standard BRST transformations in Chapter 3.
Consider in fact the following shift in $B_\mu$
\begin{align}
B_\mu\to &B_\mu
-\frac12\mathcal D_\mu^{(A-\tilde A)}(b-\tilde b)+\frac14[\mathcal D_\mu^{(A-\tilde A)}(\bar c-\tilde {\bar c}),(c-\tilde c)]\nonumber\\
&-\frac14[\mathcal D_\mu^{(A-\tilde A)}(c-\tilde c),(\bar c-\tilde {\bar c})]
\end{align}
Apply now this shift to $\bar s\psi_\mu$ and to $sA_\mu^*$ and what we get is
\begin{align}
\bar s \psi_\mu&=-B_\mu-\frac12\mathcal D_\mu^{(A-\tilde A)}(b-\tilde b)+\frac14[\mathcal D_\mu^{(A-\tilde A)}(\bar c-\tilde {\bar c}),(c-\tilde c)]
-\frac14[\mathcal D_\mu^{(A-\tilde A)}(c-\tilde c),(\bar c-\tilde {\bar c})]
\nonumber\\
sA_\mu^*&=
B_\mu
-\frac12\mathcal D_\mu^{(A-\tilde A)}(b-\tilde b)+\frac14[\mathcal D_\mu^{(A-\tilde A)}(\bar c-\tilde {\bar c}),(c-\tilde c)]
-\frac14[\mathcal D_\mu^{(A-\tilde A)}(c-\tilde c),(\bar c-\tilde {\bar c})]
\end{align}
which tell us how the anti-field of the gauge-field behaves as the anti-ghost for the ghost field associated to the shift symmetry, as well as
in ordinary BRST, $\bar c$ is the anti-ghost of $c$. The only difference in the geometric interpretation of the BRST operators as
differential operators in the superspace lies in the different sign with respect to the auxiliary field $b$ (Note we have generated
the same difference in sign with respect to $B_\mu$).
The same approach can be adopted to all the other remaining auxiliary fields associated with the shift symmetry
\begin{align}
\bar s \epsilon&=B-\frac12[c-\tilde c,b-\tilde b]-\frac18[[c-\tilde c,c-\tilde c],\bar c-\tilde{\bar c}]\nonumber\\
\bar s \bar \epsilon&=\overline{B}
-\frac12[\bar c-\tilde{\bar c},b-\tilde b]-\frac18[[\bar c-\tilde{\bar c},\bar c-\tilde{\bar c}],c-\tilde c], \nonumber\\
\bar s\epsilon_b&=-B_b-\frac14[[\bar c-\tilde{\bar c}, c-\tilde c],b-\tilde b]
+\frac1{16}[[[\bar c-\tilde{\bar c},\bar c-\tilde{\bar c}],c-\tilde c],c-\tilde c],\nonumber\\
\end{align}
and the anti-fields
\begin{align}
sc^*&=B+\frac12[c-\tilde c,b-\tilde b]-\frac18[[c-\tilde c,c-\tilde c],\bar c-\tilde{\bar c}]\nonumber\\
s\bar c^*&=\overline{B}+\frac12[\bar c-\tilde{\bar c},b-\tilde b]-\frac18[[\bar c-\tilde{\bar c},\bar c-\tilde{\bar c}],c-\tilde c]\nonumber\\
sb^*&=B_b-\frac14[[\bar c-\tilde{\bar c}, c-\tilde c],b-\tilde b]
+\frac1{16}[[[\bar c-\tilde{\bar c},\bar c-\tilde{\bar c}],c-\tilde c],c-\tilde c].
\end{align}
We notice a geometric feature in the transformations for the two fields associated respectively to the Y-M ghost and anti-ghost:
in this case, the difference in sign is not with respect to the auxiliary fields, $B$ and $\overline B$, but in the other of
the transformations. Thus is due to the fact that the anti-fields associated with $c$ and $\bar c$ have an even Grassmann parity,
reflected in the BRST and anti-BRST transformations. The anti-fields $b^*$ behaves as usual.
The transformations for the four Nakanishi-Lautrup fields are more complicated: let us see in details $sB_\mu$
\begin{align}
sB_\mu&=s\left\{ -\frac12\mathcal D_\mu^{(A-\tilde A)}(b-\tilde b)+\frac12[\mathcal D_\mu^{(A-\tilde A)}(\bar c-\tilde {\bar c}),(c-\tilde c)]
\right\}\nonumber\\
&=-\frac12\left\{\mathcal D_\mu^{(A-\tilde A)}s(b-\tilde b)-[\mathcal D_\mu^{(A-\tilde A)}(c-\tilde c), b-\tilde b]\right\}\nonumber\\
&\quad+\frac12[s\left(\mathcal D_\mu^{(A-\tilde A)}(\bar c-\tilde {\bar c})\right),(c-\tilde c)]
-\frac12[\mathcal D_\mu^{(A-\tilde A)}(\bar c-\tilde {\bar c}),s(c-\tilde c)]
\end{align}
The term $s\left(\mathcal D_\mu^{(A-\tilde A)}(\bar c-\tilde {\bar c})\right)$ vanishes and so we obtain
\begin{align}
sB_\mu&=
-\frac12\mathcal D_\mu^{(A-\tilde A)}\left\{-\frac12[c-\tilde c,b-\tilde b]-\frac18[[c-\tilde c,c-\tilde c],\bar c-\tilde{\bar c}]\right\}\nonumber\\
&\quad+\frac12[\mathcal D_\mu^{(A-\tilde A)}(c-\tilde c),b-\tilde b]
+\frac14[\mathcal D_\mu^{(A-\tilde A)}(\bar c-\tilde {\bar c}),[c-\tilde c,c-\tilde c]].
\end{align}
Using the Leibnitz rule for the covariant derivative $sB_\mu$ assumes the more compact form
\begin{align}
sB_\mu&=\frac12
\end{align}
\\
\begin{align}
sB&=s\left\{\frac12[c-\tilde c,b-\tilde b]+\frac18[[c-\tilde c,c-\tilde c],\bar c-\tilde{\bar c}] \right\}\nonumber\\
&= \\
\bar s B&=
\end{align}
\begin{align}
s{\overline B}&=s\left\{\frac12[\bar c-\tilde{\bar c},b-\tilde b]-\frac18[[\bar c-\tilde{\bar c},\bar c-\tilde{\bar c}],c-\tilde c]\right\}\nonumber\\
&= \\
\bar s\overline{B}&=
\end{align}
\begin{align}
s B_b&=s\left\{-\frac14[[\bar c-\tilde{\bar c}, c-\tilde c],b-\tilde b]
+\frac1{16}[[[\bar c-\tilde{\bar c},\bar c-\tilde{\bar c}],c-\tilde c],c-\tilde c] \right\}\nonumber\\
&= \\
\bar sB_b&=.
\end{align}
We may be naively tempted to symmetrise also the ghosts associate with the shift symmetry, but that would cause an ambiguity in the definition
of the covariant, one of the essential features of the BV formalism, as pointed out previously.
As a final part of this section, we show how to write the full Lagrangian as double BRST-exact quantity.
We remind the reader that in the last section we wrote the full Lagrangian as
\begin{align}
\mathcal L&=\mathcal L_0+s\overline{\Psi}\nonumber\\
&=\mathcal L_0+s\Big\{-A_\mu^* \tilde A^\mu+\bar c^* \tilde c+c^*\tilde{\bar c}-b^*\tilde b\Big\}.
\end{align}
By demanding the BRST symmetry to be unbroken, we can then generate the gauge-fixing anti-fermion as
\begin{align}
\mathcal L&=\mathcal L_0+s\overline{\Psi}=\mathcal L_0-(-1)^{\epsilon(\Phi)}s\Big(\Phi^*\tilde\Phi\Big)\nonumber\\
&= \mathcal L_0+s\bar s \Sigma
=\mathcal L_0+\frac12\, s\bar s\Big(-\tilde A_\mu\tilde A^\mu+\tilde c \tilde c+\tilde{\bar c}\tilde{\bar c}-\tilde b\tilde b \Big )\nonumber\\
&=\mathcal L_0-(-1)^{\epsilon(\Phi)}\frac12\,s\bar s\Big(\tilde \Phi\tilde\Phi\Big)
\end{align}

\thispagestyle{empty}
\cleardoublepage
\chapter{Conclusions}

This Thesis has been devoted to two main subjects:
the study of gauge fixing methods in non-Abelian gauge theory and the BRST formalism, both in perturbative and non-perturbative QCD.

We have thus found a representation for Landau gauge-fixing
corresponding to the FP trick being an actual change of variables
with appropriate determinant. The resulting gauge-fixing Lagrangian
density enjoys a larger extended BRST and anti-BRST symmetry. However
it cannot be represented rigorously as a BRST exact object, rather
the sum of two such objects corresponding to different BRST operations.
This means that some of the BRST machinery is not available
to this formulation, such as the Kugo-Ojima criterion for 
selecting physical states. 
We discuss cursorily
now the perturbative renormalisability of the present formulation
of the theory. Note that the procedure leading to Eq. (\ref{gfLag})
does not introduce any new coupling constants; only the
strong coupling constant $g$ is present in $M_F[A]$ coupling
the Yang-Mills field to both the new ghosts and scalars.
The dimensions of the new fields are 
\begin{equation}
[\varphi]=L^0, \quad [d]=[\bar d] = L^{-1}, \quad [B]= L^{-2} .
\end{equation}
Most importantly in this context, the kinetic term for the
new boson fields $\varphi^a$ is {\it quartic} in derivatives:
\begin{equation}
\mathcal{L}_{\rm kin} = \varphi^a (\partial^2)^2 \varphi^a\,,
\end{equation}
which is renormalisable, by power counting, since $\varphi^a$ are dimensionless.
Such a contribution is seemingly harmless in the ultraviolet regime: for 
large momenta propagators will vanish like $1/p^4$. 
Moreover it should play an important role in guaranteeing
the decoupling of such contributions in perturbative diagrams.
That such a decoupling should occur is clear from Eq. (\ref{moddet}): in
the perturbative regime fluctuations about $A_{\mu}=0$ will not feel the 
${\rm {sgn}}({\rm {det}}M_F[A])$,
so that the field theory constructed in this way must
be equivalent to the perturbatively renormalisable Landau gauge fixed
theory. For example in the computation of the running
coupling constant we expect that this property will lead to
a complete decoupling of the $t$-degrees of freedom so that the
known Landau gauge result emerges from just the gluon and standard
ghost sectors.  
Naturally, the new degrees of freedom will be relevant in
the infra-red regime, which will be the object of future study

Regarding the BRST formalism in non-perturbative non-Abelian theories, we showed that
the massive Curci-Ferrari model with its decontracted double BRST symmetry
can be formulated on the lattice without the 0/0 problem. 
The parameter $m^2$ is not interpreted as a physical mass but rather serves 
to meaningfully define a limit $m^2\to 0$ in the spirit of l'Hospital's
rule.  At finite $m^2$ the topological nature of the gauge-fixing
partition function seems lost.  It is possible,
however, to tune the Curci-Ferrari mass with the gauge parameter $\xi
$ so that the limit $m^2\to 0$ can be defined along a certain trajectory in
parameter space independent of $\xi $. An interesting open question
might then be the topological interpretation of the model within the
decontracted double BRST $osp(1|2)$ superalgebra framework.

In the Batalin-Vilkovisky formalism for non-linear gauges, we showed how the BRST and anti-BRST
transformations assume a more complicated form than with respect to standard linear gauges, such as Landau gauge.
We have constructed an algebraic BRST structure which still preserves the required nilpotency, allowing us to write
the complete B-V Lagrangian in the form of a coboundary term, both for BRST and anti-BRST transformations
The natural implementation of this work leads to derive the lattice algebraic structure of this theory, 
mimicking the Curci-Ferrai model we have already proposed, in which the background lattice gauge-fixing has to
be translated in the language of the anti-field formalism.



\thispagestyle{empty}
\cleardoublepage
\chapter{Appendix A}

\begin{center}
{\Large Connection on a principal bundle}
\end{center}

In this appendix we will enlist briefly the major topics of gauge theory from the geometric and topological point of view. This is necessary
to understand the rich geometric structure of Y-M theory. Moreover, its generalisation to super-space is essential to the comprehension
of super-symmetry, BRST and topological field theory (TFT). This will be covered in Appendix C. We assume the reader being familiar
with basic concepts of topology, such as manifolds, tangent and cotangent spaces. For this we remind the interested reader to \cite{Naber:2000bp}.
As previously said, Y-M theories can be regarded as the quantum theory of principal bundles, on which we construct connections, covariant derivatives
and curvature forms. To start with we define a principal bundle:
a differentiable principal fiber bundle over a manifold $M$ with group structure $G$ consists of a manifold $P$ and an action of $G$ on $P$ 
satisfying the following conditions
\begin{itemize}
\item $G$ acts freely on $P$ without fixed points, i.e. $gx=x$ implies $g=I$ (only the identity element fixes any $x$), $P\times G\to P$ 
      is denoted by $P\times G\ni(u,a)\to ua\in P$;
\item $M$ is the quotient space of $P$ by the equivalence relation induced by $G$, $M=P/G$, and the canonical projection $\pi:P\to M$ is differentiable;
\item $P$ is locally trivial ($P \cong \mathbb R^nl$).
\end{itemize}
To any element $A$ of the algebra $\algebra$ of $G$, we associate a vector $\Sigma(A)$ on $P$, the fundamental vector field corresponding to $A$. 
$\Sigma(A)$ is actually generated by the right action of $G$ on $P$
\footnote{Left and right actions of a group element are diffeomorphisms defined as
$L_g(h)=hg$ and $R_g(h)=hg$.}
: if $A\in\algebra$, then $\exp(tA)$ is a one-parameter subgroup of $G$, acting on $P$
as
\beq
\Sigma(A)_u\cdot f=\frac{d}{dt}f(u_t)\Big|_{t=0},
\eeq
where $u_t=R_{\exp(tA)}(u)$. $\Sigma(A)_u$ is a vector tangent to $P$ at $u$ (tangent to the fiber). Call $G_u$ the subspace of $T_u(P)$ 
of vectors tangent to the fiber through $u$, at $u$.
\beq
\Sigma:\algebra\to G_u \quad\textrm{is an isomorphism}.
\eeq
A connection in $P$ is a choice of a supplementary linear subspace $Q_u$ in $T_u(P)$ to $G_u$
\beq
T_u(P)=G_u\oplus Q_u
\eeq
where $Q_{ua}=(R_a)_\ast Q_u$ is a {\it push-forward} 
\footnote{Let $M$ and $N$ be two smooth manifolds, with dimension $m$ and $n$ respectively. Let $f: M\to N$ be a smooth function. Then, the 
{\it differential} or {\it push-forward} $f_{\ast}$ (or ${\rm d}f$) of $f$ in the point $p\in M$ is the application
$f_{\ast}:T_pM\to T_{f(p)}N$. The push-forward defines then a change of variables in tangent spaces.}
and depends differentially on $u$. $Q_u$ is called the horizontal space and $G_u$ the vertical space. 
Choosing a $Q_u$ amounts to
choosing a basis in $T_u(P)$, though this distribution is not, in general integrable. Geometrically, this corresponds to 
the non triviality of parallel transport using the holonoy group of the principal bundle.

\begin{center}
{\Large Connection form}
\end{center}

A connection form is a Lie-algebra valued 1-form $\omega$ such that 
\begin{itemize}
\item $\omega$ applied on any fundamental vector field $\Sigma(A)$ reproduces $A$, i.e. $\omega(\Sigma (A))=A$;
\item $(R^\ast_a\omega)(X)=\Ad_{a^{-1}}\cdot \omega(X)$. The horizontal subspace $Q_u$ is the kernel of $\omega$, that is to say that $X_u$ 
      is horizontal iff $\omega(X_u)=0$;
\end{itemize}
where $(R^\ast_a\omega)(X)$ is a {\it pull-back}
\footnote{The transpose action of the push-forward is the pull-back $f^\ast$ (or $\delta f$), defined as $f^{\ast}:T^{\ast}_{f(p)}N\to T^{\ast}_{p}M$.
Contrary to the push-forward, we cannot pass from the cotangent space $T^{\ast}_{p}M$ to $T^{\ast}_{f(p)}N$, but only
the other way round, linking a change of variables for cotangent spaces, dual of tangent spaces.}.
It's possible to express $\omega$, the connection form on $P$, by a family of local forms, each one being defined in an open subset of 
the base-space manifold $M$.
Let $\{U_\alpha\}$ be a covering of $M$, we choose in $P$ the preferred set of local sections $\sigma_\alpha$ and the corresponding transition 
functions $\psi_{\alpha\beta}$:
for each $\alpha$ and $\beta$. we define a Lie-algebra-valued 1-form on $U_\alpha$ by
\beq
\omega_\alpha=\sigma^\ast_\alpha\omega\qquad\textrm{pull-back of}\,\, \omega\,\,{\rm through}\,\,\sigma_\alpha
\eeq
where
\beq
\omega_\beta=\Ad_{\psi^{-1}_{\alpha\beta}}\cdot \omega_\alpha+\psi^{-1}_{\alpha\beta}\,{\rm d}\psi_{\alpha\beta}
\eeq
in $U_\alpha\cap U_\beta$. If $\omega$ is a connection form on $P=M\times G$, we can construct from a global section $\sigma_1$ of $P$ the form on $M$
\beq
\omega_1=\sigma^\ast_1(\omega).
\eeq
If we now use a $G$-valued function ${\rm g}$ on $M$ to transform $\sigma_2$ into $\sigma_2(x)=\sigma_1(x)\cdot{\rm g(x)}$, we can define a new 1-form on $M$
\beq
\omega_2=\sigma_2^\ast(\omega)
\eeq
we have
\beq
\omega_2=\Ad_{{\rm g}^{-1}}\cdot \omega_1+{\rm g}^{-1}\,{\rm d}{\rm g}.
\eeq

\begin{center}
{\Large Geometrical interpretation of gauge potentials}
\end{center}

On a 4-dim manifold, the connection form $\omega$, defined on an open subset of $M$, $U_\alpha$, can be expressed as
\beq
\omega_\alpha=A^\mu_{\alpha}(x){\rm d}x_\mu
\eeq
whose Lie-valued components transform as
\beq
A^{'\mu}(x)=\Ad_{{\rm g}^{-1}}\cdot A^\mu+{\rm g}^{-1}\,\partial^\mu{\rm g}
\eeq
which are the components of the transformed connection form
\beq
\omega'_\alpha=\sigma^{'\ast}_\alpha\omega=A^{'\mu}(x){\rm d}x_\mu.
\eeq
A change of $\sigma$ by the action of some $G$-valued function $\rm g$ on $M$ can be viewed as a change of coordinates in the principal 
fiber bundle $P$, and
it induces a transformation of the components $A^\mu$ similar to the usual gauge transformation of potentials. Then
the gauge potential naturally becomes the component of a geometrical object of a definite type: a connection form on ${\rm P}$.


\begin{center}
{\Large Covariant derivative}
\end{center}

The concept of covariant derivative is strongly related to the horizontal lift of the derivative $\partial_\mu$.
A vector field $\bar X$ is the lift of a vector field $X$ on $M$, which is the horizontal field on $P$, which projects onto $X$, s.t.
\beq
\pi_\ast(\tilde X_u)=X_{\pi(u)}\quad{\rm where}\quad \pi:P=M\times G\to M.
\eeq
Suppose we choose a local chart $U_\alpha$ on $M$, with local coordinates $\{x^\mu\}$. Then, we construct vector fields, with generators as 
$\partial_\mu=\frac{\partial}{\partial x^\mu}$, whose lift $\tilde \partial_\mu$ lies on $\pi^{-1}(U_\alpha)=U_\alpha\times G$.
If $\sigma_\alpha$ is section over $U_\alpha$, then
\beq
\omega_\alpha(\partial_\mu)=\sigma^\ast_\alpha\,\omega(\partial_\mu)=\omega(\sigma_{\alpha\ast}\partial_\mu)=(A^\nu_{\alpha}(x){\rm d}x_\nu,\partial_\mu)=
A_{\alpha\mu}=\omega(\Sigma(A_{\alpha\mu}))
\eeq
Hence
\beq
\omega(\sigma_{\alpha\ast}\partial_\mu-\Sigma(A_{\alpha\mu}))=0,
\eeq
where $\sigma_{\alpha\ast}\partial_\mu-\Sigma(A_{\alpha\mu})$ is evidently horizontal. Then 
\beq
\tilde\partial_\mu\Big|_u=\sigma_{\alpha\ast}\partial_\mu-\Sigma(A_{\alpha\mu}) \quad {\rm with}\quad u=\sigma_\alpha(x).
\eeq
We can identify $\sigma_{\alpha\ast}\partial_\mu$ with $\partial_\mu$ and $-\Sigma(A_{\alpha\mu})$ with the Lie-algebra-valued element $A_\mu$, to recover the usual
covariant derivative
\beq
\mat D_\mu=\partial_\mu-A_\mu.
\eeq
So, any point on the local section $\sigma_\alpha$, defined by $\pi^{-1}(U_\alpha)=U_\alpha\times G$, can be thought of as
\beq
u_0=\sigma_\alpha(x_0)=(x_0,e)=\sigma_{\alpha\ast}\partial_\mu\oplus \Sigma(A_{\alpha\mu}).
\eeq
This point $u_0$ is generated by the curve on the fiber $\pi^{-1}(U_\alpha)\Big|_{x_0}$
\beq
P\supseteq \pi^{-1}(x_0)\ni u_t=u_0\,\exp(tA_\mu)=(\underbrace{x_0}_{\textrm{point in $M$}},\underbrace{e^{tA_\mu}}_{\textrm{element of $G$}}).
\eeq
If $f$ is a function on $\pi^{-1}(U_\alpha)$, then the restriction of this function to $\pi^{-1}(x_0)$ is a function $F$ defined on $G$, because it's 
$e^{tA_\mu}$ which localises $\pi^{-1}(U_\alpha)$ to $\pi^{-1}(x_0)$. The directional derivative along $u_t$ is clearly the action of the Lie algebra element $A_\mu$
on $F$ at $e$. Thus, the covariant derivative is section-dependent.
There's also an other way to interpret the covariant derivative, which follows from the adjoint action on any element of $P$
\beq
\partial_\mu \psi=\partial_\mu \psi-\lim_{t\to0}\frac1t\left[e^{-tA_\mu}\psi(u_0)e^{tA_\mu}-\psi(u_0)\right]
\eeq
for any function $\psi$ on $P$, s.t. $\psi(ua)=\Ad_{a^{-1}}\psi$. It is also important to notice that while the commutator of two fundamental vectors is still a fundamental
vector, showing that this map preserves space and algebra structure, it is not true that the commutator of two 
horizontal vector fields is still horizontal. In particular
\beq
[\mat D_\mu,\,\mat D_\nu]=-(\partial_{[\mu}A_{\nu]}+[A-\mu,A_\nu])=-F_{\mu\nu}
\eeq
is a fundamental vector field on the bundle space written as a Lie algebra element, and moreover
\beq
F_{\beta\mu\nu}=\Ad_{\psi^{-1}_{\alpha\beta}}\cdot F_{\alpha\mu\nu}\quad \to \quad F'_{\mu\nu}(x)=\Ad_{{\rm g}^{-1}}\cdot F_{\mu\nu}(x)
\eeq
on $U_\alpha\cap U_\beta$.

\begin{center}
{\Large Curvature form}
\end{center}

From the commutator of two covariant derivatives, which is a fundamental vector on $P$, s.t. $\omega([\mat D_\mu,\,\mat D_\nu])=[\mat D_\mu,\,\mat D_\nu]$,
we can construct a Lie-algebra valued 2-form$\Omega$. Locally, on $U_\alpha\cap U_\beta$
\beq
\Omega_\alpha=\frac12F_{\alpha\mu\nu}\,{\rm d}x^\mu\wedge{\rm d}x^\nu
\eeq
with
\beq
\Omega_\beta=\Ad_{\psi^{-1}_{\alpha\beta}}\Omega_\alpha.
\eeq
To connect the curvature form to the connection form, we need to introduce the covariant exterior derivative ${\rm d}_\omega$, as
\beq
\Omega={\rm d}_\omega\omega={\rm d}\omega +\frac12[\omega,\omega].
\eeq
If $X$ and $Y$ are two tangent vector to the bundle, then
\beq
\Omega(X,Y)={\rm d}\omega(X,Y) +\frac12[\omega(X),\omega(Y)].
\eeq
Let's decompose $X$ and $Y$ into their vertical and horizontal components
\beq
X=hX\oplus vX\qquad Y=hY\oplus vY
\eeq
then, what we get is
\begin{align}
\Omega(X, Y)&={\rm d}\omega(hX,hY)+{\rm d}\omega(vX,vY)+{\rm d}\omega(hX,vY)+{\rm d}\omega(vX,hY)\nonumber\\
&+\frac12[\omega(vX),\omega(vY)]+\frac12[\omega(hX),\omega(hY)]\nonumber\\
&={\rm d}\omega(hX,hY).
\end{align}
Though ${\rm d}^2=0$, $\mat D^2\neq 0$, whereas $\mat D\Omega=0, \forall \omega$ (Bianchi ideintity), using the Jacobi identity.

\begin{center}
{\Large Group of gauge transformations}
\end{center}

Gauge transformations are equivariant automorphisms of some $G$-bundle $P$. The 1-forms of connections are the physical interesting objects, 
whose components
are the gauge potentials. Choosing a particular $G$-bundle automatically defines the set of Chern classes. ($P_k, k\in \mathbb Z$).
In this context, the gauge transformations assume the role of elements of an infinite-dimensional Lie group, called $\mat G$, 
whose group composition is smooth
\beq
\Phi:P\to P,\qquad \Phi\in \mat C^\infty(\Ad P).
\eeq
This group composition can be expressed as follows
\be
\bea{ccc}
\forall g\in \mat G|g:P\to P \Rightarrow \exists \gamma\in {\rm Map}(\Ad P), \gamma:P\to G,\\
\\
g(u)=u\cdot \gamma(u),\quad u\in P,\,\gamma(ua)=a^{-1}\gamma(u)a,\\
\\
\forall g,h\in \mat G, g\circ h (u)=u\cdot(\gamma_h\cdot \gamma_h)(u).
\eea
\ee
Locally, the mapping $\gamma: P\to G$ can be written as
\beq
\gamma_\beta(x)=\psi^{-1}_{\alpha\beta}(x)\,\gamma_\alpha(x)\,\psi_{\alpha\beta}(x), \forall x\in U_\alpha\cap U_\beta, \{U_{i\in I}\}\subseteq P.
\eeq
This representation is in 1-1 correspondence with the sections of the bundle $\mat B$ associated 
with $P$ with standard fiber $G$, $G$ acting on itself by the adjoint
map ($a(g)=aga^{-1}$). The group $\mat G$ of gauge transformations can be identified with the set 
$\Gamma(\mat B)$ of sections of $\mat B$, which is not a principal bundle
though, because the action of $G$ is not free. $\mat B$ will have global sections and unit element $(x,e)$.
The Lie algebra of $\mat G$, ${\rm Lie}\mat G$. As we know, elements of sections of tangent and cotangent bundles are respectively vector fields and forms.
Consider the constant unit section $s$ of $\mat B$: through any point of 
$\mat B$ passes one fiber. Using the local triviality of $mat B$ over patches $U_\alpha$,
we may identify the fiber with the group $G$. Tangent vectors to the fiber 
$s$ follow immediately, as well as parallel transport and all the operations on vector
fields. These fields are elements of the algebra of $G$, vectors to a 
fiber $\pi^{-1}_{\\mat B}(x)$, with $x\in U_\alpha$. On the transition $U_\alpha\cap U_\beta$,
the map is of course
\beq
A_\beta=\Ad_{\psi^{-1}_{\alpha\beta}}\cdot A_\alpha.
\eeq
The field we have just determined on $\mat B$ can be identified as a section of an associated bundle $E$ to $P$, 
where the fiber is $\algebra$ and the adjoint action
of $G$ on $\algebra$.
Then, $\Gamma(E)$ is the Lie algebra of $\mat G\equiv \gamma(\mat B)$. $\Gamma(E)$ is an infinite-dimensional 
module. Any section of $\mat B$ can be written as
\be
\bea{rr}
\mat C^\infty(\Ad P)=\mat G\equiv\Gamma(\mat B)\ni s=\exp(\sigma),\,\sigma\in \Gamma(E)\\
\\
\sigma:U_\alpha\to \algebra.
\eea
\ee
At last, there's a particular class of gauge transformations, those which have values in the center $Z$ of $G$: for such a transformation we have on some 
local chart $U_\alpha$
\beq
{}^g\!A^\alpha_\mu=g_\alpha^{-1}A^\alpha_\mu g_\alpha+g_\alpha^{-1}\partial_\mu g_\alpha=A^\alpha_\mu.
\eeq
This can be also written as
\beq
\partial_\mu g_\alpha+[A^\alpha_\mu,g_\alpha]=\mat D^\alpha_\mu g_\alpha=0,
\eeq
i.e. $\nabla g_\alpha=0$. Then, $g_\alpha$ belongs to the center of the holonomy group of the connection under consideration. In fact
\beq
\mat D_{[\mu}\mat D_{\nu]}g=[F_\mu\nu,g]=0.
\eeq

\begin{center}
{\Large Covariant derivative in background}
\end{center}

As an exercise, consider the covariant derivative, whose connection is a pure gauge, 
acting on a generic function $\omega$ in the adjoint representation
\beq
\mat {D}_\mu[g^\dagger \partial_\mu g]\,\omega\, = \partial_\mu \,\omega\,+[g^\dagger \partial_\mu g,\,\omega\,].  
\eeq
Explicitly, it becomes
\begin{align}
\partial_\mu \,\omega\,+[g^\dagger \partial_\mu g,\,\omega\,]&=\partial_\mu \,\omega\,+g^\dagger \partial_\mu g\,\omega\,-\,\omega\, 
g^\dagger \partial_\mu g\nonumber \\ 
&=g^\dagger g \partial_\mu \,\omega\, g^\dagger g +g^\dagger \partial_\mu g \,\omega\, g^\dagger g 
- g^\dagger g \,\omega\, g^\dagger \partial_\mu g g^\dagger g \nonumber \\
&=g^\dagger [g\partial_\mu \,\omega\, g^\dagger +\partial_\mu g \,\omega\, g^\dagger -g\,\omega\, g^\dagger \partial_\mu g g^\dagger]g\nonumber \\
&=g^\dagger [g\partial_\mu \,\omega\, g^\dagger +\partial_\mu g \,\omega\, g^\dagger 
+g\,\omega\, \partial_\mu g^\dagger]g\nonumber\\
&=g^\dagger [\partial_\mu(g \,\omega\, g^\dagger) ]g,
\end{align}
where we have used the identity
\beq
g^\dagger g=\mathbb I \qquad \Rightarrow\qquad\partial_\mu(g^\dagger g )={\bf 0}.
\eeq
To evaluate the operator $\mat D^2$, we use then this compact expression
\be
\mat {D}_\mu[g^\dagger \partial_\mu g]=g^\dagger [\partial_\mu(g \,\omega\, g^\dagger) ]g \nonumber 
\ee
\be
\Downarrow \nonumber
\ee
\beq 
\mat {D}^2(\,\omega\,)= g^\dagger [\partial_\mu(g [g^\dagger [\partial_\mu
(g \,\omega\, g^\dagger) ]g] g^\dagger) ]g= g^\dagger (\Box[g \,\omega\, g^\dagger])g.
\eeq


\thispagestyle{empty}
\clearpage
\chapter{Appendix B}

\begin{center}
{\Large Hubbard-Stratonovich transformations to linearize the quartic-ghost interaction}
\end{center}

The massive Curci-Ferrari gauge-fixing Lagrangian density presents, 
being a specific example of a broader class of non-linear gauges, the feature of a quartic-ghost interaction: this is necessary in a non-linear gauge
to preserve renormalizability. This non-linearity also contributes to prevent from applying straightforwardly Grassmann integration,
which would turn out into the more common form of a functional determinant of the Faddeev-Popov operator, as it happens for instance in Landau gauge.
To avoid such a problem, we will perform a linearisation of the quartic term, in the framework of path integral linearisation technique,
making using of the Hubbord-Stratonovich transformations. To begin with, we choose a $SU(N)$ Lagrangian density, ghost/anti-ghost symmetric,
quantized in the massive Curci-Ferrari gauge
\begin{align}
\label{lagrangian}
\mat L_{\rm mCF}&=\tr\left\{\frac\xi2b^2+ibF[{}^g\!A]+\frac{m^2}{2}({}^g\!A)^2+\frac i2 \bar c\{\partial,\mat D\}c
-im^2\xi\bar cc+\frac{{\rm g}^2}{8}\xi(\bar c\times c)^2\right\},
\end{align}
which is left invariant under the following BRST and anti-BRST matrix transformations
\begin{align}
\label{BRST}
s A_\mu&=-\mat D_\mu c                        & \bar sA_\mu &=-\mat D_\mu \bar c\nonumber\\
sc&=- \frac {\rm g}2 c\times c                & \bar s\bar c&=- \frac {\rm g}2 \bar c\times \bar c \nonumber\\
s{\bar c}&=b -\frac {\rm g}2 \bar c\times c   & \bar s{c}&=-b -\frac {\rm g}2 \bar c\times c\nonumber\\
sb&=im^2 c-\frac {\rm g}2 c\times b           & \bar sb&=im^2 \bar c-\frac {\rm g}2 \bar c\times b\nonumber\\
&-\frac {{\rm g}^2}8 (c\times c)\times\bar c  & &+\frac {{\rm g}^2}8 (\bar c\times \bar c)\times c.
\end{align}
The two transformations relative to the Nakanishi-Lautrup field $b$ are responsible of the presence of the quartic ghost interaction in 
(\ref{lagrangian}).
In the ghost/anti-ghost symmetric case, the Faddeev-Popov is $1/2\{\partial,\mat D\}$ rather than just $\partial\mat D$ as in standard linear gauges.
As pointed out in \cite{Nakanishi:1990qm}, the presence of $m^2$ in the BRST transformations and consequently 
in Eq. (\ref{lagrangian}) spoils the nil-potency
of both the BRST operators, such that their mutual anti-commutativity is given by \cite{Thierry-Mieg:1979kh, Curci:1976ar, Nakanishi:1990qm}
\be
\bea{ccc}
\{s,s\}=-m^2 \delta=-m^2\delta^\dagger \qquad \{\bar s,\bar s\}=-m^2 \bar \delta=-m^2\bar \delta^\dagger\\
\\
\{s,\bar s\}=-m^2\delta_{\rm FP},
\eea
\ee
with $\delta,\bar \delta$ and the Faddeev-Popov  ghost number operator $\delta_{\rm FP}$ generating a $SL(2,\mathbb R)$ 
algebra \cite{Dudal:2002ye, Schaden:1999ew}
\be
\bea{ccc}
{}[\delta,\delta_{\rm FP}]=-2\delta \qquad [\bar\delta,\delta_{\rm FP}]=-2\bar\delta\\
\\
{}[\delta,\bar \delta]=\delta_{\rm FP}.
\eea
\ee
In \cite{Delbourgo:1981cm} and \cite{Nakanishi:1990qm} it was argued that the 5 charges, obtained from the 5 operators we just showed,
constituted a super-symmetric Lie algebra $OSp(4|2)$: though
in\cite{Thierry-Mieg:1979kh}, it was actually found that the $b$ field broke down such a symmetry, and therefore its super-symmetric algebra.
To introduce the Hubbord-Stratonovich transformations, we simplify the above Lagrangian employing the case of $SU(2)$ (the generalisation to $SU(N)$ 
only invokes the introduction of $f^{abc}$ as structure constants), such that the quartic interaction becomes
\beq
\label{quartic}
\tr\frac{{\rm g}^2}{8}\xi(\bar c\times c)^2=\frac{{\rm g}^2}{8}\epsilon^{abc}\bar c^bc^c\epsilon^{dmn}\bar c^mc^n\ \tr  X^aX^d.
\eeq
Adopting anti-Hermitian algebra generators and a normalisation $\tr  X^aX^d=-\frac12 \delta^{ad}$, the Lagrangian density of 
Eq. (\ref{lagrangian}) becomes
\begin{align}
\label{lagrangian1}
\mat L_{\rm mCF}&=-\frac12\left\{\frac\xi2(b^a)^2+ib^aF^a[{}^g\!A]+\frac{m^2}{2}({}^g\!A^a)^2+\frac i2 
\bar c^a\{\partial,\mat D\}^{ab}c^b\right.\nonumber\\
&\left.-im^2\xi\bar c^ac^a+\frac{{\rm g}^2}{8}\epsilon^{abc}\bar c^ac^b\epsilon^{cmn}\bar c^mc^n\right\}.
\end{align}
In \cite{Schaden:1999ew, Dudal:2002ye} the quartic interaction was linearized in the light of Maximal Abelian gauge, though the BRST formulation of the 
corresponding Lagrangian wasn't explicitly revealed. In particular, it was only showed how to find the relative BRST transformation 
for the $\phi$ field in order to preserve the
invariance of the Lagrangian under BRST and in that gauge, the coupling involves $\pm2$ scalar fermions rather than scalar bosons. This is because of
the maximal abelian decomposition of the various couplings and consequently of the structure constants. We will further on see how the complete
$SU(N)$ structure constants will play an important role as far as their convolution is concerned.
Moreover, it wasn't showed again in \cite{Dudal:2002ye} how to generate the Lagrangian density over the two BRST operators, which a crucial thing
to achieve in order to prove the topological nature of a BRST-based theory.
Therefore, our objective here is to demonstrate that in the case of non-linear gauges, such as the massive Curci-Ferrari, it is possible
to re-write Eq. (\ref{lagrangian1}) as a total BRST--anti-BRST variation, and we will then present the extensive BRST transformations. 
To begin with, let's
use Hubbard-Stratonovich transformations to linearize Eq. (\ref{quartic})
\begin{align}
e^{-\int\frac{{\rm g}^2}{8}\epsilon^{abc}\bar c^ac^b\epsilon^{cmn}\bar c^mc^n}&=\mat C\int \mat D\phi\,
e^{-\int\frac\xi2\phi^a\phi^a-i{\rm g}\frac\xi2\phi^a\epsilon^{abc}\bar c^bc^c},
\end{align}
with $\mat C=\left(\det\frac{2\pi}{\xi}\right)^{1/2}$. The scalar field $\phi^a$ has vanishing ghost number and is required to be hermitian to mantain the
total Hermiticity of the Lagrangian density. It is then left invariant under FP charge operator $\delta_{\rm FP}$. 
Both ghosts and anti-ghosts functions are chosen to be hermitian, 
such that $(c^a)^\dagger=c^a$ and $(\bar c^a)^\dagger=\bar c^a$, see {\it e.g.} \cite{Alkofer:2003jr, Ojima:1980da, Kugo:1979gm} and references therein.
The local Lagrangian density becomes then
\begin{align}
\label{lagrangian2}
\mat L_{\rm mCF}&=-\frac12\left\{\frac\xi2b^ab^a+ib^aF^a[{}^g\!A]+\frac{m^2}{2}({}^g\!A^a)^2
+\frac i2 \bar c^a\{\partial,\mat D\}^{ab}c^b-im^2\xi\bar c^ac^a\right.\nonumber\\
&\left.+\frac{\xi}{2}\phi^a\phi^a-i{\rm g}\frac\xi2\phi^a\epsilon^{abc}\bar c^ac^b\right\},
\end{align}
which implies that we need $3$ additional $\phi$ fields in $SU(2)$ and $N^2-1$ in $SU(N)$.
The partition function in Euclidean space-time, which will be a functional depending on a certain background gauge field $A_\mu$ reads
\begin{align}
Z_{\rm mCF}[A]&=\mat C\int\mat Dg\mat D\bar c\mat Dc\mat Db\mat D \phi\,e^{2\tr S[{}^g\!A,b,\bar c,c,\phi]}\nonumber\\
&=\mat C\int \mat D\mu\,e^{2\int \mat L_{\rm mCF}}.
\end{align}
Before performing the integration in the ghost fields, which is defined over real ghost fields as 
$\mat D\bar c\mat Dc\equiv \prod_{x}\prod_{a}i\,\bar c^a(x)c^a(x)$, 
we wish to separate the contribution of the ghost zero and non-zero modes with respect to the eigenvalue equation of the Faddeev-Popov operator.
In non-linear gauges, the presence of a quartic ghost term allows us to absorb two zero modes 
without causing any harm as far the Grassmann integration is concerned. In addition to that, the diagonal quadratic ghost term,
$-im^2\bar c^ac^a$ can absorb an additional zero mode: this is the reason why in the massive Curci-Ferrari gauge the corresponding
Euler character does not vanish,
as it would be the case in $SU(N)$ \cite{Ghiotti:2005ih}, but it will depend on $m^2$. The eigenvalue equation is then
\begin{align}
\{\partial,\mat D[A]\}^{ab}\lambda_{(n)}^b[A]\equiv\mat M_{\rm FP}^{ab}\lambda_{(n)}^{b}[A]&=\varepsilon_{(n)}^{ab}[A]\lambda_{(n)}^b[A],\nonumber\\
\{\partial,\mat D[A]\}^{ab}\lambda_{(0)}^b[A]\equiv\mat M_{\rm FP}^{ab}\lambda_{(0)}^{b}[A]&=0.
\end{align}
In this way, we will not worry of singularities once we will deal with the inverse of the Faddeev-Popov operator.
The resulting partition function is
\begin{align}
Z_{\rm mCF}[A]&=\mat C\int\mat D\bar c_{(0)}\mat Dc_{(0)}\,e^{\int im^2\xi\bar c_{(0)}^ac_{(0)}^a+
i{\rm g}\frac\xi2\phi^a\epsilon^{abc}\bar c_{(0)}^ac_{(0)}^b}\nonumber\\
&\int\mat Dg\mat D\bar c_{(n)}\mat Dc_{(n)}\mat Db\mat D \phi\,e^{2\mat L^{(n)}_{\rm mCF}}.
\end{align}
Integrating out both ghost zero and non-zero modes, we obtain two functional determinants which both contain the auxiliary field $\phi$, 
the standard feature
in effective Meson theory \cite{Alkofer:2000wg}
\begin{align}
\label{effective}
Z_{\rm mCF}[A]&=\mat C\int\mat Dg\mat Db\mat D \phi\,e^{-\int\frac\xi2b^ab^a+ib^aF^a+\frac{m^2}{2}({}^g\!A^a)^2+\frac{\xi}{2}\phi^a\phi^a}\nonumber\\
&\times\det\left\{\int\left(\xi m^2\delta^{ab}+{\rm g}\frac\xi2\phi^c\epsilon^{abc}\right)\right\}_{(0)}\nonumber\\
&\times\det\left\{-\int\left(\mat M_{\rm FP}^{ab}-\xi m^2\delta^{ab}-{\rm g}\frac\xi2\phi^c\epsilon^{abc}\right)\right\}_{(n)}.
\end{align}
Using the formula $\det A=e^{\tr \log A}$, we can write the effective action as a non-polynomial function in $\phi$ as
\begin{align}
\label{effective1}
S_{\rm eff}[\phi]&=\int\left(\frac{\xi}{2}\phi^a\phi^a-\tr\log\left\{\int\left(\xi m^2\delta^{ab}+
{\rm g}\frac\xi2\phi^c\epsilon^{abc}\right)\right\}_{(0)}\right.\nonumber\\
&\left.-\tr\log\left\{-\int\left(\mat M_{\rm FP}^{ab}-\xi m^2\delta^{ab}-{\rm g}\frac\xi2\phi^c\epsilon^{abc}\right)\right\}_{(n)}\right).
\end{align}
Following \cite{Alkofer:2000wg}, variation of the effective action $S_{\rm eff}[\phi]$ (\ref{effective1}) yields the 
Dyson-Schwinger equations in terms of the classical
fields
\beq
\label{VEV1}
\phi_{(0)}^m(x)=\frac1\xi\tr\left\{G^{ab}_{\phi_{(0)}}(x,x){\rm g}\frac\xi2\epsilon^{abc}\delta^{cm}\right\}=\phi_{(0),{\rm cl}}^m(x),
\eeq
and
\beq
\label{VEV2}
\phi_{(n)}^m(x)=\frac1\xi\tr\left\{G^{ab}_{\phi_{(n)}}(x,x){\rm g}\frac\xi2\epsilon^{abc}\delta^{cm}\right\}=\phi_{(n),{\rm cl}}^m(x).
\eeq
The solution of these two equations determine the vacuum expectation value (VEV)
of the boson fields $\phi_{(0)}$ and $\phi_{(n)}$. $G_{\phi_{(0)}}(x,x)$ and $G_{\phi_{(n)}}(x,x)$ 
are the ghost propagators in the background respectively of the fields $\phi_{(0)}$ and $\phi_{(0)}$. They are defined, in matrix notation, as
\begin{align}
G^{-1}_{\phi_{(0)}}(x,y)&=\left(\xi m^2+{\rm g}\frac\xi2[\phi_{(0)},\cdot]\right)\delta(x-y),\nonumber\\
G^{-1}_{\phi_{(n)}}(x,y)&=\left(\mat M_{\rm FP}-\xi m^2-{\rm g}\frac\xi2[\phi_{(n)},\cdot]\right)\delta(x-y).
\end{align}
The non-zero mode effective potential $V_{\rm eff}(\phi_{\rm cl})$ for the space-time independent classical field $\phi_{\rm cl}$ is obviously
\begin{align}
\label{potential}
V_{\rm eff}[\phi_{\rm cl}]&=\frac{\xi}{2}\phi\phi-\log\left\{-\int\left(\mat M_{\rm FP}-\xi m^2-{\rm g}\frac\xi2[\phi,\cdot]\right)\right\}\nonumber\\
&=\frac{\xi}{2}\phi\phi-\log \left(G^{-1}[\phi_{\rm cl}]{\rm g}\frac\xi2\right)
\end{align}
In this semiclassical approximation, the boson field is being shifted by $\phi\to\phi_{\rm cl}+\tilde\phi$, such that 
the classical field coincides with the VEV
$\phi_{\rm cl}\equiv\langle\phi\rangle$ and the quantum fluctuation $\tilde\phi$ has a vanishing VEV.
Assuming from (\ref{VEV2}) a non-vanishing VEV for $\phi$, this would imply a non-vanishing ghost condensate: from the 
equations of motions of $\phi$, we generate  
a gap equation for the ghosts as
\beq
\phi_{\rm cl}\equiv\langle\phi^a\rangle=\frac i2{\rm g}\langle\epsilon^{abc}\bar c^bc^c\rangle=M^2.
\eeq
which can be solved by Fourier transform as
\beq
\frac{\rm g}{2}\int\frac{{\rm d}^4p}{(2\pi)^4}\frac{1}{p^2\Gamma^2(p^2)+m^2\xi+{\rm g}\frac\xi2 M^2}=M^2.
\eeq
With an ansatz for the $\Gamma$ function as $\Gamma^2(p^2)=p^{2\kappa}$, the gap equation assumes the form
\beq
\label{gap}
\int\frac{{\rm d}^4p}{(2\pi)^4}\frac{1}{p^{2(1+\kappa)}+m^2\xi+{\rm g}\frac\xi2 M^2}=
\frac1{16\pi^2}\int_0^\Lambda{\rm d}p\frac{p^3}{p^{2(1+\kappa)}+\Delta},
\eeq
with $\Delta$ being a mass function $\Delta=m^2\xi+{\rm g}\frac\xi2 M^2$ and $\Lambda$ a momentum cut-off. The solution to Eq. (\ref{gap}) 
is expressed in
terms of the Lerch's Phi function, defined as $\Phi(z,s,a)=\sum_{j=0}^\infty\frac{z^j}{(a+j)^s}$
\begin{align}
\label{gap1}
\langle\phi\rangle&=\frac{\rm g}{32\pi^2}\int_0^\Lambda{\rm d}p\frac{p^3}{p^{2(1+\kappa)}+\Delta}\nonumber\\
&=\frac{{\rm g}\Lambda^4}{32\pi^2(1+\kappa)\Delta}
\Phi\left(-\frac{\Lambda^{2(1+\kappa)}}{\Delta},1,\frac{2}{1+\kappa}\right)\nonumber\\
&=_{m^2\to 0}\frac{\Lambda^4}{16\pi^2\xi M^2}\left(\frac12-\frac{2}{2+\kappa}\frac{\Lambda^{2(1+\kappa)}}{{\rm g}\xi M^2}+
{\it O}({\rm g}^{-1})^2\right)\nonumber\\
&=M^2.
\end{align}


\begin{center}
{\Large BRST formalism in Hubbard-Stratonovich transformations}
\end{center}

After performing the Hubbard-Stratonovich transformation to linearize the quartic ghost interaction, the various BRST transformations of 
Eq. ({\ref{BRST})
will change. To generate the Lagrangian density of Eq. (\ref{lagrangian2}) as a double extended BRST variation we consider the following 
matrix transformations,
(we employ from now on $SU(2)$ as a Lie group, whose generalisation to $SU(N)$ is obtained by substituting the structure constants 
$\epsilon^{abc}$ with $f^{abc}$ in the exterior product $\times$)
\begin{align}
\label{BRST1}
s A_\mu&=-\mat D_\mu c                  & \bar sA_\mu &=-\mat D_\mu \bar c\nonumber\\
sc&=- \frac {\rm g}2 c\times c                & \bar s\bar c&=- \frac {\rm g}2 \bar c\times \bar c \nonumber\\
s{\bar c}&=b -\frac {\rm g}2 \bar c\times c   & \bar s{c}&=-b -\frac {\rm g}2 \bar c\times c\nonumber\\
sb&=im^2 c-\frac {\rm g}2 c\times b           & \bar sb&=im^2 \bar c-\frac {\rm g}2 \bar c\times b\nonumber\\
s\phi&=2\Psi                                  & \bar s \phi&= 2\bar \Psi\nonumber\\
s \Psi&=-\frac{\rm g}2\Psi\times \Psi         & \bar s\bar \Psi&=-\frac{\rm g}2\bar\Psi\times \bar\Psi\nonumber\\
s\bar \Psi&=\phi -i{\rm g}\bar c\times c      & \bar s \Psi&= \phi-i{\rm g}\bar c\times c
\end{align}
such that
\begin{align}
\label{lagrangiana}
\mat L_{\rm mCF}&=\frac i2s\bar s\left(({}^g\!A^a)^2-i\xi\bar c^ac^a-i\frac \xi2\phi^a\phi^a\right)\nonumber\\
&\qquad\qquad\qquad\qquad\qquad\qquad{}+\frac{m^2}2\left(({}^g\!A^a)^2-i\xi\bar c^ac^a\right)\nonumber\\
\nonumber\\
&{}=\frac\xi2b^ab^a+ib^aF^a[{}^g\!A]+\frac{m^2}{2}({}^g\!A^a)^2+\frac i2 \bar c^a\{\partial,\mat D\}^{ab}c^b\nonumber\\
&{}-im^2\xi\bar c^ac^a+\frac{\xi}{2}\phi^a\phi^a-i{\rm g}\frac\xi2\phi^a\epsilon^{abc}\bar c^ac^b-\xi\bar\Psi^a\Psi^a.
\end{align}
It is worth noting that the transformations involving the Nakanishi-Lautrup field $b^a$ change with respect to Eq. (\ref{BRST}), in the way that
the triple ghost term is no longer present in neither of $sb^a$ nor $\bar sb^a$, respectively 
$-\frac {{\rm g}^2}8 (c\times c)\times\bar c$ and $+\frac {{\rm g}^2}8 (\bar c\times \bar c)\times c$.
The auxiliary field $\phi$ has vanishing ghost number, ${\rm Gh}(\phi)=0$, whereas the two additional fermionic fields, introduced in (\ref{BRST1}) to
generate the coupling of $\phi$ with the quadratic-ghost term in (\ref{lagrangian}), $\Psi$ and $\bar \Psi$ have respectively
${\rm Gh}(\Psi)=1$ and ${\rm Gh}(\bar\Psi)=-1$. The field $\phi$ thus plays the role of an additional $b$-field in standard BRST, and 
$\Psi$ and $\bar\Psi$
the role of $c$ and $\bar c$.
The appearance of the term $\xi\Psi^a\bar \Psi^a$ in (\ref{lagrangiana}) produces only a multiplicative 
overall factor, because of Grassmann integration
\beq
\int\mat D\Psi\mat D\bar\Psi\,e^{\int\xi\bar\Psi^a\Psi^a}=\det(\xi),
\eeq
which will be absorbed into the overall constant $\mat C=(\det2\pi\xi)^{-1/2}$. An other interesting aspect of these non-linear gauge 
BRST transformations is that
the Lagrangian density (\ref{lagrangiana}) so generated is, at $m^2=0$ a true topological Lagrangian. In fact
\beq
Z_{\rm mCF}[A]=\mat C\int\mat Dg\mat Db\mat Dc\mat D\bar c\mat D\phi\,e^{\frac i2s\bar s\int\left(({}^g\!A^a)^2-i\xi\bar c^ac^a-
i\frac \xi2\phi^a\phi^a\right)}
\eeq
conserves its topological nature, which can be seen by rescaling the fields as
\begin{align}
b&\to \frac{b}{\sqrt\xi}\quad \phi\to \frac{\phi}{\sqrt\xi}\quad c\to \frac{c}{{}^4\!\sqrt\xi}\quad \bar c\to \frac{\bar c}{{}^4\!\sqrt\xi}\nonumber\\
&\Psi\to \frac{\Psi}{\sqrt\xi}\quad \bar \Psi\to \frac{\bar \Psi}{\sqrt\xi}
\end{align}
and noticing that $Z_{\rm mCF}[A]$ will remain unchanged.
Furthermore, demanding the nil-potency of the BRST transformations we notice that 
\beq
s^2\bar \Psi=2\Psi -i{\rm g}b\times c\neq 0   \qquad \bar s^2 \Psi= 2\bar\Psi-i{\rm g}\bar c\times b\neq 0.
\eeq
Yet, the nil-potency is restored on-shell once we use the equations of motions of the $\phi$-field. In fact, on-shell, 
$s\phi=i{\rm g}\frac\xi2s(\bar c\times c)=i{\rm g}\frac\xi2b\times c$, and therefore
\beq
s^2\bar \Psi=_{\rm on-shell}\,2i{\rm g}\frac\xi2b\times c-i{\rm g} b\times c=0.
\eeq
This is the reason why in the BRST transformations (\ref{BRST1}) we have a factor 2 upfront both $s\phi$ and $\bar s\phi$.
Also the invariance of the Lagrangian density (\ref{lagrangiana}) under the transformations of Eq. (\ref{BRSTa}) is preserved on-shell
\begin{align}
s\mat L_{\rm mCF}&=s\left(\frac{\xi}{2}\phi^a\phi^a-i{\rm g}\frac\xi2\phi^a\epsilon^{abc}\bar c^ac^b-\xi\bar\Psi^a\Psi^a\right)\nonumber\\
&=\xi\phi^as\phi^a-i{\rm g}\frac\xi2(s\phi^a)\epsilon^{abc}\bar c^ac^b-i{\rm g}\frac\xi2\phi^a\epsilon^{abc}b^ac^b\nonumber\\
&-\xi (s\bar\Psi^a)\Psi^a+\xi\bar\Psi^as(\Psi^a).
\end{align}
Using the equations of motion for $\phi^a$, $\phi^a=i\frac{\rm g}2\epsilon^{abc}\bar c^ac^b$, and $\Psi^a= i\frac{\rm g}4\epsilon^{abc}b^bc^c$,
we restore the invariance.
In this Appendix we will show how the BRST transformations (\ref{BRST1}) will change if we expand the convolution of the structure constants, 
$\epsilon^{abc}\epsilon^{cmn}$
and $f^{abc}f^{cmn}$. Let's start with $SU(2)$: the convolution of the structure constants is very simple and gives
\beq
\epsilon^{abc}\epsilon^{cmn}=\delta^{am}\delta^{bn}-\delta^{an}\delta^{bm},
\eeq
which, inserted in Eq. (\ref{lagrangian1}) gives
\begin{align}
\label{lagrangian3}
\mat L_{\rm mCF}&=-\frac12\left\{\frac\xi2(b^a)^2+ib^aF^a[{}^g\!A]+\frac i2 \bar c^a\{\partial,\mat D\}^{ab}c^b\right.\nonumber\\
&\left.+\frac{m^2}{2}({}^g\!A^a)^2-im^2\xi\bar c^ac^a+\frac{{\rm g}^2}{8}(\bar c^ac^a)^2\right\}.
\end{align}
After performing the linearisation of the quartic term, we obtain
\begin{align}
\label{lagrangian3}
\mat L_{\rm mCF}&=-\frac12\left\{\frac\xi2(b^a)^2+ib^aF^a[{}^g\!A]+\frac i2 \bar c^a\{\partial,\mat D\}^{ab}c^b\right.\nonumber\\
&\left.+\frac{m^2}{2}({}^g\!A^a)^2-im^2\xi\bar c^ac^a+\frac\xi2\phi\phi-i{\rm g}\frac \xi2\phi(\bar c^ac^a)\right\}.
\end{align}
We notice that now, the $\phi$ field carries no gauge index, due to the scalar nature of $(\bar c^ac^a)^2$, which implies that $\phi$ lives 
in the identity of
$SU(2)$. It is worth noting that in thise case, there is only one single $\phi$ field, whereas, in the case in which we do not employ the convolution
of the structure constants there were as many fields as the generators of the algebra ($\phi^a$, $a=1\ldots N^2-1$)
Consequently, the BRST transformations (\ref{BRST1}) will change accordingly as
\begin{align}
\label{BRST2}
s A_\mu&=-\mat D_\mu c                  & \bar sA_\mu &=-\mat D_\mu \bar c\nonumber\\
sc&=- \frac {\rm g}2 c\times c                & \bar s\bar c&=- \frac {\rm g}2 \bar c\times \bar c \nonumber\\
s{\bar c}&=b -\frac {\rm g}2 \bar c\times c   & \bar s{c}&=-b -\frac {\rm g}2 \bar c\times c\nonumber\\
sb&=im^2 c-\frac {\rm g}2 c\times b           & \bar sb&=im^2 \bar c-\frac {\rm g}2 \bar c\times b\nonumber\\
s\phi&=2\Psi                                  & \bar s \phi&= 2\bar \Psi\nonumber\\
s \Psi&=0                                     & \bar s\bar \Psi&=0        \nonumber\\
s\bar \Psi&=\phi -i{\rm g}\bar c\cdot c       & \bar s \Psi&= \phi-i{\rm g}\bar c\cdot c,
\end{align}
where the two fermionic fields $\Psi$ and $\bar\Psi$ transform trivially under $s$ and $\bar s$ because they live both in the identity of 
the group too, and so their
exterior product vanishes. Thus, we see that the structure-constant convolution gives a $U(1)$ BRST theory in the additional fields.
In $SU(N)$ the situation looks quite more complicated, because the convolution of $f^{abc}$ gives
\beq
f^{abc}f^{cmn}=\frac2{N_c}(\delta^{am}\delta^{bn}-\delta^{an}\delta^{bm})+d^{amc}d^{bnc}-d^{anc}d^{bmc},
\eeq
where $d^{abc}$ comes from the commutation relations of the $su(N)$ generators
\beq
[X_a,X_b]=f^c_{\phantom{c}{ab}}X_c\qquad \{X_a,X_b\}=-\frac1{N_c}\delta_{ab}-id^c_{\phantom{c}{ab}}X_c.
\eeq
The quartic ghost term will then be
\begin{align}
\label{coupling}
{\rm g}^2\frac\xi8f^{abc}f^{cmn}\bar c^ac^b\bar c^mc^n&={\rm g}^2\frac\xi{4N_c}(\bar c^ac^c)^2\nonumber\\
&+{\rm g}^2\frac\xi8d^{anc}\bar c^ac^n\,d^{bmc}\bar c^b c^m\nonumber\\
&-{\rm g}^2\frac\xi8d^{amc}\bar c^a\bar c^m\,d^{bnc}c^bc^n,
\end{align}
which should determine three different couplings. Yet, the third vanishes because of the anti-commutativity
of the ghost fields and the symmetry of the $d$ symbols, {\it e.g.} $d^{abc}c^bc^c=-d^{acb}c^c c^b=d^{abc}c^cc^b=0$, and the same thing for $\bar c$.
Therefore, the $SU(N)$ Lagrangian density appears not so different from the $SU(2)$ case, specifically
\begin{align}
\label{lagrangian4}
\mat L_{\rm mCF}&=-\frac12\left\{\frac\xi2(b^a)^2+ib^aF^a[{}^g\!A]+\frac{m^2}{2}({}^g\!A^a)^2+
\frac i2 \bar c^a\{\partial,\mat D\}^{ab}c^b\right.\nonumber\\
&\left.-im^2\xi\bar c^ac^a+\frac\xi2\phi^2-i{\rm g}\frac\xi{\sqrt{2N_c}}\phi\bar c^ac^a\right.\nonumber\\
&\left.+\frac\xi2\varphi^a\varphi^a-i{\rm g}\frac\xi{2}\varphi^ad^{abc}\bar c^bc^c\right\}
\end{align}
The corresponding partition function, expressed in terms of a double BRST variation is then a functional integral over the new additional fields as
\begin{align}
\mat Z_{\rm mCF}[A]&=\mat C\int\mat D g\mat D b\mat D\bar c\mat Dc\mat D\phi\mat D\Psi\mat D\bar\Psi\mat D\varphi\mat D\chi\mat D\bar\chi\,
e^{-S[\{\Phi\},A]}\nonumber\\
&=\mat C\int\mat D\mu\,e^{is\bar s\tr[\int({}^g\!A)^2-i\xi\bar cc-i\frac\xi2\varphi\varphi-i\frac\xi2\phi\phi]}\nonumber\\
&e^{-\frac{m^2}2\tr[\int({}^g\!A)^2-i\xi\bar cc]}
\end{align}
The corresponding $SU(N)$ BRST transformations are the following
\begin{align}
\label{BRST3}
s A^a_\mu&=-\mat D^{ab}_\mu c^b                                                 & \bar sA^a_\mu &=-\mat D^{ab}_\mu \bar c^b\nonumber\\
sc^a&=- \frac {\rm g}2 f^{abc}c^bc^c                                            & \bar s\bar c^a&=- \frac {\rm g}2 f^{abc}\bar c^b\bar c^c \nonumber\\
s\bar c^a&=b^a -\frac {\rm g}2 f^{abc}\bar c^b c^c                            & \bar sc^a&=-b^a -\frac {\rm g}2 f^{abc}\bar c^b c^c\nonumber\\
sb^a&=im^2 c^a-\frac {\rm g}2 f^{abc}c^b b^c                                    & \bar sb^a&=im^2 \bar c^a-\frac {\rm g}2 f^{abc}\bar c^bb^c\nonumber\\
s\phi&=2\Psi                                                           & \bar s \phi&= 2\bar \Psi\nonumber\\
s \Psi&=0                                                              & \bar s\bar \Psi&=0        \nonumber\\
s\bar \Psi&=\varphi^a -i{\rm g}\frac1{\sqrt{2N_c}}\bar c^ac^a               & \bar s \Psi&= \varphi^a-i{\rm g}\frac1{\sqrt{2N_c}}\bar c^ac^a\nonumber\\
s\phi^a&=2\chi^a                                                           & \bar s \phi^a&= 2\bar \chi^a\nonumber\\
s \chi^a&=-\frac {\rm g}2 f^{abc}\chi^b\chi^c                              & \bar s\bar \chi^a&=- \frac {\rm g}2 f^{abc}\bar\chi^b\bar\chi^c \nonumber\\
s\bar \chi^a&=\phi^a -i{\rm g}d^{abc}\bar c^bc^c                               & \bar s \chi^a&= \phi^a-i{\rm g}d^{abc}\bar c^b c^c.
\end{align}




\thispagestyle{empty}
\cleardoublepage

\appendix
\longpage


\footnotesize
\addcontentsline{toc}{chapter}{\bf Bibliography}
\bibliography{Biblio}
\bibliographystyle{alpha}

\end{document}


\title{BRST quantization}
\author{Marco}

\maketitle

\begin{center}
{{\bf APPENDIX A}\\
\Large Lie algebras and groups}
\end{center}

Not being a Thesis devouted to the study of groups and algebras, we will avoid inessential tecnhicalities,
risking sometimes being not rigourous as one should be, but focusing as much as possible on those issues relevant to BRST and Y-M theories.
Gauge theories, among many others, are endowed with a Lie-group structure. It would be then impossible to study these theories without
knowing the very basic ingredients of such important groups: the main idea of a Lie group is to be a differential manifold
such that the multiplicative and inverse applications are smooth. Being a group a set of abstract elements, the actual physics is defined
on the algebra of the group. \\
A Lie Algebra $\algebra$ is a linear space, spanned by a basis $X_k$, anti-Hermitean and traceless (real eigenvalues), possessing an antisymmetric
product  $[\cdot, \cdot]$ (The Lie brackets playing the role of Poisson brackets in canonical formalism) that obeys
\beq
\label{A1}
[X_i,X_j]=C^k_{\phantom{k}{ij}}\,X_k,
\eeq
over some field $K$, where the structure constants are real and antisymmetric, 
$C^k_{\phantom{k}{ij}}=-C^k_{\phantom{k}{ji}}$, 
generalisation of Clebsch-Gordon coefficients.\\
Classical algebras are defined as the infinitesimal algebras of matricial groups (the defining representation), and then the Lie 
product is realised as a matrix commutator.\\
The Jacobi identity follows from Eq.(\ref{A1}),
\beq
\label{A2}
[[X_i,X_j],X_k]+[[X_k,X_i],X_j]+[[X_j,X_k],X_i]=0 \qquad
\mat P\{C^l_{\phantom{l}{ij}}C^m_{\phantom{m}{lk}}\}=0,
\eeq
implying it is not a non-associative algebra because
\beq
\label{A3}
[[X_i,X_j],X_k]=-[X_i[X_j,X_k]].
\eeq
The dimensionality of the Lie Algebra is the cardinality (the measure of the number of elements) of the set $\{X_k\}$ of its basis elements. 
The set of basis change 
is the general linear group, either $\rm{GL}(n,\real)$ or $\rm{GL}(n,\mathbb C)$, depending on whether the Algebra is over a real or complex field.\\
A linear map $R$ of $L$, into the set of linear transformations on a vector space is called a {\it representation} $R$ of $L$, if the 
Lie products are mapped to commutators. $R$ maps a non-associative algebra into an associative one.
This is the reason why on groups adding to elements does not make sense, whereas on the algebra this operation is possible.\\
Lie algebras can be classified by the structure of their {\bf Cartan metric}, or {\bf Killing form}: the Cartan metric is defined by
\beq
\label{A4}
g_{ij}:=C^k_{\phantom{k}{im}}\,C^m_{\phantom{m}{kj}}=\sum_{k,m=0}^NC^k_{\phantom{k}{im}}\,C^m_{\phantom{m}{kj}}=
\sum_{m}^N\delta_{ij}\delta_{mm}-\delta_{im}\delta_{mj}=
N\delta_{ij}-\delta_{im}\delta_{mj},
\eeq
which is expressed in terms of the adjoint representation of the Algebra
\beq
\label{A5}
\ad(A)\,Y=\ad_AY=[A,Y] \qquad \ad(X_k)=[(X_k)^a_{\phantom{a}{b}},]=C^a_{\phantom{a}{kb}}.
\eeq
This representation is irrecucible for any simple Lie group. If the group is connected, the group can be parametrized as a product of 
one-parameter subgroups
\beq\label{A6}
g(\theta_1, \theta_2, \ldots,\theta_N)=
\exp(\theta_1X_1)\cdot\exp(\theta_2X_2)\cdots\exp(\theta_NX_N),
\eeq
from which follows the Hausdorff-Baker-Campbell formula
\beq
\label{A7}
\exp(\theta_jX_j)\,X_k\,\exp(-\theta_jX_j)=
\sum_{m=0}^\infty \frac{(\theta_k)^m}{m!}\,C(X_j: X_k, m)=
\sum_{m=0}^\infty \frac{(\theta_k)^m}{m!}\,\alpha^i_{\phantom{i}{}kj}(m)X_i,
\eeq
where $C(X_j: X_k, m)=\sum_{j=0}^k (-1)^j \binom{k}{j} B^{(k-j)} A B^j$ is the $m$-fold commutator, and $\alpha^i_{\phantom{i}{}kj}(m)$, for 
fixed $k$, the matrix 
elements of the adjoint representation of the group element $\exp(i\theta_kX_k)$. Then, in the adjoint representation, the matrices 
representing the basis elements
are formed from the structure constants. The {\bf Killing form} is defined as
\beq
\label{A8}
B(X,Y):=\Tr(\ad(X)\,\ad(Y))\qquad B(X_i\,X_j)=\Tr(\ad(X_i)\,\ad(X_j))=g_{ij}.
\eeq
The Cartan metric and the structure constants transform under the linear group of basis transformations as tensors 
($g_{ij}\in S_2(K)\subset T_2(K)$, and it lowers
the indeces). The Cartan metric is not singular if
\beq
\label{A9}
g_{ik}\,g^{kj}=\delta^j_i.
\eeq
A Lie algebra $\algebra$ is simple if it contains no non-zero invariant subalgebras (ideal: is a subspace $J$ in 
$\algebra$, s.t. $[J,\algebra]\in J$.). Simplicity
implies semisimplicity. We have four classical simple algebras plus five exceptional described by the following table
\be
\bea{ccc}
\label{A10}
\begin{tabular}{|c|c|c|}
\hline
Cartan name &
Notation    &
Dimension \\
\hline
$A_n$ & $SU(n+1),  n>1$ & $\dim A_n=(n+1)^2-1$\\
$B_n$ & $SO(2n+1), n>2$ & $\dim B_n=2n^2+n$   \\
$C_n$ & $Sp(2n),   n>3$ & $\dim C_n=2n^2+n$   \\
$D_n$ & $SO(2n),   n>4$ & $\dim D_n=2n^2-n$   \\
$E_2$ & $g(2)         $ & $\dim E_n=14$       \\
$F_4$ & $f(4)         $ & $\dim F_n=52$       \\
$E_6$ & $e(6)         $ & $\dim E_6=78$       \\
$E_7$ & $e(7)         $ & $\dim E_7=133$      \\
$E_8$ & $e(8)         $ & $\dim E_8=248$      \\
\hline
\end{tabular}\nonumber
\eea
\ee
Lie algebras exponentiate to Lie groups. More precisely, any linear Lie algebra can be integrated to a connected and simply connected Lie group. 
Caveats hold for some
non compact groups, like $O(n)$.\\
There's a strong relationship between Lie algebras and manifolds. In fact, Lie algebras can be identified with the set of left-invariant vector 
fields on the group. 
In the case of a semisimple Lie group, the group manifold can be endowed with an affine connection derived from the structure constants, 
with which the manifold
becomes a Riemannian manifold. The Cartan metric $g_{ij}$ extends to the Riemannian metric on the group manifold. Yet, the global topology of 
the group manifold
is not entirely determined by the local structure.\\
A Lie algebra over $\complex$ is called compact if it integrates to a Lie group, which as manifold is compact. The parameteres of the group 
then have bounded ranges.
Such algebras can be decomposed into simple and $U(1)$ (the charge of the algebra) Lie algebras each whose generators commute with all of the 
generators of the other 
algebras. A simple algebra has non-vanishing structure constants. which implies that the trace of a commutator vanishes in any representation.
\beq
\label{A11}
\Tr\{[X_a, X_b]\}=C^c_{\phantom{c}{ab}}\,\Tr(X_c)=0.
\eeq
There is a basis of generators of any simple compact Lie algebra in some representattion $\rho$ in which 
\beq
\label{A12}
\Tr_\rho(X_aX_b)=C(\rho)\delta_{ab},
\eeq
where $C(\rho)$ is called the quadratic invariant of the represenation $\rho$. In the adjoint representation we have
\beq
\label{A13}
C(\ad)=\frac{C_{abc}C^{abc}}{\dim G}=C_2,
\eeq
the quadratic Casimir of $G$. \\
The rank of an algebra $\rm{rank}(G)$ is equal to the maximal number of mutually commuting generators. The $U(1)^{\rm{rank}(G)}$ subalgebra of 
these generators is called
the Cartan subalgebra of $G$.
\begin{itemize}
\item A Lie algebra is semisimple if the Cartan metric $g_{ij}$ is non-singular;
\item If L is semisimple, then L is compact iff the Killing form is negative definite (if skew-Hermitean basis, positive if Hermitean).
\end{itemize}
Though a group is not semisimple, its algebra can still be compact.
\be
\bea{ccc}
\label{A15}
\begin{tabular}{||c|c|c|c|c|c||}
\hline\hline
G                 &
rank(G)           &
$\dim(G)$         &
$C(\ad)$          &
$\dim(\rm{fund})$ & 
$C(\rm{fund})$     \\ 
\hline\hline
$SU(N)$  &  $ N-1$ & $N^2-1    $ & $N$   & $N$  & $1/2$ \\
$SO(N)$  &  $ N/2$ & $N(N-1)/2 $ & $N-2$ & $N$  & $1$   \\
$Sp(2N)$ &  $ N  $ & $2N(N+1)$ & $N+1$ & $2N$ & $1/2$ \\
\hline\hline
\end{tabular}
\eea
\ee
A useful formula for computing the quadratic invariants of other representations is
\beq
\label{A16}
C(\rho_1)\dim(\rho_2)+\dim(\rho_1)C(\rho_2)=\sum_iC(\rho_i).
\eeq
Normalisation condition is, from the fundamental representation and the adjoint representation
\beq
\label{A17}
\rm{tr}_{\rho}=\frac{C(\rm{fund})}{C(\ad)}\Tr
\eeq
from which it implies that the true invariant are the ratios of quadratic invariants. The index of a classical group is defined as
\be
\label{A18}
T(\rho):=\frac{C(\rho)}{C(\rm{fund})}\qquad\Rightarrow\qquad
\left.
\bea{ll}
T(\ad[SU(N)])&=2N\\
T(\ad[SO(N)])&=N-2\\
T(\ad[Sp(N)])&=2N+2
\eea
\right.
\ee
\\

\\
\begin{center}
{{\bf APPENDIX B}\\
\Large Connection on a principal bundle}
\end{center}
\\
In this appendix we will enlist briefly the major topics of gauge theory from the geometric and topological point of view. This is necessary
to understand the rich geometric structure of Y-M theory. Moreover, its generalization to superspace is essential to the comprhension
of supersymmetry, BRST and topological field theory (TFT). This will be covered in Appendix C. We assume the reader being familiar
with basic concepts of topology, such as manifolds, tangent and cotangent spaces. For this we remind the interested reader to \cite{Naber:1997}.\\
As previously said, Y-M theories can be regarded as the quantum theory of principal bundles, on which we construct connections, covariant derivatives
and curvature forms. To start with we define a principal bundle:
a differentiable principal fiber bundle over a manifold $M$ with group structure $G$ consists of a manifold $P$ and an action of $G$ on $P$ 
satisfying the following conditions
\begin{itemize}
\item $G$ acts freely on $P$ without fixed points, i.e. $gx=x$ implies $g=I$ (only the identity element fixes any $x$), $P\times G\to P$ 
      is denoted by $P\times G\ni(u,a)\to ua\in P$;
\item $M$ is the quotient space of $P$ by the equivalence relation induced by $G$, $M=P/G$, and the canonical projection $\pi:P\to M$ is differentiable;
\item $P$ is locally trivial ($P \cong \nreal$).
\end{itemize}
To any element $A$ of the algebra $\algebra$ of $G$, we associate a vector $\Sigma(A)$ on $P$, the fundamental vector field corresponding to $A$. 
$\Sigma(A)$ is actually generated by the right action of $G$ on $P$
\footnote{Left and right actions of a group element are diffeomorphisms defined as
$L_g(h)=hg$ and $R_g(h)=hg$.}
: if $A\in\algebra$, then $\exp(tA)$ is a one-parameter subgroup of $G$, acting on $P$
as
\beq
\Sigma(A)_u\cdot f=\frac{d}{dt}f(u_t)\Big|_{t=0},
\eeq
where $u_t=R_{\exp(tA)}(u)$. $\Sigma(A)_u$ is a vector tangent to $P$ at $u$ (tangent to the fiber). Call $G_u$ the subspace of $T_u(P)$ 
of vectors tangent to the fiber through $u$, at $u$.
\beq
\Sigma:\algebra\to G_u \quad\textrm{is an isomorphism}.
\eeq
A connection in $P$ is a choice of a supplementary linear subspace $Q_u$ in $T_u(P)$ to $G_u$
\beq
T_u(P)=G_u\oplus Q_u
\eeq
where $Q_{ua}=(R_a)_\ast Q_u$ is a {\it push-forward} 
\footnote{Let $M$ and $N$ be two smooth manifolds, with dimension $m$ and $n$ respectively. Let $f: M\to N$ be a smooth function. Then, the 
{\it differential} or {\it push-forward} $f_{\ast}$ (or ${\rm d}f$) of $f$ in the point $p\in M$ is the application
$f_{\ast}:T_pM\to T_{f(p)}N$. The push-forward defines then a change of variables in tangent spaces.}
and depends differentiably on $u$. $Q_u$ is called the horizontal space and $G_u$ the vertical space. 
Choosing a $Q_u$ amounts to
choosing a basis in $T_u(P)$, though this distribution is not, in general intergable. Geometrically, this correpsonds to 
the non triviality of parallel transport using the holonoy group of the principal bundle.
\\
\\
\begin{center}
{\Large Connection form}
\end{center}

A connection form is a Lie-algebra valued 1-form $\omega$ such that 
\begin{itemize}
\item $\omega$ applied on any fundamental vector field $\Sigma(A)$ reproduces $A$, i.e. $\omega(\Sigma (A))=A$;
\item $(R^\ast_a\omega)(X)=\Ad_{a^{-1}}\cdot \omega(X)$. The horizontal subspace $Q_u$ is the kernel of $\omega$, that is to say that $X_u$ 
      is horizonatal iff $\omega(X_u)=0$;
\end{itemize}
where $(R^\ast_a\omega)(X)$ is a {\it pull-back}
\footnote{The transpose action of the push-forward is the pull-back $f^\ast$ (or $\delta f$), defined as $f^{\ast}:T^{\ast}_{f(p)}N\to T^{\ast}_{p}M$.
Contrary to the push-forward, we cannot pass from the cotangent space $T^{\ast}_{p}M$ to $T^{\ast}_{f(p)}N$, but only
the other way round, linking a change of variables for cotangent spaces, dual of tangent spaces.}.
It's possible to express $\omega$, the connection form on $P$, by a family of local forms, each one being defined in an open subset of 
the base-space manifold $M$.
Let $\{U_\alpha\}$ be a covering of $M$, we choose in $P$ the preferred set of local sections $\sigma_\alpha$ and the corresponding tranasition 
functions $\psi_{\alpha\beta}$:
for each $\alpha$ and $\beta$. we define a Lie-algebra-valued 1-form on $U_\alpha$ by
\beq
\omega_\alpha=\sigma^\ast_\alpha\omega\qquad\textrm{pullback of}\,\, \omega\,\,{\rm through}\,\,\sigma_\alpha
\eeq
where
\beq
\omega_\beta=\Ad_{\psi^{-1}_{\alpha\beta}}\cdot \omega_\alpha+\psi^{-1}_{\alpha\beta}\,{\rm d}\psi_{\alpha\beta}
\eeq
in $U_\alpha\cap U_\beta$. If $\omega$ is a connection form on $P=M\times G$, we can construct from a global section $\sigma_1$ of $P$ the form on $M$
\beq
\omega_1=\sigma^\ast_1(\omega).
\eeq
If we now use a $G$-valued function ${\rm g}$ on $M$ to transform $\sigma_2$ into $\sigma_2(x)=\sigma_1(x)\cdot{\rm g(x)}$, we can define a new 1-form on $M$
\beq
\omega_2=\sigma_2^\ast(\omega)
\eeq
we have
\beq
\omega_2=\Ad_{{\rm g}^{-1}}\cdot \omega_1+{\rm g}^{-1}\,{\rm d}{\rm g}.
\eeq
\\
\\
\begin{center}
{\Large Geometrical interpretation of gauge potentials}
\end{center}

On a 4-dim manifold, the connection form $\omega$, defined on an open subset of $M$, $U_\alpha$, can be expressed as
\beq
\omega_\alpha=A^\mu_{\alpha}(x){\rm d}x_\mu
\eeq
whose Lie-valued components transform as
\beq
A^{'\mu}(x)=\Ad_{{\rm g}^{-1}}\cdot A^\mu+{\rm g}^{-1}\,\partial^\mu{\rm g}
\eeq
which are the components of the transformed connection form
\beq
\omega'_\alpha=\sigma^{'\ast}_\alpha\omega=A^{'\mu}(x){\rm d}x_\mu.
\eeq
A change of $\sigma$ by the action of some $G$-valued function $\rm g$ on $M$ can be viewed as a change of coordinates in the principal fiber bundle $P$, and
it induces a transformation of the components $A^\mu$ similar to the usual gauge transformation of potentials. Then\\
\\
{\it the gauge potential naturally becomes the component of a geometrical object of a definite type: a connection form on ${\rm P}$}.
\\
\\
\begin{center}
{\Large Covariant derivative}
\end{center}

The concept of covariant derivative is strongly related to the horizontal lift of the derivative $\partial_\mu$.\\
A vector field $\bar X$ is the lift of a vector field $X$ on $M$, which is the horizontal field on $P$, which projects onto $X$, s.t.
\beq
\pi_\ast(\tilde X_u)=X_{\pi(u)}\quad{\rm where}\quad \pi:P=M\times G\to M.
\eeq
Suppose we choose a local chart $U_\alpha$ on $M$, with local coordinates $\{x^\mu\}$. Then, we construc vector fields, with generators as 
$\partial_\mu=\frac{\partial}{\partial x^\mu}$, whose lift $\tilde \partial_\mu$ lies on $\pi^{-1}(U_\alpha)=U_\alpha\times G$.
If $\sigma_\alpha$ is section over $U_\alpha$, then
\beq
\omega_\alpha(\partial_\mu)=\sigma^\ast_\alpha\,\omega(\partial_\mu)=\omega(\sigma_{\alpha\ast}\partial_\mu)=(A^\nu_{\alpha}(x){\rm d}x_\nu,\partial_\mu)=
A_{\alpha\mu}=\omega(\Sigma(A_{\alpha\mu}))
\eeq
Hence
\beq
\omega(\sigma_{\alpha\ast}\partial_\mu-\Sigma(A_{\alpha\mu}))=0,
\eeq
where $\sigma_{\alpha\ast}\partial_\mu-\Sigma(A_{\alpha\mu})$ is evidently horizontal. Then 
\beq
\tilde\partial_\mu\Big|_u=\sigma_{\alpha\ast}\partial_\mu-\Sigma(A_{\alpha\mu}) \quad {\rm with}\quad u=\sigma_\alpha(x).
\eeq
We can identify $\sigma_{\alpha\ast}\partial_\mu$ with $\partial_\mu$ and $-\Sigma(A_{\alpha\mu})$ with the Lie-algebra-valued element $A_\mu$, to recover the usual
covariant derivative
\beq
\mat D_\mu=\partial_\mu-A_\mu.
\eeq
So, any point on the local section $\sigma_\alpha$, defined by $\pi^{-1}(U_\alpha)=U_\alpha\times G$, can be thought as
\beq
u_0=\sigma_\alpha(x_0)=(x_0,e)=\sigma_{\alpha\ast}\partial_\mu\oplus \Sigma(A_{\alpha\mu})
\eeq
This point $u_0$ is generated by the curve on the fiber $\pi^{-1}(U_\alpha)\Big|_{x_0}$
\beq
P\supseteq \pi^{-1}(x_0)\ni u_t=u_0\,\exp(tA_\mu)=(\underbrace{x_0}_{\textrm{point in $M$}},\underbrace{e^{tA_\mu}}_{\textrm{element of $G$}}).
\eeq
If $f$ is a function on $\pi^{-1}(U_\alpha)$, then the restriction of this function to $\pi^{-1}(x_0)$ is a function $F$ defined on $G$, because it's 
$e^{tA_\mu}$ which localises $\pi^{-1}(U_\alpha)$ to $\pi^{-1}(x_0)$. The directional derivative along $u_t$ is clearly the action of the Lie algebra element $A_\mu$
on $F$ at $e$. Thus, the covariant derivative is section-dependent.\\
There's also an other way to interpret the covariant derivative, which follows from the adjoint action on any element of $P$
\beq
\partial_\mu \psi=\partial_\mu \psi-\lim_{t\to0}\frac1t\left[e^{-tA_\mu}\psi(u_0)e^{tA_\mu}-\psi(u_0)\right]
\eeq
for any function $\psi$ on $P$, s.t. $\psi(ua)=\Ad_{a^{-1}}\psi$. It is also important to notice that while the commutator of two fundamental vectors is still a fundamental
vector, showing that this map preserves space and algebra structure, it is not true that the commutator of two horizontal vector fields is stil horizontal. \\
In particular
\beq
[\mat D_\mu,\,\mat D_\nu]=-(\partial_{[\mu}A_{\nu]}+[A-\mu,A_\nu])=-F_{\mu\nu}
\eeq
is a fundamental vector field on the bundle space written as a Lie algebra element, and moreover
\beq
F_{\beta\mu\nu}=\Ad_{\psi^{-1}_{\alpha\beta}}\cdot F_{\alpha\mu\nu}\quad \to \quad F'_{\mu\nu}(x)=\Ad_{{\rm g}^{-1}}\cdot F_{\mu\nu}(x)
\eeq
on $U_\alpha\cap U_\beta$.
\\
\\
\begin{center}
{\Large Curvature form}
\end{center}

From the commutator of two covariant derivatives, which is a fundamental vector on $P$, s.t. $\omega([\mat D_\mu,\,\mat D_\nu])=[\mat D_\mu,\,\mat D_\nu]$,
we can construct a Lie-algebra valued 2-form$\Omega$. Locally, on $U_\alpha\cap U_\beta$
\beq
\Omega_\alpha=\frac12F_{\alpha\mu\nu}\,{\rm d}x^\mu\wedge{\rm d}x^\nu
\eeq
with
\beq
\Omega_\beta=\Ad_{\psi^{-1}_{\alpha\beta}}\Omega_\alpha.
\eeq
To connect the curvature form to the connection form, we need to introduce the covariant exterior derivative ${\rm d}_\omega$, as
\beq
\Omega={\rm d}_\omega\omega={\rm d}\omega +\frac12[\omega,\omega].
\eeq
If $X$ and $Y$ are two tangent vector to the bundle, then
\beq
\Omega(X,Y)={\rm d}\omega(X,Y) +\frac12[\omega(X),\omega(Y)].
\eeq
Let's decompose $X$ and $Y$ into their vertical and horizontal components
\beq
X=hX\oplus vX\qquad Y=hY\oplus vY
\eeq
then, what we get is
\beq
\Omega(X, Y)={\rm d}\omega(hX,hY)+{\rm d}\omega(vX,vY)+{\rm d}\omega(hX,vY)+{\rm d}\omega(vX,hY)+\frac12[\omega(vX),\omega(vY)]+
\frac12[\omega(hX),\omega(hY)]=\nonumber
\eeq
\beq
={\rm d}\omega(hX,hY).
\eeq
Though ${\rm d}^2=0$, $\mat D^2\neq 0$, whereas $\mat D\Omega=0, \forall \omega$ (Bianchi ideintity), using the Jacobi identity.
\\
\\
\begin{center}
{\Large Group of gauge transformations}
\end{center}

Gauge transformations are equivariant automorphisms of some $G$-bundle $P$. The 1-forms of connections are the physical interesting objects, whose components
are the gauge potentials. Choosing a particular $G$-bundle automatically defines the set of Chern classes. ($P_k, k\in \mathbb Z$).\\
In this context, the gauge transformations assume the role of elements of an infinite-dimensional Lie group, called $\mat G$, whose group composition is smooth
\beq
\Phi:P\to P,\qquad \Phi\in \mat C^\infty(\Ad P).
\eeq
This group composition can be expressed as follows
\be
\bea{ccc}
\forall g\in \mat G|g:P\to P \Rightarrow \exists \gamma\in {\rm Map}(\Ad P), \gamma:P\to G\\
\\
g(u)=u\cdot \gamma(u),\quad u\in P,\,\gamma(ua)=a^{-1}\gamma(u)a\\
\\
\forall g,h\in \mat G, g\circ h (u)=u\cdot(\gamma_h\cdot \gamma_h)(u).
\eea
\ee
Locally, the mapping $\gamma: P\to G$ can be written as
\beq
\gamma_\beta(x)=\psi^{-1}_{\alpha\beta}(x)\,\gamma_\alpha(x)\,\psi_{\alpha\beta}(x), \forall x\in U_\alpha\cap U_\beta, \{U_{i\in I}\}\subseteq P.
\eeq
This representation is in 1-1 correspondence with the sections of the bundle $\mat B$ associated with $P$ with standard fiber $G$, $G$ acting on itself by the adjoint
map ($a(g)=aga^{-1}$). The group $\mat G$ of gauge transformations can be identified with the set $\Gamma(\mat B)$ of sections of $\mat B$, which is not a principal bundle
though, because the action of $G$ is not free. $\mat B$ will have global sections and unit element $(x,e)$.\\
The Lie algebra of $\mat G$, ${\rm Lie}\mat G$. As we know, elements of sections of tangent and cotangent bundles are respectively vector fields and forms.\\
Consider the constant unit section $s$ of $\mat B$: through any point of $\mat B$ passes one fiber. Using the local triviality of $mat B$ over patches $U_\alpha$,
we may identify the fiber with the group $G$. Tangent vectors to the fiber $s$ follow immediately, as well as parallel transport and all the operations on vector
fields. These fields are elements of the algebra of $G$, vectors to a fiber $\pi^{-1}_{\\mat B}(x)$, with $x\in U_\alpha$. On the transition $U_\alpha\cap U_\beta$,
the map is of course
\beq
A_\beta=\Ad_{\psi^{-1}_{\alpha\beta}}\cdot A_\alpha.
\eeq
The field we have just determined on $\mat B$ can be identified as a section of an associated bundle $E$ to $P$, where the fiber is $\algebra$ and the adjoint acion
of $G$ on $\algebra$.
Then, $\Gamma(E)$ is the Lie algebra of $\mat G\equiv \gamma(\mat B)$. $\Gamma(E)$ is an infinite-dimensional module. Any section of $\mat B$ can be written as
\be
\bea{rr}
\mat C^\infty(\Ad P)=\mat G\equiv\Gamma(\mat B)\ni s=\exp(\sigma),\,\sigma\in \Gamma(E)\\
\\
\sigma:U_\alpha\to \algebra.
\eea
\ee
At last, there's a particular class of gauge transformations, those which have values in the center $Z$ of $G$: for such a trasnformation we have on some 
local chart $U_\alpha$
\beq
{}^g\!A^\alpha_\mu=g_\alpha^{-1}A^\alpha_\mu g_\alpha+g_\alpha^{-1}\partial_\mu g_\alpha=A^\alpha_\mu.
\eeq
This can be also written as
\beq
\partial_\mu g_\alpha+[A^\alpha_\mu,g_\alpha]=\mat D^\alpha_\mu g_\alpha=0,
\eeq
i.e. $\nabla g_\alpha=0$. Then, $g_\alpha$ belongs to the center of the holonomy group of the connection under consideration. In fact
\beq
\mat D_{[\mu}\mat D_{\nu]}g=[F_\mu\nu,g]=0.
\eeq
\\
\\
\begin{center}
{\Large Multiple Solutions}
\end{center}

Recall, in fact, that in the Faddeev-Popov method to remove the infinite gauge measure,
the following functional identity was inserted in the Y-M partition function
\beq
1=\int \mathcal D g(x)\, \delta(F[{}^g\!A])\left|\det\left(\frac{\delta F[{}^g\!A] }{\delta g}\right)\right|.
\eeq
This change of variables is only valid as long as the delta function is bijective, i.e. the delta function is single-valued
The multidimensional case is just a straightforward consequence of the one-dimensional case. In fact
\beq
\label{delta2}
\int d\vec x \,\delta^{(m)}(\vec f(\vec x))\,\vec g(\vec x)=\int dx_1\ldots dx_n\,\delta^{(m)}(\vec f(\vec x))\,\vec g(\vec x) 
\eeq
where 
\beq
\vec f(\vec x)=(f_1(\vec x),\ldots f_m(\vec x)) \qquad \textrm{and} \qquad \vec g(\vec x)=(g_1(\vec x),\ldots g_m(\vec x)).
\eeq
The multi-valued delta function is now
\beq
\vec x_i \,\Big | \, \vec f (\vec x_i)=0 \qquad i=1,2,\ldots N
\eeq
which implies (\ref{delta2}) becomes
\beq
\sum_{i=1}^N\int dx_1\ldots dx_n \,\delta^{(m)}(\frac{\partial \vec f}{\partial \vec x}\Big|_{\vec x_i}(\vec x-\vec x_i))\,\vec g(\vec x)
\eeq
The change of variables we shall use is
\beq
z_k=\frac{\partial \vec f}{\partial x_k}\Big|_{\vec x_k^{(i)}}(x_k-x_k^{(i)})
\eeq
where $k$ represents the k-th component. The differential is of course
\beq
dx_k =\frac{dz_k}{\left|\frac{\partial \vec f}{\partial x_k}\Big|_{x_k^{(i)}}\right|}
\eeq
and the volume element becomes
\beq
dx_1\wedge dx_2\wedge \ldots \wedge dx_n=\sum_{i=1}^N\frac{dz_1}{\left|\frac{\partial \vec f}{\partial x_1}\Big|_{\vec x_i}\right|}\wedge\ldots
\wedge\frac{dz_n}{\left|\frac{\partial \vec f}{\partial x_n}\Big|_{\vec x_i}\right|}=
\sum_{i=1}^N\frac{dz_1\ldots dz_n}{\left|\det\frac{\partial \vec f}{\partial \vec x}\Big|_{\vec x_i}\right|}
\eeq
It follows that the Faddeev-Popov identity in the case of a multi-valued gauge-fixing condition $F[{}^g\!A]=0$ is
\beq
\sum_{i=1}^N\frac{1}{\left|\det\frac{\delta F[{}^g\!A]}{\delta g}\Big|_{A_i}\right|}=\int \mat Dg\,\delta (F[{}^g\!A])
\eeq
and consequently
\beq
1=\int \mat Dg\,\frac{\prod_{i=1}^N\left|\det\frac{\delta F[{}^g\!A]}{\delta g}\Big|_{A_i}\right|}
{\sum_{j=1}^N\prod_{i=1}^N\left|\det\frac{\delta F[{}^g\!A]}{\delta g}\Big|_{A_i}\right|^{(1-\delta_{i,j})}}\,\delta (F[{}^g\!A])
\eeq
instead of
\beq
1=\int \mat Dg \,\left|\det \frac{\delta F[{}^g\!A]}{\delta g}\right|\,\delta(F[{}^g\!A])
\eeq
for a single-valued ``ideal'' gauge fixing condition.
For a single-valued delta function, we can write
\beq
\label{delta3}
1=\int_a^b dx\,\delta(f(x)) \left|\frac{df}{dx}\right|
\eeq
Suppose now we have a multi-valued delta function, such that
\beq
x_i\,\Big|\,f(x_i)=0 \qquad i=1,2,\ldots N
\eeq
Let's  divide the integration region $[a,b]$ into n sub-intervals, 
\beq
[a,b]=\sum_{j=0}^{N-1}[x_j,x_{j+1}]\qquad a\equiv x_0\leq x_1\ldots \leq x_{N-1}\leq x_N \equiv b 
\eeq
such that in each $[x_j,x_{j+1}]$ there exists only one solution for the delta function. Therefore 
\beq
\label{delta4}
\int_{x_j}^{x_{j+1}}dx\,\delta(f(x))\left|\frac{df}{dx}\right|=1
\eeq
Then (\ref{delta4}) becomes
\beq
\sum_{j=0}^{N-1}\int_{x_j}^{x_{j+1}}\delta(f(x))\left|\frac{df}{dx}\right|=\int_a^b dx\,\delta(f(x)) \left|\frac{df}{dx}\right|=N.
\eeq
So
\beq
\label{identity}
1=\frac1N\int_a^b dx\,\delta(f(x)) \left|\frac{df}{dx}\right|
\eeq
which implies that any Gribov copy on the same orbit  gives the same contribution. 
